\documentclass[12pt,twoside,openright]{book}
\usepackage[pdftex]{graphicx}
\usepackage[a4paper, tmargin=3.5cm, bmargin=2.5cm, lmargin=3.75cm, rmargin=2.5cm, twoside]{geometry}
\usepackage{latexsym,amsmath,amssymb,amsfonts,bm,bbm}             % AMS Math
\usepackage[toc,page]{appendix}
\usepackage{physics}
\usepackage{rotating}                    % Sideways of figures & tables
\usepackage{natbib}          % Cross-reference package (Natural BiB)
\usepackage{soul}                                         % Put References at the end of each chapter
\usepackage{xcolor}
\usepackage{float}
\usepackage{comment}
\usepackage{fancyhdr}                    % Fancy Header and Footer
\usepackage{setspace}
\usepackage[english]{babel}
\doublespace          % Line spacing
\usepackage{tocloft}
\usepackage{lscape}

\usepackage[small,hang]{caption}
\usepackage[compact]{titlesec}
\numberwithin{equation}{section}
\usepackage[T1]{fontenc}                    % Public Times New Roman text & math font
\makeatletter
\def\cleardoublepage{\clearpage\if@twoside \ifodd\c@page\else%
    \hbox{}%
    \thispagestyle{empty}%              % Empty header styles
    \newpage%
    \if@twocolumn\hbox{}\newpage\fi\fi\fi}
\makeatother
\clearpage{\pagestyle{plain}\cleardoublepage}
\newcommand{\be}{\begin{equation}}
\newcommand{\ee}{\end{equation}}
\newcommand{\bea}{\begin{eqnarray}}
\newcommand{\eea}{\end{eqnarray}}
\newcommand{\ba}{\begin{array}}
\newcommand{\ea}{\end{array}}
\newcommand{\bi}{\begin{itemize}}
\newcommand{\ei}{\end{itemize}}
\newcommand{\bc}{\begin{center}}
\newcommand{\ec}{\end{center}}
\newcommand{\bfr}{\begin{flushright}}
\newcommand{\efr}{\end{flushright}}

\renewcommand{\k}{\mathbf{k}}
\renewcommand{\r}{\mathbf{r}}
\DeclareMathOperator{\sgn}{sgn}

\usepackage{color}

\begin{document}
\thispagestyle{empty}

{ \renewcommand{\baselinestretch}{1.5}
\begin{center}
%\vspace*{1.0cm}
\begin{spacing}{2}
\noindent{\Large \bf INTENSE LASER-DRIVEN PHENOMENA IN\\ 
WEYL SEMIMETALS}
\end{spacing}
\end{center}

\vspace{18cm}
\begin{flushright}
{ {\LARGE\em  \textbf{AMAR BHARTI} ~~}}
%\noindent \vskip
%-1.0\baselineskip \noindent \rule[-2.5 mm]{\textwidth}{3 pt}\\}
\end{flushright}}

	\cleardoublepage
	\frontmatter
	\addcontentsline{toc}{chapter}{Title Page}
	\newpage
\thispagestyle{empty}

% ******* Title page *******
% **************************
\begin{titlepage}
\begin{center}
{
%% Ici je  lis  logoensem.jpg
%%  commenter si vous voulez faire du DVI :
%

%\vskip3cm

\textbf{\Large INTENSE LASER-DRIVEN PHENOMENA IN WEYL SEMIMETALS} \vskip1.0cm %\textbf{\large\emph{Thesis
%submitted to the\\  Indian Institute of Technology Bombay}}
{\large\emph{Submitted in partial fulfillment of the requirements}} \vskip 0.03cm 
{\large\emph{of the degree of}}
\singlespacing 
%\textbf{\large\emph{For award of the
%degree\\\vspace{5.0cm}of}}
%\textbf{\large\emph{For award of the
%degree}}\vskip 0.03cm
%\textbf{\large\emph{of}}
% \singlespacing
\textbf {\large Doctor of Philosophy}\\
\singlespacing
 {\large\emph{by}}\\
\singlespacing \textbf{\large AMAR BHARTI} \vskip0.2cm
%{\large (Roll No. 164120007)}\\\
\vskip1cm
{\large Supervisor:}\\
\singlespacing \textbf{\large{Prof. Gopal Dixit}} \vskip1cm

\includegraphics[height=42mm]{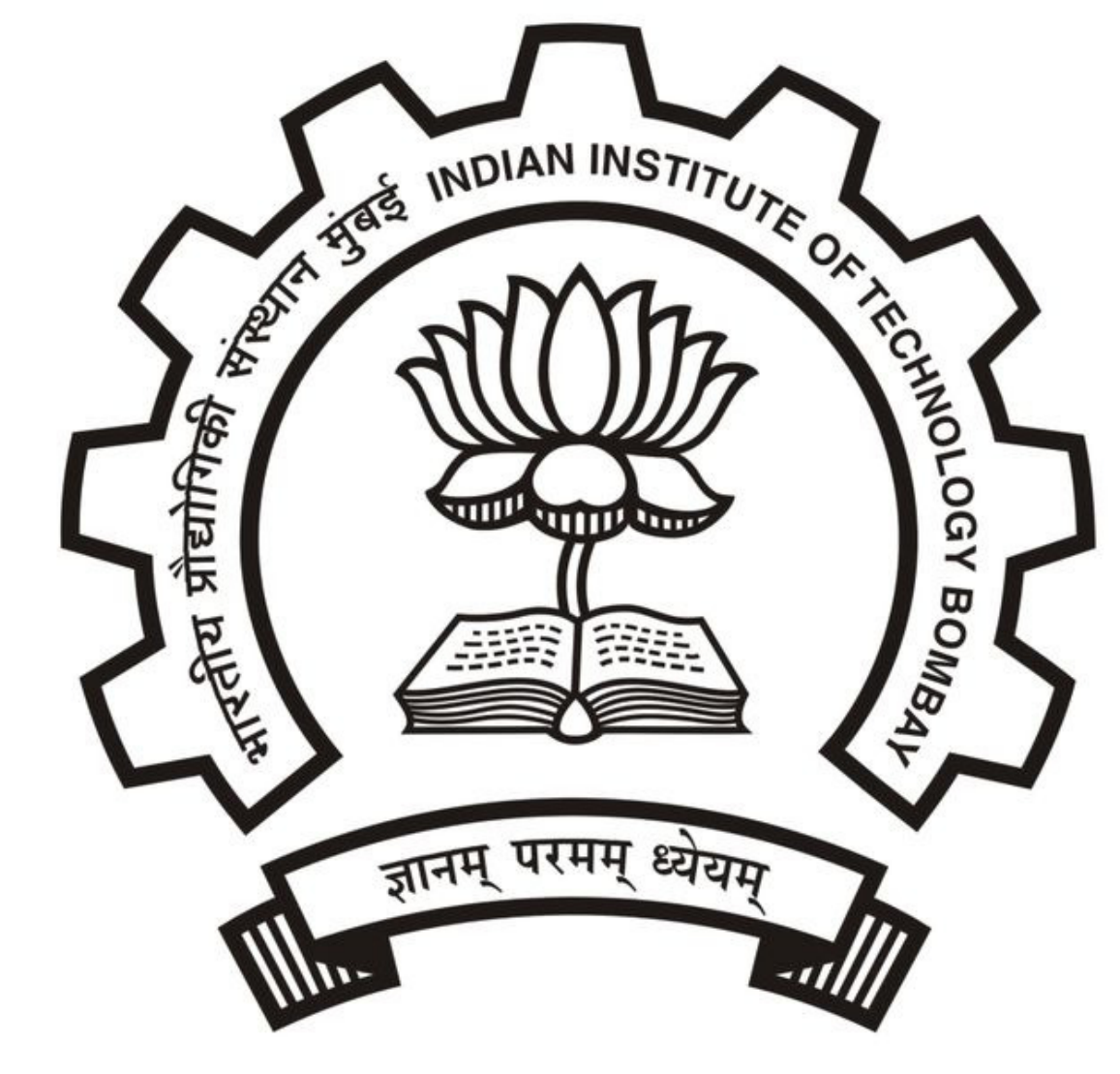}
\doublespacing
\begin{center}
\textbf{\large{DEPARTMENT OF PHYSICS\\
INDIAN INSTITUTE OF TECHNOLOGY BOMBAY\\\singlespacing
2025}}\\\singlespacing \copyright~2025 AMAR BHARTI All rights reserved.
\end{center}

\fboxsep6mm
\fboxrule1.3pt
}
\end{center}
\end{titlepage}
\pagestyle{fancy}

	\cleardoublepage
	\newpage
\thispagestyle{empty}
\setlength{\baselineskip}{32pt}
\bigskip
\bigskip
\bigskip
\bigskip
\bigskip
\vspace*{5cm}
\begin{center}
%\end{flushright}
\vspace*{0.5cm}
%\begin{flushright}
{\Large  \bf {Dedicated to my teachers} } \\
\end{center}
%\vspace*{5cm}
%\bigskip
%{\parindent 5pt \centerline{\Large \bf \it Dedicated To }}
%{\parindent 0pt \centerline{\huge \bf \it \texttt{My Mother}}
%
%\bigskip
%\bigskip

\setlength{\baselineskip}{18pt}

	\cleardoublepage \addcontentsline{toc}{chapter}{Thesis Approval}
	\include{HeadTail/certi_appro}
	\cleardoublepage
	\addcontentsline{toc}{chapter}{Declaration}
	\newpage
\thispagestyle{empty}
 \vspace*{2cm}
\begin{center}
\textbf{\LARGE{Declaration}}
\end{center}
\vspace{1cm}
\begin{spacing}{1.3}
I declare that this written submission represents my ideas in my own words and where
others' ideas or words have been included, I have adequately cited and referenced the original
sources. I also declare that I have adhered to all principles of academic honesty and integrity
and have not misrepresented or fabricated or falsified any idea/data/fact/source in my
submission. I understand that any violation of the above will be cause for disciplinary action
by the Institute and can also evoke penal action from the sources which have thus not been
properly cited or from whom proper permission has not been taken when needed.
\vspace{2.5cm}
\begin{flushright}
AMAR BHARTI \\
(Roll No. 194120016)
\end{flushright}
\begin{flushleft}
	Date : \today \\
	Place : Mumbai
\end{flushleft}
\end{spacing} 
	\cleardoublepage
	\addcontentsline{toc}{chapter}{Acknowledgements}
	\newpage
\thispagestyle{empty}
\begin{center}
\vspace*{-0.4cm}
{\LARGE {\textbf{Acknowledgements}}}
\end{center}

{\setlength{\baselineskip}{8pt} \setlength{\parskip}{2pt}
\begin{spacing}{1.5}

I offer my sincere gratitude to my supervisor, \textbf{Prof. Gopal Dixit} of the Department of Physics, IITB, for playing a vital role in my journey as a researcher. His striking insights, continuous motivation, and unwavering support for exploring uncharted territory are just a few of his contributions, for which I am thankful.
I would also like to express my utmost gratitude to the institute for providing me with fellowships, travel grants, and the necessary infrastructure to carry out my research in a fruitful way.

I would like to express my sincere thanks to my RPC members \textbf{Prof. Dipanshu Bansal} of the Department of Mechanical Engineering, IITB, and \textbf{Prof. Sumiran Pujari} of the Department of Physics, IITB, for their valuable comments and insightful questions, which helped me improve my research and its presentation.

I express my heartfelt thanks to \textbf{Prof. Misha Ivanov} of Max Born Institute for Nonlinear Optics and Short Pulse Spectroscopy, Berlin, for wonderful collaborations, which played a significant role in shaping my research. I cherish our joint endeavors to unravel the daunting problem and then narrate it as an enjoyable story. I would also like to thank him for inviting me for a scientific visit to MBI, Berlin. I sincerely appreciate the role \textbf{Dr. Mrudul M. S.} of Uppsala University, Sweden, played as my first collaborator, and I thank him for his continuous feedback and support ever since.

I extend my thanks to my labmates, Dr. Prachi Venkat, Satish Pandey and Gopika Nisha Gopalan, for their help and support. I extend my gratitude to past labmates Dr. Mrudul M. S., Dr. Sucharita Giri and Dr. Navdeep Rana for their wonderful companionship. I fondly appreciate and thank Ujjwal Upadhaya, Abhishesh Dwivedi, Midhun Krishna, Yeshma Ibrahim and Abhisek Panda for the wonderful conversations pertaining to physics and the entirety of the universe.

At last, I remember my favorite group of people for whom words would not be enough to express my gratitude. I express my utmost indebtedness to my first teachers: my parents and my grandparents. I appreciate the lessons of compassion, patience, and dedication that they instilled in me, which shaped me as a person. I extend special thanks to my mom, my sisters, and my brother-in-law for their love and support, which helped me through my research journey. I express my utmost appreciation and love for Dr. Uma, my partner in research and life. Her support made this thesis possible. Besides, listening to her talk about cell culture and gene manipulation serves as a perfect retreat for me.

\vskip2cm

\begin{flushright}
Amar Bharti\\
Department of Physics\\
IIT Bombay
\end{flushright}
%\begin{flushleft}
%	Date: \today\\
%\end{flushleft}
\end{spacing}

	\cleardoublepage

	\addcontentsline{toc}{chapter}{List of Symbols and Abbreviations} %%%%%%%%
	\markboth{List of Symbols}{List of Symbols and Abbreviations}  %%%%%%%%%%%
	\cleardoublepage
	\addcontentsline{toc}{chapter}{Abstract}
	\newpage
\thispagestyle{empty}
\begin{center}
\vspace*{-0.4cm}
{\LARGE {\textbf{Abstract}}}
\end{center}

\setlength{\baselineskip}{8pt} \setlength{\parskip}{2pt}
\begin{spacing}{1.5}
Condensed-matter provides attractive platforms to realize exotic particles, originally proposed in high-energy physics. Weyl semimetal (WSM) is one such material in which low-energy collective excitations are governed by massless Weyl fermions, which appear in pairs of opposite chirality and are topologically protected. 
Thus, the discovery of topological materials such as WSM has heralded a new era in contemporary  physics. 
Moreover, these materials offer exciting opportunities in next-generation signal processing and optoelectronics. 
This thesis explores different facets of the intense laser-driven phenomena in WSM for applications in emerging lightwave-driven Petahertz electronics and quantum technologies.

The interaction of circularly polarized laser with the Weyl node generates a chirality-sensitive current. Hence, the currents generated at the same Weyl node by the laser of opposite helicities observe mirror symmetry. We find that such mirror symmetry ceases to exist in the close proximity of the Weyl nodes. 
The reason for this is attributed to the nonlinear band energy dispersion in WSMs. 
In addition, an intense circularly polarized laser drives a helicity-sensitive photocurrent in an inversion-symmetric WSM. The generated photocurrent can be tailored by manipulating the laser parameters, such as intensity, phase delay, and ellipticity. 
Moreover, we introduce a universal method to generate photocurrent in topological as well as non-topological  materials, irrespective of their symmetries.  

The last part of this thesis focuses on investigating intriguing features of WSMs in nonperturbative high-harmonic spectra. 
It is observed that odd anomalous harmonics are generated from time-reversal-symmetry-broken WSM. 
Moreover, the harmonic spectra encode the nontrivial topology of the WSM, which is inherently linked to the Berry curvature, making it an all-optical probe of the anomalous nonlinear Hall effect. 
Higher topological charges can significantly improve the harmonics' yield and energy cutoff. Further, a comprehensive picture of high-harmonic generation from WSMs under various realistic conditions is presented. 

%\vspace{0.2in}

\end{spacing}

	\cleardoublepage
	\renewcommand{\contentsname}{Contents}  % Original name = Contents
	\begin{spacing}{1.2}  % Environment for 1.2 line spacing for contents and lists
		\tableofcontents
		\cleardoublepage
		\addcontentsline{toc}{chapter}{List of Figures} %%%%%%%%
		\listoffigures
		\cleardoublepage
		\addcontentsline{toc}{chapter}{List of Symbols and Abbreviations} %%%%%%%%
		\markboth{List of Symbols}{List of Symbols and Abbreviations}  %%%%%%%%%%%
		\chapter*{List of Symbols and Abbreviations}
\noindent {\bf Symbols}
\begin{tabbing}
aaaaaaaaaaaa \= abababababab \kill
$\mathcal{A}~(\mathbf{\Omega})$ \> Berry connection (curvature) \\
%$\mathbf{\Omega}$ \> Berry curvature\\
$\mathcal{C}$ \> Chern number \\
$\mathcal{E}$ \> Energy dispersion \\
%$\mathcal{P}$ \> Parity/inversion symmetry operator \\
%$\mathcal{T}$ \> Time-reversal symmetry operator \\
$\chi$ \> Chirality \\
$\eta$ \> Population asymmetry \\
$t_g$ \> Energy shift parameter\\
$t_c$ \> Tilt parameter\\
$n_c$ \> Residual population in conduction band \\ 
$\mathbf{A}(t)$\> Vector potential\\
$\mathbf{E}(t)$\> Electric field\\
$\mathcal{H}$\> Field-free Hamiltonian \\
$\mathcal{H}^\prime$\> Laser-Matter Interaction Hamiltonian \\
$\rho_{nn^\prime}$ \> Density matrix element \\
%$\textbf{d}_{nn^\prime}$ \> Dipole matrix element \\
%$\textbf{p}_{nn^\prime}$ \> Momentum matrix element \\
$\textrm{T}_2$ \> Dephasing time \\
$\textbf{J}(t)$ \> Total current \\
$\mathsf{J}$ \> Photocurrent\\
$\mathbf{J}_\Omega (t)$ \> Anomalous current\\
%$\mathcal{R}$ \> Amplitude ratio \\

%$\mathcal{I}(\omega)$ \> Intensity of HHG \\

\end{tabbing}
\noindent {\bf Abbreviations}
 \begin{tabbing}
aaaaaaaaaaaa \= abababababab  \kill
%au \> Atomic unit  \\
WSM(s) \> Weyl semimetal(s) \\
%WSMs \> Weyl semimetals \\
m-WSMs \> Multi-Weyl semimetals \\
HHG \> High-harmonic generation\\
SBEs \> Semiconductor Bloch equations\\
TRB \> Time-reversal-symmetry broken\\
IB \> Inversion broken\\
CPL \> Circularly polarized light\\
RCP \> Right-handed  circularly polarized  light\\
LCP \> Left-handed circularly polarized light\\
\end{tabbing}
%
%\noindent {\bf Superscript}
% \begin{tabbing}
%aaaaaaaaaaaa \= abababababab  \kill
% $*$ \>  dimensional quantity
%\end{tabbing}

		\cleardoublepage

	\end{spacing}
	
%	\addcontentsline{toc}{chapter}{List of Symbols} %%%%%%%%
%	\markboth{List of Symbols}{List of Symbols}  %%%%%%%%%%%
%	%\include{HeadTail/listofsymbol}
%	\cleardoublepage
\mainmatter
\begin{spacing}{1.5}
\chapter{Introduction}\label{Chaper1}
\begin{sloppypar}
\section{Emergence of Topological Materials}
Throughout the history of mankind, the discovery of new materials has led to revolutions. From prehistoric times, archaeological periods have been recognized as the most prevalent material being used, like the Stone Age or Iron Age. Similarly, the period after the late twentieth century can be called the Silicon Age, which is synonymous with the digital age. Indeed, the discovery of silicon as a semiconductor provided the essential ingredient for the mass production of compact electronic devices~\citep{pickard1906means,ohl1946light}. However, the invention of the transistor in 1947 by  John Bardeen, Walter Brattain, and William Shockley at Bell Labs is a broadly agreed to be the beginning of our digital age. In the following years, technology was improved incessantly, so much so that the number of components on a single chip doubled every year. This led to faster and more convenient electronic devices like mobile computing devices, to the extend that every aspect of human life -  health, commerce, utilities, travel, recreation, among others - now relies on electronic chips. 
Therefore, the discovery of new materials has played a direct role in some of the most fascinating  advancements and innovations of the twentieth century, such as light-emitting diodes, sensors, solar cells and integrated chips to name a few.

By the start of the twenty-first century, the growth of conventional electronics started saturating. 
This has fueled  the vigor in search for alternative materials to build electronics that can surpass the speed of silicon-based devices.
Graphene, a two-dimensional semimetal discovered in 2001, became one of the most influential materials in recent times~\citep {novoselov2005twopnas}. The 2010 Nobel Prize in Physics went to Andre Geim and Konstantin Novoselov for its discovery. 
This discovery brought a paradigm shift in condensed-matter physics because the electronic structure of graphene shows linear dispersion, which is described by the Dirac equation. 
Due to this surprising characteristic, graphene became an ideal candidate for the first table-top experiments for relativistic quantum mechanics. 
Therefore, it is aptly known as Dirac semimetal~\citep{novoselov2005two,neto2009electronic}.
It shows exceptional electrical conductivity, approximately six times that of copper~\citep{geim2009graphene}. Such drastic enhancement in conductivity is possible because of the large intrinsic mobility and zero effective mass of the carriers in graphene. Moreover, it is the strongest known material until now, around forty times stronger than diamond~\citep{lee2008measurement}. Immediately after its discovery, low-power devices based on graphene started making significant progress~\citep{lin2019two}. Gradually, graphene was tested to provide a better alternative for ultrafast and ultrasensitive optoelectronic and signal processing~\citep{xia2009ultrafast,prechtel2012time,gan2013chip,koppens2014photodetectors,wang2015ultrafast,tielrooij2015generation,fang2020mid,chen2022control,yoshioka2022ultrafast,agarwal2023,ghosh2024monolithic}. 

The immense potential and successful applications of graphene triggered the pursuit of three-dimensional Dirac semimetals.  
The quest was bolstered by the discovery of other novel materials - topological insulators, which formed a class of materials known as topological materials~\citep{hasan2010colloquium}. 
One of the remarkable properties of these materials is the robustness of their electronic states against perturbations like weak disorder and interactions, which has catalyzed a plethora of interesting phenomena~\citep{moore2010birth,qi2011topological,yan2017topological}.
The topological insulator acts as an insulator in the bulk but as a conductor on its surface. It has Dirac cones on its surface states, which facilitate dissipationless transport. 
The journey to the discovery of topological materials is traced to the seminal work done on the quantum Hall effect in 1985 by David Thouless, Duncan Haldane, and Michael Kosterlitz, who were awarded the 2016 Nobel Prize in Physics. They showed that topological invariants describe the phase of the two-dimensional electron gas under a strong magnetic field, hence called the topological phase. 
In 2005, Charles Kane and Eugene Mele, among others, rediscovered a topological phase, but this time in a graphene-like system without an external magnetic field, which reignited the interest in topological phases of matter~\citep{kane2005z,sheng2005nondissipative}. Finally, a theoretical prediction by Bernevig group~\citep{bernevig2006quantum} led to the experimental realization of two-dimensional topological insulators in HgTe quantum wells~\citep {konig2007quantum}.
Following this discovery, the search for a three-dimensional topological insulator, which has no counterpart to the quantum Hall effect, was led again by Kane and Mele~\citep{fu2007topological}.
In 2008, the first experimental realization of a three-dimensional topological insulator was reported using angle-resolved photoemission spectroscopy in Bi$_{x}$Sb$_{1-x}$ by Hasan group~\citep{hsieh2008topological,hsieh2009observation}. 
A single Dirac cone was reported on the surface of Bi$_2$Se$_3$, Bi$_2$Te$_3$, and Sb$_2$Te$_3$ in 2009~\citep{chen2009experimental,zhang2009topological}. The successful realization of topological insulators motivated the search for a three-dimensional counterpart of graphene - Dirac semimetals and eventually, Weyl semimetals~\citep{armitage2018weyl}. Consequently, it also served as a pedestal for the launch of the hunt for the catalog of the topological materials~\citep {vergniory2019complete}.

The studies on Na$_3$Bi and Cd$_3$As$_2$ resulted in the realization of three-dimensional Dirac semimetals~\citep{liu2014discovery,neupane2014observation}.
The three-dimensional Dirac semimetals possess 
both time-reversal and inversion protection, giving four-fold degenerate Dirac points. 
However, if one of these two symmetries is broken then these Dirac points transform into Weyl points. 
The absence of the time-reversal or inversion symmetry in the Weyl semimetal (WSM) gives rise to the Berry curvature~\citep{yan2017topological, armitage2018weyl}.
Initially, inversion-broken WSMs were discovered in the family of monopnictide: TaAs~\citep{lv2015experimental,yang2015weyl,xu2015discovery1} and NbAs, NbP, TaP~\citep{xu2015discovery,liu2016evolution}.
Time-reversal-symmetry-broken WSMs were realized in Kagome crystal Co$_3$Sn$_2$S$_2$~\citep{morali2019fermi,liu2019magnetic} and in Co$_2$MnGa ~\citep{belopolski2019discovery}.
Owing to the presence of the topologically protected Weyl points,  
WSM became a focal point among the topological materials. 
Isolated points in momentum space 
at which valance and conduction bands touch with linear energy dispersion are known as Weyl points. 
These points appear in pairs with opposite chirality and can be visualized as the source and sink of the Berry curvature.
The low-energy collective excitations in  WSM can be described by  the massless Weyl fermions,
which makes an elegant connection between high-energy and condensed-matter physics.
The massless Weyl fermions are chiral in nature, and were first proposed in high-energy physics 
and are described by the Weyl equation. 
However, there is no known candidate of Weyl fermion in the standard model of particles in high-energy physics. The successful realization of the topological WSMs has revolutionized contemporary physics. 
In recent years, multi-Weyl semimetals, a class of topological WSMs, have attracted broad interest in condensed-matter physics. 
Multi-Weyl semimetals are emerging topological semimetals with nonlinear anisotropic energy dispersion, which is characterized by higher topological charges. 

The nontrivial topology of the Berry curvature in WSM results in a series of exotic phenomenons, such as chiral magnetic effect ~\citep{vazifeh2013electromagnetic, li2016chiral,kaushik2019chiral}, circular photogalvanic effect ~\citep{de2017quantized,ma2017direct,rees2020helicity,le2021topology}, anomalous Hall effect ~\citep{burkov2014anomalous,yang2015chirality, shekhar2018anomalous,meng2019large}, Adler-Bell-Jackiw anomaly~\citep{zhang2016signatures}, diffusive magnetotransport~\citep{burkov2014chiral}  to name but a few~\citep{kim2013dirac,trescher2015quantum}.
Dirac and Weyl semimetals
have fostered the prospect for efficient conversion of light to electricity~\citep{braun2016ultrafast,wang2017ultrafast,liu2020semimetals,Tian_2022,Wang_2022,Lai_2022}. 
The topologically protected states in WSMs can facilitate the dissipationless transmission of 
information -- a prerequisite for quantum technologies. 
Interaction of light with WSMs has been a frontier approach to probe various exotic phenomena associated with the Berry curvature in 
WSMs, such as optical detection of chirality~\citep{ma2017direct}, chiral Terahertz source~\citep{gao2020chiral}, harmonic generation~\citep{lv2021high,dantas2021nonperturbative} 
to name but a few~\citep{ahn2017optical, matsyshyn2021rabi,   sirica2021shaking,lv2021experimental,  bao2021light, orenstein2021topology,  tamashevich2022nonlinear, nathan2022topological}.
Chiral Weyl fermions in WSM hold promise 
to realize qubits at ambient conditions for upcoming quantum technologies~\citep{kharzeev2019chiral}. 
In addition,  WSMs also hold promise for future technologies based on the ultrafast photodetector, chiral terahertz laser source, and quantum physics~\citep{ma2017direct,osterhoudt2019colossal,gao2020chiral}.

\section{Intense Laser-Driven Dynamics in Solids}
Search for new materials has a very long history - a glimpse of which is presented in the previous section. The traditional way of synthesizing new materials will continue in perpetuity. 
Alternatively, a new frontier of research based on intense-laser-driven physics has emerged over the last three decades. It utilizes lasers to tailor the properties of a system by controlling the dynamics of the system's constituents on their natural timescales.
The 2023 Nobel Prize in Physics is awarded  to Pierre Agostini, Ferenc Krausz, and Anne L'Huillier ``for experimental methods that generate attosecond pulses of light for the study of electron dynamics in matter''
~\citep{agostini2024nobel, krausz2024nobel, l2024nobel}. The award recognizes the immense capabilities of attosecond physics to study electron dynamics in matter on the attosecond timescale. 
The harbinger of revolution is the fact that an intense laser pulse can steer the electron dynamics and 
make it possible to videograph electron motion in matter. 
Moreover, numerous studies have shown the motion of charged particles in atoms and molecules with great precision~\citep{he2018direct,heide2024petahertz}. Interestingly, after painstaking research amid some controversy, it was concluded that there is a measurable time delay in the emission of a photoelectron from atoms and molecules. It refuted the established notion that the electron is instantly emitted in the photoelectric effect~\citep{l2024nobel}. 
Moreover, the capability of studying electron dynamics at their natural timescales has already done wonders for condensed-matter systems.  
As we delve deeper, it will be evident that we need intense laser pulses for next-generation Petahertz electronics or even quantum technologies at ambient conditions~\citep{borsch2023lightwave}. 

To understand how light shapes the dynamical properties  of  solids, let us start with 
the crown jewel of two-dimensional materials - pristine graphene, which 
possesses valley degree of freedom in addition to charge and spin. 
Valleys can be manipulated for quantum information processing, and valley-based technologies are dubbed ``valleytronics''. 
A circularly polarized light, resonant with the energy gap, can selectively excite one valley at a time 
in transition metal dichalcogenide monolayers such as MoS$_2$.
By virtue of the selection rules, a left-handed  circularly polarized light selectively excites the electronic population from valence to conduction band in one valley, whereas a right-handed  
circularly polarized light leads the population to the other valley~\citep{mak2012control}.
However, a similar protocol is not applicable for pristine graphene due to 
its inversion-symmetric lattice structure with zero band gap and zero Berry curvatures 
at $\mathbf{K}$ and $\mathbf{K}^\prime$. 
Thus, it was concluded that pristine graphene is not suitable for valleytronics till 2020.  
However, intense tailored laser pulses brought a glimpse of hope, as illustrated below. 

\begin{figure}[!h]
 \centering
\includegraphics[width=0.7\linewidth]{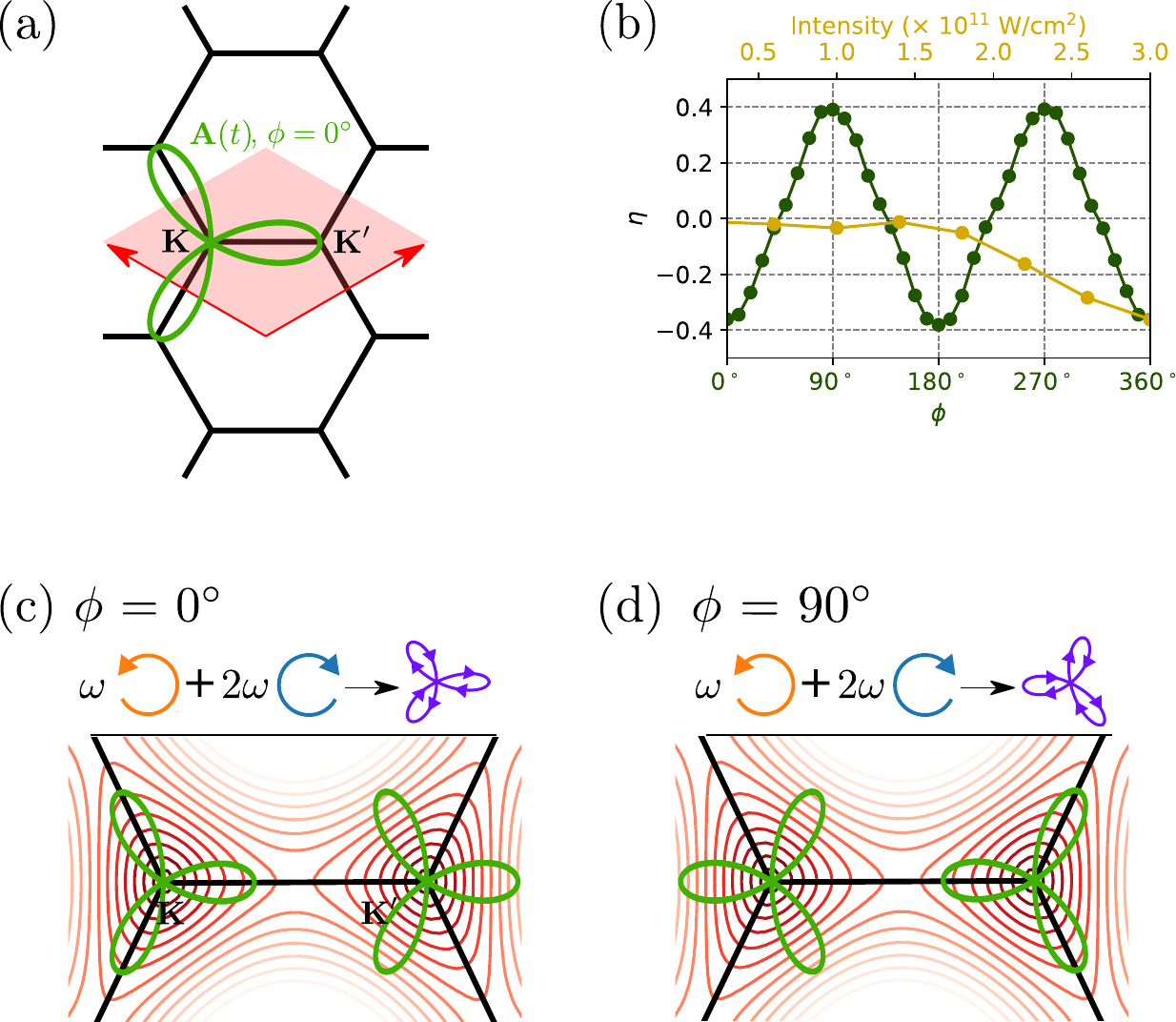}
\caption{Valleytronics in pristine graphene: (a) Trefoil structure formed by vector potential of $\omega-2\omega$ counter-rotating laser pulses shown above the Brillouin zone of the graphene. 
(b) Valley polarization ($\eta$) as the function of the subcycle phase ($\phi$) between $\omega$ and $2\omega$.  
$\eta$ becomes significant after critical intensity. 
(c) Mechanism of valleytronics: when 
the laser's trefoil structure matches the energy contours around one valley, $\mathbf{K}$, 
then the valley polarization is achieved with  $\phi=0$.  
(d) The energy contour around another valley, $\mathbf{K}^\prime$,  
matches with trefoil structure for $\phi=90^\circ$ and thus $\mathbf{K}^\prime$ is preferably populated. 
This figure is adapted from~\citep{mrudul2021light}.} \label{fig:fig1.1}
\end{figure}

A tailored light, consisting of two counter-rotating circularly polarized pulses with
frequencies $\omega$ and $2\omega$, is shown to generate valley-selective excitations in pristine graphene~\citep{mrudul2021light}. 
By changing the subcycle phase, $\phi$, between  $\omega$ and $2\omega$ pulses, one can either preferably populate $\mathbf{K}$ or $\mathbf{K}^\prime$ valley. 
If we assign ``1'' to the state when $\mathbf{K}$  valley is preferably excited, then ``0'' can be assigned for 
$\mathbf{K}^\prime$-dominated valley polarization. 
In addition, a superposition state of ``0'' and ``1'' for intermediate valley polarization
can also be achieved, allowing qubits to be realized in pristine graphene. 
Fig.~\ref{fig:fig1.1}(b) shows that the selective excitation in $\mathbf{K}$ is preferred over $\mathbf{K}^\prime$ valley for the subcycle phase between two pulses $\phi=0^\circ$. 
On the other hand, this situation changes as $\phi$ changes to  90$^\circ$ and 
$\mathbf{K}^\prime$ valley becomes the dominant one. 
To switch from one state to another, we need to achieve a state with zero polarization, which occurs for  
$\phi=45^\circ$. 
A desired control over valley polarization can be achieved by precise control over $\phi$. 
Thus, by fine-tuning laser parameters, a periodic oscillation in valley polarization can be realized in pristine graphene -- a medium where light-driven valleytronics was thought to be impossible.  
Two independent experimental groups~\citep{tyulnev2024valleytronics, mitra2024light} have  verified 
the theoretical proposal of intense tailored $\omega-2\omega$ laser-driven valleytronics~\citep{jimenez2020lightwave,mrudul2021light}. 
Therefore, it is now beyond a doubt that intense laser-shaped electron dynamics will play a critical  
role in upcoming all-optical valleytronics at Petahertz rate.

In general, photocurrent generation is forbidden in inversion-symmetric materials like graphene.
In the breakthrough work on laser-driven photocurrent in graphene, carrier-envelope phase stabilized pulses were employed to generate photocurrent that scales with the laser's intensity~\citep{higuchi2017light}. 
A carrier-envelope phase stabilized ultrashort pulse of a few-femtoseconds duration 
has a waveform which depends on its phase. 
Thus, by choosing an appropriate carrier phase, the laser waveform can be made asymmetric or symmetric
with respect to the pulse's envelope. 
The ensuing sub-cycle electron dynamics, which is on a sub-femtosecond timescale, leads to finite 
or zero photocurrent, depending on the symmetry of the waveform.  
In addition, carrier-phase stabilized pulses are recently used to successfully illustrate proof of concept 
light-driven logic gates for Petahertz electronics using graphene~\citep{boolakee2022light}.
%Later, a pair of linearly polarized pulses were shown to produce laser-driven photocurrent in graphene~\citep{heide2018coherent}.   
The underlying mechanism of generating photocurrent in an inversion-symmetric material is shaping the laser waveform in a controlled way so that it is asymmetric~\citep{heide2020sub,neufeld2021light,zhang2022bidirectional,rana2024optical}. 

\begin{figure}[!h]
    \centering
    \includegraphics[width=1\linewidth]{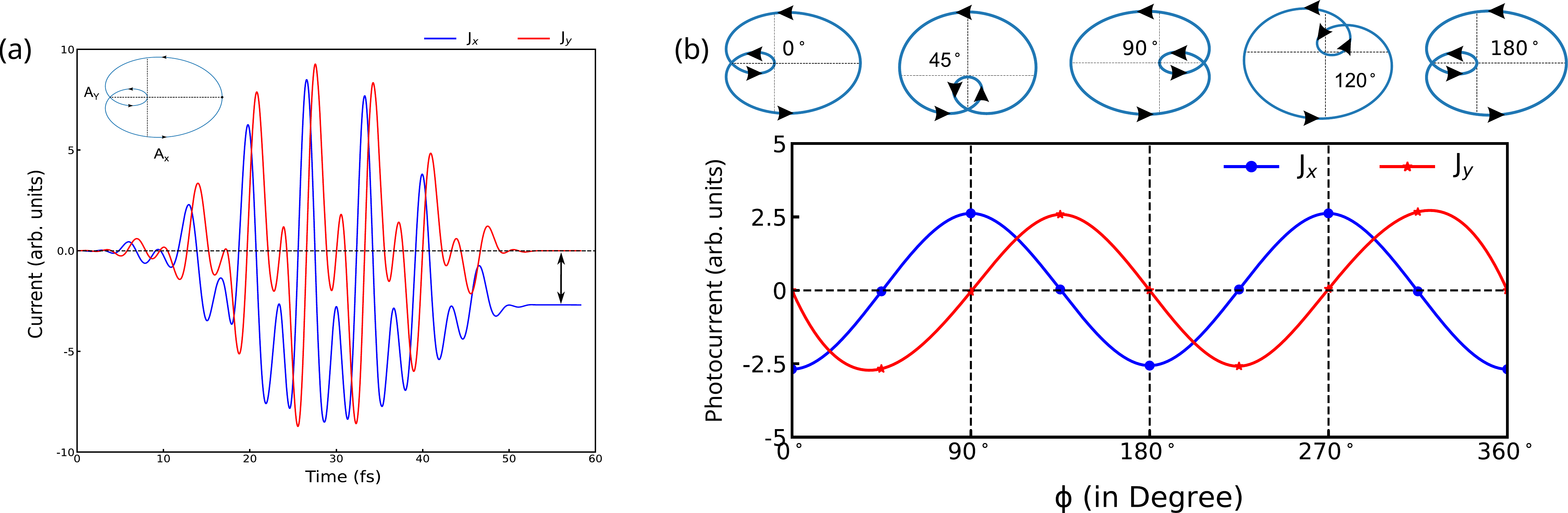}
    \caption{Laser-driven photocurrent in pristine graphene: (a) $\omega-2\omega$ co-rotating 
    circularly polarized pulses have asymmetric waveform (shown in inset) and hence lead to photocurrent 
    along $x$ direction. (b) By tuning the relative phase between $\omega$ and $2\omega$ pulses, the waveform is changed to tailor the directions and magnitude of the photocurrent in graphene. This figure is adapted from~\citep{rana2024optical}.}
    \label{fig:fig1.2}
\end{figure}

Recently, $\omega-2\omega$ co-rotating circularly polarized pulses are employed to generate intense-laser-driven photocurrent in a variety of materials~\citep{neufeld2021light}. Figure~\ref{fig:fig1.2} shows the application of $\omega-2\omega$ co-rotating pulses to graphene. 
The laser waveform of such pulses is asymmetric along  $x$ direction for $\phi = 0^\circ$, which 
results in photocurrent along the $x$ direction at the end of the laser pulse. 
By introducing subcycle phase between $\omega$ and $2\omega$ pulses, one can alter  the asymmetry of 
the waveform, making it symmetric along  $x$ direction and asymmetric along $y$ direction.
Hence,  photocurrent is generated along the $y$ direction for  $\phi = 45^\circ$, which is consistent with the asymmetry of the waveform as reflected from Fig.~\ref{fig:fig1.2}(b). 
Thus, Fig.~\ref{fig:fig1.2} (b) confirms that the direction of the photocurrent faithfully follows the asymmetry of the laser waveform.  
Analysis of the figure concludes that the material's static symmetry is not important in photocurrent generation as the symmetry of the laser waveform alters the symmetry of the entire light-driven system transiently by subcycle electron dynamics. 

By carefully studying the light-driven valleytronics and photocurrent in pristine graphene, 
one can appreciate the true capabilities of the intense-laser-driven electron dynamics on the attosecond timescale. 
An intense laser can truly shape the dynamics in solids, further enhancing their application. The laser-driven phenomena in novel materials like WSMs have enriched this research frontier  ~\citep{lv2021high,bharti2022high,bharti2023role,bharti2023tailor,bharti2023weyl,bharti2024photocurrent,bharti2024non}. 
At this point, let us discuss a cornerstone of an intense-laser driven process: 
high-harmonic generation (HHG)  from topological materials, which is at the heart of condensed-matter physics.

\section{High-Harmonic Spectroscopy of Topological Materials}
After the successful realization of HHG from a solid, a plethora of studies proliferated showcasing various  facets  of ultrafast spectroscopy of solids~\citep{ghimire2011observation,luu2015extreme,ghimire2019high,goulielmakis2022high}. 
HHG is a hallmark example of a nonperturbative nonlinear optical process, which has become the method of choice for probing various 
static and dynamical aspects of solids~\citep{zaks2012experimental,schubert2014sub,hohenleutner2015real,pattanayak2019direct,pattanayak2020influence, mrudul2021light, mrudul2021controlling,pattanayak2022role, rana2022probing, rana2022high}.   
Additionally,  HHG from topological materials has become the center  of attention as it allows one 
to investigate nonequilibrium topological features of  topological insulators, Dirac  
and Weyl semimetals in recent years~\citep{ kovalev2020non, cheng2020efficient, bai2021high, dantas2021nonperturbative,baykusheva2021all, lv2021high}. 

High-harmonic spectroscopy of monolayer MoS$_2$ established an undisputed 
signature of the Berry curvature for the first time~\citep{liu2017high}.
Following this work, Luu and W\"orner demonstrated a direct measurement of the Berry curvature of $\alpha$-quartz from the harmonic spectra~\citep{luu2018measurement}. 
Both these studies established that the signature of the Berry curvature is imprinted in ``anomalous'' harmonics, which are perpendicular to the laser's polarization direction. 
Furthermore, these studies reported the generation of even-order  ``anomalous'' harmonics due to Berry curvature, which is an essential feature of topological materials. 
Over the years, HHG has been considered as the go-to method for studying electron dynamics in topological systems on an attosecond timescale. 
A theoretical study 
of HHG from a one-dimensional lattice model showed a strong dependency on its topological edge states~\citep{bauer2018high}. 
In addition, the Ivanov group has proposed that topological phase transition in a two-dimensional Haldane model can be traced by high-harmonic spectroscopy~\citep{silva2019topological}. 
Trivial and topological phases of the Haldane insulator can be distinguished 
by measuring the helicity of the emitted harmonics. 
Parallely, the Lewenstein group showed that topological phase transition can be probed by recording 
circular dichroism of the emitted harmonics~\citep{chacon2020circular}.   
These theoretical works have motivated experimentalists to perform HHG from topological materials.  

In 2021, HHG from a three-dimensional topological insulator, Bi$_2$Se$_3$,
has been theoretically~\citep{baykusheva2021strong} and experimentally~\citep{baykusheva2021all} investigated. 
It has been observed that the yield of the emitted harmonics increases with the ellipticity of the driving laser in the mid-infrared regime.  
In the same year, the Huber group reported the generation of  the non-integer harmonics from Bi$_2$Te$_3$~\citep{schmid2021tunable}, which 
is counter-intuitive to all previous reports on the nonperturbative HHG from solids. 
Such a surprisingly efficient emission of tunable noninteger harmonics is attributed to the topological surface states in the topological insulators. 
In addition, the unique fingerprint of nontrivial topology was observed in the polarization-resolved harmonics from a topological insulator in BiSbTeSe$_2$~\citep{bai2021high}. 

In parallel, high-harmonic spectroscopy  has been employed for three-dimensional Dirac semimetals
~\citep{kovalev2020non,cheng2020efficient,lim2020efficient}. 
It has been illustrated that the ultralong scattering time and quasi-relativistic dispersion in topological semimetals are beneficial for the efficient generation of higher-order harmonics.  
Application of kinetic theory concluded that the perturbative and nonperturbative generation of the harmonics have different dependencies on the laser field strength in Dirac and Weyl semimetals~\citep{dantas2021nonperturbative}. 
HHG from inversion-broken WSM, $\beta$-WP$_2$, was reported using mid-infrared laser pulse~\citep{lv2021high}. 
Both odd- and even-order harmonics were observed, which originate from two different mechanisms:  
the even-order harmonics are attributed to ``spike-like'' Berry curvature, whereas the odd-order stem from Bloch oscillations in WP$_2$. 
The efficient generation of high-harmonics from topological semimetals warrants further studies, particularly from unexplored Weyl semimetals, such as time-reversal-symmetry-broken, type-II WSMs, and multi-Weyl semimetals with higher topological charges.  

\begin{figure}[!h]
    \centering
    \includegraphics[width=0.9\linewidth]{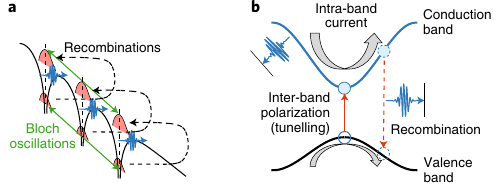}
    \caption{Mechanism of high-harmonic generation in solids: (a) A real-space view showing tunneling, Bloch oscillation, and recombination, (b) reciprocal-space picture showing the three steps: 1. tunneling (interband transition), 2. intraband motion, and 3. recombination. 
    This figure is adapted from \citep{ghimire2019high}.}
    \label{fig:fig1.3}
\end{figure}

Before proceeding to motivation and overview of this thesis, let us recap  
the underlying mechanism of HHG in solids. 
Following the three-step model of atomic HHG~\citep{corkum1993plasma}, an equivalent model for solid-state HHG is developed~\citep{ghimire2019high}. The first step is the tunnel ionization under the irradiation of an intense laser, i.e., the transfer of the Bloch electrons from valence to conduction band as visible from Fig.~\ref{fig:fig1.3}. 
This process creates a hole in the valence band and produces a Bloch electron in the conduction band. Under the influence of the intense laser, both electron and hole oscillate in their respective bands, which 
results in intraband currents within the same bands. 
As the direction of the laser field changes, there is a finite probability 
that the electron and hole can recombine, which emits interband current. 
Figure~\ref{fig:fig1.3} shows that both the interband and intraband currents generate higher-order harmonics of the driving laser field in solids. %Compared to atomic HHG, intra-band current/harmonics is a new additional pathway that is possible only in condensed matter systems. 

\section{Motivation}
Given the intriguing properties of the topological materials,  light-matter interaction yields fascinating outcomes, as already discussed in the previous sections.
%in the case of two-dimensional materials as well as three-dimensional Dirac semimetals and topological insulators
One of the essential hallmarks of WSMs is the pair of Weyl nodes, which is a topological invariant and 
carries opposite chirality.  
The theoretical predictions suggested a unique coupling of the circularly polarized light with the chirality of the Weyl node~\citep{chan2017photocurrents}.  
Thus,  the discovery of material candidates for WSM sparked a renewed interest in their linear as well as nonlinear optical responses~\citep{ahn2017optical, lv2021experimental,  bao2021light, orenstein2021topology,  tamashevich2022nonlinear, nathan2022topological, matsyshyn2021rabi,   sirica2021shaking}. 
It is anticipated that the chirality of the Weyl nodes could be optically selected analogous to ``valleytronics'' in two-dimensional materials~\citep{ma2017direct}. 
Additionally, a series of works exploited the nontrivial topology of WSMs to generate photocurrent~\citep{de2017quantized,golub2017photocurrent, rostami2018nonlinear,golub2018circular,osterhoudt2019colossal,gao2020chiral,lv2021high, hamara2023helicity}. 
However, the nonpertubative nonlinear  optical response of WSMs is a less charted territory, 
particularly for time-reversal-symmetry-broken WSMs.  

In this thesis, we aim to explore various intense-laser-driven phenomena in WSMs.
We focus on establishing signatures of the nontrivial topology of WSMs on intense-laser-driven processes, such as HHG from time-reversal-symmetry-broken WSMs.
In this endeavor, we plan to investigate WSMs with various topological charges and realistic conditions to understand how nontrivial topology and chirality are imprinted in nonperturbative nonlinear optical responses. 
Moreover, we explore avenues to exploit laser-shaped dynamics in WSMs with the idea that photocurrent 
can be generated in WSMs irrespective of their symmetries.  
We eventually aim to understand the underlying physics of the WSMs under intense laser and their technological applications in lightwave-driven Petahertz electronics and emerging quantum technologies, to name but a few.  

\section{Thesis Overview}
This thesis consists of six chapters and is organized as mentioned below.

\textbf{Chapter 1} introduces the fascinating world of materials by recounting the emergence of topological materials, especially WSMs. 
We introduce graphene and topological insulators as a foundation for unraveling novel topological semimetals and their intriguing properties. 
Later, we show how intense laser shapes electron dynamics in solids by presenting two examples - valleytronics and intense-laser-driven photocurrent in pristine graphene. 
Before moving to the next chapter,  we present a survey of HHG from topological materials, including
a brief discussion about the underlying mechanism of HHG from solids. 
The present chapter ends with a motivation and an overview of this thesis.  

We delve into their theoretical understanding after getting acquainted with topological materials and intense laser-driven phenomena. \textbf{Chapter 2} starts with a brief review of the band theory of solids. In continuation, we understand how relativistic quantum theory becomes necessary to describe certain materials, namely, topological materials. We briefly discuss key concepts of topological band theory to understand the role of topology in condensed-matter physics. 
We discuss the origin of the Berry phase in Bloch bands and in WSMs. 
At this juncture, we present minimal tight-binding models to describe WSM and its coupling with circularly polarized light. 
Toward the end, we discuss Berry curvature and the origin of the anomalous Hall effect in WSMs. 
We conclude \textbf{Chapter 2} by thoroughly discussing the semiconductor Bloch equations within the density matrix framework to describe the interaction of an intense laser pulse with WSM. 

\textbf{Chapter 3} takes up the case when WSM interacts with an intense circularly polarized laser. When the Weyl nodes interact with lasers of opposite helicity, residual electronic populations in the conduction band 
observe mirror symmetry. However, we show that this is not always the case. When the nonlinear band dispersion is considered in the vicinity of the Weyl node, the accumulated conduction band population around  
the Weyl node becomes helicity-dependent. 
This observation persists for longer wavelengths and weaker intensities of the employed laser. 
This raises a question about the massless nature of Weyl fermions in WSMs. 
Additionally, we introduce a scheme based on $\omega-2\omega$ counter-rotating circular pulses to control the helicity-dependent residual population in WSMs. 

\textbf{Chapter 4} investigates a method for generating ultrafast photocurrent in WSMs without relying on the absence of inversion symmetry. 
Our method uses a single-color circularly polarized laser to generate photocurrent in WSMs irrespective of their symmetries. 
By analyzing the total time-dependent current response, we observed that photocurrent is sensitive to 
the driving laser's helicity and ellipticity. 
In addition, our approach does not require phase stabilization of the driving laser to yield helicity-sensitive photocurrent in WSMs. 
Our proposed method remains equally applicable in various realistic conditions of WSMs, such as tilting and energy shifts of the Weyl nodes.
We further introduce a universal technique to generate photocurrent in topological and non-topological materials in two- and three-dimensional using a pair of linearly polarized pulses.  
The generated photocurrent can be easily tailored by changing the relative intensity, relative delay and angle between the two pulses. 

\textbf{Chapter 5} deals with HHG from WSMs. Firstly, we explore  
HHG from time-reversal-symmetry-broken WSM. 
We show that the emitted higher-order harmonics from WSM provide an all-optical probe of the anomalous Hall effect, which is anisotropic and nonlinear in nature. 
The ``anomalous'' odd harmonics stem from the nontrivial topology of the Berry curvature of WSM. 
We extend HHG from WSMs with various realistic conditions, such that the Weyl nodes are away from the Fermi level and tilted. It is found that the type-II WSMs lead to significant enhancement of yields of the higher-order harmonics.
We also investigate how HHG is influenced by the higher topological charge of m-WSMs. 
It is found that the scaling of the ``anomalous'' current with topological charge deviates drastically from the linear response theory.
Our studies suggest that time-reversal-broken WSMs offer a unique way to tune the polarization of emitted harmonics.

\textbf{Chapter 6} presents the conclusions and future directions of the thesis.

\end{sloppypar}
\cleardoublepage
\chapter{Theory of Weyl Semimetal and its Interaction with Intense Laser}\label{chp:Chapter2}
\newcommand{\lam}{\lambda}
\newcommand{\E}{\mathcal{E}}
 
The band theory of solids arises from the application of quantum mechanics to condensed-matter physics. The band theory of solids elucidates the quantum state of electrons inside the solid. Possessing this knowledge is essential for effectively predicting as well as explaining numerous properties of solids.  Therefore, the band theory has played a crucial role in condensed-matter physics, which deciphers the macroscopic and microscopic properties of matter. 
One such pivotal illustration is the distinctions among conductors, semiconductors, and insulators, which can only be satisfactorily explained using the band theory. 
Hence, the first step for designing and operating any electronic component is to gain a complete understanding of its band structure. 
The band structure of a material helps us 
to understand how electrons respond to external stimuli, such as light. 
The induced response pertains directly to the inherent characteristics of the material, like electric resistivity
and magnetic and optical susceptibility, to name a few.
Thus, the band theory of solids has played a crucial role in the development of various electronic and optoelectronic devices, such as light-emitting diodes, solar cells, and transistors, which 
have led revolution in electronics. Moreover, it continues to bolster the discovery of new materials, such as topological material, where it is known as topological band theory. 
Topological materials are a class of material that is beyond the Landau-Ginzburg theory of symmetry breaking. These materials lack a well-defined order parameter but are described by topological numbers, which are zero and non-zero in normal and topological phases, respectively. 

To understand how topological materials differ from ``normal'' materials, we start with a brief introduction to the band theory of solids and review its key ingredients. 
Further, we describe the emergence of a new kind of topological materials namely Weyl semimetals (WSMs). Before understanding the topological nature of WSMs, we review essential concepts of topological band theory. 
Following this, we showcase how the tight-binding Hamiltonian approach captures the topological features of WSMs. 
We move on to understand how a WSM interacts with circularly polarized light. Here, we determine how specific regimes of the band structure lead to distinguishing features in light-matter interaction.    
After this, we discuss the anomalous Hall effect and illustrate its origins due to the topological nature of WSMs. At the end of this chapter, we present a theoretical background of the semiconductor Bloch equations (SBEs), 
which is at the core of describing strong-field-driven nonperturbative light-matter interaction.
We will employ atomic units throughout this thesis unless stated otherwise. 

\section{Band Theory of Solids}

Let us briefly discuss the key element of the band theory of solids before delving into topological band theory. 
Perfectly arranged ions in a periodic lattice experience a potential with a periodicity of the Bravias lattice as
\begin{equation}
	U(\mathbf{r+R})=U(\mathbf{r}),
\end{equation}
where $\mathbf{R}$ is the Bravias lattice vector.
Time-independent Schr\"odinger equation for a single electron within an independent-electron approximation can be written as 
\begin{equation}\label{eq:perdpot}
	\left[\frac{- \nabla^2}{2} + U(\mathbf{r})  \right] \psi(\mathbf{r}) = \mathcal{E}~\psi(\mathbf{r}).
\end{equation}
Bloch theorem dictates that the solution of Eq. \eqref{eq:perdpot} is of the form:
\begin{equation}
	\psi_{n\k}(\r) = e^{i\k\cdot\r} u_{n\k}(\r),
\end{equation}
where $\k$ is wave vector (crystal momentum) and $n$ is an index for the $n^{\textrm{th}}$ eigenstate. 
The momentum operator within a non-relativistic quantum mechanics framework yields the momentum of 
the $n^{\textrm{th}}$  state as
\begin{eqnarray}\label{eq:momen}
	-i \nabla \psi_{n\k}(\r) &= &-i ~\nabla[e^{i\k\cdot\r} u_{n\k}(\r)] =  \k ~\psi_{n\k}(\r) - i e^{i\k\cdot\r}~ \nabla u_{n\k}(\r).
\end{eqnarray}
As evident from Eq. \eqref{eq:momen}, the momentum of electrons in a solid is not the same as for 
the free electron.
On substituting Eq. \eqref{eq:momen} in Eq. \eqref{eq:perdpot}, we obtain
\begin{equation}\label{eq:schro}
	\mathcal{H}_\k u_\k(\r) = \left[\frac{1}{2\omega}\left(\k - i\nabla  \right)^2 + u(\r)\right] u_\k(\r) = \mathcal{E}_\k u_k(\r).
\end{equation}
Solutions of Eq. \eqref{eq:schro} are infinitely many for each $\k$ and give rise to band structure. 
Therefore, while the free electron shows the parabolic energy dispersion, the Bloch electrons give rise to a discrete energy spectrum, as evident from Fig.~\ref{fig:bnd}. 

\begin{figure}
\centering
\includegraphics[width=\linewidth]{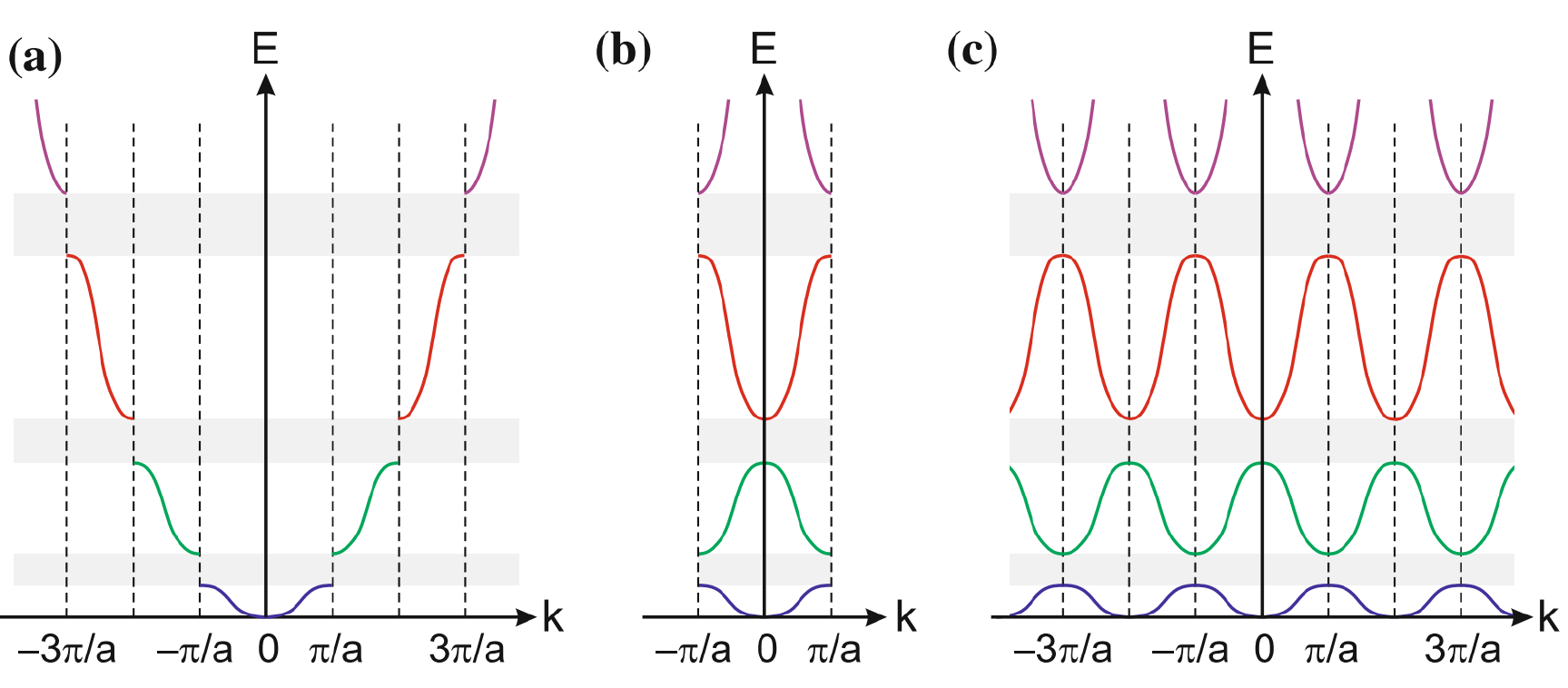}
\caption{Schematic of a band structure in various zone schemes: (a) extended zone scheme shows the band structure where bands are separated by a gap at the Brillouin zone boundary, (b) reduced zone scheme shows the band structure in the first Brillouin zone, and (c) periodic zone scheme exhibits 
bands in every Brillouin zone. This figure is adapted from Ref.~\citep{grundmann2010physics}.}
\label{fig:bnd}
\end{figure}

One of the important features of Band theory is the dressing of electrons, as we explained in 
Eq. \eqref{eq:momen} when an electron moves through a periodic potential.  The effective mass of the electron in solids is different from the mass of the electron in the vacuum and is known as  quasiparticle~\citep{madelung2012introduction}. 
The effective mass of the quasiparticle can be expressed as 
\begin{equation}
m^{*} = \left[\pdv[2]{\mathcal{E}(\mathbf{k})}{\mathbf{k}}\right]^{-1}.
\end{equation}
In general, the effective mass is a tensor and can be negative or positive depending on the curvature of 
the energy band dispersion. 
In the later part of this chapter, we will see that the quasiparticle becomes massless as the dispersion 
relation becomes linear for topological semimetals. 

\begin{figure}[H]
\centering
\includegraphics[width=0.7\linewidth]{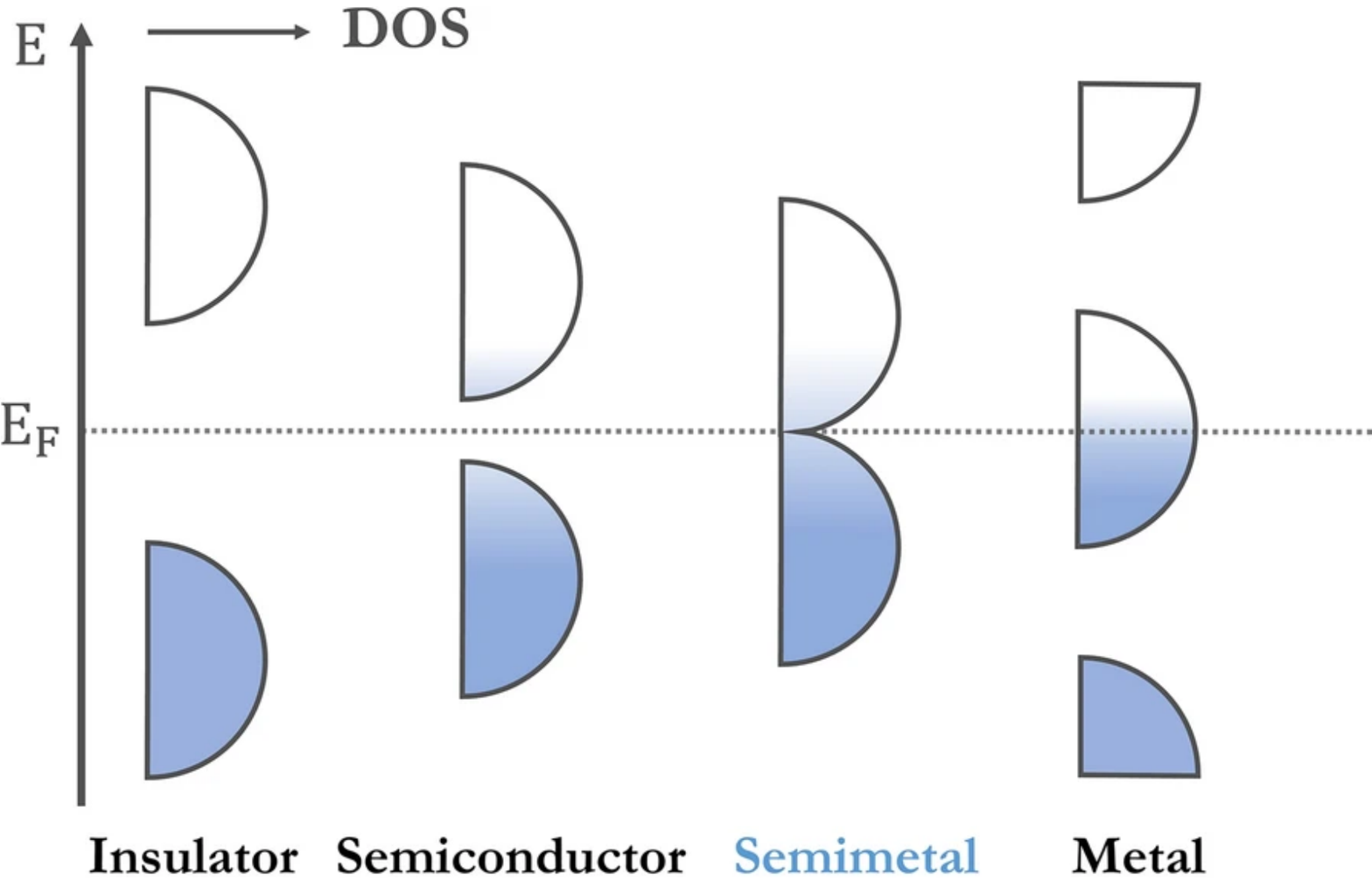}
\caption{Density of state allows the distinction among metal, semimetal, semiconductor, and insulator. While both metal and semimetal are gapless, semimetal has, in contrast, a vanishing density of states. This figure is adapted from Ref.~\citep{guo2023light}.}
\label{fig:dos}
\end{figure}

The energy dispersion allows us to estimate the available energy states per $\mathbf{k}$ and is known as the density of state as
\begin{equation}
D(\E) = \int \frac{d\mathbf{k}}{(2\pi)^3} ~\delta(\E-\E(\k)).
\end{equation}
The filling of the available energy states decides the electronic behavior of a given material. By noticing  
at the position of the Fermi level, $\mathcal{E}_f$, different electronic states can be identified as shown in Fig.~\ref{fig:dos}. While metals have at least one band inside $\mathcal{E}_f$, semiconductors and insulators have $\mathcal{E}_f$ inside the band gap.
The semimetal is an intermediate state between metal and semiconductor where the density of states goes to zero at $\mathcal{E}_f$. 

\section{Relativistic Quantum Mechanics in Condensed-Matter Physics}
In general, the energy scale involved in condensed-matter physics is typically much smaller than the rest mass of the electron. Hence, even when we consider realistic scenarios where electron-electron interaction
is significant, Schr\"odinger equation in Eq.~\eqref{eq:schro} provides a satisfactory description of a given material. 
The relativistic quantum mechanics, embodied by the Dirac equation, became essential  
after the realization of graphene in 2001. 
Let us briefly review the implications of the Dirac equation in condensed-matter physics, which plays a significant role in our understanding of topological semimetals.

Description of an electron within quantum mechanics, consistent with special relativity, can be written in terms of Dirac equation as
\begin{equation}\label{eq:diraceq}
(i\gamma^\mu \partial_\mu -m)\Psi = 0.
\end{equation}\label{gammacomu}
Here, $\gamma^\mu$ are known as gamma matrices where $\mu\in(0, 1, \ldots, d)$ for $d$-dimensions. The gamma matrices are chosen in such a way that $\gamma^\mu$ obey the given anticommutator relation as   
\begin{equation}
\{\gamma^\mu,\gamma^\nu\} = 2\eta^{\mu\nu} \mathbbm{1},
\end{equation}
where $\mathbbm{1}$ is $4\times4$ identity matrix.

Herman Weyl further simplified the Dirac equation for massless particles ($m=0$) in odd (spatial) dimensions. 
The Weyl equation is derived from the Dirac equation in a straightforward manner.
To do this, let us express the Dirac equation  in one-dimension as 
\begin{align}
	i\partial_t \Psi = (-i\gamma^0 \gamma^1 \partial_x + m\gamma_0) \Psi.
\end{align} 
After considering $m=0$ and $\hat{p}_{x} = -i\partial_x$ with $\gamma_5=i^k \gamma^0\gamma^1\gamma^2\dots\gamma^d$, we arrive as 
\begin{align} \label{eq:weyl}
	i\partial_t \Psi = \gamma^0\gamma^1~\hat{p}_{x}\Psi-\hat{p}_{x}~\gamma_5\Psi = \mp \hat{p}_{x} \Psi_\pm, 
\end{align}  
where $\Psi_\pm$ is chosen in such a way that $\gamma_5\Psi_\pm=\pm \Psi_\pm$, and 
$\Psi_\pm$ is  the eigenstate of $\gamma_5$. 
The above equation can be written in compact form as 
\begin{equation}
	i \partial_t\Psi = \pm \hat{\mathbf{p}}~\Psi.
\end{equation}
The above equation is known as the Weyl equation for massless particles. 

\begin{figure}[H]
\centering
\includegraphics[width=0.7\linewidth]{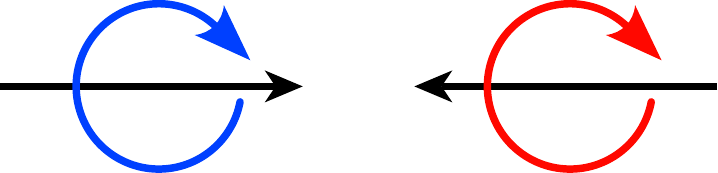}
\caption{Sense of rotation, denoted by the red or blue arrow, and the direction of momentum, denoted by the black arrow, is either parallel or anti-parallel.}
\label{fig:weyl2}
\end{figure}

The energy dispersion for the Weyl equation is $\E_\pm(\mathbf{p}) = \pm \mathbf{p}$, where the sign represents the right (left) moving particle or (chiral) Weyl fermions [see Fig.~\ref{fig:weyl2}]. 
The gamma matrices in three-dimensions are considered as 
$\gamma^0=\mathbbm{1}\otimes\sigma_x, ~ \gamma^i=\sigma^i\otimes i\sigma_y, ~\text{and,}~  \gamma^5=-\mathbbm{1}\otimes\sigma_z$.
Thus, the Weyl equation reads in compact form as
\begin{equation} \label{weylham}
i \partial_t\Psi = \mathcal{H}_{\textrm{Weyl}} \Psi = \pm (\mathbf{p}\cdot\sigma)~ \Psi,
\end{equation}
where $ \mathcal{H}_{\textrm{Weyl}}= \pm \mathbf{p}\cdot\sigma$ is known as the Weyl Hamiltonian. 

\section{Topological Band Theory}
Concepts illustrated in the band theory of solids were sufficient to distinguish different electronic phases until 2005. 
Moreover, a phase transition followed the Landau-Ginzburg theory, which describes a phase transition as the breaking of a symmetry. 
For example, the phase transition of liquid into a solid is associated with breaking the translational symmetry. 
Symmetry breaking goes hand in hand with an order parameter. 
Now, consider the phase transition of paramagnet to ferromagnet, in contrast to the paramagnetic phase,  
spins are aligned in a particular direction in the ferromagnetic phase, resulting in broken-rotational symmetry. 
Additionally, there is non-zero (zero) magnetization in the ferromagnetic (paramagnetic) phase, which is considered an order parameter. 
This ``conventional'' theory started to fail after the realization of topological insulators in 2007. 
Topological and ``normal'' insulators are two phases related by a phase transition. 
However, the corresponding phase transition is not associated with a symmetry breaking and an order parameter. 
Hence, such a phase transition from ``normal'' to a topological phase is not described by Landau-Ginzburg's theory. 
Thus, we need topological band theory to describe topological materials, which requires a new 
concept -- topological invariant.
It is zero in the ``normal'' phase but non-zero after the topological phase transition. 
To understand how topological invariant arises, let us understand the origin of the Berry phase via a simple example, which is also the central pillar of topology in condensed-matter physics. 

\subsection{Berry Phase in Bloch Bands}
In order to understand the origin of topology in condensed-matter systems, let us start by considering a generic Hamiltonian of the following form
\begin{equation}\label{eq:dham}
	\mathcal{H} = \bm{\sigma} \cdot  \mathbf{h}(\k) ,
\end{equation}
where $\sigma$ is the Pauli matrices.
Note that Eq. \eqref{eq:dham} is frequently used to describe graphene and WSMs within the two-band representation. 
Let us parameterize $\mathbf{h}(\k)$ as $\mathbf{h}= |h| [\sin(\theta) \cos(\phi), \sin(\theta)\sin(\phi), \cos(\theta)]$ in the Bloch sphere with 
 $|h|$ as the modulus of the vector ($\mathbf{h}$),  $\theta= \arccos(h_z/|h|)$, and $\phi = \arctan(h_y/|h|)$.
Two eigenstates with energies $\mp |h|$ are obtained by diagonalizing Eq. \eqref{eq:dham} as 
\begin{equation}
	\ket{v}= \begin{pmatrix}
		\sin(\theta/2) e^{-i\phi} \\
		-\cos(\theta/2)
	\end{pmatrix},~\textrm{and}~ \quad \ket{c}= \begin{pmatrix}
		\cos(\theta/2) e^{-i\phi} \\
		\sin(\theta/2)
	\end{pmatrix}.
\end{equation}

Let us calculate Berry connection for the state $\ket{v}$ in parameter space $\mathbf{R}$ using a general  definition as 
\begin{equation}
	\mathcal{A}_n(\mathbf{R}) = i \matrixelement{v}{\pdv{}{\mathbf{R}}}{v}.
\end{equation}  
The corresponding Berry connection can be written as
\begin{align}\label{dconn1}
	\mathcal{A}_\theta = \matrixelement{v}{i\partial_\theta}{v} = 0, \quad \text{ and} \quad \mathcal{A}_\phi = \matrixelement{v}{i\partial_\phi}{v} = \sin[2](\theta/2).
\end{align}
It turns out that the above Berry connections are not gauge invariant. Hence, a gauge invariant quantity, called Berry curvature, for an eigenstate $n(\mathbf{R})$ is defined as 
\begin{eqnarray}\label{bcurv}
	\Omega^n_{\mu\nu}(\mathbf{R}) &=&\pdv{\mathcal{A}_\nu^n(\mathbf{R})}{\mathbf{R}^\mu} - \pdv{\mathcal{A}_\mu^n(\mathbf{R})}{\mathbf{R}^\nu}\nonumber\\
	&= &i \left[ \braket{\pdv{n(\mathbf{R})}{\mathbf{R}^\mu}}{\pdv{n(\mathbf{R})}{\mathbf{R}^\nu}}  - \braket{\pdv{n(\mathbf{R})}{\mathbf{R}^\nu}}{\pdv{n(\mathbf{R})}{\mathbf{R}^\mu}}   \right].
\end{eqnarray}

Using the above expression of the Berry connections in Eq. \eqref{dconn1}, the Berry curvature can be calculated as 
\begin{equation}
	\Omega_{\theta\phi}= \partial_\theta \mathcal{A}_\phi - \partial_\phi \mathcal{A}_\theta = \sin(\theta)/2.
\end{equation}
As illustrated above, a generic Hamiltonian in Eq. \eqref{eq:dham} can lead to non-zero Berry curvature, which arises due to the non-trivial nature of the eigenstates.
Interestingly, the eigenstates are not smooth functions under adiabatic evolution. 
If one evolves $\ket{v}$ adiabatically to $\theta=\pi$, it becomes ill-defined. Similarly, taking $\ket{c}$ adiabatically to $\theta=0$ renders it meaningless. 
This ill-defined behavior of eigenstates is the central pillar of topology in the electronic state of 
condensed-matter systems, which will be explained in detail below.

\begin{figure}[!htb]
\centering
\includegraphics[width=0.7\linewidth]{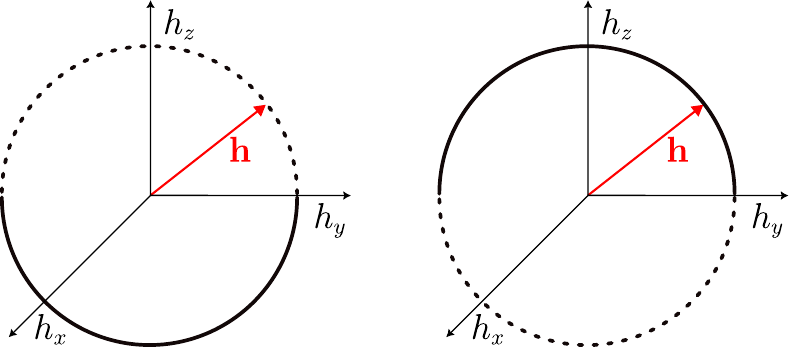}
\caption{Bloch sphere representation of eigenstates: Ill-defined eigenstates in the north or south pole.}
\label{fig:blochsphere}
\end{figure}

At this point, we attempt to gauge transform the eigenstate to make it well-defined. Let us consider $\ket{v} = e^{i\phi} \ket{v}$, such that
\begin{equation}
	\ket{v}= \begin{pmatrix}
		\sin(\theta/2)  \\
		-\cos(\theta/2)e^{i\phi}
	\end{pmatrix}.
\end{equation}
We redo the calculation of the Berry connections  for the updated $\ket{v}$, which leads to
\begin{eqnarray}
	\mathcal{A}_\theta=0,\quad \text{and}\quad\mathcal{A}_\phi=-\cos[2](\theta/2).
\end{eqnarray}
Note that the Berry curvature remains the same as $\Omega = \sin(\theta)/2$. 
Our choice of gauge transformation makes $\ket{v}$ a smooth function at $\theta=\pi$ but becomes ill-defined at $\theta=0$. 
Hence, unless we can find a gauge transformation that makes the eigenstates smooth everywhere, the gauge-independent \textit{local} quantity Berry curvature is expected to be non-zero.
In other words, under adiabatic evolution in a closed loop, a Berry phase is accumulated in addition to the dynamical phase. 
The Berry phase acquired in a cyclic adiabatic evolution over a loop $C$ is defined as 
\begin{equation}
	\gamma_n = \int_C d{\textbf{R}}\cdot \mathcal{A}_n(\textbf{R}). 
\end{equation}
In accordance with Stokes' theorem, the Berry phase can have an equivalent formulation as
\begin{equation}
	\gamma_n = \int_S   d{\textbf{S}} \cdot \mathbf{\Omega}_n(\textbf{R}). 
\end{equation}

\subsection{Topology of Weyl Semimetals}
Let us consider a  specific form of $\mathbf{h}(\k)$, which describe WSMs as 
\begin{equation}\label{eq:Weyl}
\mathcal{H}_{\textrm{Weyl}}(\k) = \bm{\sigma} \cdot \mathbf{d}(\k) = \sum_i v_i k_i \sigma_i.
\end{equation}
It is straightforward to show that the Berry curvature corresponding to $\mathcal{H}_{\textrm{Weyl}}(\k)$ is
\begin{equation}\label{eq:monopoles}
	\Omega^\pm = \mp \frac{\mathbf{d}}{2|\mathbf{d}|^3}.
\end{equation}
In earlier works, Dirac showed that the field described by Eq. \eqref{eq:monopoles} is generated by monopoles situated at $\k=0$. In this case, it is also energy degenerate. Hence, these degenerate points act as the source and sink of the Berry curvature flux, which can be seen as the \textit{pseudo} magnetic field in momentum space. These degenerate points are known as Weyl points or Weyl nodes. 

\begin{figure}[H]
\centering
\includegraphics[width=0.6\linewidth]{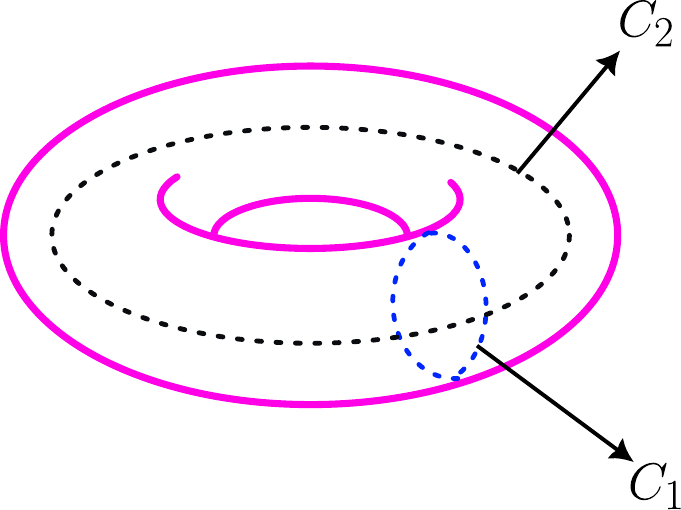}
\caption{Non-trivial topology can be described by genus number or number of holes. On the surface of a sphere, loop $C_1$ can be smoothly deformed. However, on the surface of a Torus, there exists a $C_2$ like loop that can not be deformed smoothly due to the presence of a hole.}
\label{fig:torus}
\end{figure}

On integrating Eq. \eqref{eq:monopoles} over a sphere containing the monopole, we obtain
\begin{equation}
	\mathcal{C} = \int \Omega^\pm_{\theta\phi}~\dd\theta \dd\phi = \pm~ 2\pi.
\end{equation}
Here, $\mathcal{C}$ is known as the Chern number. In general, the Berry curvature integrated over a closed manifold is quantized in the units of 2$\pi$, and is equal to the net number of monopoles inside. 
In the present context, $\mathcal{C}$ is calculated over the periodic Brillouin zone, which is topologically equivalent to a torus and also has no boundary. 
For a well-defined eigenstates, $\mathcal{C}$  obtained by integrating over the torus as
$$\mathcal{C} = \oint_S d{\textbf{S}} \cdot \Omega  = \oint_C d{\textbf{R}}\cdot \mathcal{A} = 0.$$
Hence, the non-zero $\mathcal{C}$ is a testament that there is an obstruction to the application of the Stokes theorem to the entire Brillouin zone, which is illustrated in Fig.~\ref{fig:torus}.  

At this point, it becomes interesting to ask when there is non-zero Berry curvature in a real material. 
In other words,  what is the simplest condition for the realization of topological material?  
To answer the question, let us look at the symmetries of the Berry curvature. 
Under inversion ($\mathcal{P}$) and time-reversal ($\mathcal{T}$), $\Omega_n$ transforms as 
\begin{equation*}
	\Omega_n(\k) \xrightarrow{\mathcal{P}}\Omega_n(-\k),~~~\textrm{and}~~~  \Omega_n(\k). \xrightarrow{\mathcal{T}}-\Omega_n(-\k). 
\end{equation*}
If both time-reversal and inversion symmetries are preserved in a system, then Berry curvature is zero. 
Hence, one of these symmetries must be broken to get a finite Berry curvature. 
In addition, a non-zero $\mathcal{C}$ arises when time-reversal symmetry is broken.

\section{Weyl Semimetal}
Weyl semimetal (WSM) can be realized by breaking either time-reversal or inversion symmetry or both symmetries of the corresponding Dirac semimetal phase, as shown in Fig.~\ref{fig:weyl1}. 
In either case, non-zero Berry curvature  results~\citep{yan2017topological}.
Weyl semimetal  consists of the topologically protected degenerate points known as 
Weyl points, which act as the monopoles of the Berry curvatures in momentum space~\citep{armitage2018weyl}. 

\begin{figure}[!h]
\centering
\includegraphics[width=0.8\linewidth]{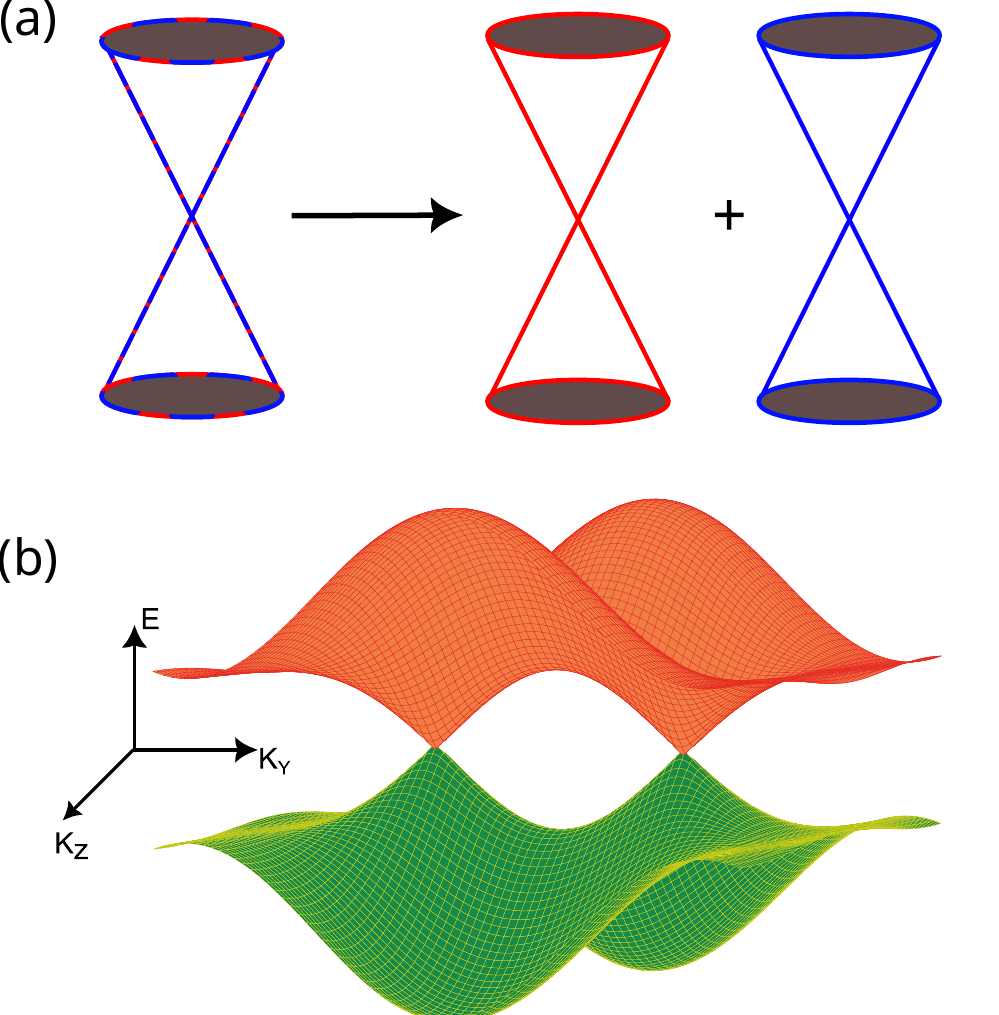}
\caption{(a) Dirac points have copies of two chiralities superimposed, which can be separated into two Weyl points by breaking inversion or time-reversal symmetry. (b) Band structure of a generic Weyl semimetal with ``at least'' two band-touching, or Weyl nodes.}
\label{fig:weyl1}
\end{figure} 

Let us consider a time-reversal symmetry broken (TRB) WSM, whose components of $\mathbf{d}(\textbf{k})$ in Eq. \eqref{eq:Weyl} are expressed as~\citep{hou2016weyl}  
\begin{equation}\label{eq:weylhamtb}
\mathbf{d}(\mathbf{k}) =   t\big[\cos(k_x a), \cos(k_y a), \{\cos(k_z a) -  \alpha(1 + \sin({k_x a}) \sin({k_y b}))\}\big]. 
\end{equation}
Here, $\alpha=0.5$ is a dimensionless quantity, and the Weyl nodes are situated at $\mathbf{k} = [0,0,\pm \pi/(2a)]$.
For spinless fermions, the representation for inversion and time-reversal symmetries have the following form: $\mathcal{P} \longleftrightarrow \sigma$ and $\mathcal{T}\longleftrightarrow \mathcal{K}$ with 
$\mathcal{K}$ as the anti-Hermitian complex conjugate operator such that  $\mathcal{K}^\dagger i \mathcal{K}=-i$.
For inversion symmetric WSM, we have 
$$\mathcal{P}^\dagger \mathcal{H}_{\textrm{Weyl}}(-\k) \mathcal{P} = \sigma_0^\dagger \mathcal{H}_{\textrm{Weyl}}(-\k) \sigma_0 = \mathcal{H}_{\textrm{Weyl}}(\k).$$ 
Thus, it can be concluded that the Hamiltonian in Eq.~\eqref{eq:weylhamtb}  is inversion symmetric. 
Similarly, we can show that $\mathcal{T}^\dagger \mathcal{H}_{\textrm{Weyl}}(-\k) \mathcal{T} \neq \mathcal{H}_{\textrm{Weyl}}(\k)$, which means that WSM represented by Eq.~\eqref{eq:weylhamtb} exhibits a breaking of 
the time-reversal symmetry. 
The energy band structure of the corresponding TRB WSM is shown in Fig. \ref{fig:model}.

%\section{Weyl Equation}
\begin{figure}[!h]
\centering
\includegraphics[width=\linewidth]{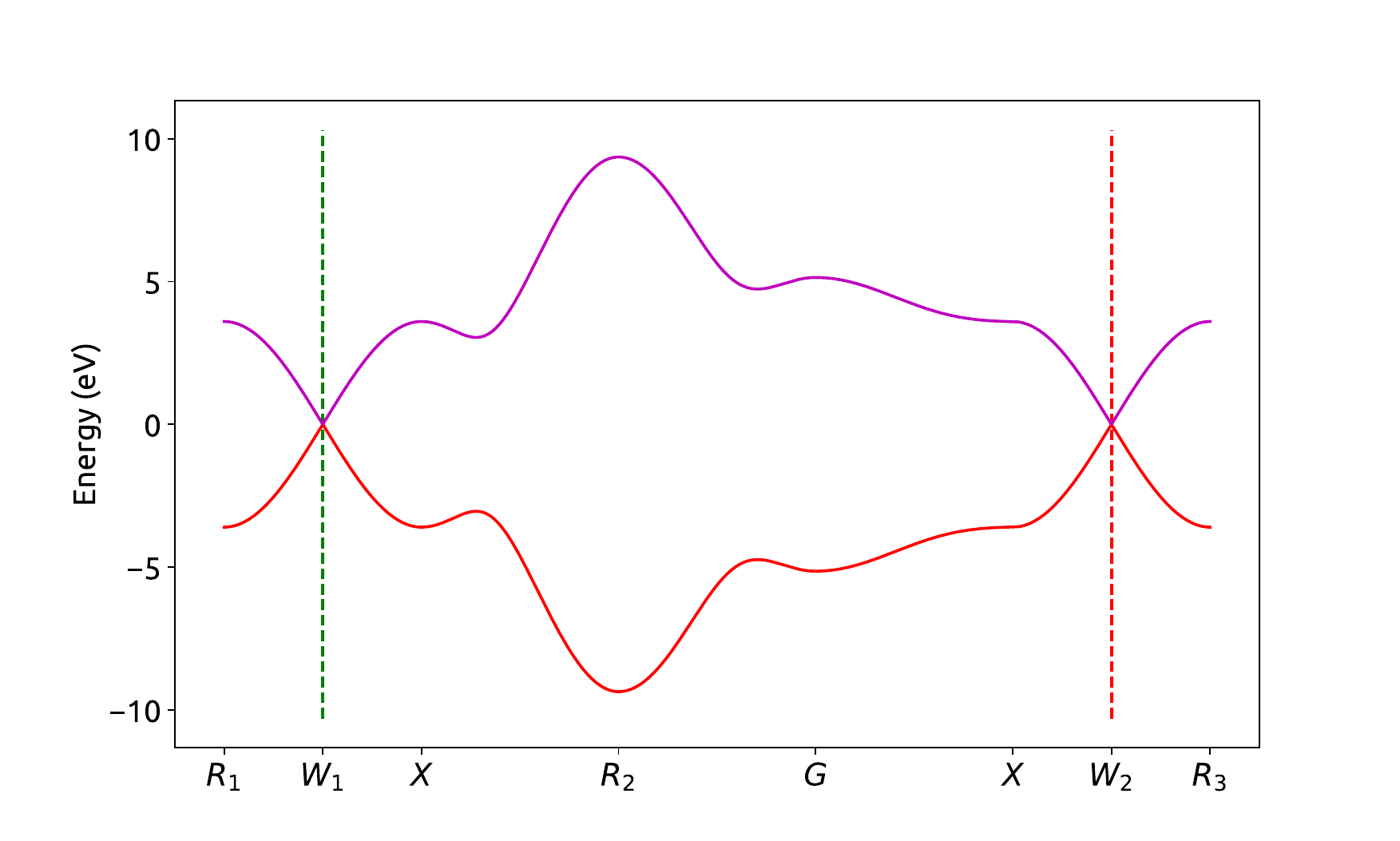}
\caption{Energy band structure of the time-reversal symmetry broken Weyl semimetal corresponding to the Hamiltonian in Eq.~\eqref{eq:weylhamtb}}.
\label{fig:model}
\end{figure}

Let us approximate the Hamiltonian near the two Weyl points 
$W_{1}$ and $W_{2}$ as shown in  Fig. \ref{fig:model}. 
Hamiltonian  near the first Weyl point $W_1$ at  $k_z = \pi/2a$ can be written as 
\begin{equation}
\mathcal{H}_{W_{1}} = t[(-k_x a) \sigma_x + (k_y a) \sigma_y - (k_z a) \sigma_z].
\end{equation} 
There are two eigenvalues associated with $\mathcal{H}_{W_{1}}$ as 
\begin{equation}
\mathcal{E}^{W_{1}}_{v/c} = \pm \nu \sqrt{k_x^2 + k_y^2 + k_z^2 } = \pm \lambda,
\end{equation}
and the corresponding eigenstate of the valence   band ($-\lambda$) is 
\begin{equation}
	\label{eq:vecH1}
	\ket{v}_{W_{1}} = \frac{1}{\sqrt{2\lambda^2 + 2\lambda k_z}}\begin{pmatrix}
		k_z + \lambda\\
		k_x - i k_y
	\end{pmatrix},
\end{equation}	
and  the corresponding eigenstate of the conduction band ($+\lambda$)  is
\begin{equation}		
	\ket{c}_{W_{1}}  = \frac{1}{\sqrt{2\lambda^2 - 2\lambda k_z}}\begin{pmatrix}
		k_z - \lambda\\
		k_x - i k_y
	\end{pmatrix}.
\end{equation}
Similarly, the other Hamiltonian  near the second  Weyl point  $W_2$ at $k_z = -\pi/2a$ is  
\begin{equation}
\mathcal{H}_{W_{2}}  = t[(-k_x a) \sigma_x + (k_y a) \sigma_y + (k_z a) \sigma_z].
\end{equation}
The eigenvalues corresponding to $\mathcal{H}_{W_{2}}$  are 
\begin{equation}
\mathcal{E}^{W_{2}}_{v/c} = \pm \nu \sqrt{k_x^2 + k_y^2 + k_z^2 } = \pm \lambda,
\end{equation}
and the associated eigenstates for valence  ($-\lambda$) and conduction ($\lambda$)  bands as 
\begin{equation}
\label{eq:vecH2}
\ket{v}_{W_{2}} = \frac{1}{\sqrt{2\lambda^2 - 2\lambda k_z}}\begin{pmatrix}
		-k_z + \lambda\\
		k_x - i k_y
	\end{pmatrix},~~\textrm{and}~~	\ket{c}_{W_{2}} = \frac{1}{\sqrt{2\lambda^2 + 2\lambda k_z}}\begin{pmatrix}
		-k_z - \lambda\\
		k_x - i k_y
	\end{pmatrix}.
\end{equation}
As one can see $\mathcal{H}_{W_{1}}$ and $\mathcal{H}_{W_{2}}$ differ by sign of $k_z$. 
Thus, the eigenstates also have a different sign of $k_z$. The mirror symmetry of the Weyl Hamiltonian also characterizes embedded chirality. As with the enantiomers, Weyl cones are superimposable if one of them is flipped about $k_z$.

\section{Coupling of Weyl Hamiltonian  with Circularly Polarized Light}

Interaction between WSM and laser light can be expressed in terms of an interaction Hamiltonian as $\mathcal{H}^{\prime}{(t)} = \mathbf{d}_{nn^\prime}\cdot\mathbf{E}(t)$, where $\mathbf{d}_{nn^\prime}$ is the dipole matrix element 
between electronic states $n$ and $n^\prime$,  and $\mathbf{E}(t)$ is the electric field associated with laser. 
The electric field within dipole approximation is $\mathbf{E}(t) = E_0 e^{i \omega t}~\hat{\textbf{e}}$ with 
$E_{0}$ is an amplitude, $\omega$ is the frequency and 
$\hat{\textbf{e}}$ as the polarization direction. 
Thus, the interaction term for circularly polarized light (CPL) with left ($\circlearrowleft$)/right ($\circlearrowright$) helicity becomes 
\begin{equation} \label{eq:interterm}
\mathbf{d}_{nn^\prime}\cdot\mathbf{E}(t)= \mathbf{d}_{nn^\prime}\cdot E_0 e^{i \omega t}~\hat{\textbf{e}}_{\circlearrowleft/\circlearrowright}\nonumber = i \matrixelement{n}{\pdv{}{\mathbf{R}}}{n^\prime} \cdot E_0 e^{i \omega t}~\hat{\textbf{e}}_{\circlearrowleft/\circlearrowright}.
\end{equation}
The above equation exhibits mirror symmetry in the helicity of light. 
In general, the left and right helicities of the CPL interact identically. 
However, two helicities of the light behave in equal and opposite manner in the case of chiral entities. 
Thus, selective response captures the chiral nature.
As WSM encompasses at least a pair of chiral Weyl nodes, it is interesting to see how they respond to CPL. 
One part of the Weyl cone can be selectively excited depending on the chirality of the said node, as evident from Fig.~\ref{fig:selectionrule}. 
In order to understand it, the detailed calculations of the dipole matrix element and how the symmetry of it explains the observed phenomenon are given in Appendix A. 

\begin{figure}[!h]
\centering
\includegraphics[width= 0.7 \linewidth]{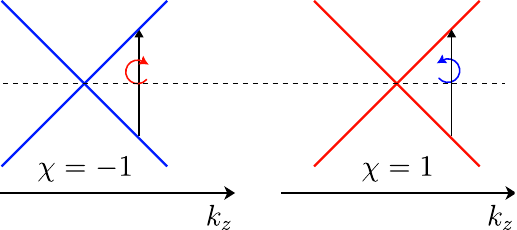}
\caption{Selective excitation of the Weyl cones by circularly polarized light. 
The left helicity of light excites the Weyl cone with chirality $\chi=-1$ in the $+k_z$ direction, and the same side of the Weyl cone with chirality $\chi=1$ can be excited by the right helicity of light.}
\label{fig:selectionrule}
\end{figure}

\section{Anomalous Hall effect}
Berry curvature corresponding to a single Weyl node with chirality $\chi$,  located at $\k =\mathbf{b}$, is written as 
\begin{equation} \label{eq:singlebc}
\Omega_\chi(\k) = \chi \frac{\k-\mathbf{b}}{2|\k-\mathbf{b}|^3}. 
\end{equation}
The contribution to the total current due to Berry curvature is termed as anomalous current:
\begin{equation}
\mathbf{J}_\Omega \propto  \int_{\k} d{\k} ( \mathbf{E} \times \mathbf{\Omega}).
\end{equation} 

\begin{figure}[!h]
\centering
\includegraphics[width= 0.8\linewidth]{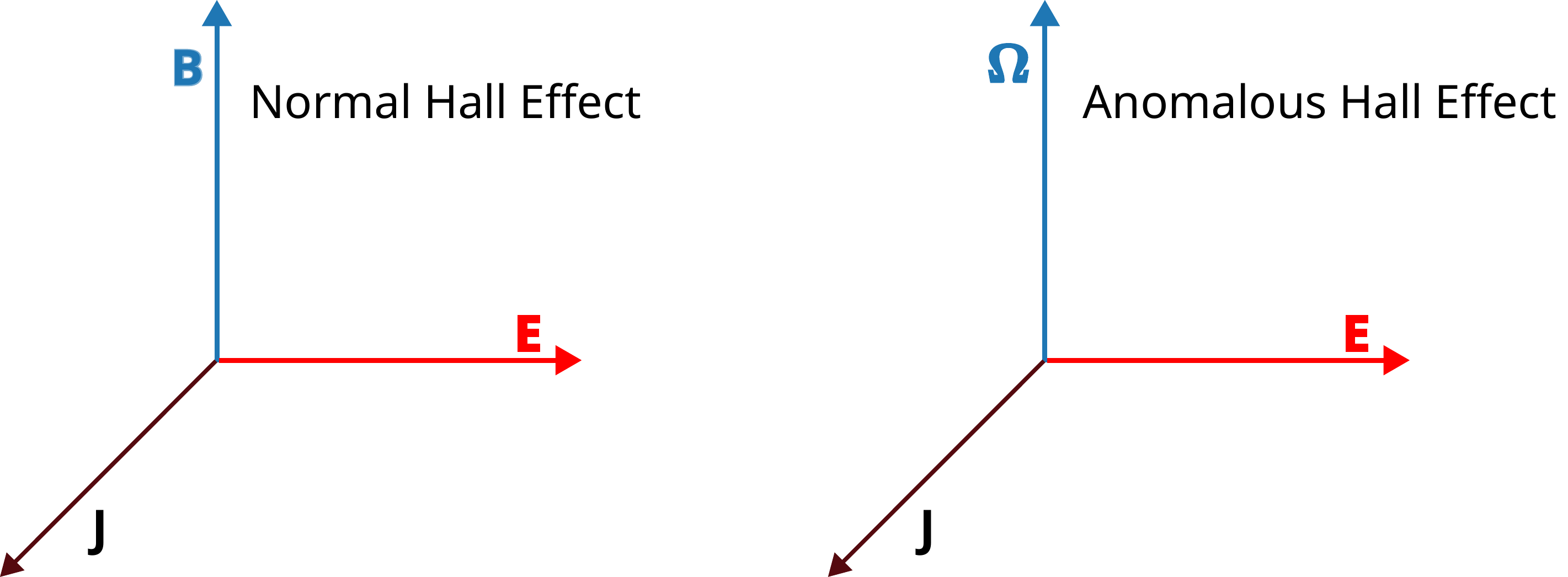}
\caption{Left panel: Orthogonal electric and magnetic fields result in perpendicular electric current in normal Hall effect. 
Right panel: the role of the magnetic field is replaced by  $\Omega$  (``pseudo-magnetic field'') in the anomalous Hall effect.} 
\label{fig:ahe}
\end{figure}

We can estimate anomalous current stemming from each Weyl node using Eq.~\eqref{eq:singlebc} by evaluating the following integral~\citep{rostami2018nonlinear}
\begin{eqnarray}
\int  d\k ~~\mathbf{\Omega}& = &\int  d\k \frac{\k - \mathbf{b}}{|\k - \mathbf{b}|^3}\nonumber\\
	&=& \int_0^\infty  d{k}~k^2 \int_{-1}^1 d({\cos(\theta)}) \int_0^{2\pi} d{\phi} \frac{(k \cos\theta-b)\hat{\textbf{b}} + k \sin \theta \hat{\mathbf{\rho}}}{(k^2 + b^2 - 2kb \cos \theta)^{3/2}} \nonumber \\
	&=& 2\pi \hat{\mathbf{b}} \int_0^\infty  d{k}~k^2 \int_{-1}^1 d{x} \frac{kx-b}{(k^2 + b^2 -2kbx)^{3/2}}\nonumber\\
	&=& -2\pi \hat{\mathbf{b}}  \frac{1}{b^2}\int_0^\infty  d{k}~k^2 \{1+ \sgn[b-k] \}\nonumber\\
	&=& -4\pi \hat{\mathbf{b}}  \frac{1}{b^2} \int_0^b  d{k}~k^2   = -\frac{4\pi}{3} \mathbf{b}. 
\end{eqnarray} 
In the above simplification, we have used $x= \cos(\theta)$. 
Thus, anomalous current in WSM is proportional to $\chi (\mathbf{E}\times \mathbf{b})$ with $\mathbf{b}$ as the separation between the Weyl nodes.

\section{Berry Curvature for Tight-Binding Model}
We start our discussion by writing the  general expression for the Berry curvature as
\begin{equation}\label{eq:berry}
	\Omega(\mathbf{k})_{\mathbf{k}, \pm, i} = \pm \epsilon_{ijl} \frac{\mathbf{d}(\mathbf{k})_\mathbf{\alpha} \cdot \left( \frac{\partial{\mathbf{d}(\mathbf{k})_{\mathbf{\alpha}}}}{\partial{k_j}} \times  \frac{\partial{\mathbf{d}(\mathbf{k})_{\mathbf{\alpha}}}}{\partial{k_l}}  \right)} 
	{4|\mathbf{d}(\mathbf{k})_{\mathbf{\alpha}}|^3},
\end{equation}
where $+ (-)$ corresponds to conduction (valence) bands,  $\epsilon_{ijl}$  
is the Levi-Civita tensor and 
$\mathbf{d}(\mathbf{k})_\mathbf{\alpha = 1, 2, 3}$ are the components of 
$\mathbf{d}(\mathbf{k})$ in $\mathcal{H}_{\textrm{Weyl}}(\mathbf{k})$.
Let us calculate the Berry curvature corresponding to the Hamiltonian  given by 
Eq~\eqref{eq:weylhamtb}. 
The components of $\mathbf{d}(\mathbf{k})_\mathbf{\alpha}$ 
as given in  Eq.~\eqref{eq:weylhamtb}   are
\begin{equation}\label{eq:dk} 
\mathbf{d}(\mathbf{k}) =   t\big[\cos(k_x a), \cos(k_y a), \{\cos(k_z a) -  \alpha(1 + \sin({k_x a}) \sin({k_y b}))\}\big]. 
\end{equation}
After substituting the expression from Eq.~(\ref{eq:dk}) to  Eq.~(\ref{eq:berry}), 
the components of the Berry curvature can be written as
\begin{eqnarray}
	\Omega(\k)_{k_x} & = &a^2t^3 \cos(a k_x)\sin(ak_y)\sin(ak_z)/\mathcal{N},\\
	\Omega(\k)_{k_y} & = &a^2t^3\cos(ak_y)\sin(a k_x)\sin(ak_z)/\mathcal{N},~~\textrm{and}\\
	\Omega(\k)_{k_z} & = & -a^2t^3 \big[\alpha \cos^2(ak_y)\sin^2(ak_x) + \alpha \cos^2(ak_x)\sin^2(ak_y) \nonumber \\
	&&- \sin(ak_x)\sin(ak_y) \{\cos(ak_z) -\alpha(1+\sin ak_x\sin ak_y)\}\big]/\mathcal{N}. 
\end{eqnarray}
Here, $\mathcal{N}= 4 |\mathbf{d}|^3$ is the normalization factor.
As evident from the above equations,  $\Omega(\k)_{k_{x}/k_{y}}$ is an odd functions of $k_z, k_y$ and $k_x$. 
This implies that the  anomalous current $\mathbf{J}_\Omega \propto  \int_{\k}  \{\mathbf{E}\times \mathbf{\Omega}\} d{\k} $ yields a non-zero contribution only from $\Omega(\k)_{k_z}$. 
Thus, a linearly polarized laser pulse along the $x$ or $y$ direction generates an anomalous Hall effect in WSM.

\section{Semiconductor Bloch Equations}
To describe the interaction of WSM with an intense laser, let us write an 
equation of motion for density matrix  $\rho$ as
\begin{equation}
\dot{\rho} = -i \left[\mathcal{H}(t), \rho\right].
\end{equation}
The equation of motion has an explicit dependence on $\k$ 
in  the case of solids.
In addition, to account for the collisions and other dissipative process during the interaction, 
the above equation modifies as~\citep{manzano2020short}
\begin{equation} \label{eq:eom3}
\dot{\rho}(\k) = -i \left[\mathcal{H}(\k, t), \rho(\k)\right] + \dv{\rho}{t}\bigg|_{\textrm{col}}.
\end{equation}
Here, $\mathcal{H}(\k, t)$ contains field-free Hamiltonian for WSM and 
$\mathcal{H}^{\prime}(t)$ for the interaction between WSM and an intense laser. 
For field-free Hamiltonian, we have $\mathcal{H}_{\k} \ket{n \k} = \mathcal{E}(\k) \ket{n\k}$ with 
$\ket{n \k}$ as the Bloch basis. 
Matrix element and density matrix elements between $\ket{n \k}$ and $\ket{n^{\prime} \k}$ are estimated as 
$\mathcal{H}_{nn^\prime}= \matrixel{n \k}{\mathcal{H}(\k, t)}{n^\prime\k}$ and $\rho_{nn^\prime} = \matrixel{n\k}{\rho}{n^\prime\k}$, respectively. 

It is known that the crystal momentum $\k$ changes to $\k_t = \k+\mathbf{A}(t)$  in the presence of a laser, where $\mathbf{A}(t)$ is the vector potential. 
In this respect, it is convenient to work in Houston basis $\ket{n\k_t}$. 
Thus, density matrix in the Houston basis becomes $\rho_{nn^\prime}=\matrixel{n\k_t}{\rho(\k, t)}{n^\prime\k_t}$. 
In order to simply the equation of motion in Eq.~(\ref{eq:eom3}), let us project it by ($\langle {n\k_t}|$) and ($| {n^{\prime} \k_t} \rangle$) from left and right as 
\begin{eqnarray}
i\matrixel{n\k_t}{\partial_t \rho(\k, t)}{n^\prime\k_t}& =& \matrixel{n\k_t}{\left[\mathcal{H}(\k_t),\rho(\k, t)\right]}{n^\prime\k_t}\\\nonumber
	&=&\mathcal{E}_{nn^\prime} \matrixel{n\k_t}{\rho(\k, t)}{n^\prime\k_t}=\mathcal{E}_{nn^\prime} \rho_{nn^\prime}. \nonumber
\end{eqnarray}
First we will focus on the left-hand side term as
\begin{eqnarray} \label{eq:leftterm}
\matrixel{n\k_t}{\partial_t \rho(\k, t)}{n^\prime\k_t} &=& \partial_t\left(\matrixel{n\k_t}{\rho(\k, t)}{n^\prime\k_t} \right)\nonumber \\
	&&- (\partial_t \bra{n \k_t})\rho(\k, t)\ket{n^\prime\k_t} \nonumber\\
	&&-\bra{n\k_t}\rho(\k, t)(\partial_t \ket{n^\prime \k_t}).
\end{eqnarray} 
We know that the identity relation $\sum_n \dyad{n\k_t} = \mathbbm{1}$, 
which follows $\partial_t\ket{n\k_t}=-\partial_t\bra{n\k_t}$. 
Further, we find that 
\begin{equation}\label{eq:vec}
\partial_t \ket{n\k_t} = - \dot{\mathbf{A}}(t) \partial_{\k_t} \ket{n\k_t} = \mathbf{E}(t)\partial_{\k_t}\ket{n\k_t}.
\end{equation}

On substituting the findings of Eqs.~\eqref{eq:leftterm} and ~\eqref{eq:vec}, 
it is straightforward to show that
\begin{eqnarray} \label{eq:lasteq}
\matrixel{n\k_t}{\partial_t \rho(\k, t)}{n^\prime\k_t} &=& \partial_t\left(\matrixel{n\k_t}{\rho(\k, t)}{n^\prime\k_t} \right) \\
	&&+ \mathbf{E}(t)\sum_{n^*} \bra{n\k_t}(\partial_{\k_t}\ket{n^*\k_t}) \bra{n^*\k_t} \rho(\k, t)\ket{n^\prime k_t}\nonumber\\
	&&-\mathbf{E}(t)\sum_{n^*} \matrixel{n\k_t}{\rho(\k, t)}{n^*\k_t}\bra{n^*\k_t}(\partial_{\k_t}\ket{n^\prime k_t}).\nonumber
\end{eqnarray}
In the above equation, $\mathbf{d}_{nn^\prime}^{\k_t} = \matrixel{n\k_t}{i\partial_{\k_t}}{n^\prime\k_t}$ is 
identified as the dipole matrix element. 
Thus, we have
\begin{eqnarray}
\matrixel{n\k_t}{\partial_t \rho(\k, t)}{n^\prime\k_t} &=& \partial_t\left(\matrixel{n\k_t}{\rho(\k, t)}{n^\prime\k_t} \right) \\
&&+ \mathbf{E}(t) \left[ \sum_{n^*}\matrixel{n\k_t}{\rho(\k, t)}{n^*\k_t} \mathbf{d}_{n^*n^\prime}^{\k_t} - \mathbf{d}_{nn^*}^{\k_t}\matrixel{n^*\k_t}{\rho(\k, t)}{n^\prime\k_t} \right].\nonumber
\end{eqnarray}

On using the above relation, the equation of motion is simplified as 
\begin{eqnarray}
\left(i\partial_t - \mathcal{E}_{nn^\prime}^{\k_t} \right) \matrixel{n\k_t}{\rho(\k,  t)}{n^\prime\k_t} &= & \mathbf{E}(t)\left[ \sum_{n^*} \matrixel{n\k_t}{\rho(\k, t)}{n^*\k_t} \mathbf{d}_{n^*n^\prime}^{\k_t} -\right.\nonumber\\
	&&\quad\quad\left.\sum_{n^*}\mathbf{d}_{nn^*}^{\k_t}\matrixel{n^*\k_t}{\rho(\k, t)}{n^\prime\k_t}\right].	
\end{eqnarray}
The above equation in compact form is expressed as 
\begin{equation} \label{eq:sbe}
\left(i\partial_t - \mathcal{E}_{nn^\prime}^{\k_t}\right)\rho_{nn^\prime}^{\k}(t)  =  \mathbf{E}(t)\cdot  \sum_{n^*}
\left[ \rho_{nn^*}^{\k}(t) \mathbf{d}_{n^*n^\prime}^{\k_t} - \mathbf{d}_{nn^*}^{\k_t}\rho_{n^*n^\prime}^{\k}(t) \right]. 
\end{equation}
The above equation is known as semiconductor Bloch equations (SBEs). 
Considering the complex nature of different dissipation processes (dephasing), rigorous treatment of them is cumbersome. Hence, we include a wide variety of dissipating processes using a phenomenological 
term as $\frac{(1-\delta_{nn^\prime})}{\textrm{T}_2}\rho_{nn^\prime}^{\k}$ with $\textrm{T}_2$ as the dephasing time. 

The total current at any $\k$-point is calculated using current operator $\hat{\mathbf{j}}=-\hat{\mathbf{p}}$ as
\begin{equation} \label{eq:totcur}
 \mathbf{J}(\k, t) = \Tr[\hat{\mathbf{j}}~\hat{\rho}] = \sum_{nn^{\prime}} \mathbf{p}_{nn^{\prime}}^{\k_t}~\rho^{\k}_{n^{\prime}n}(t),
\end{equation}
where $\mathbf{p}_{nn^{\prime}}^{\k_t} = \matrixel{n\k}{\nabla_\k \mathcal{H}(\k, t)}{n^{\prime}\k}$ is momentum matrix element. 
Total current can be decomposed into intraband and interband contributions, respectively, as
\begin{eqnarray} \label{eq:currcont}
 \mathbf{J}(\k, t) &= &\sum_{nn^{\prime}} \mathbf{p}_{nn^{\prime}}^{\k_t}~\rho^{\k}_{n^{\prime}n}(t) \nonumber\\
	&=& \sum_{n = n^{\prime}} \mathbf{p}_{nn^{\prime}}^{\k_t} \rho^{\k}_{n^{\prime}n}(t) + \sum_{n \neq n^{\prime}} \mathbf{p}^{\k_t}_{nn^{\prime}} \rho^{\k}_{n^{\prime}n}(t).
\end{eqnarray}
The total current is the integration of Eq.~\eqref{eq:totcur} over the entire Brillouin zone. Similarly, integration of the first term in Eq.~\eqref{eq:currcont} gives intraband current solely due to band dispersion, and the integration of the second term gives interband current contribution to the total current. 
  
The Berry connection  between conduction and valence bands is evaluated  as
$\mathcal{A}_{cv}(\mathbf{k}) = -i \braket{\nabla_{\mathbf{k}} u_{c}({\mathbf{k}})}{u_{v}({\mathbf{k}})} = \mathcal{A}_{vc}^{*}(\mathbf{k})$, which  can be visualized as the off-diagonal dipole matrix elements~\citep{nematollahi2020topological,ngo2021microscopic}.
Thus, the  Berry curvature between conduction and valence bands is written as 
$\mathbf{\Omega}_{cv}(\mathbf{k}) = i (\mathcal{A}_{vc} (\mathbf{k}) \times \mathcal{A}_{cv}(\mathbf{k}))$. 
The current due to the Berry curvature is evaluated as  $\mathbf{J}_\mathbf{\Omega}(\mathbf{k},t) =  \sum_{cc,vv} \mathbf{E}(t) \times \mathbf{\Omega}_{cv}(\mathbf{k})~\rho_{cc,vv}(\mathbf{k}, t)$ 	
and is commonly known as the anomalous current.
Once we have the total current, we can simulate the total harmonic spectra using Fourier transform as
\begin{equation} \label{eq:hhg}
\mathcal{I}(\omega) = \left|\mathcal{FT} \left( \frac{d}{dt}\mathbf{J}(t)\right) \right|^2.
\end{equation}

\cleardoublepage
\chapter{How Massless are Weyl fermions in Weyl Semimetals}
\begin{sloppypar}
Light-driven optical response
has played a pivotal role in understanding and probing 
exotic properties of WSMs~\citep{sirica2019tracking, ma2017direct,ma2019nonlinear, lv2021experimental, lv2021high, orenstein2021topology, bharti2022high}.
One such optical response is circularly polarized light-driven selective excitations in the vicinity of the  Weyl nodes, as shown in Fig.~\ref{fig:selectionrule}. 
The excitation process depends on  
the chirality of the Weyl fermions and  the helicity 
of CPL~\citep{yu2016determining}.
Helicity-driven selective excitations in  inversion-broken WSMs lead
to population asymmetry around the  Weyl nodes and  the circular photogalvanic effect:
the generation of current upon irradiation with circular light~\citep{konig2017photogalvanic,chan2017photocurrents,ma2017direct}. 
Broken inversion symmetry in WSMs is a prerequisite to ensure noncancellation of the contribution from a pair of chiral Weyl nodes.
Thus, when a measurement of coupling between the massless fermions and circularly polarized light is 
integrated over both nodes, the nonzero result arises only in the inversion-broken WSMs~\citep{ma2017direct}.  

Since this conclusion assumes perfectly massless Weyl fermions, 
i.e., a gapless system with a perfectly linear dispersion near the nodes, it welcomes a question: how 
quickly is this assumption violated as one moves away from the exact location of the node?
Note that deviations from linear dispersion imply that even for gapless nodes, the mass becomes nonzero as soon as one moves away from the degenerate point.
Can circularly polarized light with opposite helicity generate non-mirror-symmetric  excitations in inversion-symmetric WSMs 
once the nonlinearity of the band structure is taken into account, even near the Weyl nodes?
We show that the answer to the latter question is positive in this chapter.

We begin with the perfectly massless Weyl fermions, where the population induced around the chiral Weyl 
node with $\chi =1$ by right circularly polarized  light is superimposable with that of $\chi = -1$ induced 
by the left circularly polarized light and vice versa. 
These populations provide the reference for the more general case of nonlinear band dispersion. 
Once quadratic corrections to the Weyl equation 
are included, helicity-sensitive asymmetric excitations become nonzero and significant already at the Weyl nodes. 
That is, the excitation generated with one helicity at the $\chi=1$  node is no longer superimposable with that 
generated by the opposite helicity at the $\chi=-1$ node, and the excitations at a given node for light with opposite helicities are not mirror symmetric. 
The same result is obtained for the more general  
inversion-symmetric Hamiltonian of a WSM.
While the induced asymmetry reduces when decreasing the light frequency, so that
the resonant excitations are located very close to the node, it still remains substantial.  
Last but not least, we devise a scheme based on two-color counterrotating CPL to 
control the helicity-sensitive asymmetric excitation. 
Our control scheme can tailor the asymmetry from positive to zero to negative.

\section{Theoretical Methodology}
We start our discussion by writing the components of $\mathbf{d}(\mathbf{k})$ corresponding to the Hamiltonian given in Eq.~(\ref{eq:Weyl}) for a type-I WSM as~\citep{hou2016weyl}
%\begin{equation}\label{eq:eqham}
%\mathcal{H}(\mathbf{k}) = 2t_{x}\cos(k_{x} a)\sigma_{x} + 2t_{y}\cos(k_{y} a)\sigma_{y} 
%+ 2t_{z}\left[\cos(k_{z} a) - \alpha - \beta \sin(k_{x} a)\sin(k_{y} a) \right] \sigma_{z},
%\end{equation} 
\begin{equation}\label{eq:eqham}
\mathbf{d}(\mathbf{k})  = [2t_{x}\cos(k_{x} a),  2t_{y}\cos(k_{y} a), 
 2t_{z}\left \{\cos(k_{z} a) - \alpha - \beta \sin(k_{x} a)\sin(k_{y} a) \right\} ],
\end{equation} 
where $t$'s are hopping parameters, and  $\alpha$ and $\beta$ are dimensionless parameters. 
%The Hamiltonian corresponds to an inversion-symmetric WSM with broken time-reversal symmetry~\citep{hou2016weyl}. 
To make our discussion simple, we have considered 
$t_{x, y, z} = t$ and $\alpha = \beta$. 
%Diagonalization of Eq.~(\ref{eq:eqham}) yields the band structure shown in Fig.~\ref{fig:fig13}(a). 
The two Weyl nodes are positioned at 
$W_{1, 2} = (\frac{\pi}{2a}, -\frac{\pi}{2a}, \pm \frac{\pi}{2a})$,  i.e., $(0.5,0,\pm0.25)$ in reduced coordinates,  and are at the Fermi level as shown in Fig.~\ref{fig:fig13}(a).
The energy contours in their vicinity in the $k_{x} - k_{y}$ plane are isotropic [see Fig.~\ref{fig:fig13}(b)], so that
light-induced excitation should yield a symmetric population. 

\begin{figure}[h!]
\centering
\includegraphics[width=\linewidth]{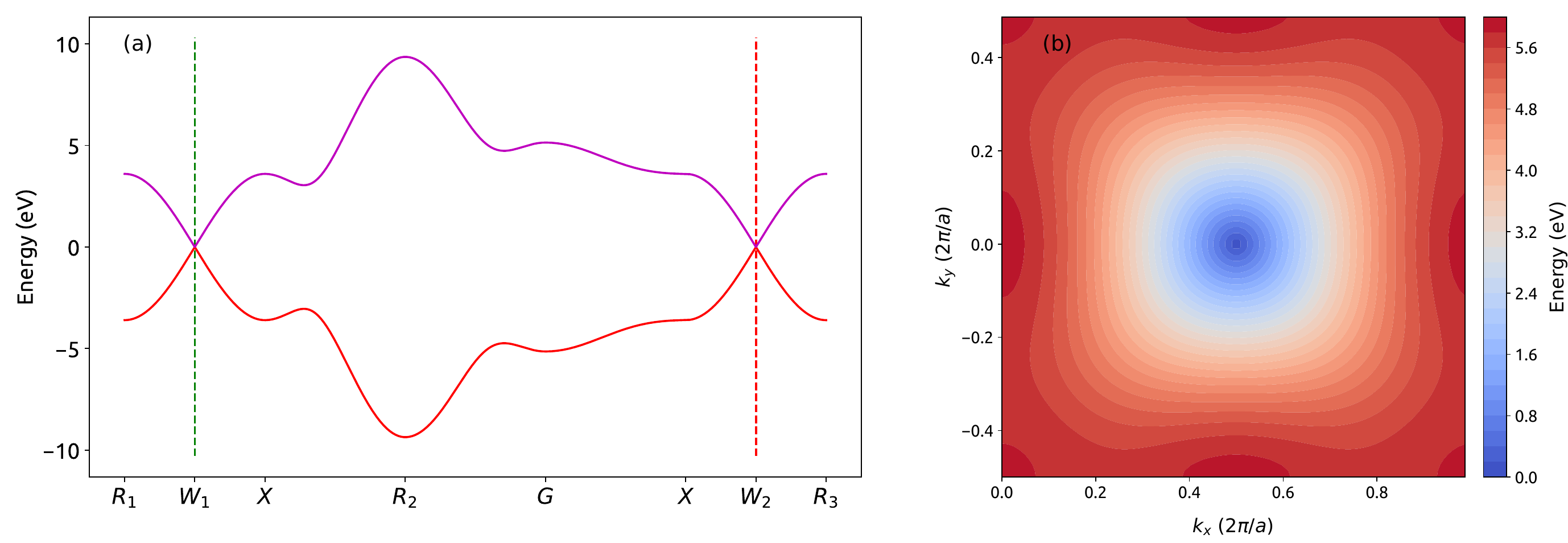}
\caption{(a) Energy dispersion along high-symmetry points of an inversion-symmetric 
Weyl semimetal as given in Eq.~(\ref{eq:eqham}). (b) Energy contour around one of the Weyl nodes in $k_x-k_y$ plane (Weyl planes). 	The hopping parameter  is $t =1.8$ eV, and the lattice parameters 
are  $a = 6.28~\mathring{\text{A}}$  \ and $\beta = 0.8$.
The lattice vectors are $a_1 = (a,-a,0)$,  $a_2 = (a,a,0)$,  $a_3 =(0,0,a)$, and the
reciprocal vectors are $b_1= ( \pi/a, -\pi/a, 0)$,  $b_2= ( \pi/a, \pi/a, 0)$, $b_3= ( 0, 0 , 2\pi/a)$, leading 
	to reduced coordinates for the high-symmetry points as follows: 
	$R_1 (\pi/2a,-\pi/2a,-\pi/a)$, $X (\pi/2a,-\pi/2a,0)$, $R_2 (0,\pi/a,-\pi/a)$, $G (0,0,0)$, and  $R_3 (\pi/2a,-\pi/2a,\pi/a)$. } \label{fig:fig13}
\end{figure}	
Let us first focus on the linear part of the band dispersion. Expanding the terms of Eq.~(\ref{eq:eqham}) up to linear terms 
near the Weyl nodes, Hamiltonians near the Weyl nodes can be written as
\begin{subequations}\label{eq:Eqlin1}
\begin{align}
	\mathcal{H}_{1}(\mathbf{k}) & = d_{1, x}(\mathbf{k})\sigma_{x} + d_{1, y}(\mathbf{k})\sigma_{y} + d_{1, z}(\mathbf{k})\sigma_{z}, \\
	\mathcal{H}_{2}(\mathbf{k}) & = d_{2, x}(\mathbf{k})\sigma_{x} + d_{2, y}(\mathbf{k})\sigma_{y} + d_{2, z}(\mathbf{k})\sigma_{z}. 
\end{align}
\end{subequations}
Here, $\mathbf{k}$ denotes the deviation from the  Weyl node [for both nodes, Eqs.~(\ref{eq:Eqlin1}a) and (\ref{eq:Eqlin1}b)], 
$d_{1(2), x}(\mathbf{k}) = v \left(-k_{x} a\right), d_{1(2), y}(\mathbf{k}) = v \left(k_{y} a \right)$, and 
$d_{1(2), z}(\mathbf{k}) = v \left[-(+) \tilde{k}_{z} a \right]$, where  
$-\tilde{k}_{z}(+\tilde{k}_{z})$  is measured relative to the Weyl node 1 (2), and $v = 2t$.  
The above  Hamiltonian in Eq.~(\ref{eq:Eqlin1}) 
represents  the Weyl equation and can be written as $\mathcal{H}_{\textrm{Weyl}} = v  ~\mathbf{k} \cdot  \bm{\sigma}$. 
As pointed above, the two Weyl 
nodes described by $\mathcal{H}_{1}(\mathbf{k})$ and $\mathcal{H}_{2}(\mathbf{k})$ are degenerate and only differ by chirality, 
which is defined   as $\chi = \textrm{sgn}(d_{x}\cdot d_{y} \times d_{z}$).
The Weyl nodes 1 and 2  have $\chi= 1$ and -1, respectively.
	
Light-driven electronic excitation is simulated using the semiconductor Bloch equations framework as discussed in the previous chapter. 
To account for the decoherence between electron and hole during the excitation process, 
a phenomenological dephasing term with 1.5 fs is introduced.  Our findings are robust against the 
dephasing term ranging from 1.5 to 10 fs.
The conduction band population  is obtained by integrating 
the density matrix in the conduction band after  the end of the laser pulse; 
the population is integrated over $k_x$ and $k_y$, and is shown along the $\tilde{k}_z$ direction, where 
$\tilde{k}_z = 0$ is the Weyl plane which contains both chiral Weyl nodes, for Eq. \eqref{eq:Eqlin1}.  
We used  100 fs long circularly polarized pulses with intensity  10$^{11}$ W/cm$^2$ and  
wavelength 3.2 $\mu$m (i.e., $\omega$ = 0.39 eV); different wavelengths upto 10.6 $\mu$m 
(i.e., $\omega$ = 0.12 eV) were also studied, with the results described below.

\section{Results and Discussion}
\begin{figure}[h!]
	\includegraphics[width=\linewidth]{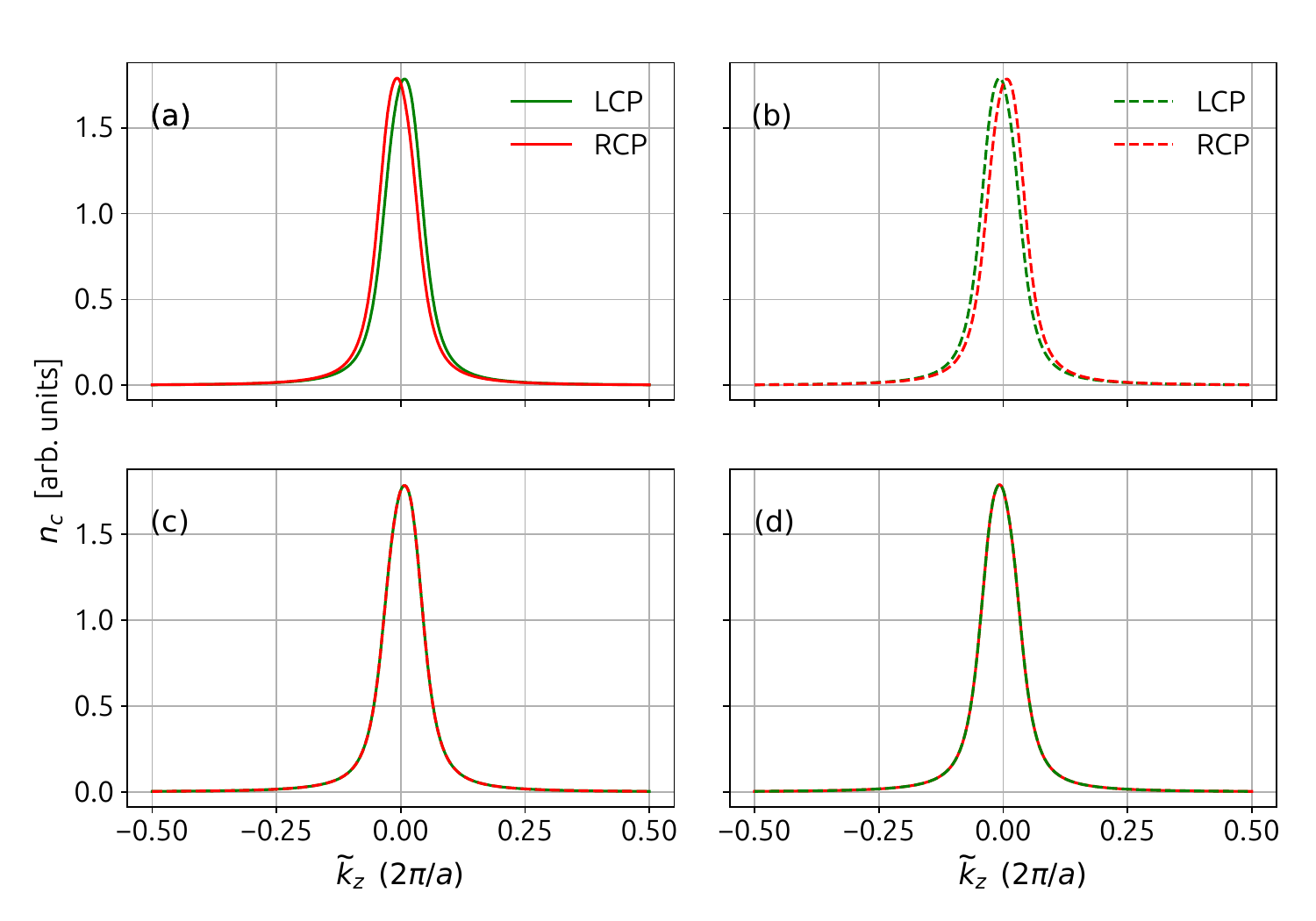}
	\caption{Residual population in the conduction band ($n_{c}$) after the end of the left-handed circularly polarized (LCP)  and right-handed circularly polarized (RCP) 
		light around a Weyl node with (a)  $\chi = -1$  and (b) $\chi = 1$.  
		(c) Comparison of the residual populations from a Weyl node with $\chi = -1$  due to LCP, and from a Weyl node with $\chi = 1$  due to RCP. (d) Same as (c) for a Weyl node with $\chi = -1$  due to RCP, and from a Weyl node with $\chi = 1$  due to LCP.  The Weyl nodes 
		with  $\chi = -1$  and $\chi = 1$ are described by  Eq.~(\ref{eq:Eqlin1}). }\label{fig:fig23}
\end{figure}

Figure~\ref{fig:fig23} shows  
the residual  population around the two Weyl nodes in the conduction band after the end of the pulse 
for  $\chi = -1$ and $\chi = 1$ calculated for Hamiltonians in Eqs.~(\ref{eq:Eqlin1}a) and (\ref{eq:Eqlin1}b), respectively.    
As expected, the population asymmetry is zero at $\tilde{k}_z = 0$ and is mirror symmetric with respect to 
changing either 
the light helicity or the chirality of the node. In particular, the population at $\chi = -1$ induced by 
the left circularly polarized (LCP) pulse is the same as that induced at $\chi=+1$ by 
the right circularly  polarized (RCP) pulse [see Fig.~\ref{fig:fig23}(c)]. 
The same is true for the population induced by the RCP at  $\chi = -1$  compared to the population induced by  LCP near  $\chi = 1$ [see Fig.~\ref{fig:fig23}(d)]. 
	
Having established this reference, 
we now go beyond the linear approximation and 
expand Eq.~(\ref{eq:eqham}) to the second order, resulting in the following expression,
\begin{subequations}\label{eq:Eqquad1}
\begin{align}
\tilde{\mathcal{H}}_{1}(\mathbf{k}) & = d_{1, x}\sigma_{x} + d_{1, y}\sigma_{y} +  \tilde{d}_{1, z}\sigma_{z},  \\
\tilde{\mathcal{H}}_{2}(\mathbf{k}) & = d_{2, x}\sigma_{x} + d_{2, y}\sigma_{y} +  \tilde{d}_{2, z}\sigma_{z},  
\end{align}
\end{subequations}
where  $\tilde{d}_{1(2), z}(\mathbf{k}) = v \left[-(+) \tilde{k}_{z} a \right] - \tilde{v} \left[\frac{(k_{x} a)^2 + (k_{y} a)^2}{2} \right]$ with $\tilde{v} = 2t \alpha$. 
The $\tilde{d}_{z}$ component now contains additional terms quadratic in $k_{x}$ and  $k_{y}$, whereas  $d_{x}$ and $d_{y}$ remain identical in both the equations.  

\begin{figure}[h!]
	\includegraphics[width=\linewidth]{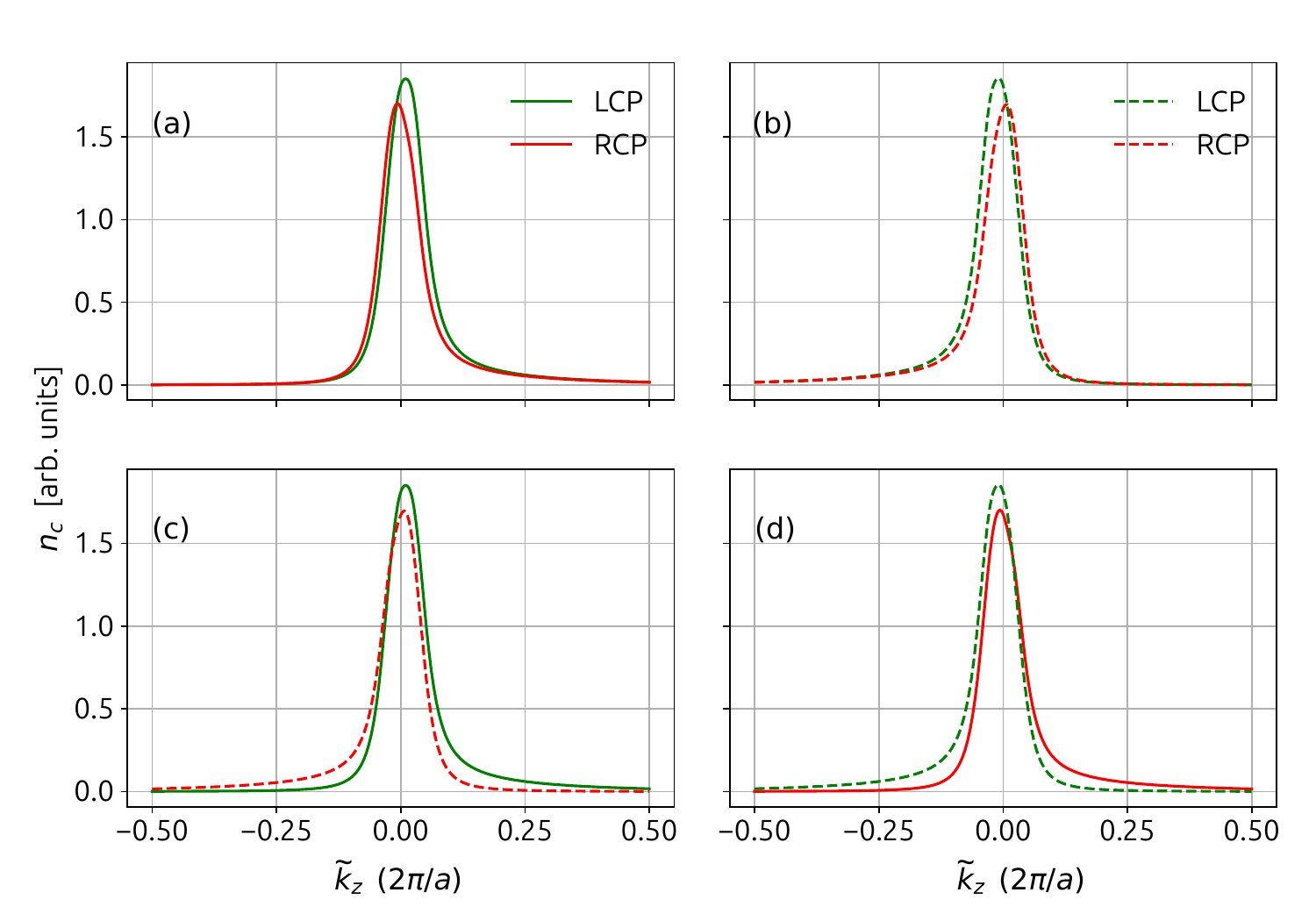}
	\caption{Same as in Fig.~\ref{fig:fig23} for the Weyl nodes with $\chi = -1$  and $\chi = 1$, but now 
		using the  Hamiltonian Eq.~(\ref{eq:Eqquad1}) which includes the quadratic terms.}\label{fig:fig33}
\end{figure}

The quadratic terms affect the final population already in the immediate vicinity of the 
Weyl nodes 
as visible from 
Figs.~\ref{fig:fig33}(a) and ~\ref{fig:fig33}(b). 
The mirror symmetry upon changing the handedness of the Weyl node is, of course, preserved: the 
population near $\chi=-1$ is mirror symmetric with that near $\chi=+1$ with respect to
changing $\tilde{k}_z \rightarrow - \tilde{k}_z$. However, for a given Weyl node, the peaks of the populations induced by RCP and LCP pulses 
do not coincide. Similarly, the excitation induced near the $\chi=-1$ node by LCP pulse does not overlap
with the excitation induced near the $\chi=+1$ node by RCP pulse [see Fig.~\ref{fig:fig33}(c)]. Likewise, 
the excitation induced near the $\chi=+1$ node by the LCP pulse does not overlap
with the excitation induced near the $\chi=-1$ node by RCP pulse  [see Fig.~\ref{fig:fig33}(d)].
This stands in stark contrast with Fig.~\ref{fig:fig23}.
The fact that this asymmetry, associated with
the deviations from the linear dispersion, arises in the immediate vicinity of the nodes, 
i.e., in what is supposed to be the zero-mass region, raises the question posed in the title of this chapter: 
How massless are the Weyl fermions under practical conditions of typical laser wavelengths and intensities?  
\begin{figure}[h!]
\centering
\includegraphics[width=\linewidth]{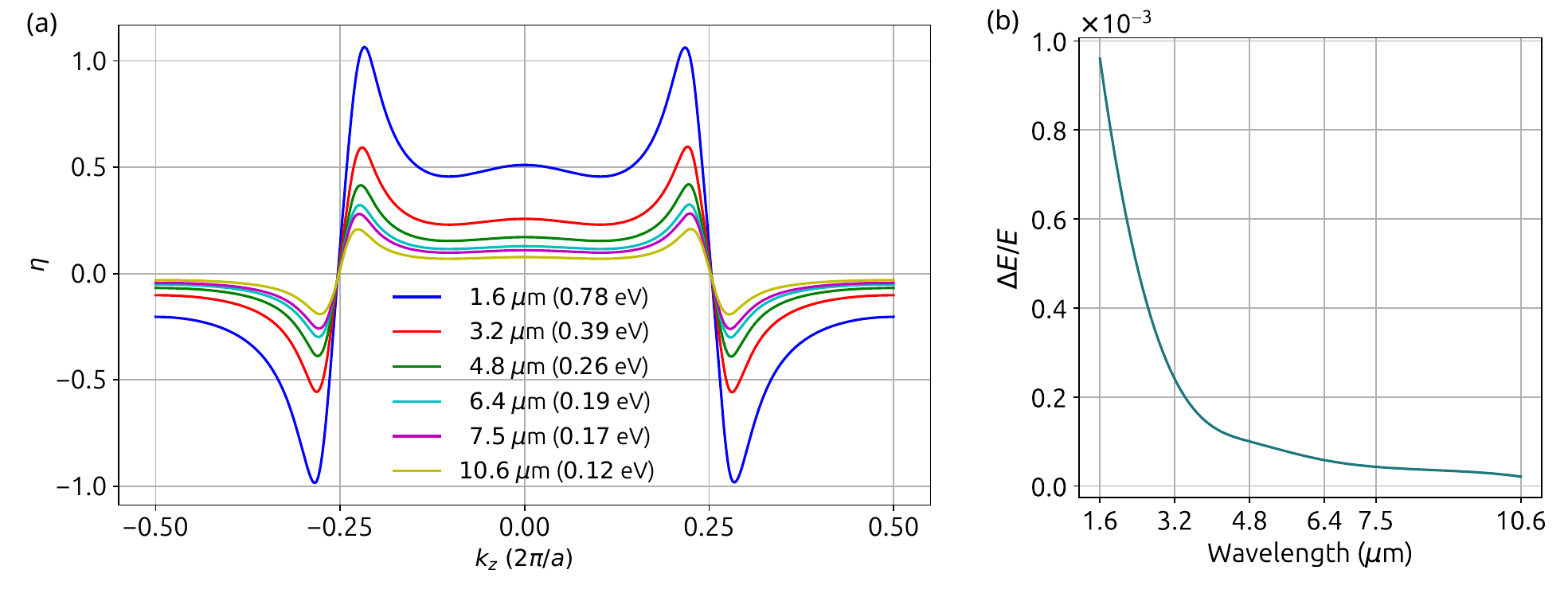}
\caption{(a) Normalized population asymmetry ($\eta$)  as a function of the wavelength of the circularly 
polarized pulse with intensity $5\times 10^{9}$ W/cm$^2$. (b) Nonlinear correction ($\Delta E$) 
to the energy ($E$) obtained from the linear dispersion.
The simulations use the full Hamiltonian given in Eq.~(\ref{eq:eqham}). } \label{fig:fig43}
\end{figure}

Since the deviations from the massless behavior could have come from our specific 
choice of the laser wavelength and intensity, which could have forced the electrons to explore the nonlinear 
parts of the dispersion, we will scan the laser intensity and wavelength while 
using the full Hamiltonian given in Eq.~(\ref{eq:eqham}). 
In the following, we shall use 
normalized population asymmetry  defined as
\begin{equation}\label{eq:eqasym}
\eta = \frac{n_{c}^{\circlearrowright} - n_{c}^{\circlearrowleft}}
{(n_{c}^{\circlearrowright} + n_{c}^{\circlearrowleft})/2},
\end{equation}
where $n_{c}^{\circlearrowright}  (n_{c}^{\circlearrowleft})$ is the final population 
due to LCP (RCP) pulse along $k_z$, integrated in the $k_x-k_y$ plane. 

Figure~\ref{fig:fig43}(a) shows $\eta$ for the driving wavelengths $\lambda$ 
from $1.6~\mu$m to  $10.6~\mu$m, which allows one to access 
different parts of energy dispersion during the excitation. 
We see that for all $\lambda$ the asymmetry
$\eta$ is nonzero around the Weyl nodes at $k_z=\pm 0.25$. While the 
asymmetry reduces with $\lambda$, even for the 
longest wavelength substantial values of $\eta$ at the levels $\sim 10$\% 
arise in the immediate vicinity of the Weyl nodes.  
We note that the deviation from linear dispersion for the wavelength studied is below 0.001 as shown in 
Fig.~\ref{fig:fig43}(b), 
while the circular dichroism asymmetry induced is several orders of magnitude higher as in Fig.~\ref{fig:fig43}(a).

\begin{figure}[h!]
	\includegraphics[width= \linewidth]{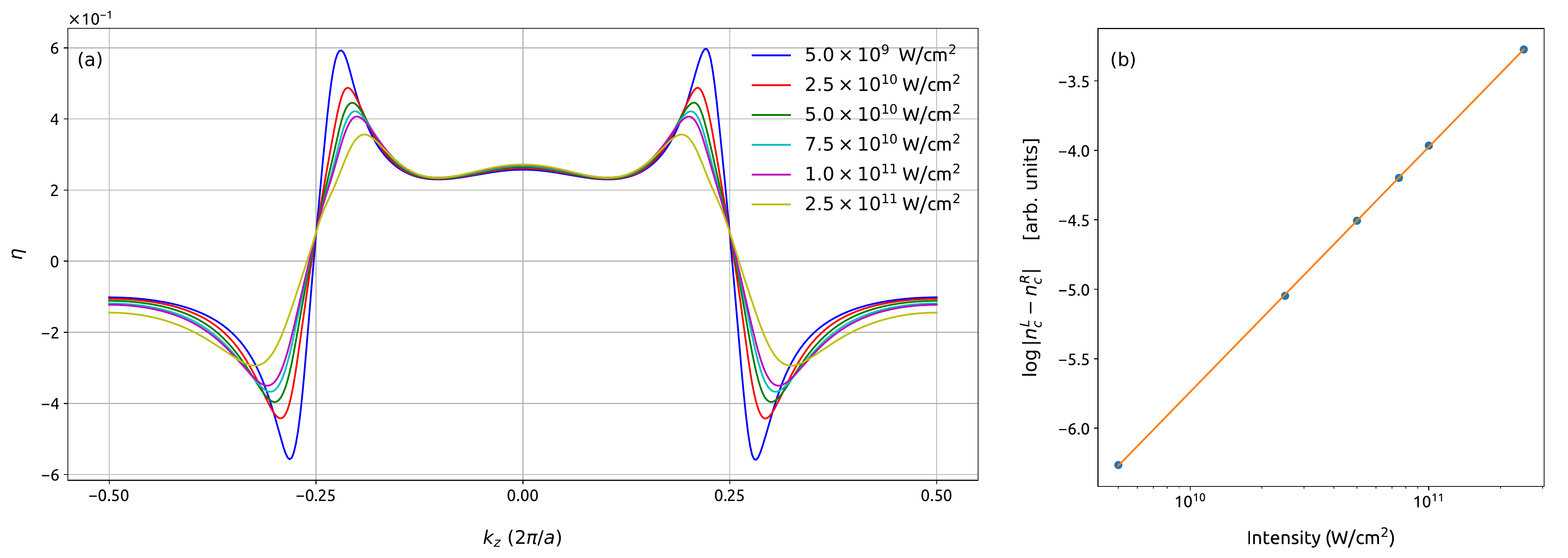}
	\caption{ (a) Variations of $\eta$ 
		for different intensities of the circularly polarized light. 
		(b) Logarithm of the difference in the population around a given Weyl node excited by
		the left- or right-handed circularly polarized light as a function of the laser's  intensity. 
		The slope of the fitted line is $0.8$, i.e., is below unity expected for linear processes.  
		The driving light wavelength is  $3.2~\mu$m.  
		The simulations use 
		the full Hamiltonian given in  Eq.~(\ref{eq:eqham}).}
	\label{fig:fig53}
\end{figure}

\begin{figure}[h!]
\centering
\includegraphics[width= 0.8\linewidth]{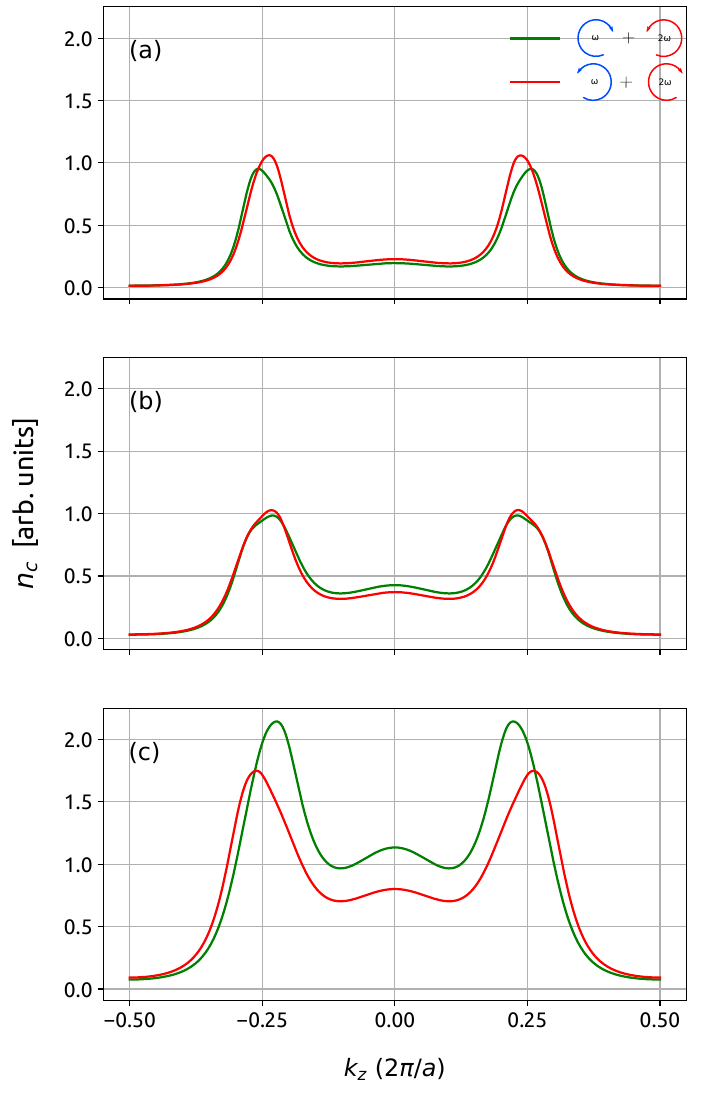}
\caption{Residual population for different ratio  ($\mathcal{R}$)   of the  two-color $\omega-2\omega$ laser pulses:   
(a) $\mathcal{R}=0.2$, (b) $\mathcal{R}=0.5$, and (c) $\mathcal{R}=1.0$. 
Intensity and wavelength of the $\omega$ pulse are $5\times10^{10}$ W/cm$^2$ and 
$3.2~\mu$m, respectively.} \label{fig:fig63}
\end{figure}

Figure~\ref{fig:fig53}(a) shows the dependence of $\eta$  on laser intensity, for 
$\lambda=3.2 ~\mu$m.  Notably, we find that the asymmetry is nonzero 
exactly at the Weyl node, where the dispersion is linear. This is true for all laser intensities,  
with the position of the zero asymmetry moving away from the node with increasing intensity.
The another surprise is that the asymmetry decreases 
with increasing intensity, i.e., when the electron is driven to explore a wider range of the 
Brillouin zone, where the dispersion nonlinearity is stronger.  This observation is
supported by Fig.~\ref{fig:fig53}(b), which shows that the intensity dependence of the 
non-normalized asymmetry is sublinear, with the slope $0.8$.

At this point, it is natural to explore the possibilities to control the ratio of the asymmetry induced by LCP and RCP pulses at a given node. 
To this end, we apply $\omega-2\omega$ counterrotating CPL with the total vector potential given by 
\begin{equation}\label{eq:eqtwocolor}
\mathbf{A}(t) = \frac{A_0 f(t)}{\sqrt{2}} \left( \left[ \cos(\omega t + \phi) + \mathcal{R} \cos(2 \omega t) \right] \hat{\mathbf{e}}_{x} \right. + \left.\left[ \sin(\omega t + \phi) - \mathcal{R} \sin(2 \omega t)\right] \hat{\mathbf{e}}_{y} \right).
\end{equation}
Here, $A_{0}$ is the amplitude of the total vector potential, and $f(t)$ is the envelope. 
The ratio between $\omega$ and $2 \omega$ fields is controlled by $\mathcal{R}$ and $\phi$ describes  the subcycle  phase between the $\omega$ and $2 \omega$ pulses. 

The  population excited by $\omega-2\omega$ counterrotating pulses is shown in Fig.~\ref{fig:fig63}, with 
the fundamental wavelength $\lambda=3.2~\mu$m.
For  $\mathcal{R} = 0.2$, the RCP-LCP combination generates more excitation
than the LCP-RCP combination [see Fig.~\ref{fig:fig63}(a)]. 
Moreover, the peak induced by RCP-LCP combination leans toward the center of the Brillouin zone.  
As  $\mathcal{R}$ changes from  0.2 to 0.5, both combinations yield almost similar populations. 
However, the peaks due to LCP-RCP combination change its direction and peaked toward the center. 
The situation reverses for $\mathcal{R} = 1$ where the LCP-RCP combination generates higher excitation than the
RCP-LCP combination [see Fig.~\ref{fig:fig63}(c)]. 
The reason behind such behavior is the interplay of the two competing resonant processes driven by LCP and RCP pulses, 
which is controlled by changing  $\mathcal{R}$. 
Thus, the ratio and the behavior of the residual population can be controlled by tailoring the value of 
$\mathcal{R}$ in $\omega-2\omega$ counterrotating pulses.  

\section{Conclusion}
In summary, we have demonstrated the generation of a helicity-sensitive population in an inversion-symmetric Weyl semimetal, which
is not symmetric with respect to the helicity of the driving circular light. The effect is general
and persists for different wavelengths and intensities. Even for the longest wavelengths and weakest 
intensities studied, it is triggered by the deviations of the Weyl fermion 
mass from zero, even in the immediate vicinity of the Weyl node.  
The origin of this phenomenon is embedded in the Berry connection, which remains unaffected by any modifications in the Hamiltonian of the Weyl semimetal~\citep{sadhukhan2021role}. 
We have proposed  a way to control and manipulate  the asymmetric population using 
counterrotating bicircular light, which allows tailoring the asymmetry 
from positive to negative via nearly zero.
%The asymmetric residual population  can be probed  via time- and angle-resolved photoemission spectroscopy in a pump-probe setup \citep{weber2021ultrafast}.  
\end{sloppypar}
\cleardoublepage
\chapter{Generating and Tailoring Photocurrent in Solids}\label{Chapter4}
\begin{sloppypar}
In the previous chapter, we  
have witnessed how an intense circularly polarized pulse leads to helicity-sensitive  residual population
in the conduction band of the time-reversal symmetry-broken (TRB) Weyl semimetal (WSM). 
Continuing on the same line, the present chapter focuses on the generation of photocurrent, which can be used to probe the unique coupling of light's helicity with WSMs. 
The application of intense laser pulses holds potentials of signal processing at Petahertz rate~\citep{boolakee2022light}. 
Thus,  employing ultrafast intense laser on topological materials is an emerging avenue for converting light into electricity efficiently.
In the later part of this chapter, we will extend the idea of photocurrent generation and its control to topological and non-topological as well as two- and three-dimensional systems.

Asymmetric electronic population distribution in the conduction band after the end of the laser pulse  in WSM renders a finite photocurrent -- photogalvanic effect -- 
which can be realized in various ways, such as the chiral magnetic effect~\citep{taguchi2016photovoltaic,kaushik2019chiral,kaushik2020transverse}, 
via transfer of angular momentum of light to the Weyl nodes~\citep{de2017quantized}, 
and nonlinear optical responses in the perturbative regime~\citep{ishizuka2016emergent,golub2017photocurrent,golub2018circular,zhang2018photogalvanic,wang2020electrically,watanabe2021chiral, gao2021intrinsic,heidari2022nonlinear,sirica2022photocurrent}.  
It has been shown that the inversion-broken  (IB) WSM  with gyrotropic symmetry produces 
second-order nonlinear optical responses -- 
injection and shift currents, which lead to a colossal photocurrent in TaAs ~\citep{morimoto2016topological, osterhoudt2019colossal,orenstein2021topology,ma2021topology}. 
In addition, photocurrent can exhibit a sign flip with the change in the helicity of the circularly polarized light~\citep{de2017quantized,chan2017photocurrents,konig2017photogalvanic}. 
The broken mirror-symmetry of Weyl nodes is a key reason 
behind helicity-sensitive photoresponse in the IB WSMs
~\citep{ma2017direct,rees2020helicity,ni2021giant}. 
So far, the majority of the work on photocurrent is focused on the IB WSMs with various tilts and crystal symmetries, as in the recent one in an inversion-symmetric  WSM~\citep{hamara2023helicity}. 
Thus, a universal method to generate photocurrent from both inversion-symmetric and IB WSMs that does not rely on such materials' symmetry details is missing.

It is a commonly accepted notion that the single-color CPL fails to generate 
photocurrent in WSMs with mirror-symmetric Weyl nodes. 
While each Weyl node generates current depending on its chirality, the currents in a chiral pair of the mirror-symmetric Weyl nodes cancel each other. 
In contrast to this widely accepted notion, we unequivocally demonstrate that a single-color CPL is able to generate photocurrent in mirror-symmetric WSMs.
A schematic of our idea is shown in Fig.~\ref{fig:fig4.1}(a), where a circularly polarized pulse with polarization in $xy$ plane is shined on an inversion-symmetric WSM.
Moreover, the present approach does not rely on the system's symmetry as it is equally applicable to both inversion-symmetric and IB WSMs with isotropic band dispersion 
and even with all Weyl nodes at the Fermi level.
Recently, bicircular laser pulses have been proposed for generating photocurrent in two- and three-dimensional materials, including  WSM, described by a linear anisotropic Hamiltonian~\citep{neufeld2021light,ikeda2023photocurrent}. 
However, a single-color laser based photocurrent, with a relatively easy experimental setup, is highly desirable for practical purposes. 

Few-cycle carrier-envelope phase stabilized linearly polarized laser can induce photocurrent as shown for graphene~\citep{higuchi2017light,zhang2022bidirectional}. 
However, such photocurrent cancels out if the phase is not stabilized, outweighing its applicability.
Our approach is also robust against such carrier-envelope phase stabilization.
We employ three- and six-cycle laser pulses in the mid-infrared regime to generate photocurrents whose direction and magnitude can be tailored by the phase of the circular pulse.
Moreover, it is observed that the photocurrent in an inversion-symmetric WSM is sensitive to the helicity and ellipticity of the laser pulse.
Two of the three proposed methods, counter-rotating $\omega-2\omega$ field as well as multicycle circular pulse, fail to generate photocurrent in two-dimensional materials.
Thus, a universal method to transform light into electricity applicable to topological and ``normal'' materials in two- and three-dimensions is lacking. 
In the second part of this chapter, we introduce a universal way to generate photocurrent in topological and non-topological materials based on a set of linearly polarized light. 

A schematic setup of our universal idea is shown in Fig.~\ref{fig:fig4.1}(b) where a pair of  linearly polarized pulses with frequencies $\omega$ and $2\omega$ is shined on an inversion-symmetric WSM, 
which is extended to other solids. 
The polarization of the $\omega$ pulse is fixed along   $x$ axis, whereas  $2\omega$ pulse  is along an 
arbitrary direction making an angle $\theta$ with respect to $x$ axis in the $xy$ plane. 
When both $\omega-2\omega$ pulses are in collinear configuration ($\theta = 0$), 
the resultant laser pulse is asymmetric
along $x$ direction with no component along $y$ direction. 
In addition, there are more oscillations on the negative side than the positive side in the resultant laser waveform 
as evident from the top panel of Fig.~\ref{fig:fig4.12}(b). 
On the other hand, the orthogonal configuration of the $\omega-2\omega$ pulses, i.e., $\theta = \pi/2$, results in a symmetric laser waveform {as reflected from the bottom panel of Fig.~\ref{fig:fig4.12}(b)}. 

\begin{figure}
\centering
\includegraphics[width=\linewidth]{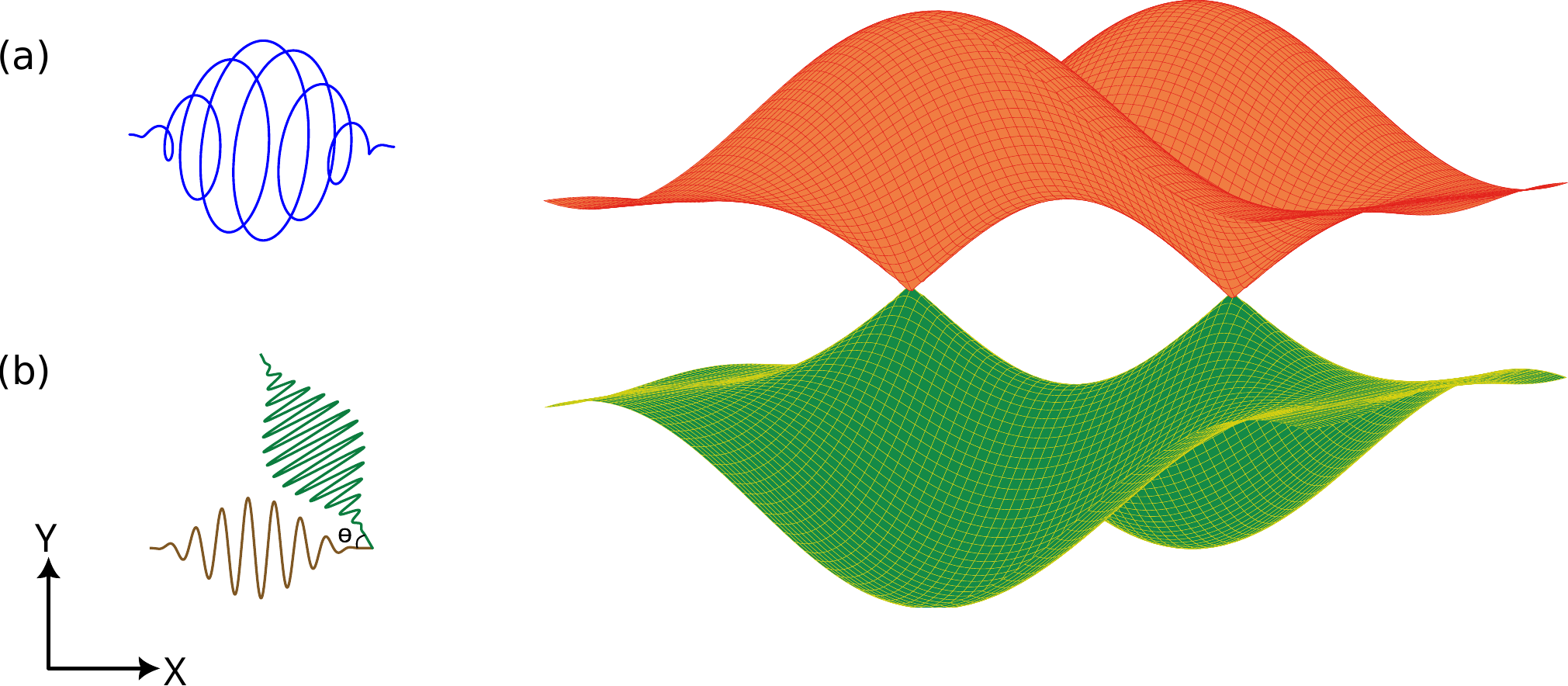}
\caption{Schematic setup for ultrafast photocurrent generation in a solid, which is represented by an     
inversion-symmetric Weyl semimetal. (a) A circularly-polarized laser with frequency $\omega$ is interacting to a solid. (b) Alternatively, two linearly-polarized pulses with frequencies $\omega$ and $2 \omega$ with polarization directions at an angle $\theta$ are interacting with the solid.} 
\label{fig:fig4.1}
\end{figure}

In the following, we will illustrate that the photocurrent in WSMs is maximum (minimum) when the setup is in a collinear (orthogonal) configuration. 
Furthermore, it has been found that the presence of a weak $2 \omega$ pulse is enough to generate photocurrent, 
which can be further manipulated by tuning the intensity of the $2 \omega$ pulse with respect to the $\omega$ pulse. 
In addition, the photocurrent can be further tuned by controlling the interplay of $\omega-2\omega$ pulses through variations in angle $\theta $, amplitude ratio, and time delay between them.

To demonstrate the universality and robustness of our idea, we will extend our study to inversion-symmetric and inversion-broken two-dimensional materials with trivial topology.
Asymmetry of the carrier-envelope-phase stabilized  few-cycle laser waveform has been utilized 
to generate photocurrent in  two-dimensional materials, where 
the photocurrent was controlled by tuning the carrier-envelope-phase of the pulse~\citep{higuchi2017light, heide2020sub}. 
Further, the resultant photocurrent gives a lower bound on coherence time in graphene~\citep{heide2021electronic}.
In contrast to previous studies, our approach does not require carrier-envelope-phase stabilization to generate photocurrent. 
Moreover, our approach shows a universal application of intense light to engender photocurrent in both topological and non-topological materials in two- and three-dimensions.

\section{Theoretical Methodology}
We start our discussion by writing the expressions of  the  components of $\mathbf{d}(\mathbf{k})$
for an inversion-symmetric WSM as~\citep{sadhukhan2021role,menon2021chiral} 
\begin{equation}\label{eq:trb4}
	\mathbf{d}(\mathbf{k})  =   \big[t\sin(k_x a), t\sin(k_y a),
	t\{\cos(k_z a) - \cos(k_0 a) +2- \cos(k_x a) - \cos(k_y a)\}\big],
\end{equation}
and for an IB WSM read as 	 
\begin{eqnarray}\label{eq:invb4}
	&\mathbf{d}(\mathbf{k})  =  & \big[t\{\cos(k_0 a)-\cos(k_y a) + \mu[1-\cos(k_z a)]\}, t\sin(k_z a),\nonumber \\
	&& t\{\cos(k_0 a)-\cos(k_x a) + \mu[1-\cos(k_z a)]\}\big].
\end{eqnarray}
Here $k_0$ determines the position of the Weyl nodes, which are considered as $\pi/(2a)$ for both WSMs. 
The Weyl nodes for inversion-symmetric and inversion-broken systems are situated at $\mathbf{k} = [0,0,\pm \pi/(2a)]$ and $\mathbf{k}=[\pm\pi/(2a),\pm\pi/(2a),0]$, respectively.  
A simple cubic crystal structure is considered with lattice parameter $a = 6.28$ $\mathring{\text{A}}$~and isotropic hopping parameter $t=1.8$ eV  
in Eqs.~(\ref{eq:trb4}) and~(\ref{eq:invb4}). 
Moreover, a dimensionless parameter $\mu = 2$ is used in Eq.~(\ref{eq:invb4}). 
Energy band dispersions corresponding to Eqs.~(\ref{eq:trb4}) and  (\ref{eq:invb4}) are shown in Fig.~\ref{fig:4.2}(a) and (b), respectively. Similar tight-binding Hamiltonians for pristine graphene and  MoS$_2$ are adopted from 
Refs.~\citep{mrudul2021high} and ~\citep{rana2023all}, respectively.

The  photocurrent originates from the residual population asymmetry and can be written as
~\citep{soifer2019band}
\begin{equation}\label{eq:4.1}
	\mathbf{J}(t) = \int_\mathbf{k}  d\mathbf{k}~\left[ \rho(\mathbf{k}) - \rho(-\mathbf{k}) \right] \pdv{\mathcal{E}(\mathbf{k})}{\mathbf{k}},
\end{equation}
where $\mathbf{J}(t)$ is the total current, $\rho$ is the residual  population  density  
after the end of the laser pulse, 
and  $\mathcal{E}(\mathbf{k})$ is the energy dispersion in a solid.     
A nonzero photocurrent, $\mathbf{J}(t)$, arises if there is an asymmetric electronic population in the conduction band 
after the end of the laser pulse, i.e., $\rho(\mathbf{k}) \neq \rho(-\mathbf{k})$~\citep{bharti2023tailor,soifer2019band}. In this chapter, we analyze the photocurrent for various configurations of the laser pulse to ascertain suitable configurations for producing the asymmetric population and, hence, in turn, photocurrent in solids.

\begin{figure}[h!]
\centering
\includegraphics[width=\linewidth]{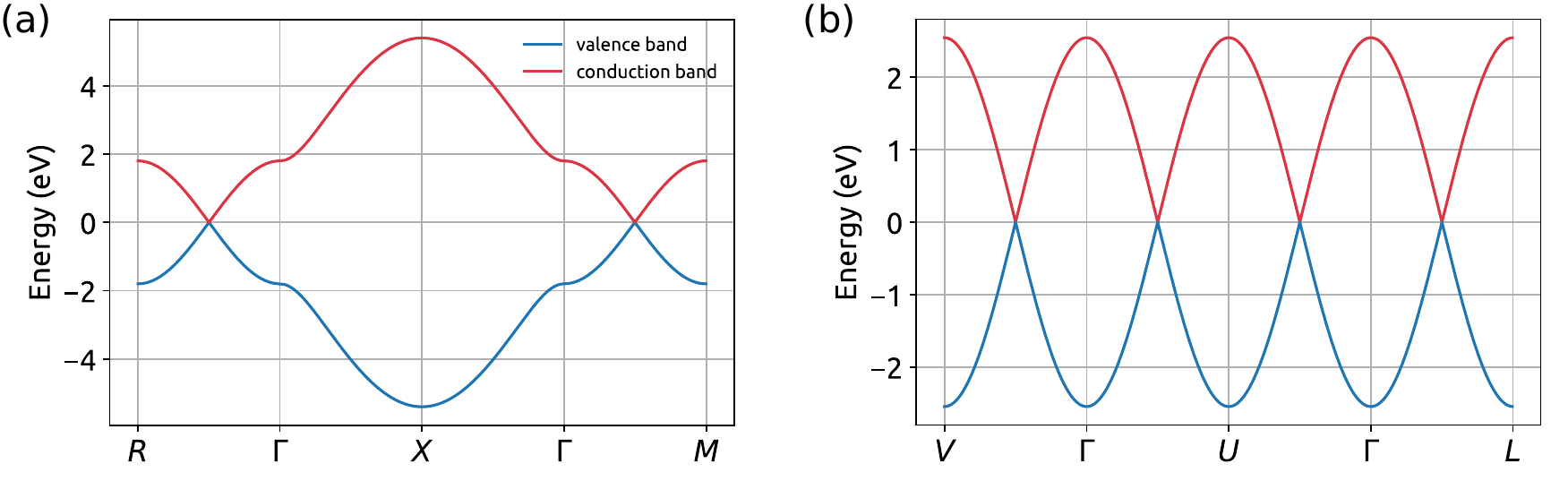}
\caption{(a) Energy dispersion of an inversion-symmetric Weyl semimetal, described by Eq.~(\ref{eq:trb4}) 
along  high-symmetry path $R(0,0,\pi/a)\rightarrow \Gamma(0,0,0)\rightarrow X(\pi/a,0,0)\rightarrow M(0,0,-0.5)$. (b) Energy dispersion of an inversion-broken Weyl semimetal, described by Eq.~(\ref{eq:invb4})  
along high-symmetry path $V(\pi/a,-\pi/a,0)\rightarrow \Gamma (0,0,0)\rightarrow U(\pi/a, \pi/a,0) \rightarrow L(-\pi/a,-\pi/a,0).$} \label{fig:4.2}
\end{figure}

For the first part of the chapter, the vector potential of the circularly polarized laser is written as 
$\mathbf{A}(t) = A_0 f(t) \Re{e^{i \omega t + \phi} ~\hat{\mathbf{e}}_{\pm}}$,  
where $\hat{\mathbf{e}}_{\pm} = (\hat{\mathbf{e}}_{x} \pm i \epsilon\ \hat{\mathbf{e}}_{y})$ corresponds to left- and right-handed circularly polarized laser pulse with ellipticity $\epsilon=1$,  $A_0$ is the amplitude
and $f(t)$ is the $\sin^2$ envelope. 
The subcycle phase of the laser pulse is denoted by $\phi$, which controls the orientation of the Lissajous profile of the laser. 
For the second part, the total vector potential of a pair of linearly polarized laser pulses is written as 
\begin{equation}\label{eq:laser2}
	\textbf{A}(t) = A_0 f(t)   \left[ \cos(\omega t) \hat{e}_{x} + \mathcal{R} \cos(2\omega t) \{ \cos(\theta) \hat{e}_{x}  + 
	\sin(\theta) \hat{e}_y  \} \right],      
\end{equation}
where  
$\mathcal{R}$ is a dimensionless parameter to tune the intensity of the $2\omega$ pulse with respect to the $\omega$ pulse 
and  $\theta$ is the angle between the polarization of two linearly polarized pulses as shown in Fig.~\ref{fig:fig4.1}(a). 
A laser with  wavelength $3.2 ~\mu$m  
and pulse duration ranging from $\sim$ 35 to 100 fs is employed to generate photocurrent.

\section{Circularly Polarized Laser-Driven Photocurrent}

\begin{figure}[!htb]
	\centering
	\includegraphics[width=\linewidth]{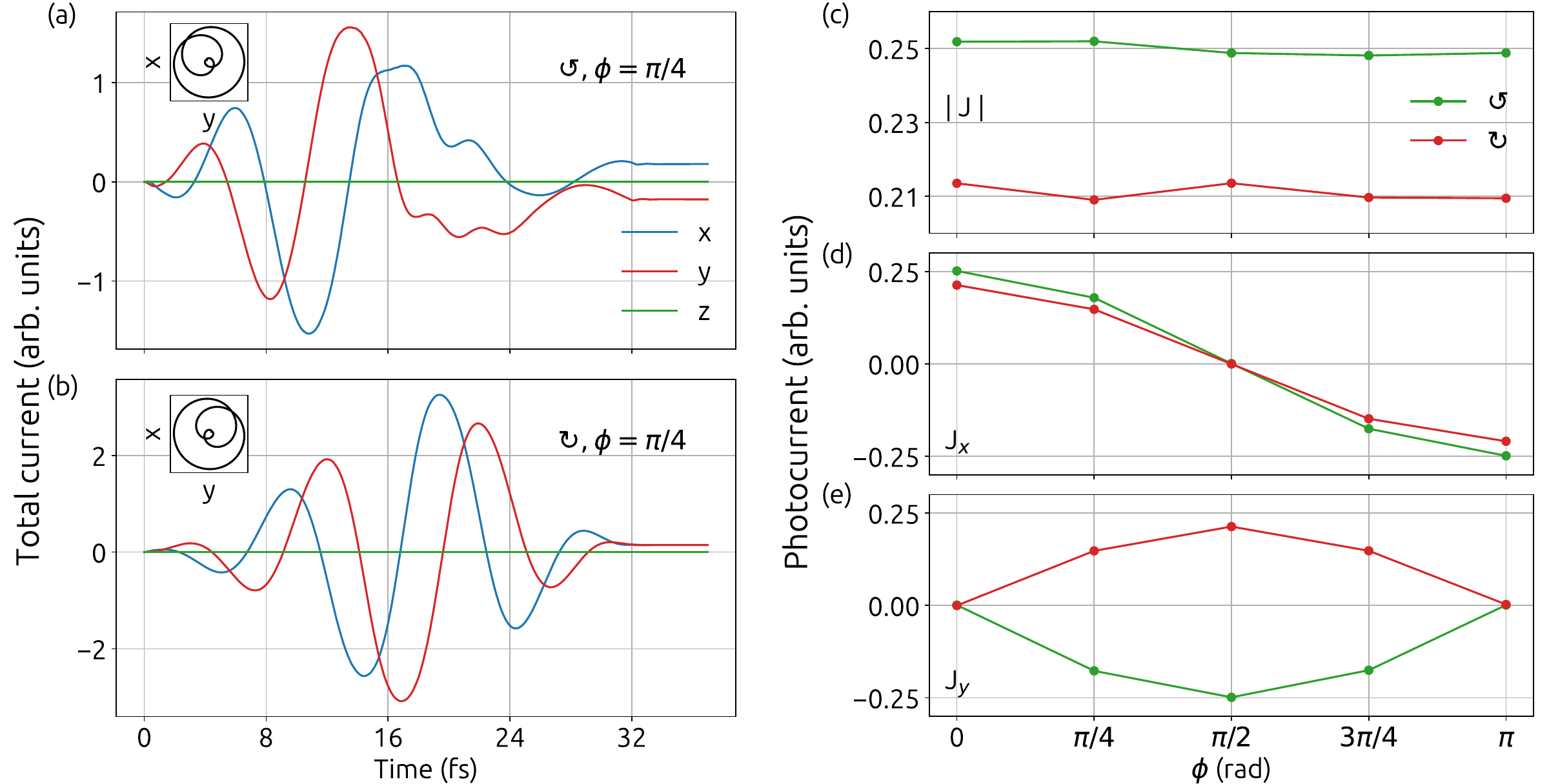}
	\caption{Total photocurrent in an inversion-symmetric Weyl semimetal 
		induced by (a)  left-handed and (b) right-handed circularly polarized lasers with  phase 
		$\phi = \pi/4$. The Lissajous curve in the $xy$ plane is shown in respective insets. 
		Variations in (c) the   photocurrent, (d) its $x$  ($\mathsf{J}_{x}$), and (e) $y$  
		($\mathsf{J}_{y}$) components with respect to the phase. 
		These results are obtained for a three-cycle circularly polarized laser pulse with  $\sim$ 32 fs duration, 3.2 $\mu m$ wavelength, and $10^{11}$ W/cm$^2$ intensity.}
	\label{fig:fig4.3}
\end{figure}

Let us analyze results for an inversion-symmetric WSM, which exhibits 
a finite photocurrent  along $x$ and $y$ directions after the end of the laser pulse 
as shown in  Fig.~\ref{fig:fig4.3}(a). 
As the helicity of the laser changes from right to left, the sign of the photocurrent  along the $y$ direction  flips  
from negative to positive as evident from Fig.~\ref{fig:fig4.3}(b). 
To unravel the underlying mechanism  for the flip, we analyzed the Lissajous profile of the vector potential 
in the polarization plane as shown in the insets. 
The change in the Lissajous curve with a change in the helicity is a primary 
reason for the sign flips along  $y$ direction. 
This shows that the photocurrent is susceptible to the profile of the laser pulse. 
At this juncture, it is pertinent to know how the photocurrent is sensitive to the phase of the laser pulse. 
To this end, we investigate variation in the total photocurrent and its components with respect to  the phase.

\begin{figure}[!htb]
	\centering
	\includegraphics[width=0.8\linewidth]{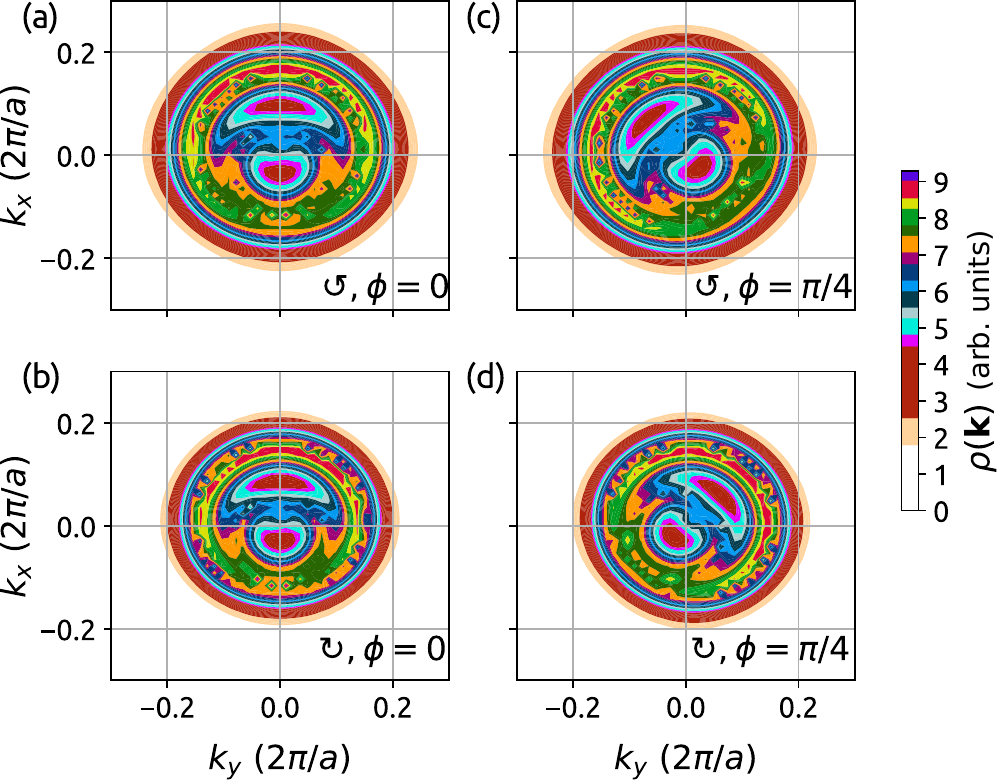}
	\caption{Residual population in the conduction band  of an inversion-symmetric Weyl semimetal 
		after the end of the  (a) left-handed, and (b) right-handed circularly polarized lasers with $\phi =0$. 
		(c) and (d) are, respectively the same as (a) and (b) for $\phi = \pi/4$. 
		Weyl nodes are situated at 
		$\mathbf{k}=[0,0, \pm \pi/(2a)]$. 
		Population along $k_z$ direction is integrated in all cases. Laser parameters are identical to  Fig. \ref{fig:fig4.3}. }
	\label{fig:fig4.4}
\end{figure}

\begin{figure}[!htb]
	\centering
	\includegraphics[width=\linewidth]{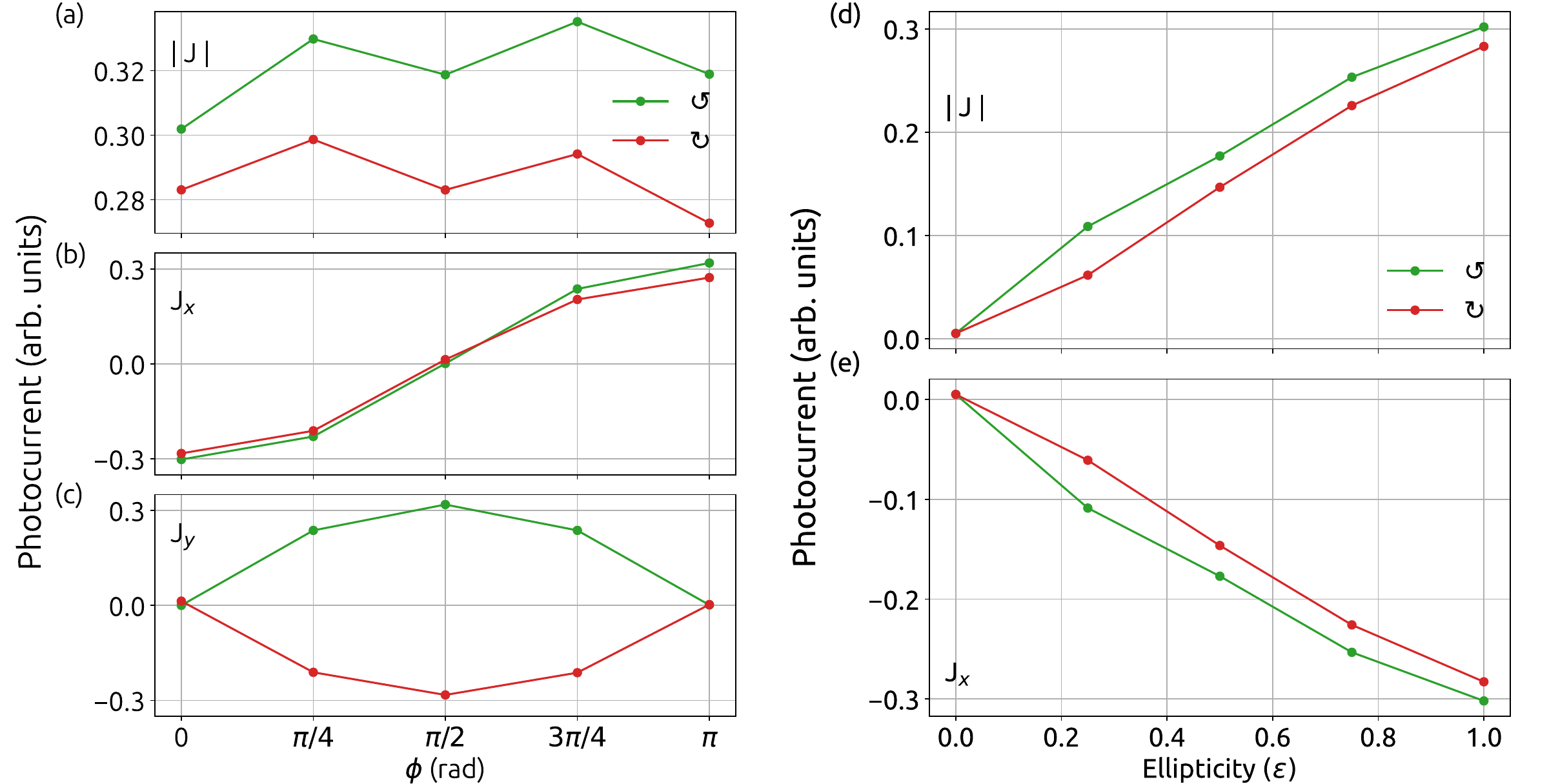}
	\caption{Variations in (a) the  total photocurrent,  
		(b) its $x$  ($\mathsf{J}_{x}$), and (c) $y$  
		($\mathsf{J}_{y}$) components with respect to $\phi$ for an inversion-symmetric Weyl semimetal.  
		Sensitivity of (d) the photocurrent and (e)   $\mathsf{J}_{x}$  with respect to the ellipticity of the laser for
		$\phi=0$.  The laser pulse consists of six cycles with  $5\times10^{11}$ W/cm$^2$ intensity and 3.2 $\mu m$ wavelength.}
	\label{fig:fig4.5}
\end{figure}

Figure~\ref{fig:fig4.3}(c) shows the insensitivity of the photocurrent with respect to $\phi$, 
which concludes that phase stabilization is not a prerequisite to generate photocurrent in WSM.  
However,  the $x$ component ($\mathsf{J}_{x}$) changes from a  positive  to 
a negative value as $\phi$ changes from 0 to $\pi$, including zero at $\phi = \pi/2$ [see Fig.~\ref{fig:fig4.3}(d)]. 
Both helicities display similar behavior for $\mathsf{J}_{x}$, whereas 
the $y$ component ($\mathsf{J}_{y}$) exhibits an opposite trend as the helicity is reversed 
from left and right, except $\phi = 0$ and  $\pi$ where it is zero [see Fig.~\ref{fig:fig4.3}(e)].  
Analysis of Fig.~\ref{fig:fig4.3} raises a crucial question 
about factors determining the nonzero photocurrent and its components.  

The residual population in the conduction band around a Weyl node after the end of the laser is presented in Fig.~\ref{fig:fig4.4}. 
Owing to the zero band-gap nature of the Weyl node, the region around the node is significantly populated, which decreases rapidly as we move away from the origin. 
Population about  the $k_{x} = 0$ plane is significantly asymmetric, which in results nonzero 
photocurrent along this direction as 
$\rho({k_{x}}) \neq \rho(-{k_{x}})$ for both helicities. 
However, the population exhibits mirror symmetry about the $k_{y} = 0$ plane for $\phi = 0$, which in results in zero 
photocurrent for both helicities as evident from Figs.~\ref{fig:fig4.3}(e),  
\ref{fig:fig4.4}(a) and \ref{fig:fig4.4}(b). 
A change in  $\phi$  from 0 to $\pi/4$ induces asymmetry along $k_{y} = 0$, which generates finite photocurrent as 
reflected from Figs.~\ref{fig:fig4.4}(a) and \ref{fig:fig4.4}(b). 
In addition, the direction of the induced asymmetry along $k_{y} = 0$ flips as we change the helicity from left and right, which results in a sign change in  $\mathsf{J}_{y}$ as shown in Fig.~\ref{fig:fig4.3}(e). 
Thus, observations 
in Figs.~\ref{fig:fig4.3}  and \ref{fig:fig4.4} are consistent  with Eq.~(\ref{eq:4.1}).

One of the striking features of Fig.~\ref{fig:fig4.4} is the extent of the asymmetries along $k_{x} = 0$ and  $k_{y} = 0$ planes, which are significantly different for both helicities. 
Recently, it has been shown that the electronic excitation from the nonlinear part of the band dispersion 
can effectuate the helicity-dependent population in an inversion-symmetric WSM~\citep{bharti2023weyl}. 
Therefore, owing to the unique coupling of the circularly polarized laser with the WSM,  
the residual population along $k_z$, integrated along other directions, is sensitive to the laser's helicity~\citep{bharti2023weyl}. 
Thus, the helicity-sensitive population asymmetry leads to different photocurrent for the left- and right-handed laser pulses, as shown in Fig.~\ref{fig:fig4.3}. 

\begin{figure}[!htb]
	\centering
	\includegraphics[width=\linewidth]{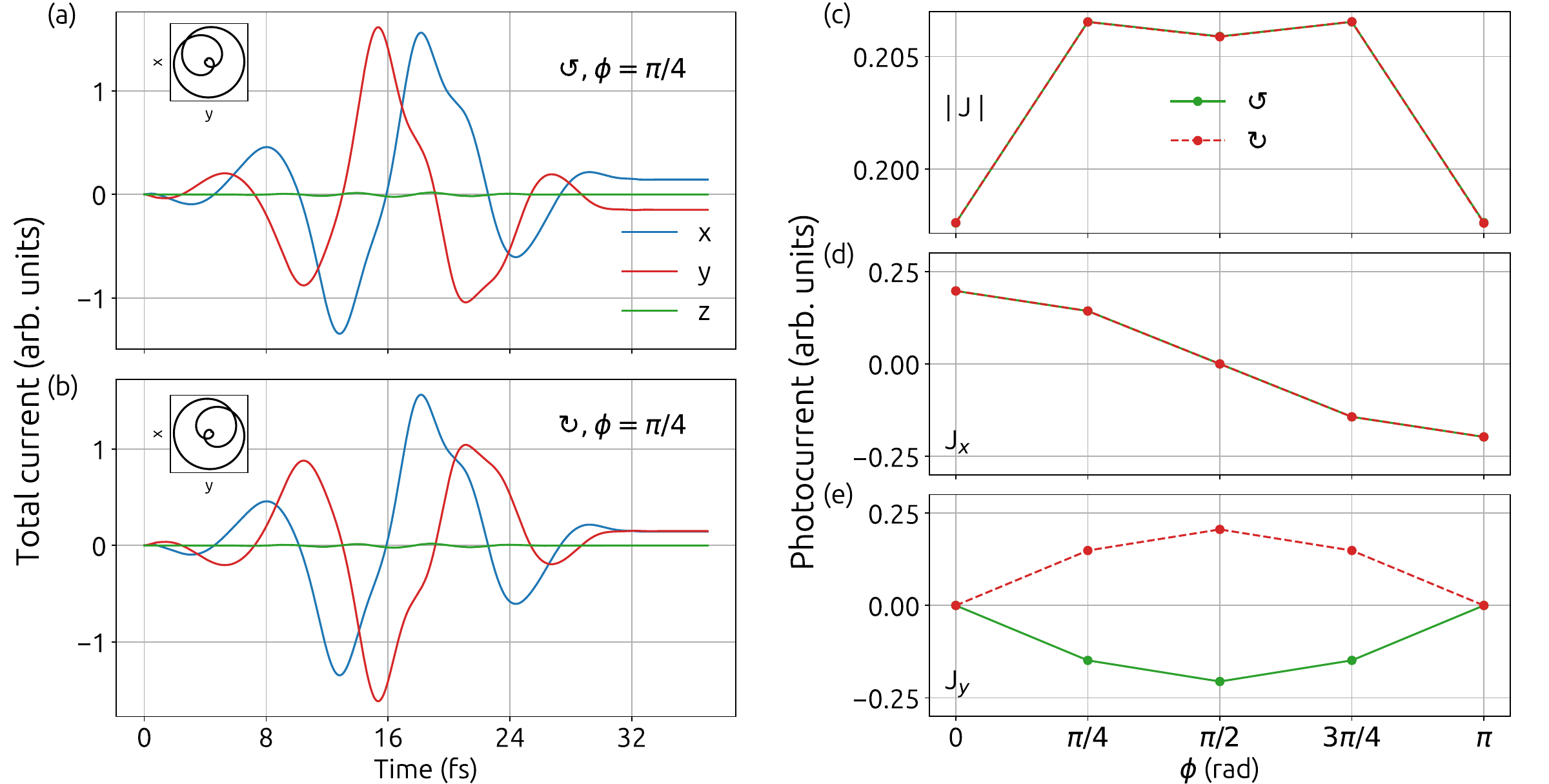}
	\caption{Same as Fig. \ref{fig:fig4.3} for an inversion-broken Weyl semimetal as described  
		by the Hamiltonian in Eq. \eqref{eq:invb4}.}
	\label{fig:fig4.6}
\end{figure}

So far, we have discussed the results of the three-cycle laser pulse. 
It is known that the vector potential can be nonzero when the electric field is  zero for a  
few-cycle laser pulse with stabilized carrier-envelope phase. 
The nonzero vector potential can induce asymmetric population and photocurrent in graphene, as discussed in Refs.~\citep{higuchi2017light,zhang2022bidirectional}. 
Thus, it is natural to ask about the robustness of our results with the pulse duration.
Generating photocurrent in WSMs via relatively long laser pulse in mid-infrared regime is highly desirable for numerous practical  applications~\citep{ishizuka2016emergent,golub2017photocurrent,golub2018circular, wang2020electrically,watanabe2021chiral,heidari2022nonlinear,golub2017photocurrent,sirica2022photocurrent}. 

Towards that end, let us  increase the pulse duration from $\simeq$   30 to 65 fs by changing 
the number of cycles from three to six while keeping the intensity constant.   
In this case, a finite photocurrent with a relatively smaller magnitude is observed.  
It is found that the intensity needs to be increased by five times 
to make the magnitude of the photocurrent comparable for three- and six-cycle pulses [see Fig.~\ref{fig:fig4.5}(a)]. 
On comparing Figs.~\ref{fig:fig4.3}(c) and \ref{fig:fig4.5}(a), it is evident 
that an increase in intensity leads to a reduction in the contrast between the photocurrent for different helicity. 
The reduction in the contrast can be attributed to the underlying mechanism of the helicity-dependent asymmetric population, which relies on the resonant excitation at various $\mathbf{k}$ and thus reduces the asymmetry with an increase in intensity. 

\begin{figure}[!htb]
	\centering
	\includegraphics[width=\linewidth]{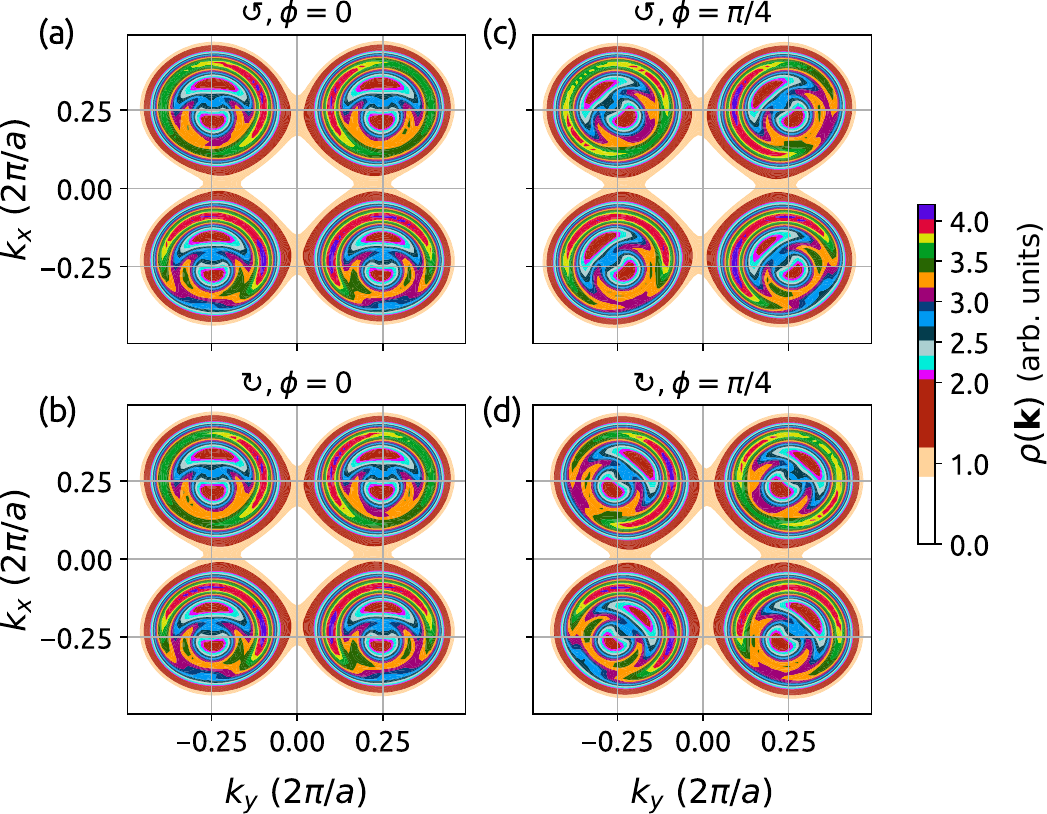}
	\caption{Same as Fig. \ref{fig:fig4.6} for an inversion-broken Weyl semimetal.}
	\label{fig:fig4.7}
\end{figure}

In contrast to the three-cycle pulse, $\mathsf{J}_{x}$ transits from negative to positive magnitude as $\phi$ changes from 
0 to $\pi$, whereas $\mathsf{J}_{y}$ exhibits similar behavior for three- and six-cycle pulses [see Figs.~\ref{fig:fig4.5}(b) and \ref{fig:fig4.5}(c)]. 
Note that the photocurrent 
can be positive or negative based on whether $-\mathbf{k}$ or $\mathbf{k}$ is more populated, which 
depends on the intensity and pulse duration~\citep{zhang2022bidirectional}. 
The photocurrent is not only sensitive to the pulse duration but also to the ellipticity of the laser pulse, as shown in Fig.~\ref{fig:fig4.5}(d) for $\phi =0$. 
Photocurrent monotonically reduces to zero as the ellipticity changes from one (circular)  to zero (linear) for both helicities. Similar observations can be made for $\mathsf{J}_{x}$ from Fig.~\ref{fig:fig4.5}(e). 
Note that $\mathsf{J}_{y}$ is zero for $\phi =0$. Our analysis establishes that a laser pulse with definite chirality, but nonzero ellipticity, 
is able to engender photocurrent in 
an inversion-symmetric WSM, which also encapsulates a unique coupling of chiral light with WSM~\citep{bharti2023weyl}.

After demonstrating the photocurrent generation in an inversion-symmetric WSM, 
let us focus our discussion to an IB WSM. 
Figure~\ref{fig:fig4.6} presents finite photocurrent in an IB WSM driven by a circularly polarized laser. 
By the virtue of the Lissajous profile flip, the photocurrent along $y$ direction flips its sign as the laser's helicity changes for $\phi = \pi/4$ [see Figs.~\ref{fig:fig4.6}(a) and \ref{fig:fig4.6}(b)]. 
The total photocurrent does not change significantly with variation in $\phi$ and is identical for both helicities as shown in  Fig.~\ref{fig:fig4.6}(c). 
Similar to an inversion-symmetric case, 
$\mathsf{J}_{x}$ changes its magnitude from positive to negative as 
$\phi$ changes from 0 to $\pi$ [see Fig.~\ref{fig:fig4.6}(d)], and $\mathsf{J}_{y}$ remains either positive or negative depending on the helicity 
except at $\phi = 0$ and $\pi$ [see Fig.~\ref{fig:fig4.6}(e)]. 
Thus, the behavior of the photocurrent and its components are robust with respect to $\phi$. 
Note that there is a finite photocurrent in the plane of polarization for other polarization directions of the laser, and can be tailored by changing $\phi$.

We also analyze the residual  
population in the conduction band to corroborate the photocurrent's results in Fig.~\ref{fig:fig4.6}. 
Significant  population around four Weyl nodes at $\mathbf{k}=[\pm\pi/(2a),\pm\pi/(2a),0]$ is observed 
as shown in Fig.~\ref{fig:fig4.7}. 
Population is asymmetric in nature with respect to $k_x=0$ plane for $\phi=0$ 
and exhibits $k_y = 0$ as a plane of reflection, which results in nonzero (zero)  photocurrent along $x$ ($y$) axis.   
Reflection symmetry  about $k_y = 0$  plane is lost as $\phi$ changes  to $\pi/4$ 
[see Figs.~\ref{fig:fig4.7}(c) and \ref{fig:fig4.7}(d)],  
which results in nonzero photocurrent along this direction as evident from Fig.~\ref{fig:fig4.6}(e). 
In addition, the populations corresponding to both helicities are identical, which is in contrast to the one observed for  
an inversion-symmetric WSM.

\subsection{Scaling of Photocurrent with Laser's Intensity}

\begin{figure}[!htb]
\centering
\includegraphics[width=0.8\linewidth]{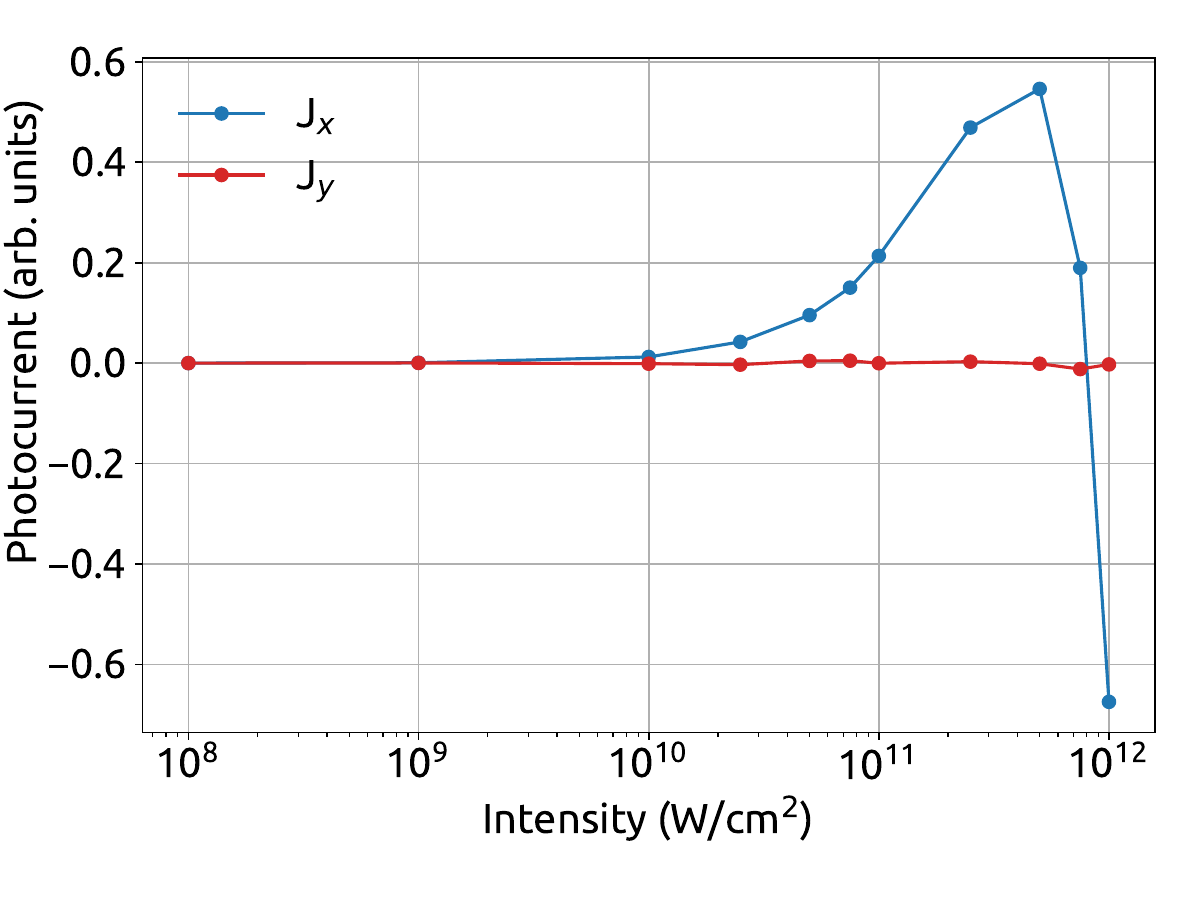}
\caption{Scaling of the photocurrent in an inversion-symmetric Weyl semimetal with the laser's intensity of the right-handed circularly polarized light with $\phi = 0$.  The other laser parameters are same as Fig.~\ref{fig:fig4.3}.} \label{fig:fig4.8}
\end{figure}

The generated photocurrent is nonperturbative in nature as evident from its scaling with laser's intensity, see Fig.~\ref{fig:fig4.8}. We discuss the variation of the photocurrent in an inversion-symmetric WSM with the laser's intensity.  
A right-handed circularly polarized laser consisting of three cycles with $\phi = 0$ is used to study the intensity scaling.  
It is observed that the photocurrent increases rapidly with an increase in the intensity. 
A reversal of the photocurrent at very large intensity is seen, which 
is consistent with previous observations in graphene~\citep{higuchi2017light,zhang2022bidirectional}.

\begin{figure}[!h]
	\centering
	\includegraphics[width=  \linewidth]{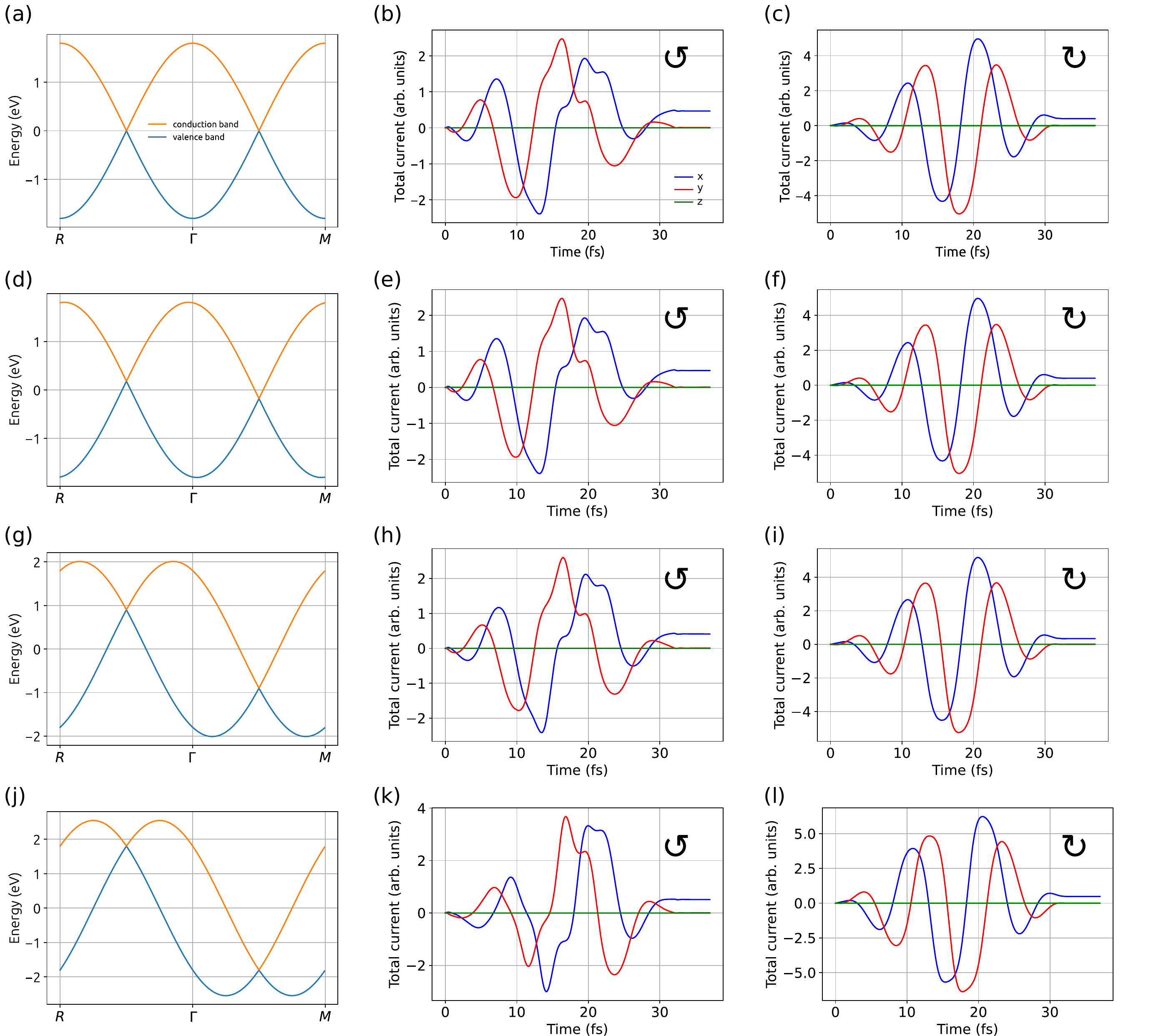}
	\caption{Variation in the photocurrent in an inversion-symmetric Weyl semimetal with Weyl nodes  
		situated at different energies (various energy splits) with respect to the Fermi level. 
		(a) Energy dispersion for $t_g=0$, i.e., both Weyl nodes are at Fermi level, total current due to the 
		(b)  left-handed, and (c)  right-handed circularly polarized light with $\phi = 0$.  
		(d), (e) and (f) are the same as (a), (b), and (c) for $t_g = 0.1t$, respectively.  
		(g), (h) and (i) are the same as (a), (b), and (c) for $t_g = 0.5t$, respectively.
		(j), (k) and (l) are the same as (a), (b), and (c) for $t_g = 1.0t$, respectively.
		The other laser parameters are same as Fig.~\ref{fig:fig4.3}.}
	\label{fig:fig4.9}
\end{figure}

\subsection{Role of Energy Split of the Weyl Nodes from Fermi Level and  Tilt of the Weyl Nodes}

Our approach is equally applicable to realistic situations when the Weyl nodes are nondegenerate [see Fig.~\ref{fig:fig4.9}], situated at different energies, and have tilt along a certain direction [see Fig.~\ref{fig:fig4.10}]. 
The Hamiltonian in Eq.~\eqref{eq:trb4} is modified 
to include tilt and chemical potential in the WSM 
by adding $d_{0}\sigma_0$ term in $  \bm{\sigma} \cdot \mathbf{d}(\mathbf{k})$ with 
$d_{0} = t_c [\cos(k_z a) + \cos(k_x a) - 1]+ t_g \sin(k_z a)$.  
Here, $t_c$ decides the tilt of the Weyl nodes, and $t_g$ decides the energy split between the non-degenerate Weyl nodes~\citep{menon2021chiral}. 
We consider different tilt of the Weyl nodes with tilt parameters as $t_c = 0.5t, 0.8t$ and $1.2t$ in Fig.~\ref{fig:fig4.10}.
Note that $ \mathbf{d}(\mathbf{k})$ for an inversion-symmetric WSM is given in Eq.~\eqref{eq:trb4}.

\begin{figure}[!h]
	\centering
	\includegraphics[width=  \linewidth]{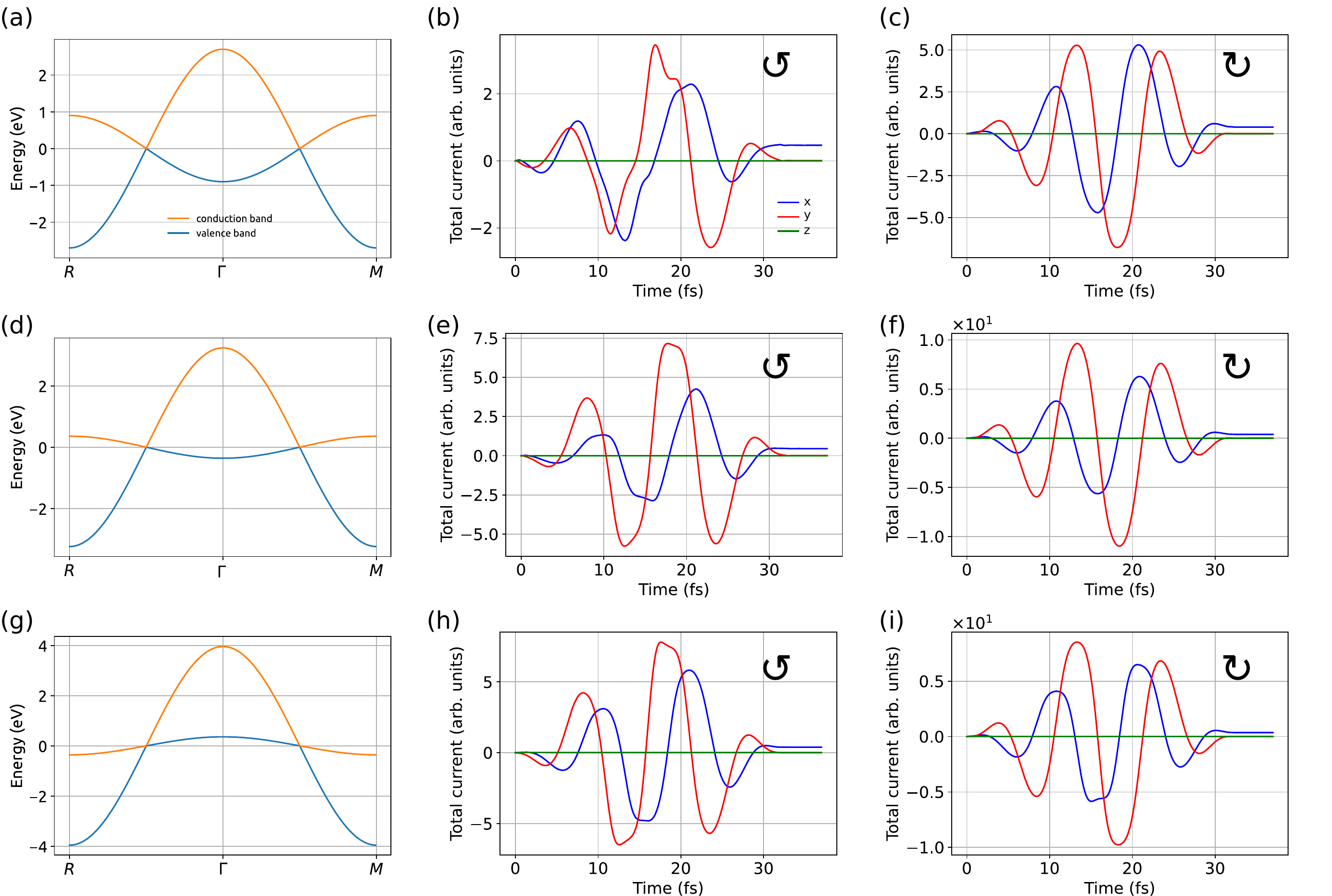}
	\caption{Variation in the photocurrent in an inversion-symmetric Weyl semimetal with Weyl nodes  having finite 
		tilt along a certain direction. 
		(a) Energy dispersion for $t_c=0.5t$, total current due to the 
		(b) left-handed, and (c)  right-handed circularly polarized light with $\phi = 0$. 
		(d), (e) and (f) are the same as (a), (b), and (c) for $t_c=0.8t$, respectively.
		(g), (h) and (i) are the same as (a), (b), and (c) for $t_c = 1.2t$, respectively.
		The other laser parameters are same as Fig.~\ref{fig:fig4.3}.}
	\label{fig:fig4.10}
\end{figure}

\subsection{Photocurrent via  Bicircular Counter-Rotating Laser Fields}
Further, we discuss photocurrent in an inversion-symmetric WSM driven 
by two-color $\omega-2\omega$ bicircular counter-rotating laser pulses. 
Recently, Morimoto and co-workers have used $\omega-2\omega$ bicircular counter-rotating laser pulses to generate perturbative photocurrent in inversion-symmetric WSM, described by anisotropic linear Hamiltonian~\citep{ikeda2023photocurrent}. In this chapter, an inversion-symmetric WSM is described by the ``realistic'' lattice Hamiltonian, which goes beyond the linear model. 

\begin{figure}[!h]
	\centering
	\includegraphics[width= 0.8\linewidth]{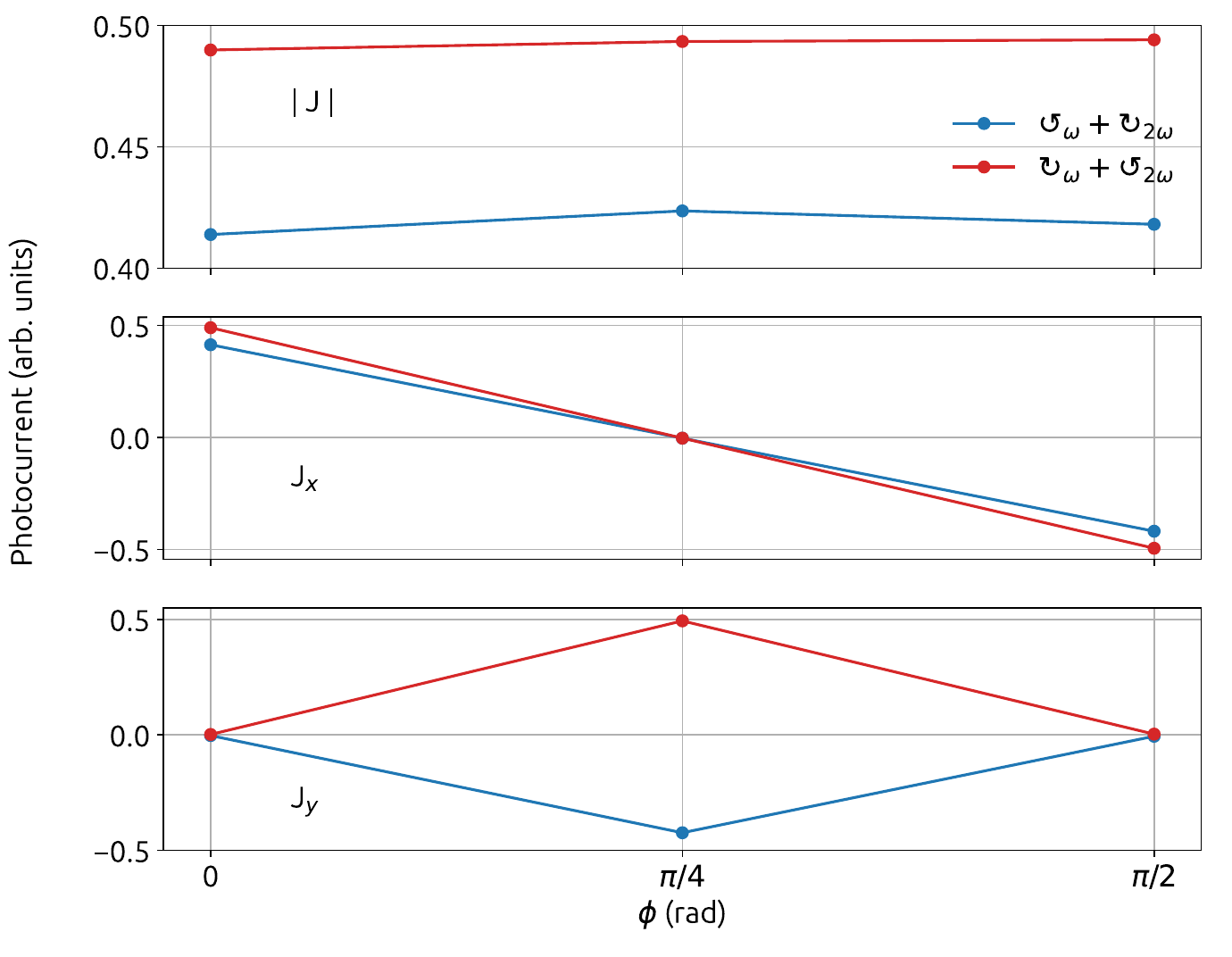}
	\caption{Variation in the photocurrent and its components in an inversion-symmetric Weyl semimetal, described by Eq.~\eqref{eq:trb4} with respect to the sub-cycle phase between $\omega-2\omega$ fields. 
		Eight-cycle laser pulse with its fundamental wavelength of 3.2 $\mu$m and  intensity  of $0.5\times10^{10}$ W/cm$^2$ is used.}
	\label{fig:fig4.11}
\end{figure}

The total vector potential of $\omega-2\omega$ bicircular counter-rotating laser pulses  in $x-y$ plane 
is written as 
\begin{equation}
	\mathbf{A} = A_0 f(t) [\{\cos(\omega t + \phi) + \cos(2\omega t)\}\hat{e}_x  + \mathcal{R}\{\sin(\omega t + \phi) - \sin(2\omega t)\}\hat{e}_y]. 
\end{equation}
Here,
$\phi$ is the sub-cycle phase between the two color. 
$\mathcal{R} = +1$ represents $\omega(2 \omega)$ field is left(right)-handed, whereas 
$\mathcal{R} = -1$ means  $\omega(2 \omega)$ field is right(left)-handed. 
To compare the results for the single and two colors on equal footing, the intensity of the two color is reduced to half of the single setup. 

From the above figure, it is evident that the photocurrent generated by the two-color bicircular counter-rotating laser pulses and single-color circular pulse is of the same order as evident from the figure. 
The photocurrent is insensitive to the sub-cycle phase, whereas the magnitude of the photocurrent 
slightly depends on the combination of the helicities of the $\omega-2\omega$  fields, which is consistent with 
our previous observation of the residual electronic populations in the conduction band~\citep{bharti2023weyl}.
In addition,  our results  show that similar control over the photocurrent's direction 
can be achieved by tuning the sub-cycle phase. 

\section{Linearly Polarized Laser-Driven Photocurrent}

\subsection{Photocurrent in Weyl Semimetals}

We start our discussion by analyzing the sensitivity of the photocurrent with respect to $\theta$ for an inversion-symmetric WSM, as shown in Fig.~\ref{fig:fig4.12}(c). 
It is evident that there is nonzero photocurrent along the laser polarization for $\theta = n \pi$ with $n$ as an integer. 
The origin of the finite photocurrent can be attributed to the asymmetric laser waveform [see Fig.~\ref{fig:fig4.12}(a)], 
which results in an asymmetric residual electronic population in the conduction band, i.e.,  $\left[ \rho(\mathbf{k}) - \rho(-\mathbf{k}) \right] \neq 0$. 
The asymmetry of the laser waveform can be flipped by changing  $\theta = 0$ or $2 \pi$ to $\pi$. 
As a result,  the magnitude of the photocurrent tunes from negative to positive by changing the collinear configuration from parallel to antiparallel.
In addition, the waveform asymmetry reduces as  $\theta$ deviates from the collinear configuration, such as at 
$\theta=\pi/4$, which leads to the reduction of the photocurrent. 
The orthogonal configuration of the $\omega - 2 \omega$ pulses renders 
a symmetric waveform along the $x$ direction, which results in zero photocurrent. 
Thus, the underlying mechanism for the photocurrent generation is the asymmetric laser waveform, which is imprinted in the excitation processes, leading to an asymmetric electronic population.
Interestingly, there is also a small photocurrent along the $y$ direction for $\theta \neq n \pi$ as reflected from the figure.
The photocurrent along the $y$ direction arises due to the asymmetric waveform in the $y$ direction [see inset of the bottom panel of Fig.~\ref{fig:fig4.12}(a)].
Thus, the analysis of Fig.~\ref{fig:fig4.12}(b) establishes that the magnitude and direction of the 
photocurrent is tunable with  $\theta$.

\begin{figure}[!h]
	\centering
	\includegraphics[width=\linewidth]{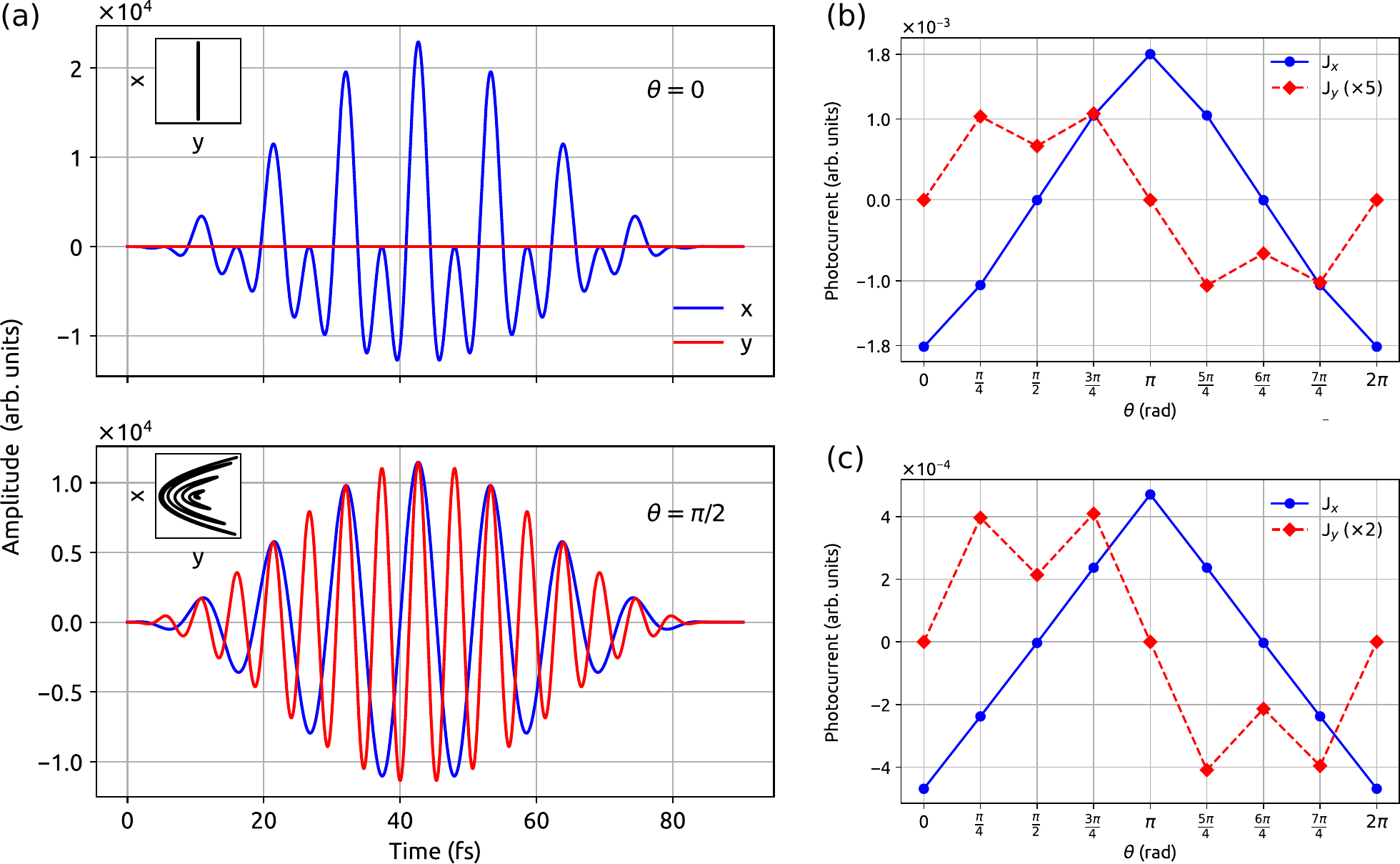}
	\caption{ (a)  Amplitude of the vector potential of the two pulses when both pulses are in collinear (top panel) and orthogonal (bottom panel) configurations. 
		The Lissajous curve of the total vector potential in the $xy$ plane is shown in respective insets. 
		(b) Variations in the photocurrent with respect to $\theta$ in an inversion-symmetric Weyl semimetal. (c) Same as (b) for
		an inversion-broken Weyl semimetal. 
		Wavelength of the $\omega$ pulse is 3.2 $\mu$m with pulse length $\sim$ 100 fs. 
		Laser intensity equal to $5\times10^{10}$ W/cm$^2$  with $\mathcal{R}  = 1$ is used for both 
		inversion-symmetric and inversion-broken Weyl semimetals.} \label{fig:fig4.12}
\end{figure}

Figure~\ref{fig:fig4.12}(c) presents the variation in the photocurrent with $\theta$ for an IB WSM. 
The collinear configuration results in a finite photocurrent in this case. 
This observation, together with the sensitivity of the photocurrent's magnitude for other $\theta$, 
exhibits similar behavior as in the case of the inversion-symmetric WSM discussed above.
In addition, the trend in the photocurrent is universal in the sense that it does not depend on the inversion symmetry of the material.
Thus, Figs.~\ref{fig:fig4.12}(b) and \ref{fig:fig4.12}(c) establish that the $\omega - 2 \omega$ pulse setup 
generates a finite photocurrent, which emanates by imprinting asymmetry of the laser waveform on the electronic population. 
The overall magnitude of the photocurrent is tunable with  $\theta$ and critically depends on the material and intensity employed.
So far, we have investigated photocurrent generation within the $\omega - 2 \omega$ setup with identical intensity of the laser pulses. 
At this junction, it is worth wondering how the photocurrent changes when the intensity ratio of the two pulses varies.

\subsection{Role of the Intensity Ratio}
To address this pertinent issue, we will consider two  cases where the waveform's asymmetry 
is extremum, i.e., $\theta = 0$  (collinear) and $\pi/2$ (orthogonal) configurations. 
Moreover, an inversion-symmetric WSM is chosen for further discussion from here onward as both inversion-symmetric and IB WSMs behave similarly. 
Figure~\ref{fig:fig4.13} presents the sensitivity of the photocurrent as a function of the amplitude ratio  
$\mathcal{R}$ [see Eq.~\eqref{eq:laser2}]. 
The residual electronic population in the conduction band, after the end of the laser pulse,  
exhibits asymmetry along $k_x=0$ and the asymmetry increase with  $\mathcal{R}$ [see Fig.~\ref{fig:fig4.14}].

\begin{figure}[!h]
	\centering
	\includegraphics[width=\linewidth]{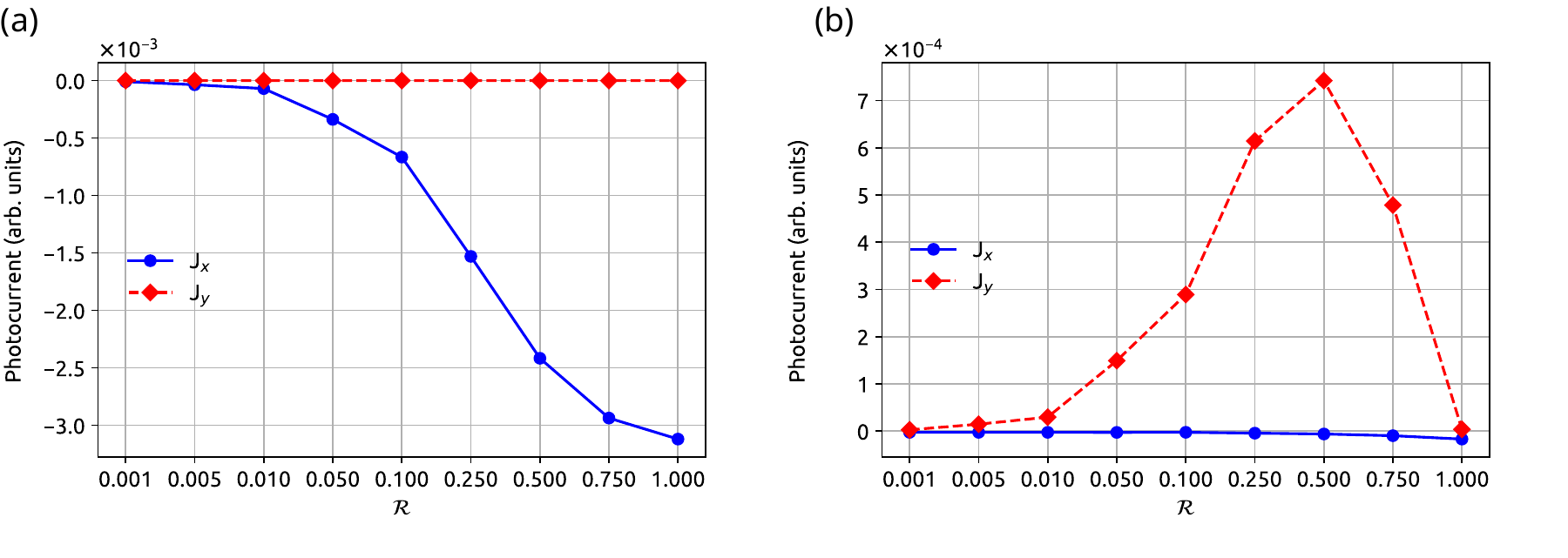}
	\caption{Variation in photocurrent with the amplitude ratio  ($ \mathcal{R}$) 
		of the $2\omega$ pulse with respect to the $\omega$ pulse for an 
		inversion-symmetric Weyl semimetal in the  (a) collinear and (b) orthogonal configurations. 
		Laser parameters are the same as in Fig.~\ref{fig:fig4.12} with intensity of the $\omega$ pulse as $10^{11}$ W/cm$^2$.} \label{fig:fig4.13}
\end{figure}

Consequently, the photocurrent becomes significant as $\mathcal{R}$ is increased, as evident from Fig.~\ref{fig:fig4.13}(a). 
Further, it is notable that the presence of a weaker $2 \omega$ pulse is enough to
generate photocurrent of the same order of magnitude.
Also, in comparison  to $\mathcal{R} = 1$, there is an appreciable photocurrent even when the intensity of the 
$2\omega$ field is  one-tenth of the $\omega$ field. 
In general, the generation of the $2\omega$  pulse from $\omega$, say using a beta barium borate crystal, reduces the intensity of the $2\omega$  pulse drastically in typical experimental setups. 
Thus, the presence of a 
weaker $2\omega$ pulse in the $\omega - 2\omega$  setup in our approach 
is sufficient for the photocurrent generation to provide flexibility to the experimentalist.
Note that the spatial distribution of the residual population resembles the laser waveform as  
 $\mathbf{k}$ alters to $\mathbf{k}_{t} = \mathbf{k} + \mathbf{A}(t)$.  
Thus, the laser waveform controls the asymmetry of  the residual population and therefore photocurrent.

\begin{figure}[]
	\centering
	\includegraphics[width=\linewidth]{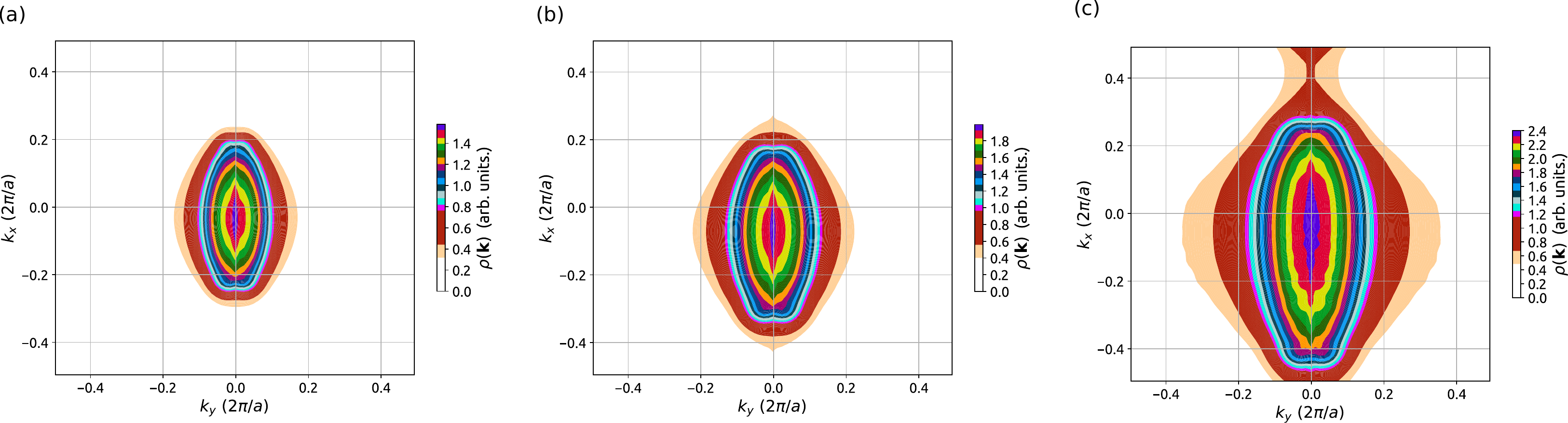}
	\caption{Residual electronic population in the conduction band after the end of the laser pulse for 
	 (a)  $\mathcal{R} = $ 0.1, 
		(b) $\mathcal{R} = $ 0.25 and (c) $\mathcal{R} = $ 1.0. 
		The population along $k_{z}$ direction is integrated as the $\omega - 2\omega$ field setup lies on the $k_{x} - k_{y}$ plane. Note that the scale of the color bar is different in all cases. Rest of the parameters are same as Fig.~\ref{fig:fig4.12}.}
	\label{fig:fig4.14}
\end{figure}

Analysis of Fig.~\ref{fig:fig4.12}(b) indicates that there is insignificant photocurrent along the $y$ direction  
when $\theta = \pi/2$ for $\mathcal{R} = 1$.
However, photocurrent along the $y$ direction can be boosted by an order of magnitude by tuning $\mathcal{R} = 1$ to 0.5 as reflected from Fig.~\ref{fig:fig4.13}(b). 
Similar to the collinear configuration, the residual population in the 
conduction band is fairly symmetric along $k_x = 0$ for $\mathcal{R}=0.1$. 
On the other hand, the population is asymmetric along $k_y=0$, which results in photocurrent along the $y$ direction 
[see Fig.~\ref{fig:fig4.15}].
Moreover, the orthogonal configuration exhibits the non-monotonic behavior of the photocurrent, which is in contrast to the observation in the collinear configuration.
The nonmonotonic behavior can be attributed to the laser-driven  nonperturbative  electron dynamics in the conduction bands. 
A further increase in the intensity, by increasing $\mathcal{R}$, leads to the sign change of $\left[\rho(\mathbf{k})-\rho(-\mathbf{k})\right]$, which results in the reversal of the photocurrent's direction. Our observation about the direction reversal with intensity is 
consistent with previous reports~\citep{wachter2015controlling,wismer2016strong,higuchi2017light,zhang2022bidirectional}.
The same observations hold true for an IB WSM qualitatively.  
Thus, $\mathcal{R}$ adds another knob to tune the photocurrent in WSMs along with $\theta$. 

\begin{figure}[!h]
	\centering
	\includegraphics[width=\linewidth]{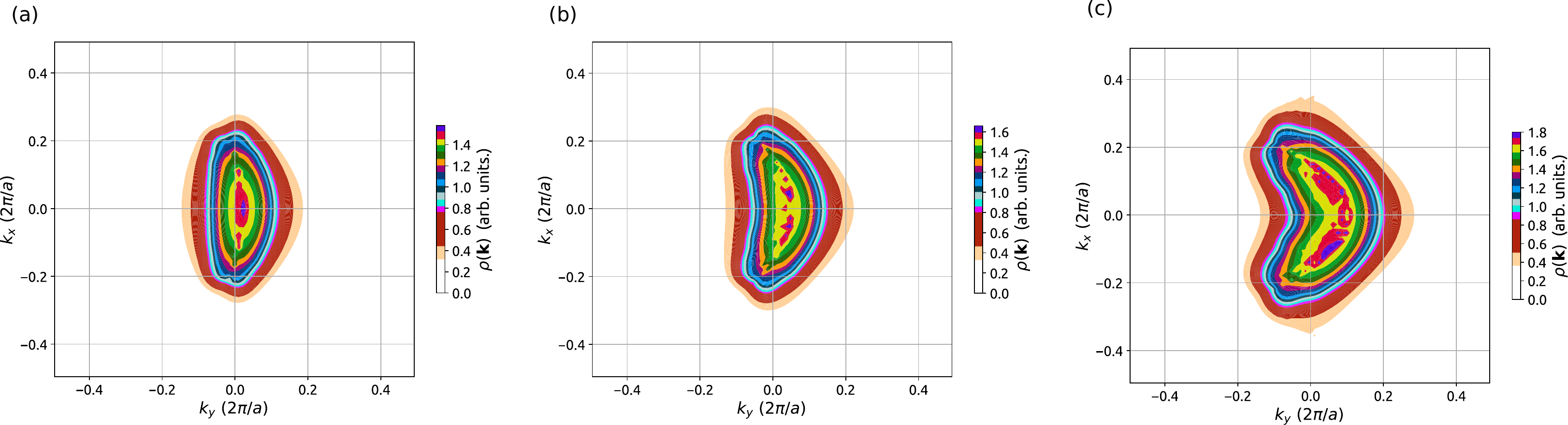}
	\caption{Same as  Fig.~\ref{fig:fig4.14}  for  the orthogonal configuration.}
	\label{fig:fig4.15}
\end{figure}

\subsection{Photocurrent in Two-Dimensional Materials} 

Until now, we have observed that the photocurrent is insensitive to the symmetries and topology  
of the WSMs 
and exhibits similarities with $\theta$  and $\mathcal{R}$ variations. 
This conclusion raises a crucial question about the universality of our observation. 
To address this important question, we transit from three-dimensional topological to two-dimensional trivial materials, namely inversion-symmetric graphene and inversion-broken molybdenum disulfide (MoS$_2$). 
These two-dimensional materials have been the center of exploration for photodetection and other optoelectronic applications in recent years~\citep{koppens2014photodetectors,agarwal2023, mak2016photonics, bussolotti2018roadmap}. 

\begin{figure}[!h]
	\centering
	\includegraphics[width=\linewidth]{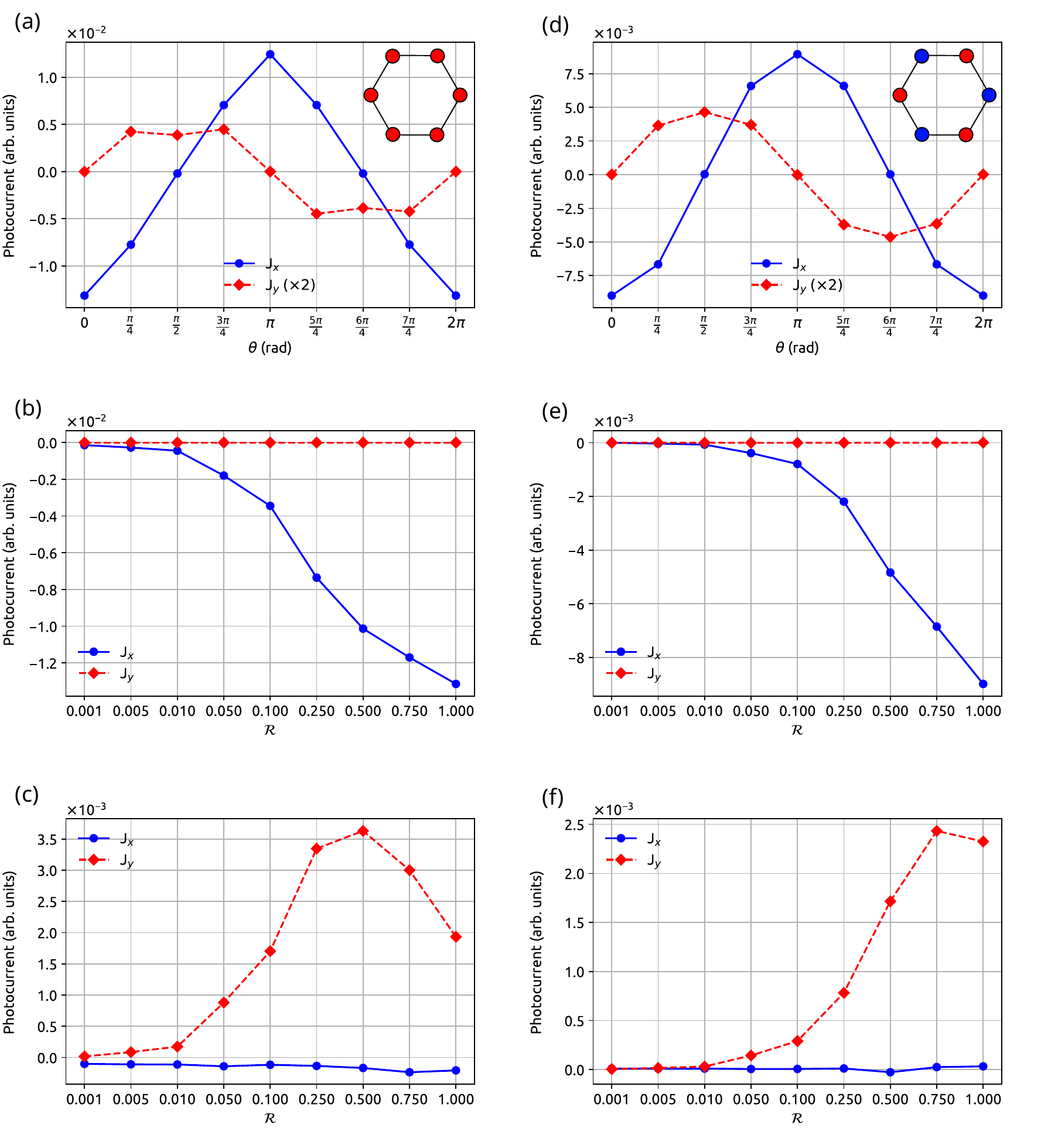}
	\caption{(a) Sensitivity of  the photocurrent in an inversion-symmetric graphene with respect to 
		(a) $\theta$, the amplitude ratio ($ \mathcal{R}$)  in (b) collinear and (c) orthogonal configurations of the $\omega-2\omega$ setup. Panels (d), (e), and (f) are the same as (a), (b), and (c) for MoS$_2$, respectively. 
		Laser parameters are the same as in Fig.~\ref{fig:fig4.12} with intensity of the $\omega$ pulse as $10^{11}$ W/cm$^2$.}
	\label{fig:fig4.16}
\end{figure}

There is a finite photocurrent along the $x$ direction for graphene in the collinear configuration 
($\theta = 0, \pi$ and $2 \pi$)  as evident from Fig.~\ref{fig:fig4.16}(a). 
On the other hand, the photocurrent's magnitude reduces as $\theta$ changes and reaches a minimum  
for  $\theta = \pi/2$ and $3\pi/2$. 
A similar trend in the photocurrent with $\theta$  is visible for MoS$_2$ [see Fig.~\ref{fig:fig4.16}(d)].
Interestingly, the photocurrent's magnitude in MoS$_2$ is reduced in comparison to graphene, 
which can be attributed to the finite band gap of MoS$_2$.
Overall, other features of the photocurrent along $x$ and $y$ directions remain robust with a variation in $\theta$.

Sensitivity of the photocurrent with $\mathcal{R}$ 
is presented in Figs.~\ref{fig:fig4.16}(b) and ~\ref{fig:fig4.16}(e) for graphene and MoS$_2$, respectively. 
It is noted that the small presence of the $2 \omega$ component in collinear configuration is sufficient to generate photocurrent along the $x$ direction. 
On the other hand, the value of $\mathcal{R}$ crucially depends on the material's nature to maximize the photocurrent along the $y$ direction in the orthogonal configuration, as evident 
from Figs.~\ref{fig:fig4.16}(c) and ~\ref{fig:fig4.16}(f) for graphene and MoS$_2$, respectively.  
Observations of Figs.~\ref{fig:fig4.12} - \ref{fig:fig4.16} confirm  that the  photocurrent can be tuned by varying  the laser's parameters in the $\omega-2\omega$ setup irrespective of the materials and their underlying symmetry, which  
establishes the universality of our approach.

\subsection{Role of Time-Delay between $ \omega - 2 \omega$  Pulses} 
In this subsection,  we investigate how the photocurrent can be additionally controlled by introducing a relative time delay between $\omega$ and $2 \omega$ pulses. So far, we have considered zero delays between the pulses.  
Figure~\ref{fig:fig4.17}(a) presents the variation in the photocurrent as a function of the delay for an inversion-symmetric WSM in a collinear configuration ($\theta = 0$) 
with $\mathcal{R} = 1$ at which the photocurrent is maximum [see Fig.~\ref{fig:fig4.13}(a)].  
The photocurrent's amplitude along the $x$ direction can be modulated from negative to positive value by changing the relative delay in units of a quarter of the fundamental ($\omega$ pulse) time period. 
The modulation in the photocurrent can be attributed to the change in the asymmetry of the laser waveform caused by the time delay.  
Figure~\ref{fig:fig4.17}(b) shows a similar control over the photocurrent's amplitude along the $y$ direction  
in the  orthogonal  configuration ($\theta = \pi/2$) with $\mathcal{R} = 0.5$ at which the photocurrent is maximum [see Fig.~\ref{fig:fig4.13}(b)].
Thus, the photocurrent can be modulated by merely introducing the delay, which adds another convenient control knob to tailor photocurrent in materials.   

\begin{figure}[!h]
	\centering
	\includegraphics[width=\linewidth]{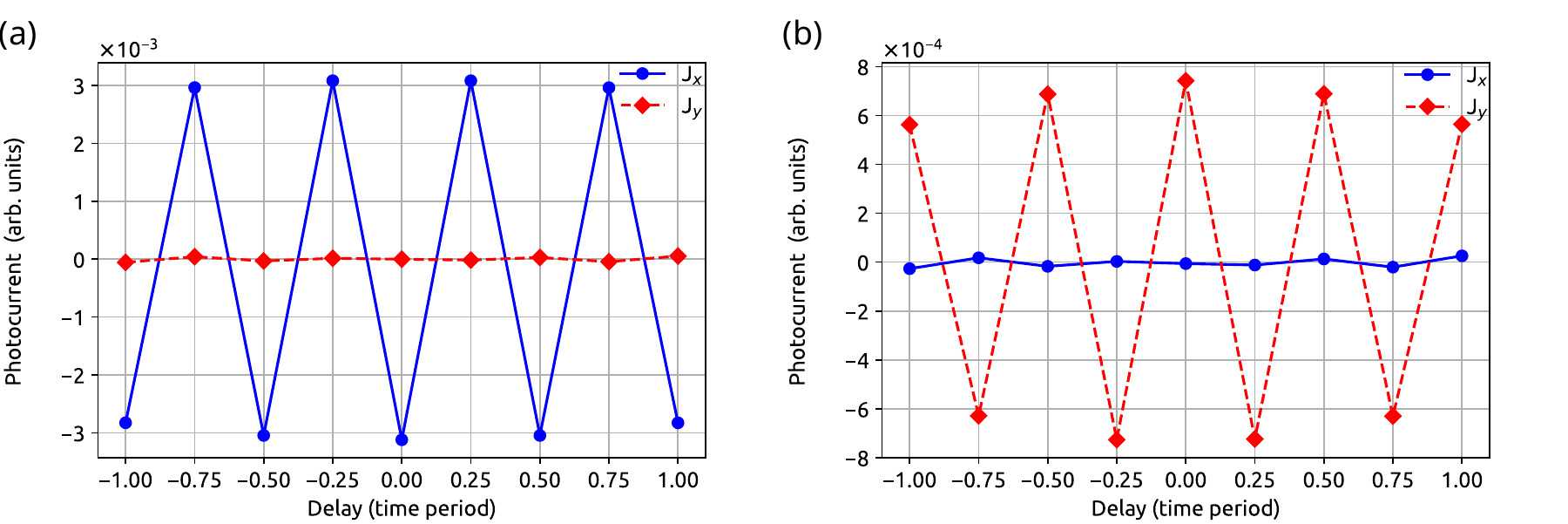}
	\caption{Effect of the time-delay between  of $\omega$ and $2\omega$ pulses on the photocurrent in  
		an inversion-symmetric Weyl semimetal. 
		The $\omega$ and $2\omega$ pulses are in (a) collinear and (b) orthogonal configurations 
		with $ \mathcal{R} = 1$ and 0.5, respectively. 
		The laser parameters are the same as in Fig.~\ref{fig:fig4.12}.} \label{fig:fig4.17}
\end{figure}

\subsection{Role of the Laser's Intensity} 

So far, we have limited our discussion on the photocurrent for a fixed intensity of the $\omega$ pulse.  
At this point, it is worth knowing how photocurrent scales with the intensity.
Figure~\ref{fig:fig4.18} discusses how the photocurrent scales with the intensity   
in collinear and orthogonal configurations. 
The intensity of both $\omega$ and $2\omega$ pulses 
are varied in a fixed ratio, which corresponds to the maximum photocurrent as in Fig.~\ref{fig:fig4.13}.
It is evident that the photocurrent becomes appreciable at $10^{10}$ W/cm$^2$ and exhibits nonmonotonic behavior 
in the collinear configuration as shown in Fig.~\ref{fig:fig4.18}(a). 
The photocurrent peaks at $10^{11}$ W/cm$^2$ and starts decreasing with an increase in intensity, which  
results in the reversal of the photocurrent's direction as discussed above. 
The asymmetry in the residual electronic population increases along $k_{x}$ as the intensity increases, which leads to an increase in photocurrent. However, after reaching  a maximum, the asymmetry starts reducing as the    
residual  population migrates from positive  to negative $k_{x}$ region and vice versa, 
which results in the reduction of the photocurrent's magnitude, and can be understood by analyzing
the residual population as shown in Fig.~\ref{fig:fig4.19}. 

\begin{figure}[!h]
	\centering
	\includegraphics[width=\linewidth]{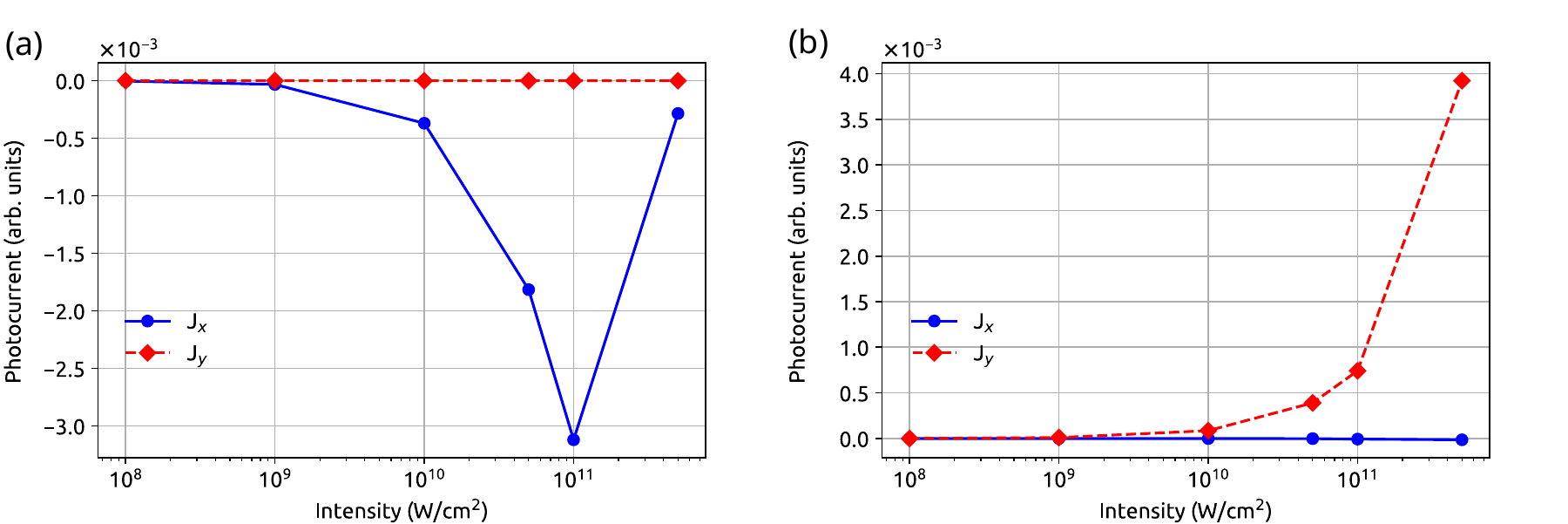}
	\caption{Scaling of the photocurrent  with the intensity of the 
		$\omega$ and $2\omega$ pulses in (a) collinear and (b) orthogonal configurations 
		with $ \mathcal{R} = 1$ and 0.5, respectively. 
		The laser parameters are the same as in Fig.~\ref{fig:fig4.12} for an inversion-symmetric Weyl semimetal.}
	\label{fig:fig4.18}
\end{figure}

The  minimum intensity required to generate photocurrent  and its nonmonotonic nature  indicate that the generated  photocurrent is nonperturbative in nature. 
This observation is consistent with an earlier report for graphene exposed to a few-cycle phase-stabilized laser pulse~\citep{higuchi2017light,zhang2022bidirectional} and WSM~\citep{bharti2023tailor}.
In contrast,  the orthogonal configuration at the same intensity yields minuscule photocurrent as reflected from Fig.~\ref{fig:fig4.18}(b). 
Photocurrent in the orthogonal  configuration increases monotonically with the laser's intensity studied.
The residual populations in the conduction band  in the parallel and orthogonal configurations for different laser's intensity are, respectively,  presented in  Figs.~\ref{fig:fig4.19} and~\ref{fig:fig4.20}, which can be analyzed in a similar fashion as discussed above.
Note that the maxima and directional reversal of photocurrent will appear at intensity different from Fig.~\ref{fig:fig4.18}, 
if we choose a different value of $\mathcal{R}$ for any configuration.
Nonetheless, above a threshold intensity, the $\omega-2\omega$ field can produce photocurrent which can be optimized by tuning the ratio of amplitude.

\begin{figure}[!h]
	\centering
	\includegraphics[width=\linewidth]{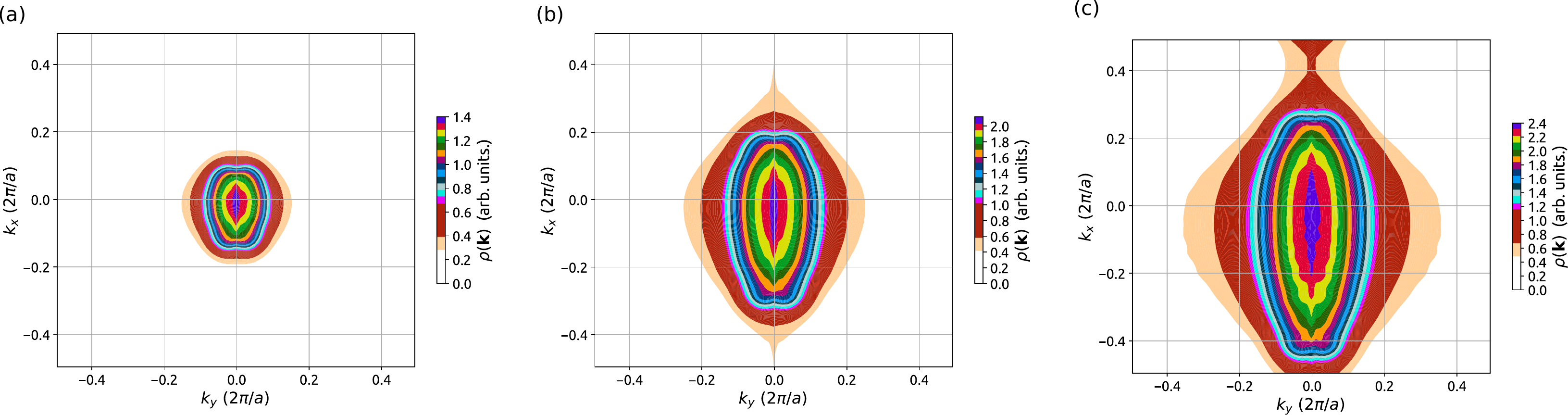}
	\caption{Residual electronic population in the conduction band after the end of laser pulse for different  intensity of the $\omega$ field in the parallel configuration: (a) $10^{10}$, (b) $5\times10^{10}$  and (c) $10^{11}$ W/cm$^2$. The population along $k_{z}$ direction is integrated as the $\omega - 2\omega$ field setup lies  on the $k_{x} - k_{y}$ plane. Note that the scale of the color bar is different in all cases. Rest of the parameters are same as Fig.~\ref{fig:fig4.12}.}
	\label{fig:fig4.19}
\end{figure}

\begin{figure}[!h]
	\centering
	\includegraphics[width=\linewidth]{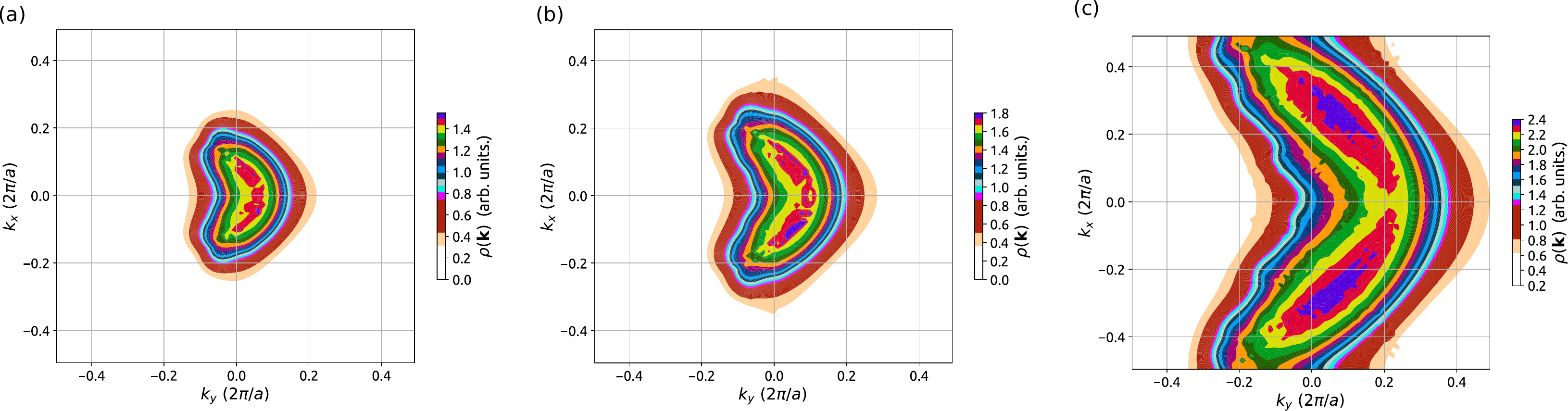}
	\caption{Same as  Fig.~\ref{fig:fig4.19} for different intensity of the $\omega$ field for orthogonal configuration.}
	\label{fig:fig4.20}
\end{figure}

\section{Conclusion}
In summary, we introduce a robust and universal  method to generate photocurrent in both 
inversion-symmetric and IB WSMs using a single-color circularly polarized light.  
Both WSMs have degenerate Weyl nodes at the Fermi level. 
We unequivocally demonstrate   that phase stabilization is  not a prerequisite to generate  photocurrent in 
both types of WSMs as the generated photocurrent is insensitive to the phase of the laser pulse. 
Photocurrent in an inversion-symmetric  WSM is sensitive to the helicity of the laser as the left-handed circularly polarized laser yields more photocurrent in comparison to the right-handed laser. 
Moreover, the components of the photocurrent in an  inversion-symmetric  WSM are also 
sensitive to the helicity, whereas 
only the $y$ component exhibits sensitivity in the case of an IB WSM. 
In addition, the strength of the photocurrent reduces as the ellipticity of the laser changes from circular to linear. 
It is anticipated that the measurement of the photocurrent can quantify the coupling of the spin-angular momentum of light with nonlinear band dispersion in WSMs. 

Further, we present a universal method to generate and tailor photocurrent in normal and topological materials, namely graphene, MoS$_2$, and WSMs using a set of linearly polarized light.
A pair of linearly polarized pulses comprised of $\omega$ and $2\omega$ frequencies  can produce a highly asymmetric laser 
waveform, which steers electrons on an attosecond timescale to generate photocurrent in materials with trivial and 
nontrivial topology. 
In this regard, we find that the presence of a comparatively weak $2\omega$ pulse is sufficient for the photocurrent generation.
Interestingly, the generated photocurrent can be tailored by simply varying the laser parameters of the $\omega-2\omega$ setup. 
The photocurrent is found to be sensitive to the variation in the angle between the polarization directions, amplitude ratio, and relative time delay of the two pulses.
Even orthogonal linearly polarized pulses drive asymmetric population for a certain amplitude ratio and thus give 
rise to comparable photocurrent as the collinear pulses. 
Our proposed methods showcase various ways to tailor laser waveform to generate photocurrent for optoelectronic and photodetection applications in a nonmaterial specific manner.
%-- thus a universal way.
\end{sloppypar}

\cleardoublepage
\chapter{High-Harmonic Spectroscopy of Weyl Semimetals}\label{chp:chapter5}
\begin{sloppypar}
In the previous chapters, we have explored intense laser-driven residual electronic population in the 
conduction band and resultant photocurrent in solids. 
Present chapter focuses on another intriguing aspect of an intense laser-driven process, namely nonperturbative high-harmonic generation (HHG), which is a nonlinear  frequency up-conversion process
~\citep{ghimire2011observation, ghimire2019high}. 
Numerous static and dynamic properties of solids have been probed  by analyzing 
the emitted radiation during HHG~\citep{zaks2012experimental,schubert2014sub,luu2015extreme,hohenleutner2015real,langer2018lightwave, pattanayak2020influence,  mrudul2020high,mrudul2021high,  mrudul2021light,  neufeld2021light}.   

In recent years, topological materials have turned out to be the center of attention for HHG~\citep{bauer2018high,reimann2018subcycle, silva2019topological,chacon2020circular, bai2021high,  dantas2021nonperturbative}. 
It is experimentally found that the bulk  and the topological surface 
play different roles during HHG from a topological insulator~\citep{schmid2021tunable}. 
The interplay of  the time-reversal symmetry protection and the spin-orbit coupling in a topological insulator leads to anomalous dependence of harmonic yield on the polarization of the driving laser~\citep{baykusheva2021all}. 
Berry curvature plays an important role  in determining the behavior of high-harmonic spectra in both  cases. 
In three-dimensional Dirac semimetal, coherent dynamics of the Dirac electrons plays the central role in HHG ~\citep{kovalev2020non, cheng2020efficient}.  Moreover, it has been reported that the 
nonlinear responses of the three- and two-dimensional Dirac semimetals are significantly different~\citep{lim2020efficient}.  
In all cases, time-reversal symmetry (TRS) is inherently preserved in topological insulators and Dirac semimetal.  
Therefore, it is natural to envision exploring how the breaking of the TRS affects HHG from topological materials. 

It is known that the energy band dispersion is linear in the vicinity of the Weyl nodes at which valence and 
conduction bands touch each other.
Ideally, the band dispersion forms an upright cone at these nodes, which are energy degenerate at the Fermi energy.  
This kind of WSM is known as the type-I~\citep{yan2017topological}. 
However, Weyl nodes in real materials usually appear away from the Fermi level 
and the energy dispersion does not form an upright cone but is tilted in nature~\citep{lv2021experimental}. 
The earliest discovery of transition-metal monopnictides WSM, 
among others, has Weyl nodes that exhibit tilted cones away from the Fermi level~\citep{xu2015discovery,xu2015discovery1,lv2015experimental}. 
Moreover, significant tilting of the Weyl cones can alter the sign of the Weyl fermions' velocity near the nodes. 
In this scenario,  WSM is in the type-II phase, which is different from the type-I phase with small or no tilting~\citep{soluyanov2015type,deng2016experimental,li2017evidence,jiang2024revealing}.

The first part of this chapter focuses on addressing some crucial questions such as how TRS breaking and resultant modifications in Berry curvature affect HHG  and how the separations of the Weyl points  influence HHG in WSM. In the following, we  demonstrate that non-zero Berry curvature in TRS-broken WSM leads to anomalous current in a direction perpendicular to the electric field and anomalous odd harmonics -- analogous to the anomalous Hall effect. Moreover, we  discuss 
that the directions of the emitted anomalous odd harmonics are related to the nature of the 
Berry curvature's components. 
The appearance of the anomalous odd harmonics allows us to probe non-trivial topology of the TRS-broken WSM by measuring the polarization of the emitted anomalous odd harmonics.
Recently,  HHG from an inversion-symmetry broken WSM was explored experimentally in which linearly  polarized pulse leads to the generation of even harmonics, related to non-zero Berry curvature~\citep{lv2021high}.  
Present findings are in contrast to previously reported works where Berry curvature mediated  
anomalous electron's velocity leads to the generation of even harmonics~\citep{schubert2014sub, liu2017high, hohenleutner2015real, luu2018measurement}.
Consequently, we find that the HHG spectra are sensitive to the intricate geometrical properties of the nontrivial Berry curvature and the Weyl nodes in WSM. 
In addition, the emitted harmonics from WSM can be a versatile source of chiral light with tunable   
energy.  

Despite preliminary research,  HHG from WSM with realistic  environments  is relatively scarce, e.g., 
it is not obvious how  HHG from WSM is affected when WSM transits from type-I to type-II phase or how the tilt and energy splitting of the Weyl nodes affect HHG. 
The second part of this chapter focuses on the systematic study of HHG from WSM in different realistic situations. 
Two symmetry classes of WSM, namely, time-reversal broken (TRB) 
WSM and inversion-broken (IB) WSM are considered.
We start our discussion with  WSM, in which upright Weyl nodes (cones) are energetically degenerate at Fermi energy.  
It is demonstrated that the finite tilt of the Weyl nodes leads to gradual enhancement in the harmonic yield.
Enhancement of as large as two orders of magnitude in the yield is achievable when WSM transits from type I to type II. 
However, finite energy splitting between the Weyl nodes with zero tilting results in a nominal boost in the yield. 
Moreover, significant splitting between the Weyl nodes in the type-II phase facilitates the generation of even-order harmonics.

The earlier realizations of the WSMs, as considered above, have a  unit topological charge,  $n =1$ ~\citep{lv2015experimental,xu2015discovery,morali2019fermi,liu2019magnetic,belopolski2019discovery}. 
The sign of the  Berry-curvature  monopole determines the sign of the topological charge of the  WSMs. 
The magnitude of the topological charge is $n = \int_{\textrm{BZ}} \nabla_{\mathbf{k}}\cdot\Omega(\mathbf{k})~ d\mathbf{k}$. 
A class of  WSMs, multi-Weyl semimetals (m-WSMs), with  
topological charges higher than one came into existence in recent years.  
The m-WSMs with topological charges two and three are, respectively,  manifested by 
quadratic and cubic energy dispersions along particular directions, whereas they exhibit linear dispersion along other directions~\citep {xu2011chern,fang2012multi,huang2016new, liu2017predicted}. 
This contrasts with ``conventional'' WSMs with   topological unit charges, 
which exhibit isotropic linear energy dispersion. 
Weyl points in m-WSMs can be formed by 
annihilating Weyl points of the same chirality in WSMs with lower topological charges~\citep{roy2022non}. 
The upper bound on the topological charge in m-WSMs is limited to three by 
certain crystalline symmetries~\citep{ xu2011chern,fang2012multi,yang2014classification}.

The m-WSMs  manifest unique quantum response due to the higher topological charges and resultant anisotropic energy dispersions~\citep{nag2020thermoelectric}. 
It has been shown that the chirality accumulation is possible without a magnetic field 
due to higher topological charges~\citep{huang2017topological}.
A distinct chiral anomaly-induced nonlinear Hall effect, associated with the 
different topological charges in m-WSMs, has been discussed~\citep{nandy2021chiral}. 
It has been found that the anomalous Hall current  in the nonperturbative regime 
saturates, and the anomalous Hall conductivity scales linearly with the topological charges~{\citep{nandy2019generalized,dantas2021nonperturbative}.  
		
In the third and the last part of this chapter, we investigate how the nonlinear optical responses in m-WSMs  
are sensitive to the topological charges for different laser intensities.   
In the following, we show that  the  current, parallel to the laser's polarization, and 
the anomalous current, perpendicular to the laser's polarization, exhibit distinct behavior as a function 
of the laser's intensity for different topological charges. 
In addition, we demonstrate that the  current is anisotropic in nature, which strongly depends 
on the direction of the laser polarization. 
The  parallel current increases linearly below a critical laser's intensity. 
However, above a critical intensity, the parallel current 
shows a nonlinear increment for different topological charges, and it starts saturating after another critical intensity. 
On the other hand, the anomalous current displays linear behavior for relatively larger intensity.  
The Berry-curvature driven anomalous current approaches  similar values for 
different topological charges at large intensity limits. 
We employ high-harmonic spectroscopy to probe the distinct behaviors of the normal and anomalous currents in m-WSMs. 
It has been found that the topological charge drastically alters the harmonics' yield and energy cutoff.

%\section{Theoretical Methodology}

\section{Results And Discussion}

\subsection{Generation of Anomalous Odd-Order  High-Harmonics}

\begin{figure}
	\centering
	\includegraphics[width=\linewidth]{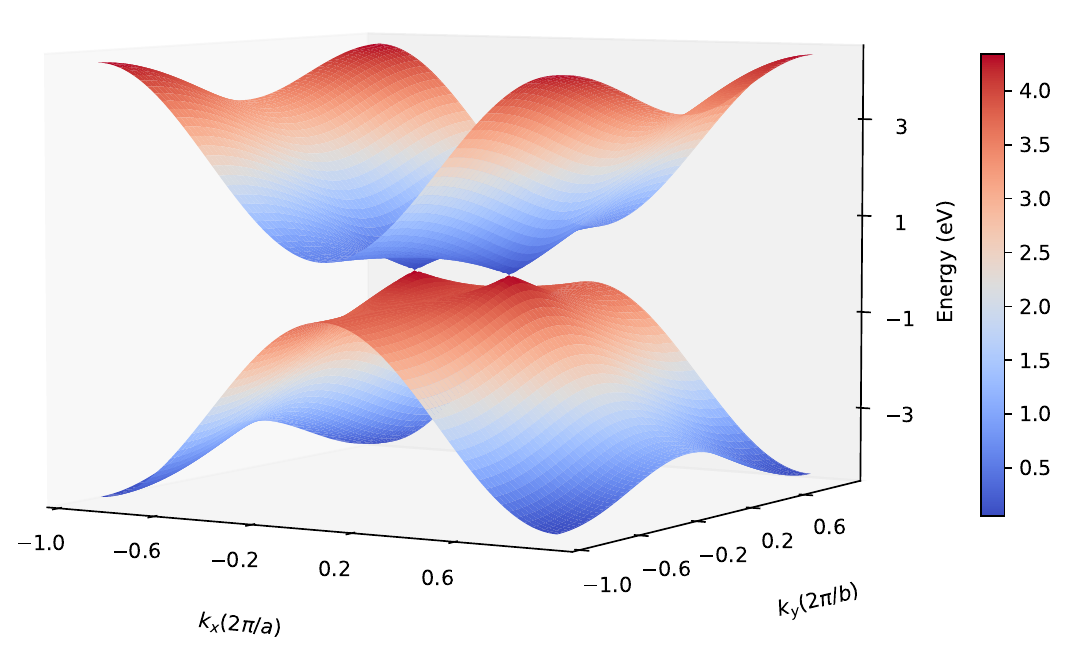}
	\caption{Energy band structure of the time-reversal symmetry broken 
	Weyl semimetal. Two Weyl points are present in the band structure at 
	($k_{x},  k_{y}, k_{z} = \pm 0.2, 0, 0$) reciprocal lattice units, 
	and energy dispersion is linear around these points.}
	\label{fig:fig5.0}
\end{figure}

Components of $\textbf{d}(\k)$ for the TRB WSM can be written as 
$[t_x \{ \cos(k_x a) - \cos(k_0 a)\} + t_y \{\cos(k_y b) -1\} + t_z \{\cos(k_z c) -1\}, 
t_y \sin(k_y b), t_z \sin(k_z c)]$ with 
$t_{x,y,z}$ as the hopping parameters and $a, b, c$ are lattice parameters. 
Here, we assume ferromagnetic WSM with tetragonal crystal structure, i.e., $a = b \neq c$~\citep{meng2019large}.  
$a = b = 3.437~\mathring{\text{A}}, c = 11.646 ~\mathring{\text{A}}$ and $t_{x} = 1.88~\textrm{eV}, t_{y} = 0.49~\textrm{eV}, 
t_{z} = 0.16~\textrm{eV}$ are considered. 
The parameters used here are in accordance with the ones used in Ref.~\citep{nematollahi2020topological}. 
After diagonalizing the  Hamiltonian, 
energy dispersion can be obtained as $\mathcal{E}_\pm(\mathbf{k}) = 
\pm \sqrt{d_1^2 + d_2^2 + d_3^2}$. We have considered $k_0$ = 0.2 rad/au.
The corresponding band-structure in $k_z=0$ plane is presented in Fig.~\ref{fig:fig5.0}.

\begin{figure}[!h]
	\includegraphics[width= \linewidth]{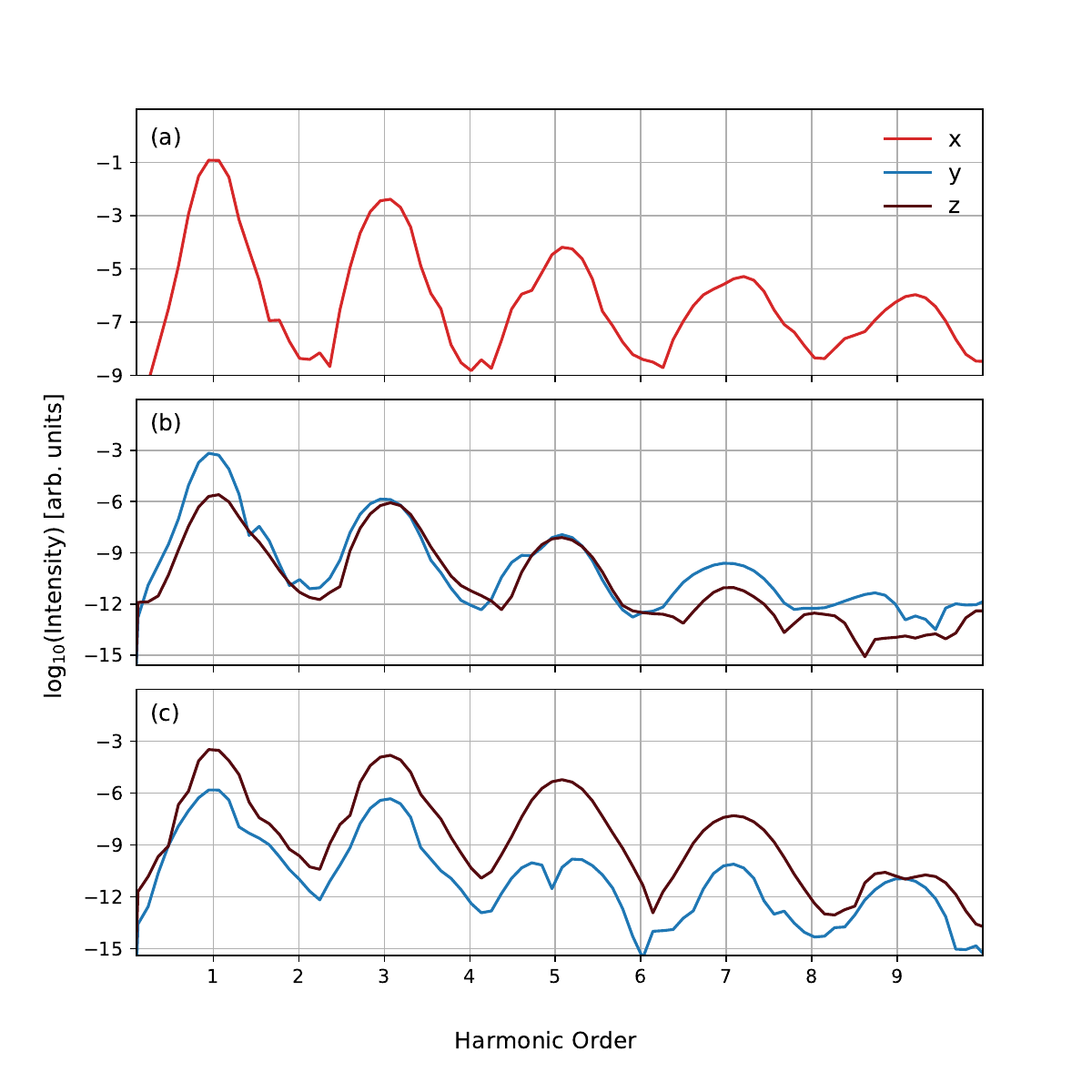}
	\caption{High-harmonic spectra corresponding to a linearly  polarized pulse. The pulse is polarized  along 
		(a) $x$, (b) $y$, and (c) $z$ directions. 
		Driving pulse is $\simeq$ 100 fs long with intensity $1 \times 10^{11}$ W/cm$^2$, and wavelength 3.2 $\mu$m.  Decoherence time of 1.5 fs is added phenomenologically.}  \label{fig:fig5.1}
\end{figure}

Figure~\ref{fig:fig5.1} presents high-harmonic spectra corresponding to linearly polarized pulse.
When the pulse is polarized along $x$ direction, odd harmonics are generated  along the laser polarization  as evident from Fig.~\ref{fig:fig5.1}(a). 
However, results become intriguing  when the pulse is  polarized along $y$ or $z$ direction. 
In both  cases, odd harmonics are generated along the laser polarization. 
Moreover, anomalous odd harmonics  along perpendicular directions are also generated.  
As reflected from Figs.~\ref{fig:fig5.1}(b) and \ref{fig:fig5.1}(c), when laser is polarized 
along $y$ or $z$ direction, anomalous odd harmonics along  
$z$ or $y$ direction, respectively, are generated.  
However, the yield of the anomalous  harmonics is relatively weaker  in comparison to the parallel  harmonics. 

\begin{figure}[h!]
	\includegraphics[width= 1.05 \linewidth]{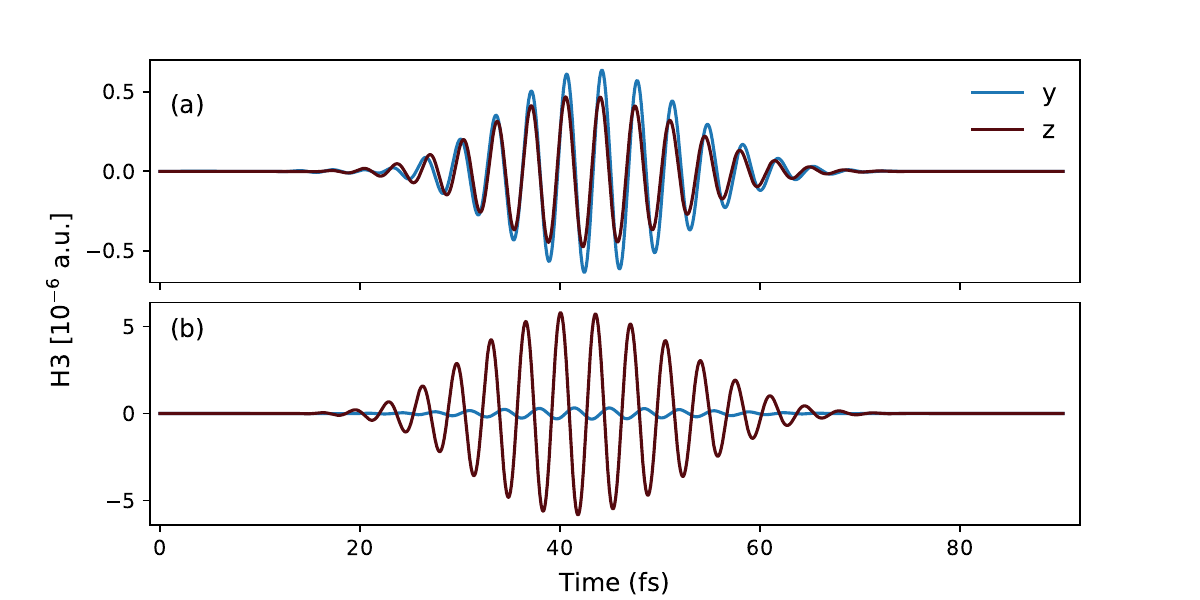}
	\caption{Third harmonic (H3) in time domain. 
		Driving laser pulse is linearly polarized along (a) $y$  and (b) $z$ directions. The driving pulse has the same parameters as in Fig.~\ref{fig:fig5.1}.}  \label{fig:fig5.2}
\end{figure}

To understand why linearly polarized driving  pulse leads to parallel and anomalous odd harmonics, and how these 
findings are related to TRS breaking in WSM, we employ  
semiclassical equation  
of Bloch electrons in an external electric field $\textbf{E}(t)$. Within this approach, expression of the anomalous current is written as
$\textbf{J}_\Omega(t) = -  \int \{ \textbf{E}(t) \times \boldsymbol{\Omega}_\mu(\mathbf{k}) \}~\rho_{ \mu}(\mathbf{k},t)~ d \mathbf{k}$ with 
$\boldsymbol{\Omega}_\mu$ and $\rho_\mu(\mathbf{k},t)$ as the Berry curvature and band-population of $\mu^{\textrm{th}}$ energy band, respectively~\citep{liu2017high}. 
We can assume that the initial band population is symmetric under inversion as 
$\rho(\mathbf{k}, 0) = \rho(-\mathbf{k}, 0)$. 
In the presence of laser, momentum of an electron changes 
from $\mathbf{k}$ to $\mathbf{k_t}$, which leads 
to the change in the  band population as $ \rho(\mathbf{k}, t) = \rho(\mathbf{k_t},0)$. 
Under time-translation of the laser $t\rightarrow t + \textrm{T}/2$ , anomalous current can be expressed as 
\begin{eqnarray}
\textbf{J}_\Omega(t+\textrm{T}/2)  & = & - \int \{ \mathbf{E}(t+ \textrm{T}/2) \times  \boldsymbol{\Omega}_\mu(\mathbf{k})\}~\rho_\mu\big(\mathbf{k}_{\mathbf{t}+ \textrm{T}/2},0\big) ~d \mathbf{k} \nonumber \\
	& = &  \int \{ \mathbf{E}(t) \times \boldsymbol{\Omega}_\mu(\mathbf{k})\} ~\rho_\mu(\mathbf{k}+\mathbf{A}(t), 0\big) ~d \mathbf{k}  \nonumber \\
	& = &   \int \{  \mathbf{E}(t) \times \boldsymbol{\Omega}_\mu(\mathbf{k})\} ~\rho_\mu\big(\mathbf{k_t}, 0\big) ~d \mathbf{k} \nonumber \\
	& = &  - \textbf{J}_\Omega(t). 
\end{eqnarray}
In the above equations, we have used $\mathbf{E}(t+ \textrm{T}/2) = - \mathbf{E}(t)$ and $ \mathbf{A}(t+ \textrm{T}/2) = -\mathbf{A}(t)$,  $\rho(\mathbf{k}, 0) = \rho(-\mathbf{k}, 0)$; 
and changed the dummy variable  $\mathbf{k}\rightarrow-\mathbf{k}$ in the integral. 
Also, Berry curvature for an inversion-symmetric system with  broken TRS  obeys  
$ \boldsymbol{\Omega}(\mathbf{k})= \boldsymbol{\Omega}(-\mathbf{k})$. 
The contribution of $\textbf{J}_\Omega (t)$ to the $n^{\textrm{th}}$ harmonic is given by 
$  \textbf{J}^n_\Omega(\omega) \propto \int_{-\infty}^\infty \textbf{J}_\Omega(t) e^{in\omega t}~dt$. 
By changing  $t \rightarrow t +  \textrm{T}/2$ in the integral, we obtain  
$ \textbf{J}^n_\Omega(\omega) \propto  \int_{-\infty}^\infty \textbf{J}_\Omega(t+ \textrm{T}/2) e^{in\omega (t+ \textrm{T}/2)}~ dt = 
-e^{in\pi}\int_{-\infty}^\infty \textbf{J}_\Omega(t) e^{in\omega t}~dt$, which implies that only odd harmonics are allowed 
as $\exp(in\pi) = -1$. 
Thus, TRS-broken systems lead to anomalous odd harmonics, 
which is in contrast to the case of TRS-preserving systems with broken-inversion symmetry in which anomalous current leads to the generation of even harmonics~\citep{schubert2014sub, liu2017high, hohenleutner2015real, luu2018measurement}. 

\begin{figure}[h!]
	\centering
	\includegraphics[width= 0.9 \linewidth]{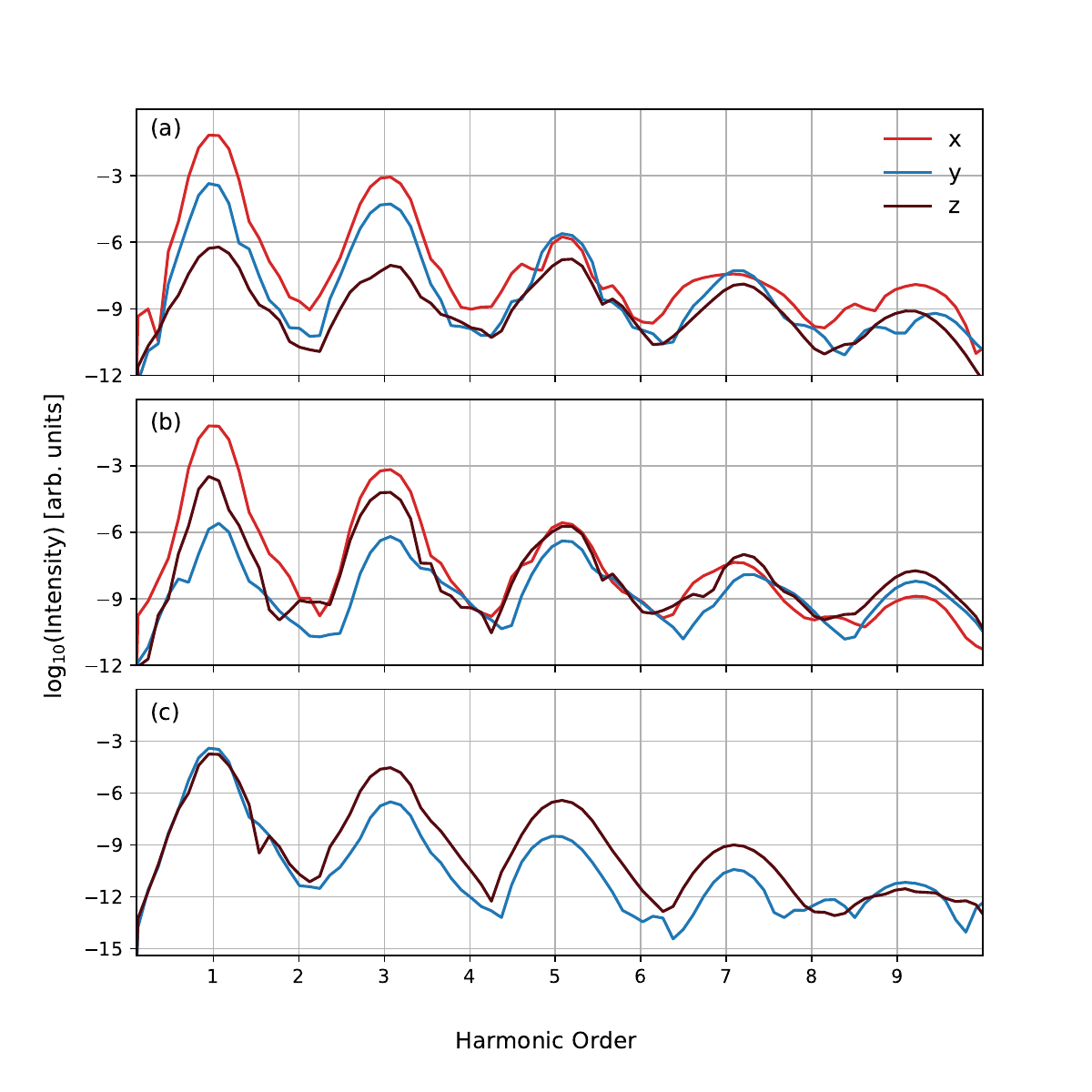}
	\caption{High-harmonic spectra generated by right-handed circularly polarized pulse in 
		(a) $xy$, (b) $xz$, and (c) $yz$- planes. The parameters of the laser and decoherence time are same as given in 
		Fig.~\ref{fig:fig5.1}.}  \label{fig:fig5.3}
\end{figure}

In order to discern the directions of the anomalous current, we need to understand the distinct   
role  of the Berry curvature's components, which is written as 
$\boldsymbol{\Omega}(\mathbf{k}) = \Omega_{k_{x}}(\mathbf{k}) \hat{e}_{k_{x}} + 
\Omega_{k_{y}}(\mathbf{k}) \hat{e}_{k_{y}} + \Omega_{k_{z}}(\mathbf{k}) \hat{e}_{k_{z}}$. 
The  expressions of the Berry curvature's components are given in Appendix B.  
The direction of the anomalous current is given by $\mathbf{E} \times \mathbf{\Omega}$ and the integral is perform over entire Brillouin zone. 
Moreover, $\mathbf{E}$ is a function of time and $\boldsymbol{\Omega}$ is function of $\mathbf{k}$, so their product does 
not changes the parity. 

If the laser is polarized along $x$ direction, then it is straightforward to see that anomalous current along 
$y$ and $z$ directions turn out to be zero as $\Omega_{k_{y}}$ and $\Omega_{k_{z}}$ are odd functions in the two directions. 
On the other hand, $\Omega_{k_{x}}$ is an even function in all the directions, contributing to the anomalous current when the laser is polarized along $y$ or $z$ direction. 
Thus, present theoretical analysis is consistent with numerical results shown in Fig.~\ref{fig:fig5.1}, which unequivocally establishes that non-trivial topology of the Berry curvature leads to nonlinear anomalous odd harmonics -- light-driven nonlinear anomalous Hall effect.

At this point it is natural to investigate what determines the phase between the 
parallel  and anomalous harmonics. To address this issue, we focus on 
the third harmonic (H3) in time domain.
When the laser is polarized along $y$ direction, H3 along $y$ and $z$ directions is in phase as evident from  Fig.~\ref{fig:fig5.2}(a). However, it becomes out of phase in the case of  $z$ polarized pulse.  The reason behind in phase or out-of-phase of H3 can be attributed to the sign of 
$\textbf{J}_\Omega (t)  \propto \int \{ \mathbf{E}(t) \times \mathbf{\Omega}(\mathbf{k}) \} d\mathbf{k}$, which yields positive (negative) sign when laser is along $y~(z)$ direction. 
%Therefore our numerical results are consistent with our theoretical findings. 

\begin{figure}[!h]
\centering
\includegraphics[width= \linewidth]{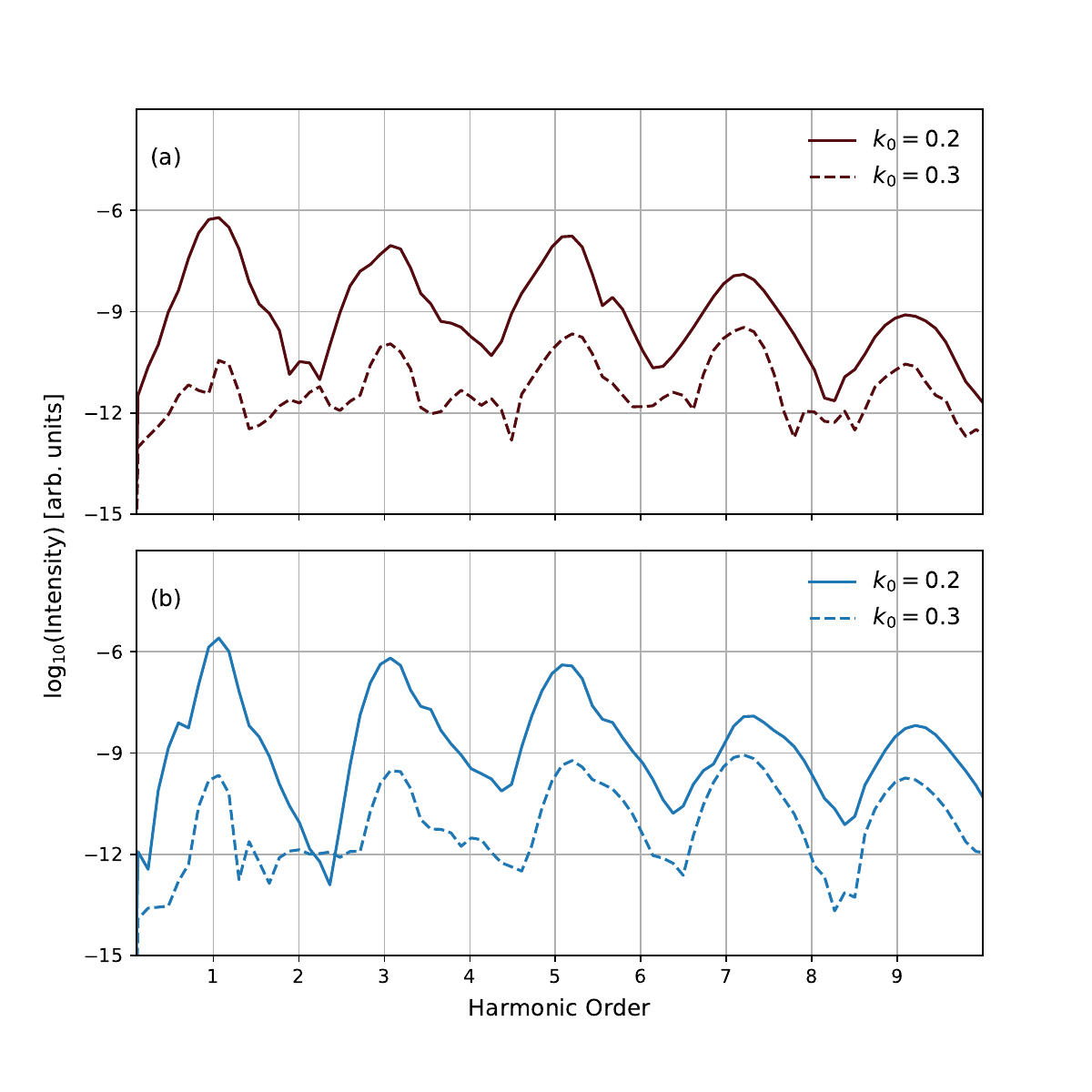}
\caption{Comparison of the anomalous high-harmonic yield for different values of $k_0$,  which is proportional to the distance between the two Weyl points. 
Odd anomalous harmonics along (a) $z$ direction when polarization of the pulse is in $xy$- plane, and  (b) 
$y$ direction when polarization of the pulse is in $xz$- plane.}  \label{fig:fig5.4}
\end{figure} 

To corroborate  our findings about the generation of the  anomalous odd 
harmonics and its relation with non-trivial topology of 
the Berry curvature's component, high-harmonic spectra generated by circularly polarized pulse 
are presented in Fig.~\ref{fig:fig5.3}. In agreement with the two-fold rotation symmetry of the Hamiltonian, only odd harmonics are generated.
When the pulse is in $xy$- plane, odd harmonics along $x$ and $y$ directions are generated as 
$x$ and $y$ components of the 
driving electric field are non-zero.  
Moreover,  due to the non-zero $y$ component of the driving field, 
anomalous odd harmonics are generated along $z$ direction 
[see Fig.~\ref{fig:fig5.3}(a)]. In this case,  the mechanism   
is same as it was in the case of linearly polarized pulse along $y$ direction. 
Same is applicable in the case of circularly polarized  pulse in $xz$- plane. In this case, 
parallel odd harmonics are generated along $x$ and $z$ directions, whereas anomalous odd harmonics are generated along $y$ direction [see Fig.~\ref{fig:fig5.3}(b)]. However, when the pulse is polarized in $yz$- plane, only parallel harmonics along $y$ and $z$ directions are generated, and no anomalous harmonics along $x$ direction are generated. This is expected due to even and odd natures of $ \Omega_{k_{x}}$ and $ \Omega_{k_{z}/k_{z}}$ (see Appendix B)}.  

\begin{figure}[h!]
\centering
\includegraphics[width=0.8\linewidth]{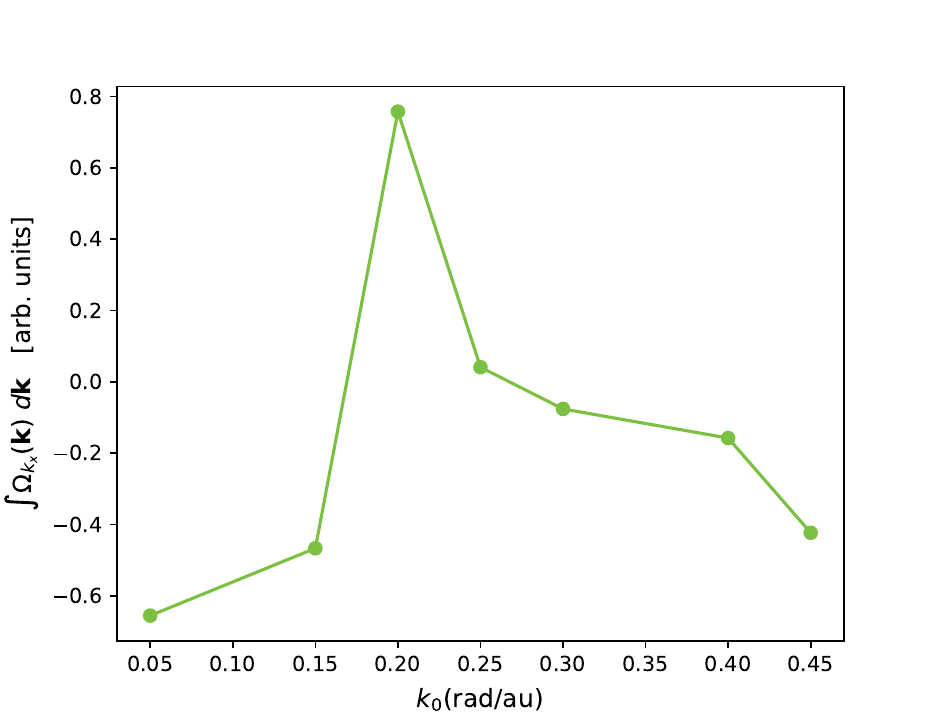}
\caption{Value of the integral of the Berry curvature's component as a function of $k_{0}$.}
\label{fig:fig5.5}
\end{figure}

After establishing the non-trivial role of the 
Berry curvature's components  and their parity,  
let us explore  how their strengths affect the yield of the anomalous odd harmonics. 
We know that  the magnitude of the  anomalous current depends on $\mathbf{E}\times \mathbf{\Omega}$. Moreover, the magnitude of the Berry curvature's components depend on $k_0$, as shown in Fig.~\ref{fig:fig5.5}.
Therefore, as we change the value of  $k_0$ from $0.2$ to $0.3$, the strength of the Berry curvature's components reduces, which lead to the reduction in the strength of the anomalous current. 
Fig.~\ref{fig:fig5.4}  presents a comparison of the  yield of the anomalous harmonics  for two different values of 
$k_0$. 

\begin{figure}[h!]
\centering
\includegraphics[width=\linewidth]{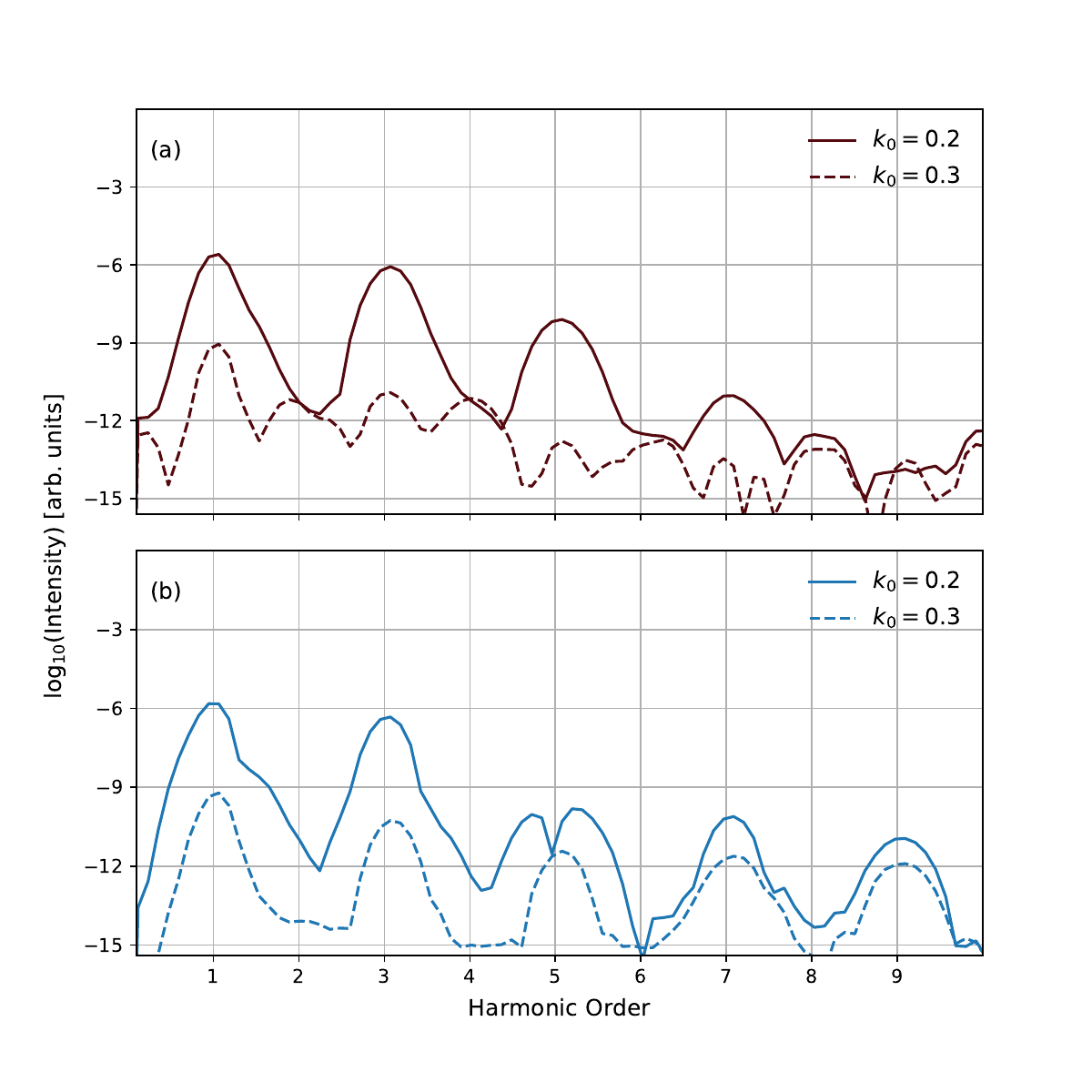}
\caption{Comparison of the anomalous harmonic yield for two different values of $k_{0}$.  
Odd anomalous harmonics along (a) $z$ direction when laser is linearly polarized along $y$ direction, 
and  (b) along $y$ direction when laser is linearly polarized along $z$ direction.
The value of $k_{0}$ is varied from $\pm 0.2$ to  $\pm 0.3$ rad/au.}  \label{fig:fig5.6}
\end{figure} 

In the case of circularly polarized pulse in $xy$- plane, the anomalous harmonics  along $z$ direction reduces drastically  as we change $k_0$ from $0.2$ to $ 0.3$ rad/au [see Fig.~\ref{fig:fig5.4}(a)]. 
The same is true for the pulse in $xz$- plane and anomalous harmonics  along $y$ direction [see Fig.~\ref{fig:fig5.4}(b)].   

\begin{figure}[h!]
\centering
\includegraphics[width=  \linewidth]{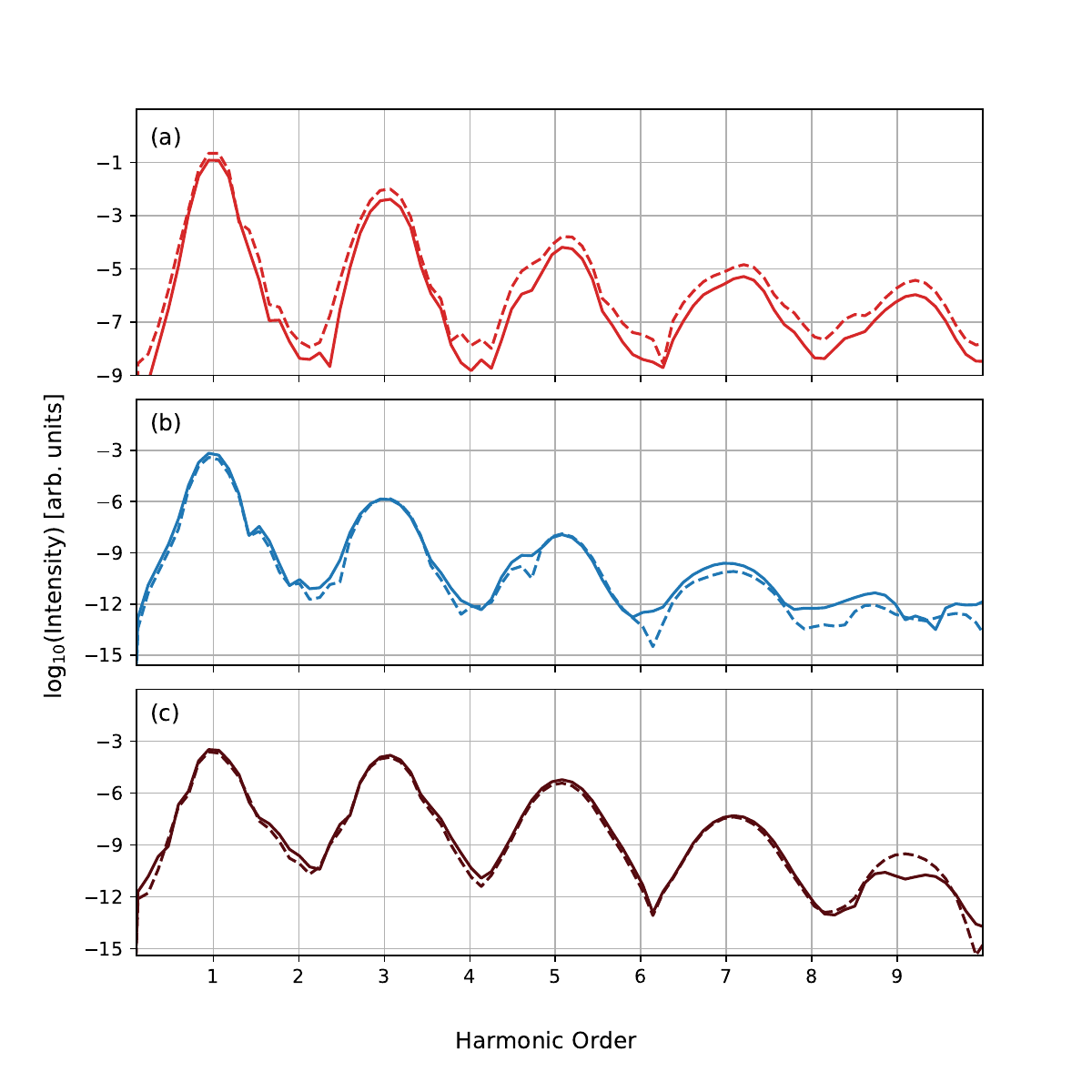}
\caption{Comparison of the parallel  harmonic yield for two different values of $k_{0}$. 
The laser is polarized along (a) $x$, (b) $y$, and (c) $z$ directions. 
Red,  blue and brown correspond to $x$, $y$ and $z$ polarization, respectively. 
Solid (dotted) line represents the value of $k_{0} = \pm 0.2 (\pm 0.3)$.  
The value of $k_{0}$ is varied from $\pm 0.2$ to  $\pm 0.3$ rad/au.}  \label{fig:fig5.7}
\end{figure} 

Therefore, the yield of the anomalous harmonics reduces drastically as 
the value of $k_0$ increased from 0.2  to 0.3 rad/au. 
Similar conclusions can be drawn in the case of HHG from linearly polarized pulse (see Fig.~\ref{fig:fig5.6}). 
However, the yield of the parallel harmonics is insensitive to the change in the value of $k_0$  
(see Figs.~\ref{fig:fig5.7} and \ref{fig:fig5.8}).
Our findings are similar to anisotropic anomalous Hall effect in which the magnitude of the current depends on the integral of the Berry curvature~\citep{yang2021noncollinear}.

\begin{figure}[h!]
	\centering
	\includegraphics[width=  \linewidth]{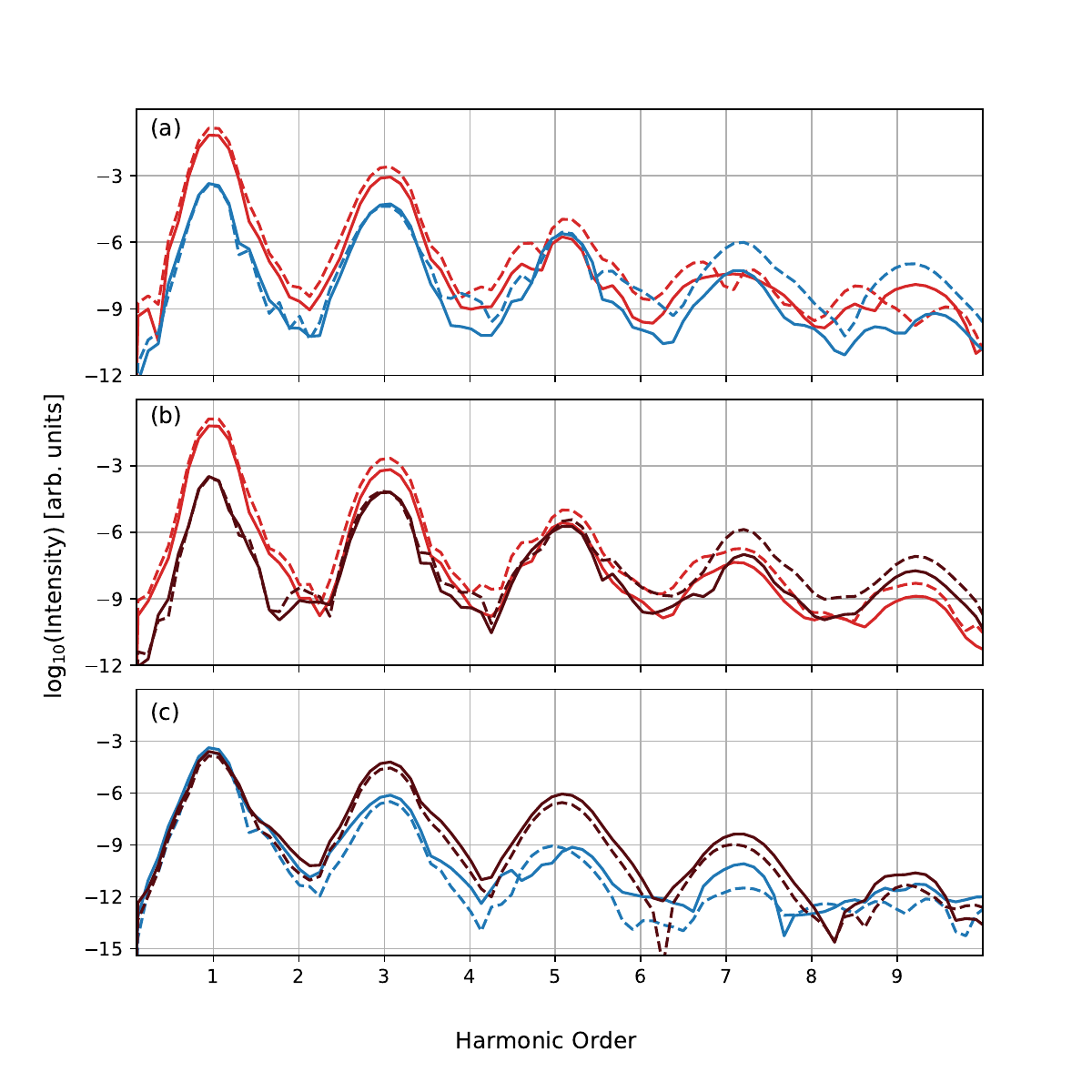}
	\caption{Comparison of the parallel  harmonic yield for two different values of $k_{0}$. 
		The laser is circularly polarized  in (a) $xy$, (b) $xz$, and (c) $yz$- planes. 
		Red,  blue and brown correspond to $x$, $y$ and $z$ polarization, respectively. 
		Solid (dotted) line represents the value of $k_{0} = \pm 0.2 (\pm 0.3)$.  
		The value of $k_{0}$ is varied from $\pm 0.2$ to  $\pm 0.3$ rad/au.}  \label{fig:fig5.8}
\end{figure}

Not only anomalous current and harmonics encode the non-trivial symmetry and the magnitude of the Berry curvature's components but also tailor the polarization of the emitted harmonics, which offers an elegant way  to probe non-trivial topological properties of the Berry curvature by all-optical way. 
As evident from Fig.~\ref{fig:fig5.2},  $y$ and $z$ components of H3 are in-phase 
and out-of-phase when the driving laser is polarized along $y$ and $z$ directions, respectively, which gives two different polarization of H3 [see Fig.~\ref{fig:fig5.9}(a)]. 
Thus, by measuring the polarization of H3, 
the non-trivial topology of the Berry curvature can be probed as it controls 
the strength and the phase between  
$y$ and $z$ components of H3. 
The same observations are true for other higher-order harmonics corresponding to  linearly polarized driver [see Figs.~\ref{fig:fig5.1} and ~\ref{fig:fig5.9}]. 
Similar conclusions can be made when CPL is used for HHG 
[see Figs.~\ref{fig:fig5.3} and ~\ref{fig:fig5.10}].

\begin{figure}[!h]
	\centering
	\includegraphics[width= 0.7 \linewidth]{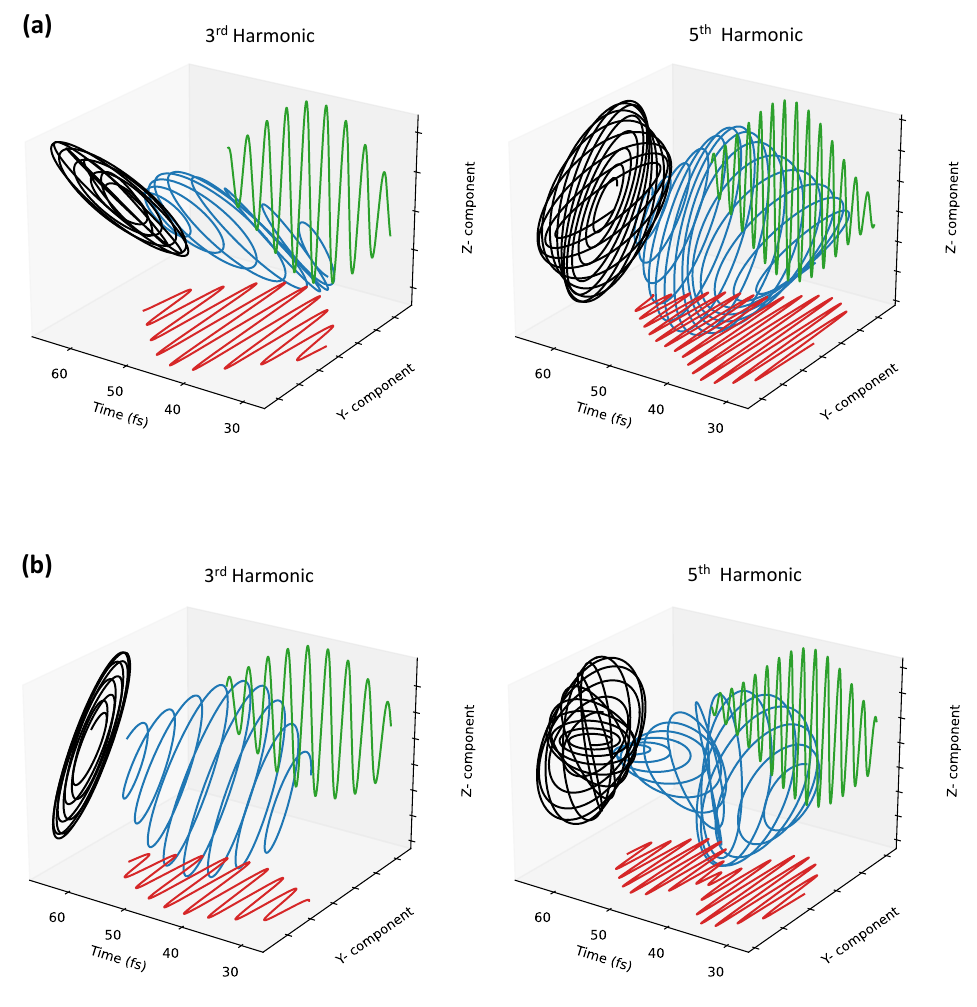}
	\caption{The total time-dependent electric field (in blue) corresponding to third and fifth harmonics in time-reversal broken Weyl semimetal. The $y$-component (in red), $z$-component (in green), and Lissajous figure (in black). The harmonics in Weyl semimetal are generated using linearly polarized pulse along (a) $y$-axis and (b) $z$-axis as shown in Figs.~\ref{fig:fig5.1}(b) and \ref{fig:fig5.1}(c), respectively.}  \label{fig:fig5.9}
\end{figure} 
\begin{figure}[!h]
	\centering
	\includegraphics[width=  0.7\linewidth]{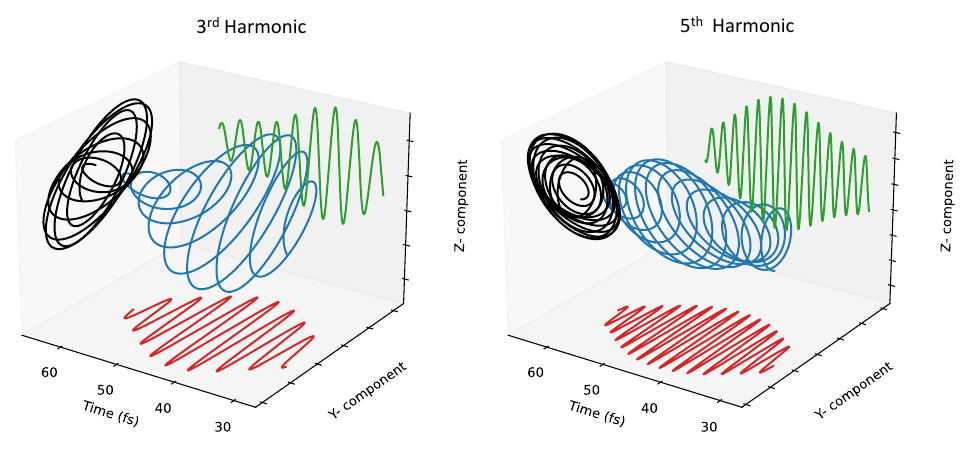}
	\caption{Same as in Fig.~\ref{fig:fig5.9}. The harmonics are generated using circularly  polarized pulse in  $xy$- plane as the harmonic spectrum is shown in Fig.~\ref{fig:fig5.3}(a).}  \label{fig:fig5.10}
\end{figure}

\clearpage

\subsection{Role of Tilt and Chemical Potential}
To address the role of tilt and chemical potential, let us add $d_0$ term in $\textbf{d}(\textbf{k})$. 
The components of $\textbf{d}(\textbf{k})$ for  TRB  WSM and IB WSM are, respectively,  expressed as~\citep{mccormick2017minimal, menon2021chiral} 
\begin{eqnarray}\label{eq:trb}
	&\textbf{d}(\textbf{k}) =  & t\big[t_c \{\cos(k_z a) + \cos(k_x a) -1\} + t_g \sin(k_z a), \sin(k_x a),\sin(k_y a), \nonumber \\
	&&   \{\cos(k_z a) - \cos(k_0 a) +2- \cos(k_x a) - \cos(k_y a)\}\big]
	%&&~~\textrm{and} 
\end{eqnarray}
and
\begin{eqnarray}\label{eq:invb}
	&\textbf{d}(\textbf{k})  =  & t \big[ t_c\{\cos(2k_x a) - \cos(k_0 a)\} \{\cos(k_z a) - \cos(k_0 a)\} \nonumber\\
	&& + t_g \{\sin(k_x a) + \sin(k_z a) \},  -2 \{1- \cos^2(k_z c) -\cos(k_y a)\} \nonumber \\
	&&+ 2\{\cos(k_x a) - \cos(k_0 a) \},-2\sin(k_y a), -2\cos(k_z a)\big].
\end{eqnarray}
Here, $t_c$ and $t_g$ describe the tilt and energy splits of the Weyl nodes, respectively. 
The Weyl nodes in the TRB  WSM and IB WSM are 
situated at $\mathbf{k} = [0,0,\pm \pi/(2a)]$ and $\mathbf{k}=[\pm\pi/(2a),0,\pm\pi/(2a)]$, respectively. 
We have considered a cubic lattice $a = 6.28~\mathring{\text{A}}$ ~and isotropic hopping parameter $t=1.8$ eV.

\begin{figure}[h!]
	\centering
	\includegraphics[width=\linewidth]{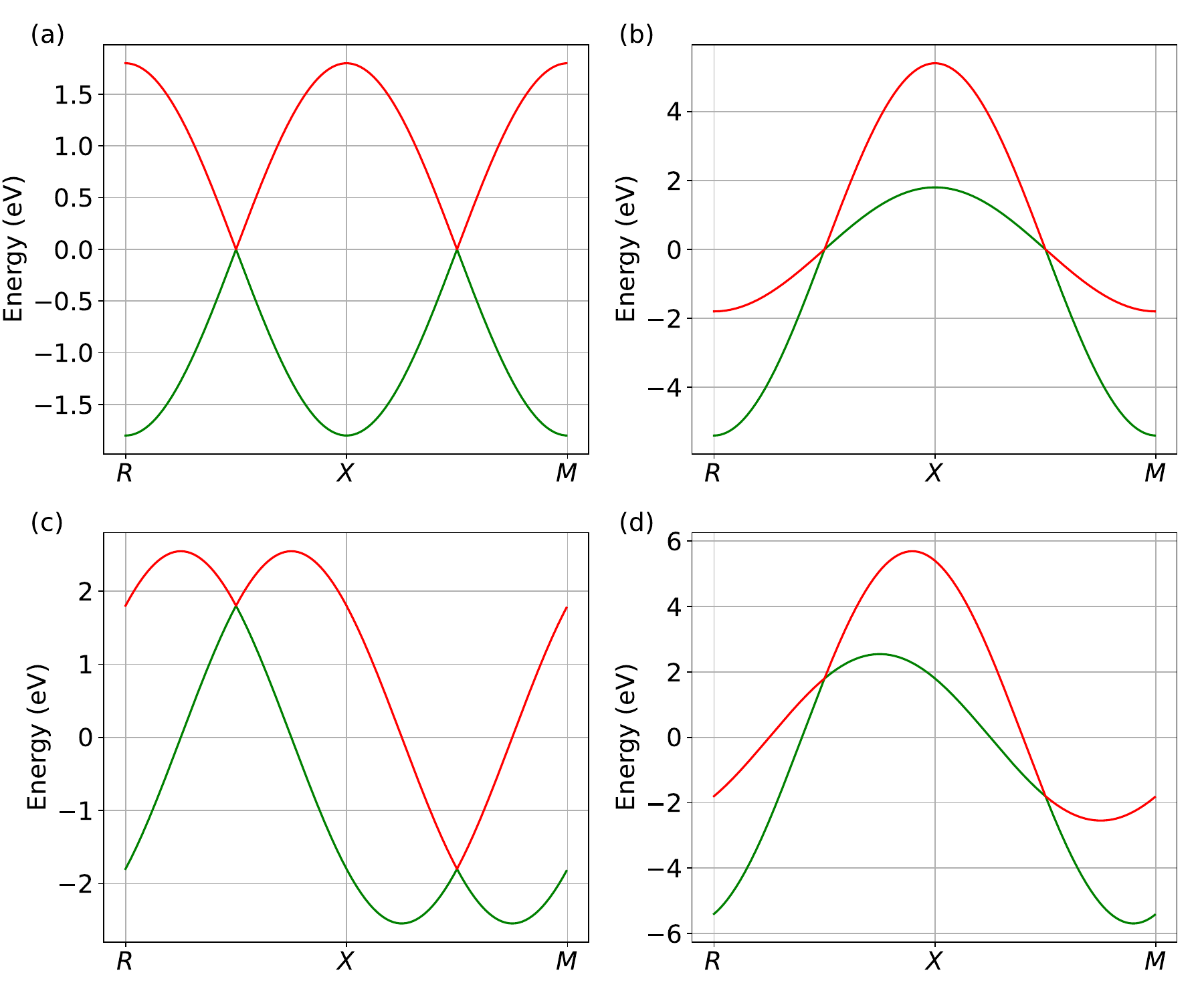}
	\caption{Energy bands corresponding to time-reversal symmetry broken Weyl semimetal  
		along  high-symmetry directions $R (0,0,\pi/a)\to \Gamma (0,0,0) \to M(0,0,-\pi/a)$ 
		for (a) $t_{c} = t_{g} = 0$, (b) $t_{c} = 2$ and  $t_{g} = 0$, (c) $t_{c} = 0$ and  $t_{g} = 1$ and (d) $t_{c} = 2$ and  $t_{g} = 1$.  The Fermi level is set at zero energy in all cases.}
	\label{fig:fig5.11}
\end{figure}

An ideal type-I WSM with degenerate Weyl nodes corresponds to $t_{c} = t_{g} = 0$ as evident from 
Fig.~\ref{fig:fig5.11}(a).
As  we tune $t_{c}$ from zero to $t_{c} > 1$,  untilted Weyl nodes transit to the tilted one, type-II phase 
[see Fig.~\ref{fig:fig5.11}(b)].   
Both cones at the Weyl nodes are tilted in opposite directions for $t_{c} = 2$ as evident from Fig.~\ref{fig:fig5.11}(c), resulting in electron and hole pockets. 
Weyl cones are upright (type-I) with the finite energy difference between them ($t_{g} = 1$) in Fig.~\ref{fig:fig5.11}(c). 
The combined effect of $t_{c}$ and $t_{g}$ on the Weyl nodes is shown in Fig.~\ref{fig:fig5.11}(d), where 
$t_{g} = 1$ shift  one Weyl node at $k_{z} = -\pi/(2a)$ to a higher energy while pushing the other 
node at $k_{z} = \pi/(2a)$ to a negative energy. Together with this,  $t_{c} = 2$ makes the Weyl cone tilt. 
These modifications in the energy band structure also alter the momentum matrix elements significantly. 
However, Berry connections remain unaffected 
as the tilt and energy split parameters appear in $d_{0}$ [see Eqs.~(\ref{eq:trb}- \ref{eq:invb})]. 

We start analyzing high-harmonic spectra for TRB WSM by varying model parameters. 
The harmonic spectrum of the WSM  without any tilt and energy split 
displays odd harmonics as evident  from  Fig.~\ref{fig:fig5.12}(a). 
The appearance of the odd-order harmonics is due to the presence of an inversion symmetry in WSM. 
In addition, an absence of the time-reversal symmetry  allows the generation of anomalous odd harmonics, 
perpendicular to the incident laser polarization, which is akin to the anomalous Hall effect and consistent with earlier reports~\citep{bharti2022high,avetissian2022high, medic2024high}.  
As the finite tilt ($t_c =2$)  is introduced, the harmonic yield is boosted by more than two-orders in magnitude as 
visible from Fig.~\ref{fig:fig5.12}(b). 
Such a significant boost in the harmonic yield for the finite tilt establishes that type-II Weyl semimetals are superior over type-I for a given driving laser. 
The strength of the tilt is not sufficient to break the inversion symmetry of WSM, which results in odd harmonics only. 
This is also evident from Eq.~\eqref{eq:trb} that any nonzero value of $t_c$ preserves the inversion symmetry of the Hamiltonian. 

\begin{figure}[h!]
\centering
\includegraphics[width=\linewidth]{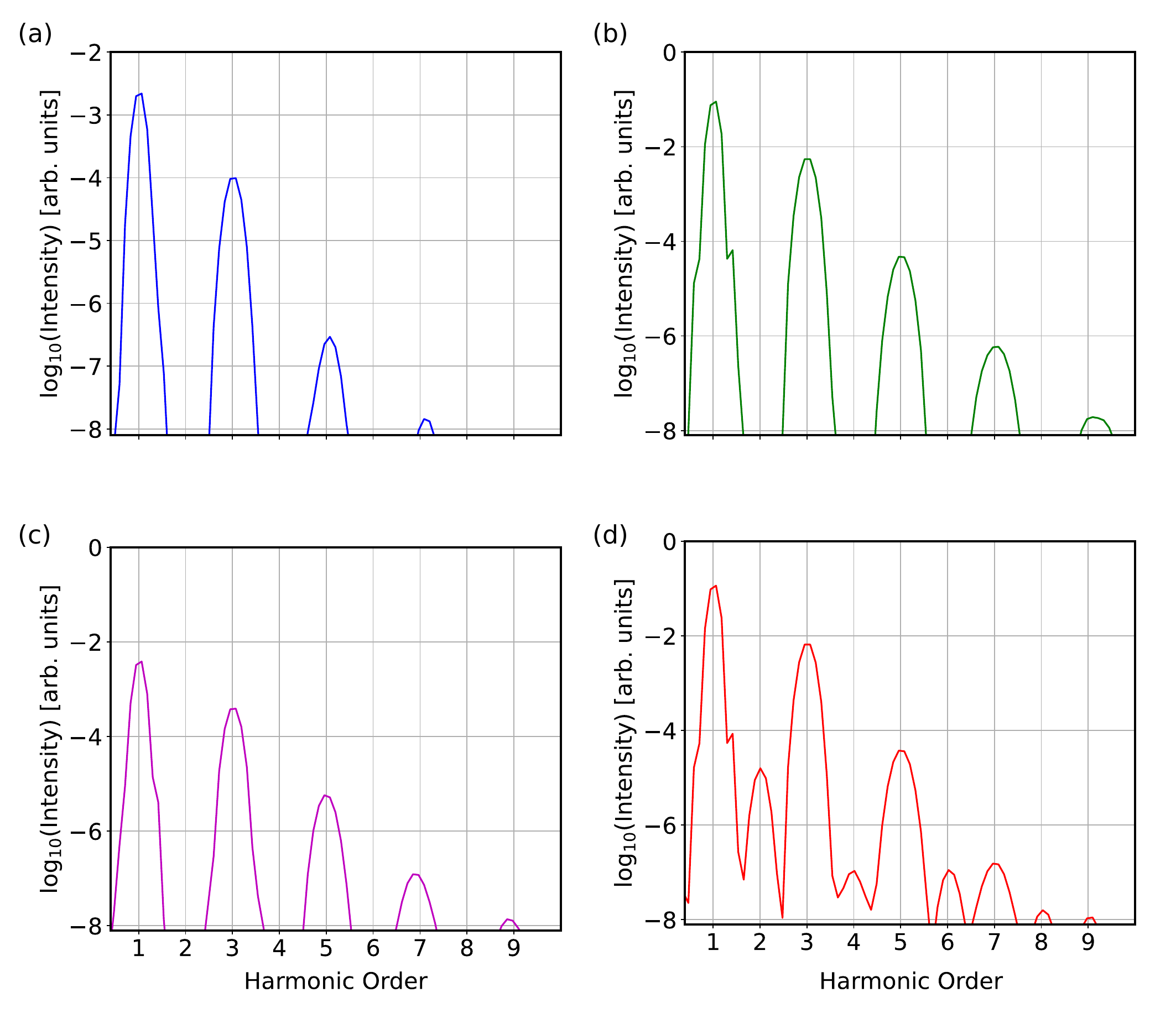}
\caption{High-harmonic spectra of  time-reversal symmetry broken Weyl semimetal  
for (a) $t_{c} = t_{g} = 0$, (b) $t_{c} = 2$ and  $t_{g} = 0$, (c) $t_{c} = 0$ and  $t_{g} = 1$ and (d) $t_{c} = 2$ and  $t_{g} = 1$. The peak intensity of the laser is $10^{11}$ W/cm$^2$ with 3.2 $\mu$m wavelength and about 80 fs long. }\label{fig:fig5.12}
\end{figure}

Let us turn our discussion about the impact of the finite energy split of the Weyl nodes on HHG. 
The presence of a finite split leads to an enhancement in the harmonic yield by an order of magnitude [see  
Fig.~\ref{fig:fig5.12}(c)  for $t_g = 1$]. 
In this case, also, only odd-order are allowed, which is in accordance with the fact that
a finite $t_g$ does not break the inversion symmetry [see Eq.~\eqref{eq:trb}].
However, a combined effect of a finite tilt and the energy split breaks inversion symmetry, which results
in the appearance of the even-order harmonics as shown in  Fig.~\ref{fig:fig5.12}(d) for $t_{c} = 2~\text{and}~t_{g} = 1$. 
In this case, the yield of the odd-order harmonics is enhanced drastically.  
The analysis of Fig.~\ref{fig:fig5.12} establishes that WSM with finite tilt and energy split is best suited for 
HHG that is advantageous for numerous technological applications.  

\begin{figure}[h!]
\centering
\includegraphics[width= 0.9\linewidth]{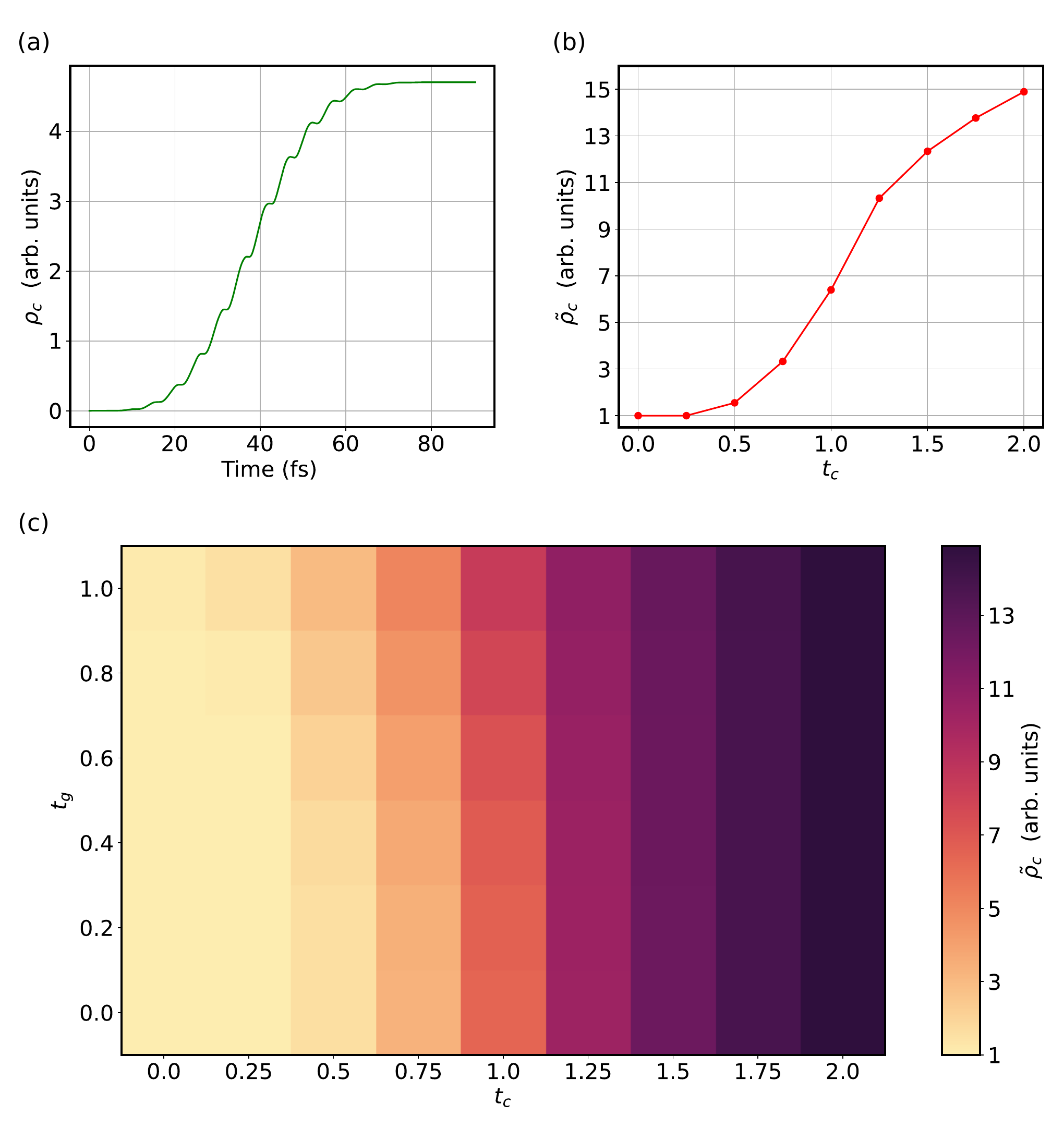}
\caption{Temporal evolution of the conduction band population in time-reversal symmetry broken Weyl semimetal  for 
(a) $t_{c} = t_{g} = 0$, (b) as a function of $t_{c}$ for $t_{g} = 0$ and (c) as a function of $t_{c}$ for $t_{g}$. 
The maximum population for $t_{c} \neq 0$ and $t_{g} \neq 0$  ($\tilde{\rho}_c$) is normalized by the maximum population for $t_{c} = 0$ and $t_{g} = 0$ given in  Fig.~\ref{fig:fig5.13}(a).}
	\label{fig:fig5.13}
\end{figure}

In order to understand the underlying mechanism responsible for significant enhancement  in  the  yield for finite tilt and energy
split, we investigate the electronic population in the conduction band during the laser pulse ($\tilde{\rho}_c$). 
The temporal evolution of the population in WSM  for $t_{c} = t_{g} = 0$  ($\rho_c$) is presented in Fig.~\ref{fig:fig5.13}(a). 
The conduction band is empty in the beginning, and it starts filling up during the subcycle timescale of the laser. 
At the end of the pulse, the conduction band is maximally populated. 
Figure~\ref{fig:fig5.13}(b) shows how the maximum population is sensitive to $t_{c}$. 
The population increases drastically as WSM transits from type-I to type-II, and
the population becomes significantly close to the transition point at  $t_{c} = 1$. 
The increase in the population is slow for type-I WSM  for $t_{c} < 1$, which 
exhibits an important distinction between type-I (untilted) and type-II (over-tilted) phases.
An introduction of the finite $t_{g}$ leads to a minuscule increase in the 
normalized population, as evident from Fig.~\ref{fig:fig5.13}(c). 
However, the combined effect of  $t_{c}~\text{and}~t_{g}$ results in a drastic increase in the population. 
In addition,  the impact of the energy split becomes insignificant for type-II WSM. 
The significant enhancement in the harmonic yield for finite $t_{c}~\text{and}~t_{g}$  can be attributed to the 
sensitivity of the normalized population with $t_{c}~\text{and}~t_{g}$. 

\begin{figure}[h!]
\centering
\includegraphics[width=\linewidth]{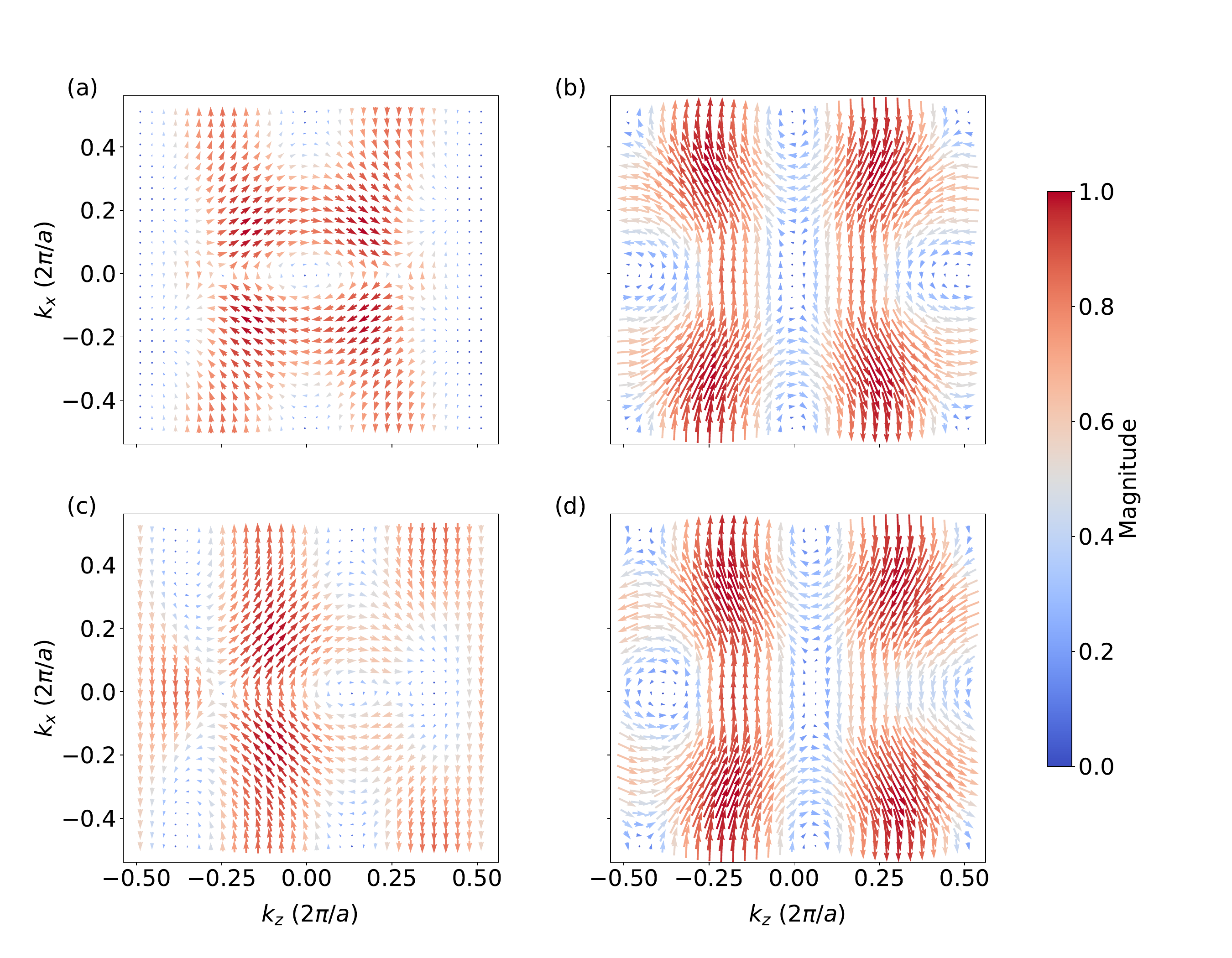}
\caption{Momentum matrix element $\mathbf{p}_{cc}^{\mathbf{k}} = \langle c, \mathbf{k} | 
\nabla_\textbf{\textbf{k}} \mathcal{H}_{\textrm{Weyl}}(\k) | c,\textbf{k}\rangle$ with $| c,\textbf{k}\rangle$ as the eigenstate of  the conduction band of time-reversal symmetry broken Weyl semimetal  for (a) $t_{c} = t_{g} = 0$, (b) $t_{c} = 2$ and  $t_{g} = 0$, (c) $t_{c} = 0$ and  $t_{g} = 1$ and (d) $t_{c} = 2$ and  $t_{g} = 1$.} \label{fig:fig5.14}
\end{figure}

To delve further into the responsiveness of HHG with  $t_{g}$ and  $t_{c}$, 
we analyze the momentum matrix element $\mathbf{p}_{cc}^{\mathbf{k}}$ -- an important ingredient in semiconductor Bloch equation framework. 
To simplify our discussion, we approximate the Hamiltonian in the vicinity of the Weyl nodes as 
$\mathcal{\tilde{H}}(\mathbf{k}) =  v_{0}(\mathbf{k}) \sigma_{0} + \sum_{i = 1}^{3} v_{i}(\mathbf{k}) \sigma_{i}$, where $v_{0}(\mathbf{k})$ 
includes the essence of $t_{g}$ or $t_{c}$, which is consistent with the expression for ${d}_{0}$ in Eq.~\eqref{eq:trb}. 
The velocity near the Weyl nodes alters as 
the value of $t_{g}$ or $t_{c}$  changes, 
which also impacts the energy band structure as reflected in Fig.~\ref{fig:fig5.11}.
Divergence of the momentum vector  flux in the $k_{x} - k_{z}$ plane is clearly  visible around the Weyl nodes
[see Fig.~\ref{fig:fig5.14}(a)]. 
The strength of the vector flux is largest in the vicinity of the nodes with mirror symmetry along the $k_{x}$ and $k_{z}$ directions. The flux diminishes away from the nodes.  
The vector flux's magnitude  increases drastically and becomes significant away from the Weyl nodes for type-II WSM
[see Fig.~\ref{fig:fig5.14}(b)].
As the finite energy splitting is introduced, the vector flux loses its mirror symmetry along the $k_{z}$  
direction as evident from  Fig.~\ref{fig:fig5.14}(b). 
However, there is no significant increase in the magnitude as compared to the case with $t_{g} = 0$. 
The vector flux's magnitude increases significantly  
when both tilt and energy split is present in  WSM [Fig.~\ref{fig:fig5.14}(d)]. 
Overall, the momentum vector shows a larger magnitude in the type-II phase, which, together with ample conduction band population, leads to enhanced non-linearity and, thus, higher-order harmonics with significant yield. 

\begin{figure}[h!]
\centering
\includegraphics[width=\linewidth]{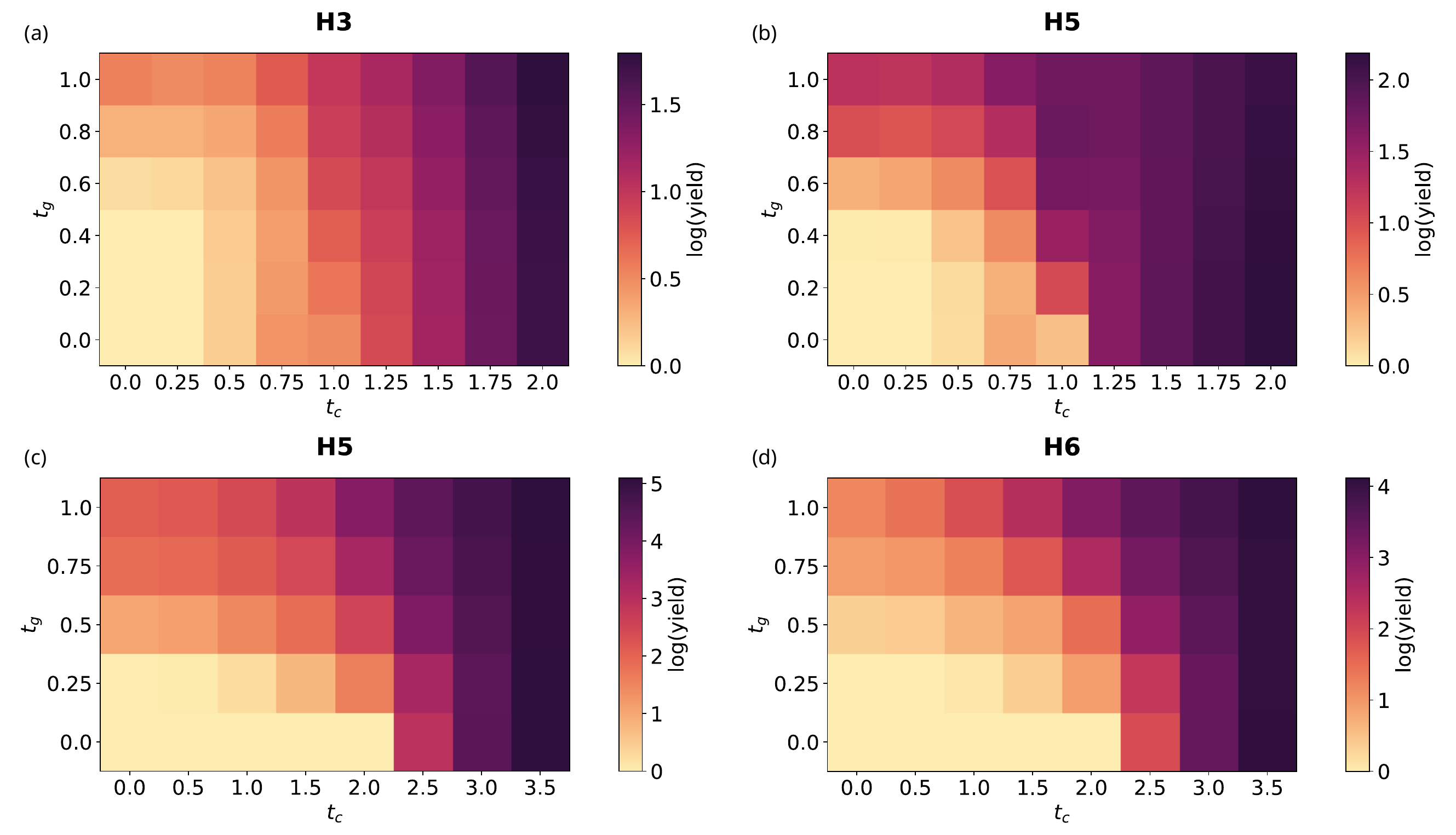}
\caption{Sensitivity of the integrated relative harmonic yield of (a) third (H3) and  (b) fifth (H5) harmonics as a function of $t_{c}$ and $t_{g}$ in time-reversal symmetry broken Weyl semimetal. 
(c) and (d) is the same as (a) and (b) for the fifth (H5) and sixth (H6) harmonics in  inversion-symmetry broken Weyl semimetal, respectively. 
The harmonic yield for $t_{c} \neq 0$ and/or  $t_{g} \neq 0$ is normalized with respect to the harmonic with 
$t_{c} =  t_{g}  = 0$.} 
	\label{fig:fig5.15}
\end{figure}

Before moving to next section, let us investigate how the yield of different harmonics responds with
$t_{g}$ and $t_{c}$. 
Figures~\ref{fig:fig5.15}(a) and \ref{fig:fig5.15}(b) present a comprehensive picture of the relative yield of the third (H3) and fifth (H5) harmonics in TRB  WSM. 
The yield of the H3 remains unaffected until $t_{g}$ becomes 1  as visible from Fig.~\ref{fig:fig5.15}(a) [see also Fig.~\ref{fig:fig5.12}]. 
However, the yield become comparatively significant for $t_{c} = 1.25$ with $t_{g}=0$, 
which is also an indication of the transition to type-II phase. 
In addition, H5 is more responsive to the tilt in comparison with H3, and the relative yield of H5 is more pronounced.   
The IB WSM allows the generation of even harmonics, which are further enhanced by tuning $t_{c}$ as evident from Fig.~\ref{fig:fig5.15}(d) [also see Eq.~\eqref{eq:invb}]. 
In contrast, the yield of even harmonics in the TRB case is zero in type-I, which become non-zero in type-II when there is significant $t_g$ [see Fig.~\ref{fig:fig5.12}].
The sensitivity of the yield of H5 with $t_{c}$ and $t_{g}$ remains qualitatively the same as the TRB case. 
However, we observe a sharper transition in the relative yield in IB WSM transits from type-I to type-II. 

%In conclusion, we studied high-harmonic generation from WSM that mimics the realistic conditions, which go beyond ideal WSM with degenerate,  untilted Weyl nodes. We found that by suitably selecting the WSM, with a moderate to large tilting, yields of higher harmonics can be improved by a few orders of magnitude with given parameters of driving laser pulse.  
%We anticipate similar results for higher topological charges in WSM~\citep{bharti2023role}. 
%Alternatively, the energy split between Weyl nodes can also significantly improve the yield of higher harmonics. 
%Hence, it is expected that unconventional WSM, which exhibits higher topological charge and finite energy split, would serve as a better source for higher-order harmonics~\citep{flicker2018chiral,chang2017unconventional}. 
%In addition, our study will be useful in generating tailored laser pulses with tunable polarization and energy~\citep{bharti2022high,wang2024table}.

\subsection{Role of Topological Charges}

Inversion-symmetric m-WSMs with broken time-reversal symmetry for different 
topological charges $n$ can be collectively written as 
$\mathcal{H}^{(n)}(\mathbf{k}) = \bm{\sigma} \cdot  \mathbf{d}^{(n)}(\mathbf{k})$~\citep{dantas2020non}.
The full expressions of  the three components of $\mathbf{d}^{(n)}(\mathbf{k})$ for  topological charges $n = 1, 2,$ and 3 are  
written as  	 
\begin{eqnarray}\label{eq:n_1}
	\mathbf{d}^{(1)}(\mathbf{k}) & =  & \left[t\sin(k_x a), t\sin(k_y a)\right.,\\\nonumber
	&&\left. t\{\cos(k_z a) - \cos(k_0 a) +2- \cos(k_x a) - \cos(k_y a)\}\right],
\end{eqnarray}
\begin{eqnarray}\label{eq:n_2}
	\mathbf{d}^{(2)}(\mathbf{k}) & = & \left[t\{\cos(k_x a)-\cos(k_y a)\}, t\sin(k_x a)\sin(k_y a),\right. \\\nonumber
	&&  \left. t\{\cos(k_z a) - \cos(k_0 a) +2- \cos(k_x a) - \cos(k_y a)\}\right],
\end{eqnarray}
and 
\begin{eqnarray}\label{eq:n_3}
	\mathbf{d}^{(3)}(\mathbf{k}) & = & \left[t\sin(k_x a)\{3\cos(k_y a)-\cos(k_x a) -2\},\right.\\\nonumber
	&&\left. t\sin(k_y a)\{3\cos(k_x a)-\cos(k_y a) - 2\}, \right.\\\nonumber
	&&\left. t\{\cos(k_z a) - \cos(k_0 a) +2- \cos(k_x a) - \cos(k_y a)\}\right].
\end{eqnarray}
Positions of Weyl points for Eqs.~(\ref{eq:n_1}-\ref{eq:n_3}) are $(0,0,\pm k_0)$ as shown in 
Fig.~\ref{fig:fig5.18}. 
	
Let us apply  a laser polarized along the $z$ direction to induce electron dynamics in m-WSMs. 
Figure~\ref{fig:fig5.16} presents the total current in m-WSMs with different  topological charges. 
The current in a WSM with $n = 2$ is approximately one-order higher in magnitude  
in comparison to the current for $n = 1$. 
However, the total current  is comparable for $n = 2$ and 3 as evident from  Fig.~\ref{fig:fig5.16}.
The overall shape of the currents in all cases  are significantly distorted with respect to the shape
of the laser's electric field as shown in the inset. 
This distortion indicates  the involvement of nonlinear optical processes during  electron dynamics.  
Note that  the laser polarized along the $z$ direction does not yield current along any direction other than the polarization direction, which will be discussed later.

\begin{figure}[h!]
\centering
\includegraphics[width=\linewidth]{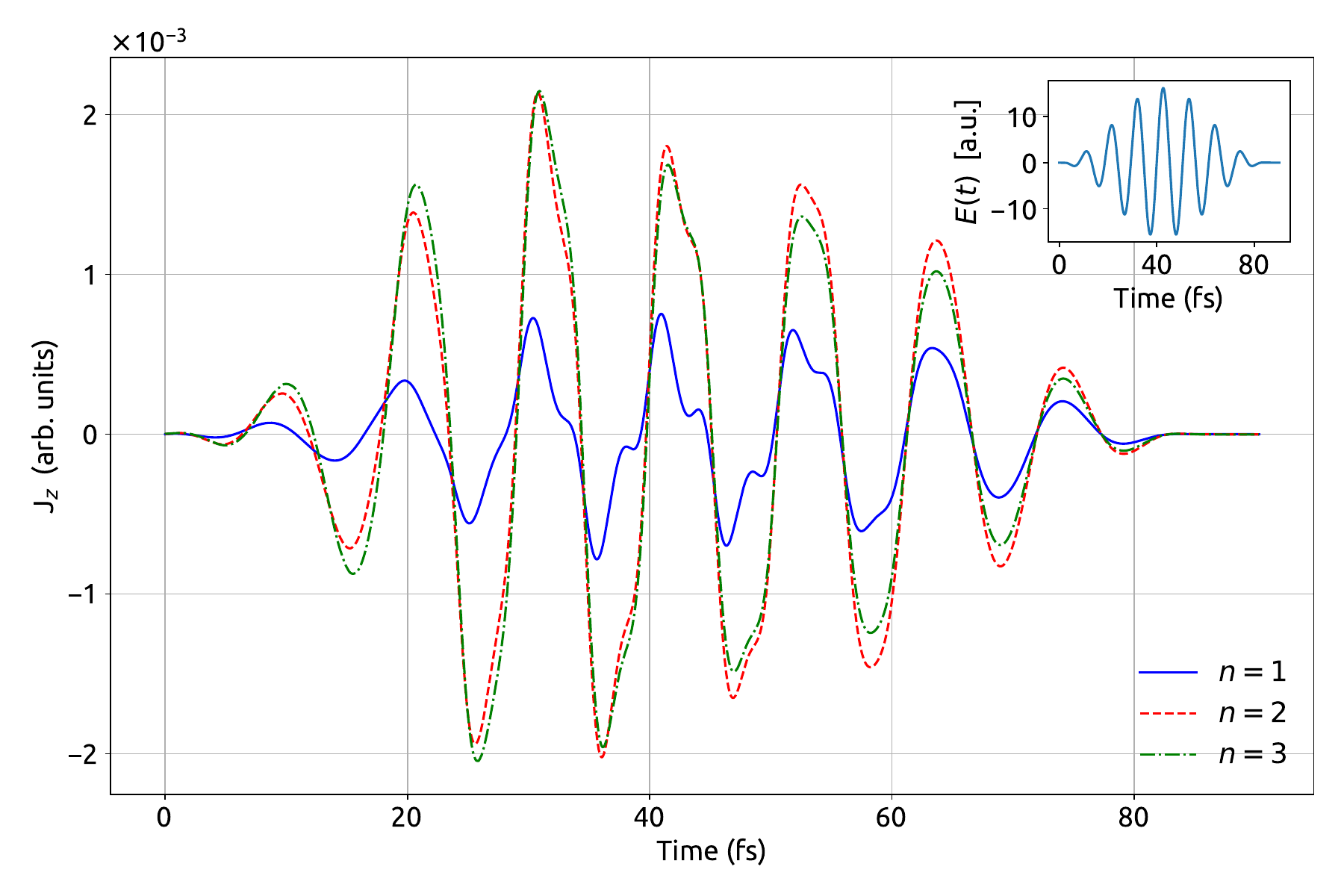}
\caption{Total current along the $z$ direction in m-WSMs for different  topological charges $n$. 
The linearly polarized laser pulse along the $z$ direction is  employed to generate the current and  is 
shown in the inset. 
The laser is approximately 100 fs long with a  wavelength of $3.2~\mu$m and 
an intensity of  $10^{11}$ W/cm$^2$. 
The value of the phenomenologically  decoherence time is 1.5 fs.}\label{fig:fig5.16}
\end{figure}

To understand the unusual behavior in the strength of the total current for different $n$ values, let us first analyze
the energy  band structures of the m-WSMs.     
Owing to a similar structure of $\mathcal{H}^{(n)}(\mathbf{k})$, the eigenvalues can be succinctly written as $\mathcal{E}_{c,v}(\k) = \pm \sqrt{d_{x}^{2} + d_{y}^{2} + d_{z}^{2}}$, where $\mathcal{E}_{c,v}(\k)$ corresponds to conduction and valence bands, respectively. 
Figure~\ref{fig:fig5.17}(a) shows the energy band structures along the $k_{z}$ direction for different $n$ values.  
The band structures  with two Weyl nodes of opposite chirality at  $\pm k_0$ 
are identical for $n = 1, 2$, and 3 as evident from the figure. 
The reason behind the identical energy dispersion along $k_{z}$  can be attributed to the 
same hopping parameter $t$ and lattice parameter  $a$ for different $n$'s. 
However, the band structures  near one of the Weyl nodes 
are significantly different  along $k_{x}$ for  different $n$'s as documented in Fig.~\ref{fig:fig5.17}(b). 
Note that the band structures  along  $k_{x}$ and $k_{y}$ are identical. 
The different energy  dispersions along $k_{x}$ are directly related to the topological charges 
of the Weyl nodes. 

\begin{figure}[h!]
\centering
\includegraphics[width=\linewidth]{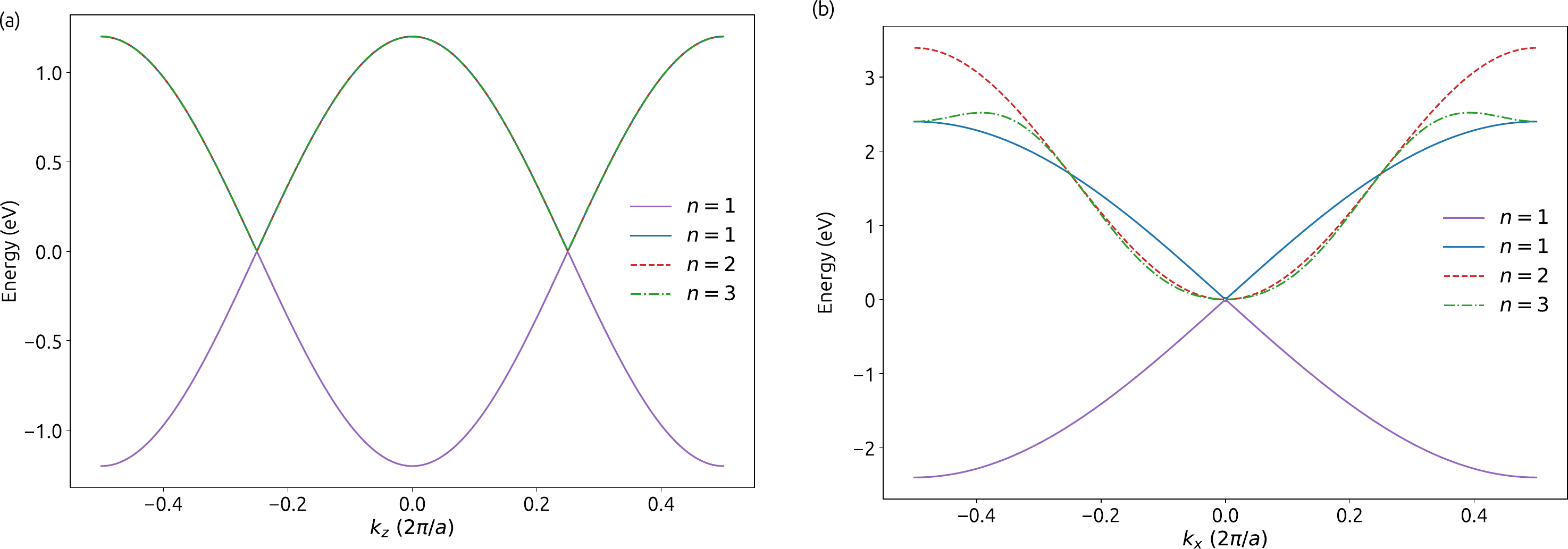}
\caption{Energy bands corresponding to m-WSMs with different topological 
charges along (a) the $k_{z}$ direction, and (b) the $k_{x}$  direction around a Weyl node. 
The band structures along the $k_{y}$ and  $k_{x}$ directions are identical.
The band structure is obtained by diagonalizing  $\mathcal{H}^{(n)}$  with the hopping parameter  $t = 1.2$ eV and the lattice parameter $a = 6.28 ~\mathring{\text{A}}$. 
The purple line corresponds to the valence band in all cases.}  \label{fig:fig5.17}
\end{figure}
The relation between the energy dispersion and the topological charge is apparent from  
the low-energy Hamiltonian near a Weyl node as 
$\mathrm{H}^{(n)}(\k) = v\left(k_x^n \sigma_x + k_y^n\sigma_y + k_z \sigma_z\right)$ for $n=1, 2$, 
and 3. 
It is straightforward to see that  
the effective low-energy band structures are linear, quadratic, and cubic for $n =1$, 2, and 3, 
respectively. 
Furthermore, it can be shown explicitly by calculating Berry curvature 
that the Chern numbers corresponding to linear, quadratic and cubic  dispersions are one, two, and three, respectively, as calculated elsewhere~\citep{nag2022distinct}.

Analysis of  Fig.~\ref{fig:fig5.17}  indicates that the current should be comparable for 
the m-WSMs with different $n$'s as the energy dispersions are identical along the direction of laser polarization.  
However, the energy gap between valance and conduction bands is smaller for $n = 2$ and 3 in comparison 
to $n =1$ around the Weyl node as evident from the energy dispersion along $k_{x}$, which translates 
into a higher probability of electron excitation for $n = 2$ and 3. 
Moreover, the energy dispersions for $n = 2$ and 3 are similar.
The distinct energy dispersion naturally leads to the distinct gradient of the group velocity. 
Thus, the cubic band dispersion offers a larger intraband current  compared to the quadratic one. 
Similarly, the linear dispersion has the smallest current flow out of the three. 
Both these facts explain why the total current exhibits similar features 
for $n = 2$ and 3 and is higher in magnitude in comparison to $n =1$.   
Note that the entire Brillouin zone contributes to the total current in three-dimensional systems like WSMs~\citep{gu2022full}.  

\begin{figure}[h!]
\centering
\includegraphics[width=\linewidth]{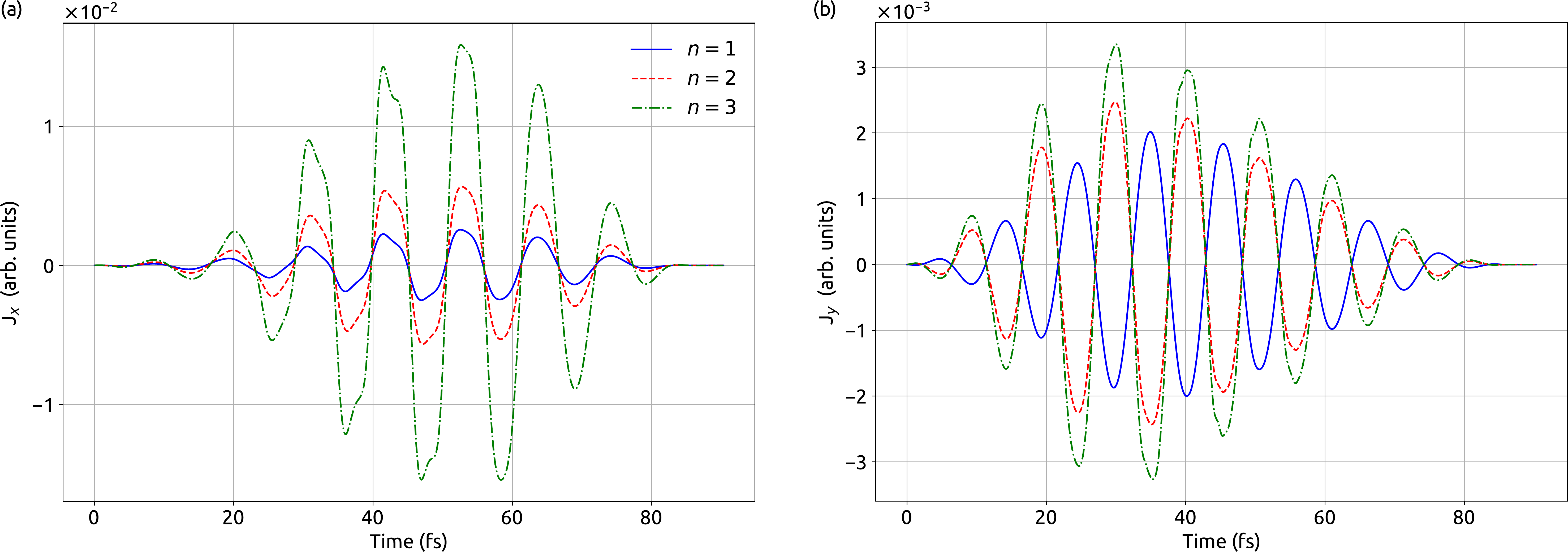}
\caption{Total current along (a) the direction of the laser polarization, i.e., normal current, and (b) 
along the perpendicular $y$ direction,  i.e., anomalous current.  
The laser is polarized along the $x$ direction. 
Details of the simulation  and the laser parameters are the same as those in Fig.~\ref{fig:fig5.16}.}
\label{fig:fig5.18}
\end{figure}

So far, we have witnessed that the laser polarized along the $z$ direction does not generate current along any 
perpendicular direction. 
Let us see how this observation is modified  when the polarization of the laser is tuned  
from the $z$ direction to the $x$ direction. 
Figure~\ref{fig:fig5.18}(a) presents the current parallel to the laser polarization for different 
topological charges.  The strength of the current is drastically different for different $n$'s, which
is expected from the band structures along $k_{x}$ as shown in Fig.~\ref{fig:fig5.16}(b). 
The current increases monotonically as the band dispersion goes from linear to quadratic and cubic with 
the increase in  $n$. 
This observation is in contrast with the previous one where the current was comparable for $n = 2$ and 3.   
Before we delve into the detailed reason for the increase in current as  a function of $n$, 
let us focus on the perpendicular component of the total current along the $y$ direction 
as shown in Fig.~\ref{fig:fig5.18}(b). 
A similar perpendicular component appears along the $x$ direction when the laser is polarized along the $y$ direction. 
In the following, let us use the terminology normal and anomalous currents for the currents along and perpendicular to the laser polarization, respectively.  

\begin{figure}[h!]
\centering
\includegraphics[width=\linewidth]{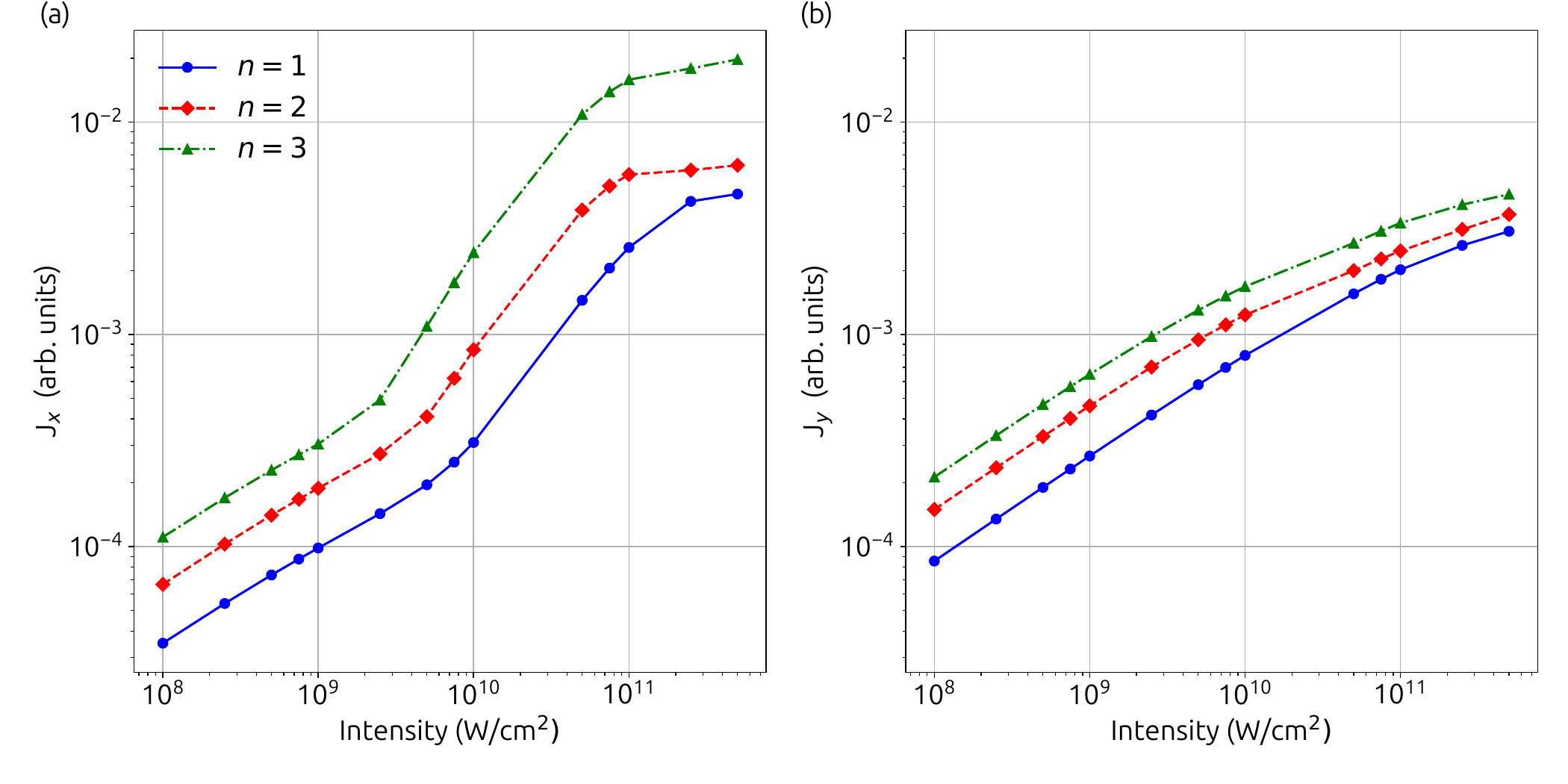}
\caption{Variations of the maximum amplitude for (a) the normal current along $x$ and (b) the anomalous current along $y$ as a function of the laser's intensity for m-WSMs with topological charges $n = 1, 2,$ and 3. The other parameters are the same as those in Fig.~\ref{fig:fig5.18}.}
\label{fig:fig5.19}
\end{figure}

We employ a semiclassical equation of motion for laser-driven electron dynamics 
to understand why the $x$ polarized laser generates  anomalous current, whereas $z$ polarization does not. 
It is known that the WSM has a non zero Berry curvature, which results  
an anomalous velocity as $\mathbf{E}(t) \times \mathbf{\Omega}$. 
Moreover, the Berry curvature in an  inversion-symmetric WSM follows 
$\mathbf{\Omega}(-\k) = \mathbf{\Omega}(\k)$, which leads  to nonzero current due to the anomalous velocity of  $\int \{\mathbf{E }\times \mathbf{\Omega}(\k)\}  \rho(\k) ~d\k$. 
The parity of the Berry curvature's components in the present case is 
such that there is no anomalous velocity  if the electric field is along the 
same direction as the line connecting the Weyl nodes as discussed in Ref.~\citep{bharti2022high} [see Appendix C].
Moreover, the anomalous current in the time-reversal broken WSM produces  the 
anomalous Hall effect~\citep{xiao2010berry}. 
Thus, the anomalous current is zero when the laser is polarized along  the $z$ direction, 
the direction along which the two Weyl nodes are situated. 
The situation changes drastically as the polarization direction changes from $z$ to $x$ or $y$.  

The anomalous current is proportional  to $\mathbf{\Omega}$, which means it is also proportional to $n$. 
Thus, the strength of the anomalous current increases as $n$ increases [see Fig.~\ref{fig:fig5.18}(b)]. 
However, the anomalous currents due to $n = 2$ and 3 are out of phase with respect to $n = 1$. 
This behavior is due to  the sign of the integral of the Berry curvature's components as shown in Ref.~\citep{bharti2022high}. 
The sign of $\int \{\mathbf{E}\times \mathbf{\Omega}(\k)\}  \rho(\k) ~d\k$
is positive for $n$ = 1 and 3 and negative for $n$ = 2, which leads the anomalous
current for $n$ = 2 out of phase.
Thus, the strength and the phase of the  anomalous current encode the information about the nontrivial topology of  the Berry curvatures in m-WSMs. 
Note that the anomalous current is one-order weaker  than the normal current as evident from  Fig.~\ref{fig:fig5.18}. 
At this juncture, it is interesting to wonder how these features in the normal and anomalous currents 
alter with the laser's intensity.

\begin{figure}[h!]
\centering
\includegraphics[width=0.9\linewidth]{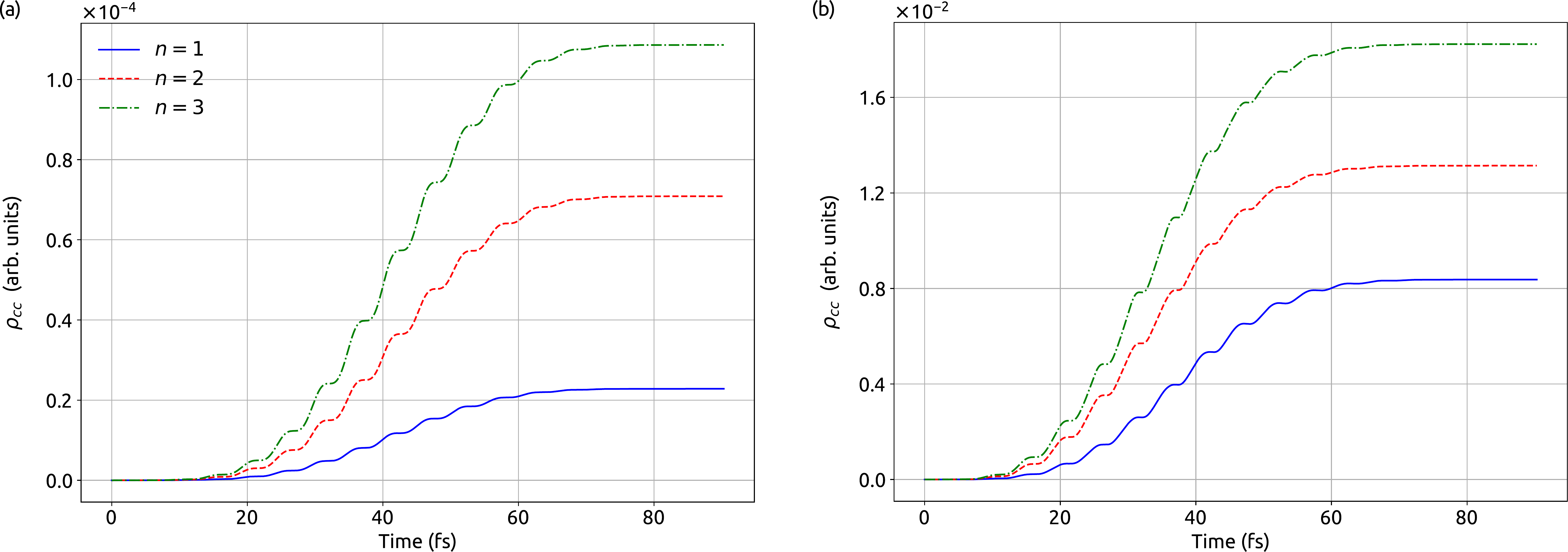}
\caption{Temporal evolution of the conduction band population, $\rho_{cc}$,  for different $n$  values 
during the laser for two intensities:  (a) $10^8$ W/cm$^2$ and (b) $10^{11}$ W/cm$^2$. The other parameters are the same as those in  Fig.~\ref{fig:fig5.18}.}
\label{fig:Fig5.20}
\end{figure}

Figure~\ref{fig:fig5.19} presents variation in the peak current for different $n$ values as a function of the intensity ranging from $10^8$  to $10^{11}$ W/cm$^2$. 
In the beginning, the anomalous current dominates over the normal current for each $n$ at  $10^8$ W/cm$^2$. 
However,  the normal current takes over the anomalous current  at some critical intensity.  
On comparing the normal and anomalous currents for each $n$, we find that this  critical intensity 
gets lower as  $n$ increases.  
In addition, the normal current starts to grow exponentially  
at a much lower intensity for higher $n$ values. 
The  peak current increases linearly as the intensity increases from $10^8$ to $10^{10}$ W/cm$^2$ [see Fig.~\ref{fig:fig5.19}(a)].   
Moreover, the rate of increase is much higher for the normal current as compared to the anomalous current, which 
starts to saturate at an intensity much lower than that of the 
normal current. 
There is no exponential growth region in the anomalous current compared to the normal current, which grows exponentially in the intensity window of $10^{10}$ to $10^{11}$ W/cm$^2$ [see Fig.~\ref{fig:fig5.19}(b)].
It is expected that the laser drives electrons further away in the energy band as the intensity increases, 
and therefore the normal current increases. 
However, the comparatively large anomalous current at lower intensity is interesting.
Moreover, the subdued increment in the anomalous current needs further investigation. 
To understand  these interesting observations, we analyze the laser-driven electronic population in the conduction band. 

\begin{figure}[!h]
\centering
\includegraphics[width=\linewidth]{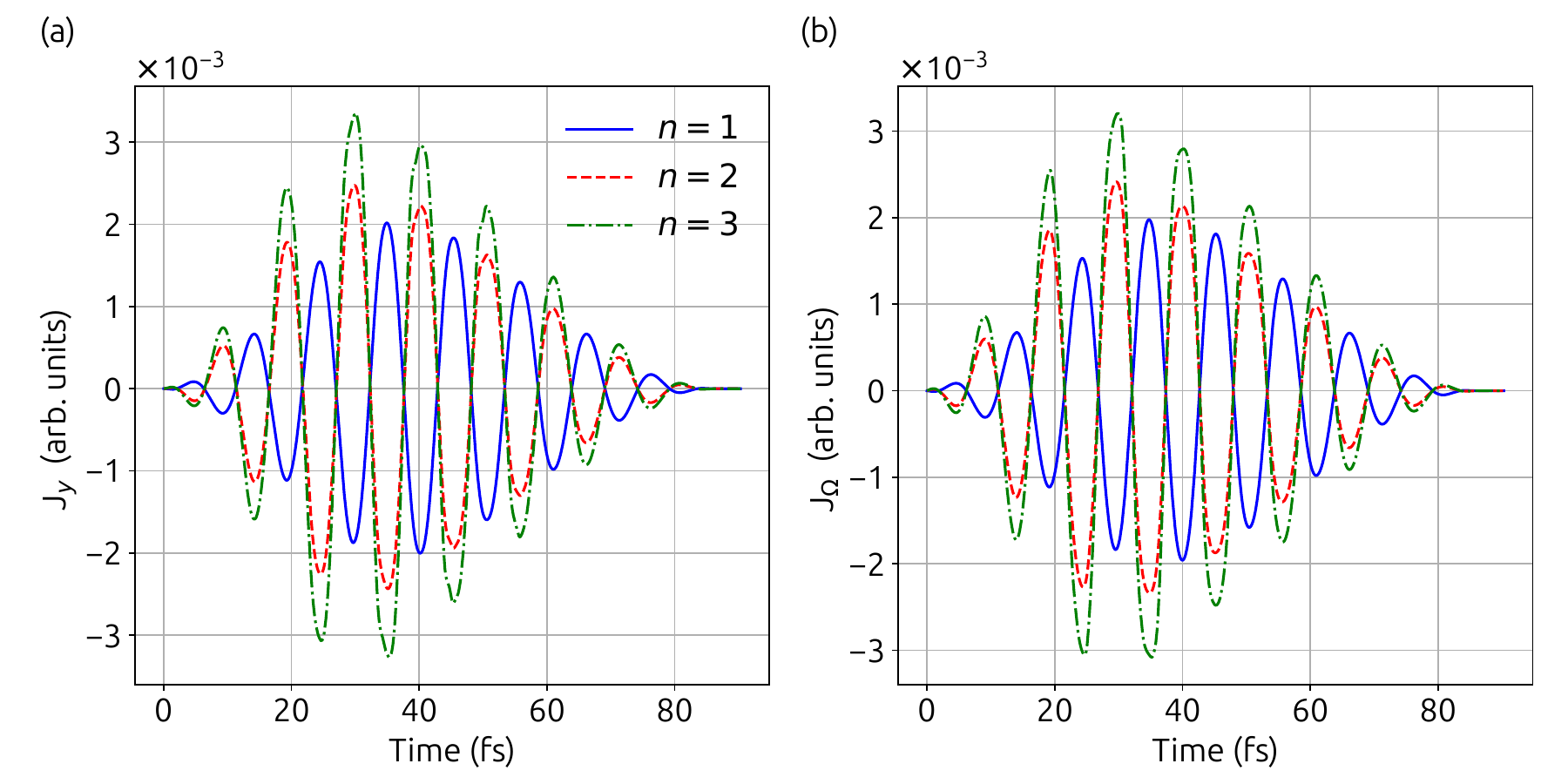}
\caption{(a) Total current along $y$ direction, (b) current solely due to the Berry curvature along 
$y$ direction for different topological charges $n$.
The laser  is approximately 100 fs long with  wavelength of $3.2~\mu$m and  intensity of  $10^{11}$ W/cm$^2$. The laser pulse is linearly polarized along $x$ axis.} \label{fig:fig5.22}
\end{figure}

Electronic population in the conduction band for  $n = 1$  is one-oder in magnitude weaker  than 
the population  for  $n = 3$ as shown in Fig.~\ref{fig:Fig5.20}(a). 
Moreover, the conduction band  is sparsely populated for  $n = 1$ at $10^8$ W/cm$^2$.
As the intensity increases from $10^8$ to $10^{11}$ W/cm$^2$, the overall  
population increases by two-orders in magnitude [see Fig.~\ref{fig:Fig5.20}]. 
The conduction band population for  $n = 3$ remains substantially more significant than that of the 
lower topological charges 
even at $10^{11}$ W/cm$^2$ as evident from Fig.~\ref{fig:Fig5.20}(b).
There are  two key factors governing   the overall behavior of the  conduction band population for different $n$ values at two intensities: 
the first one is the reduction in the energy gap between valence and conduction bands around a Weyl node
as $n$ increases [see Fig.~\ref{fig:fig5.17}(b)]. This results in higher probability of excitation and thus more population as $n$ increases.     
The other important factor is the dipole matrix amplitude and its relation with $n$ [see Eq. (\ref{eq:sbe})], 
which manifests higher electronic population as $n$ increases. 
Thus, it can be concluded from Fig.~\ref{fig:Fig5.20} that the population increases monotonically with 
$n$ irrespective of the laser's intensity. 

\begin{figure}[h!]
\includegraphics[width= \linewidth]{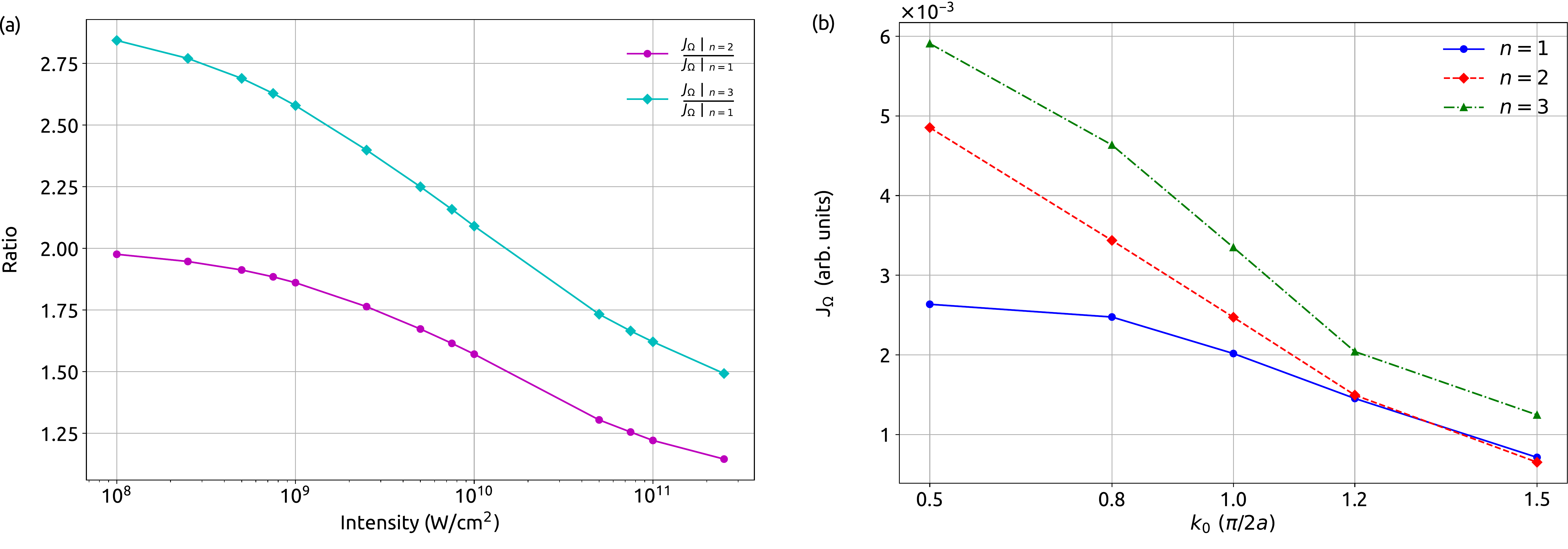}
\caption{(a) Ratio of the anomalous current's peak, $\textbf{J}_{\Omega}$, 
as a function of the laser's intensity, and 
(b) variation in the anomalous current's peak, $\textbf{J}_{\Omega}$,  
as a function of the separation between two Weyl nodes, $k_{0}$, at 
intensity $10^{11}$ W/cm$^2$ for different topological charges $n$.} \label{fig:fig5.21}
\end{figure}

Returning back to the considerable anomalous current at the lowest intensity $10^8$ W/cm$^2$, 
let us explore how the topological charge impacts  the anomalous current.
By following the  analysis of the linear response from the WSM, 
the anomalous current is proportional to $n$
as $\mathbf{J}_\mathbf{\Omega} \propto n (\mathbf{b}\times\mathbf{E})$, 
where $\mathbf{b}$ is
the vector joining the Weyl nodes~\citep{nandy2019generalized}. 
The quantity $\mathbf{b}\times\mathbf{E}$ determines the direction of the anomalous current.
Note that the same reasoning
we have used earlier to explain why the laser polarized along the direction of the line connecting the 
Weyl nodes results  in no  anomalous current. 
If we consider the expression of $\mathbf{J}_\mathbf{\Omega}$ to be true for the considered laser intensities    
then,  the anomalous current should increase monotonically with intensity as 
$\mathbf{J}_\mathbf{\Omega} \propto \mathbf{E}$. 
However, the anomalous current  deviates significantly from this expectation  at higher intensity, signaling a nonlinear optical response from m-WSMs. 
It is important to emphasis 
that the anomalous current  along the perpendicular direction is mainly driven by the Berry curvature as shown in Fig.~\ref{fig:fig5.22} .

\begin{figure}[!h]
	\centering
	\includegraphics[width=0.9\linewidth]{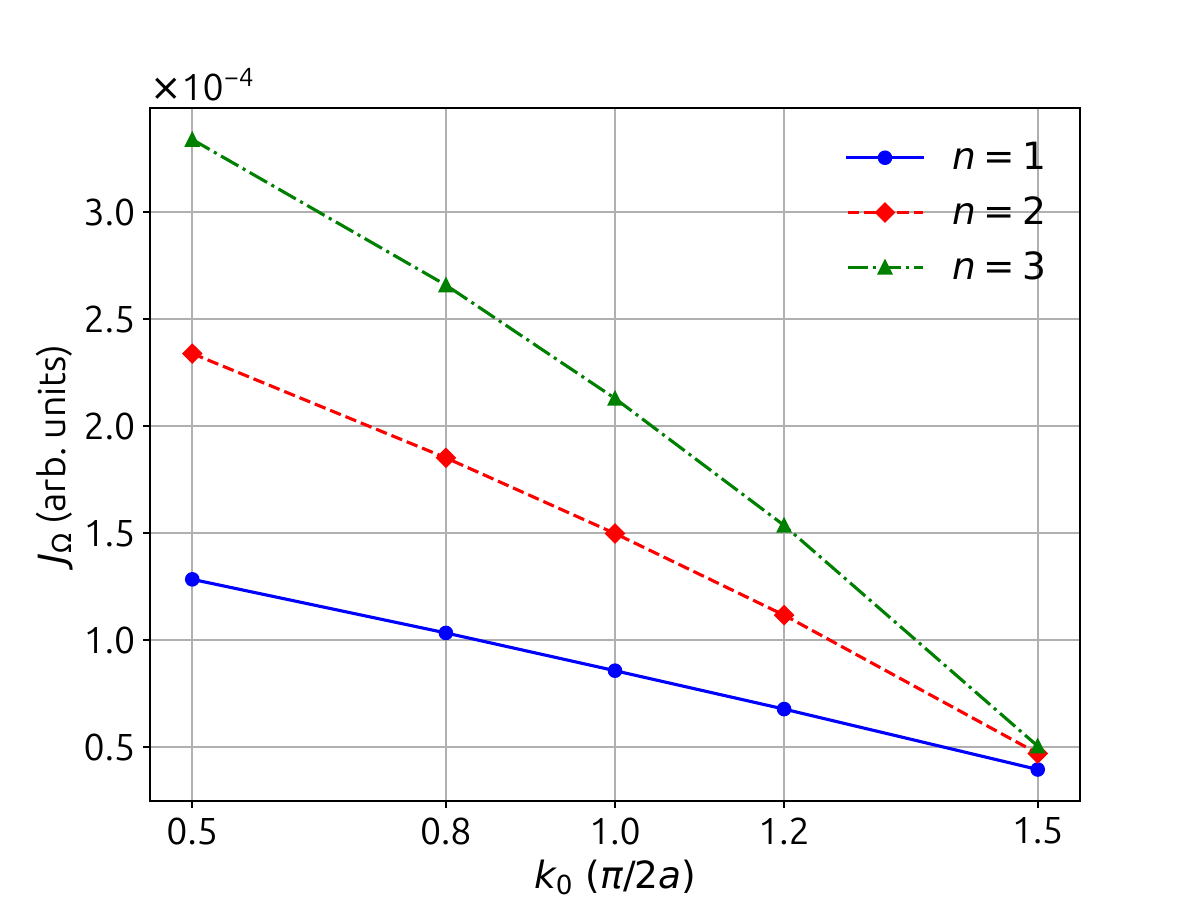}
	\caption{Variation in the anomalous current's maximum amplitude along $y$ axis, which is solely due  to the Berry curvature, $\textbf{J}_{\Omega}$,  
		as a function of the separation between two Weyl nodes, $k_{0}$, for different topological charges $n$.   
		The laser parameters are identical as in Fig.~\ref{fig:fig5.22} except the  intensity $10^{8}$ W/cm$^2$.}
	\label{fig:fig5.23}
\end{figure}

To corroborate our claim about the nonlinear optical response, 
let us analyze the ratio of the anomalous current's peak for different $n$ values 
as a function of the laser's intensity.
The ratios of the  peak current for $n=3$ to $n=1$, and $n=2$ to $n=1$ are presented in
Fig.~\ref{fig:fig5.21}(a). 
Within the linear response framework, it is expected that the ratios 
$\mathbf{J}_\mathbf{\Omega}|_{n = 3} /\mathbf{J}_\mathbf{\Omega}|_{n = 1}$ and 
$\mathbf{J}_\mathbf{\Omega}|_{n = 2} /\mathbf{J}_\mathbf{\Omega}|_{n = 1}$ should be 3 and 2, 
respectively as 
$ \mathbf{J}_\mathbf{\Omega} \propto  n$. 
This expectation is true at   $10^{8}$ W/cm$^2$ where the ratios are  
close to 3 and 2 as evident from Fig.~\ref{fig:fig5.21}(a). 
However, both ratios decrease monotonically as intensity increases, albeit at different rates. 
The ratios $\mathbf{J}_\mathbf{\Omega}|_{n = 3} /\mathbf{J}_\mathbf{\Omega}|_{n = 1}$ and 
$\mathbf{J}_\mathbf{\Omega}|_{n = 2} /\mathbf{J}_\mathbf{\Omega}|_{n = 1}$ reach approximately  2 and 1.5 at   
$10^{10}$ W/cm$^2$, respectively --  a drastic deviation from the expectation of linear response theory. 
The ratio $\mathbf{J}_\mathbf{\Omega}|_{n = 3} /\mathbf{J}_\mathbf{\Omega}|_{n = 1}$
decreases with a much faster rate compared to 
$\mathbf{J}_\mathbf{\Omega}|_{n = 2} /\mathbf{J}_\mathbf{\Omega}|_{n = 1}$. 
This implies that the anomalous  current due to the Berry curvature decreases faster with an increase 
in $n$. 
Thus, $\mathbf{J}_\mathbf{\Omega}$ originating from two different topological charges 
may be comparable at a certain intensity. 
However, it is practically not feasible to keep increasing 
the intensity as it can go above the damage threshold of the material. 

\begin{figure}[h!]
\centering
\includegraphics[width=0.9\linewidth]{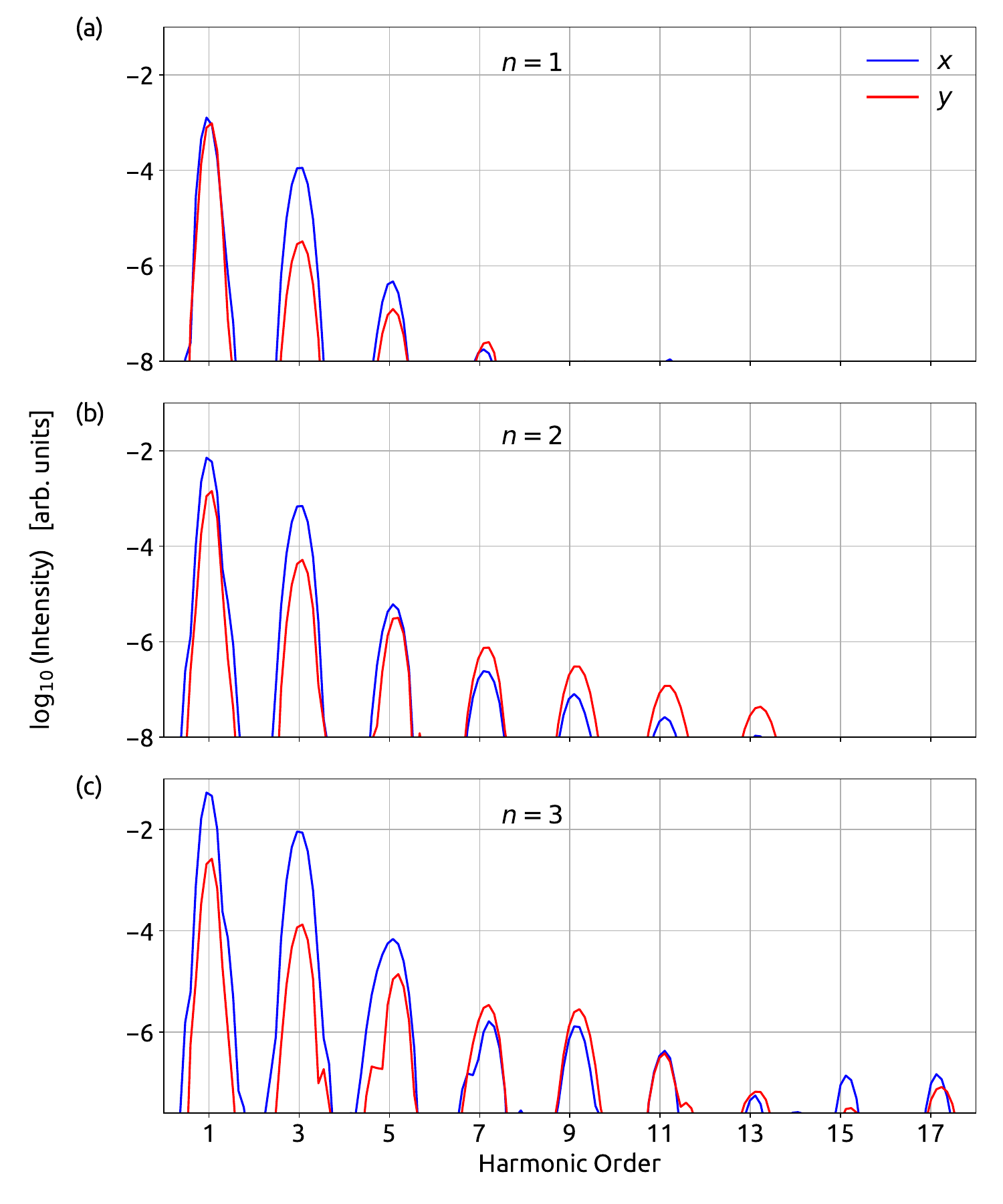}
\caption{High-harmonic spectrum corresponding to m-WSMs with topological charge 
(a) $n = 1$, (b) $n = 2$, and (c) $n =3 $.  The driving laser is 100 fs long 
with  wavelength of $3.2~\mu$m and intensity of  $10^{11}$ W/cm$^2$. 
The laser pulse is linearly  polarized along $x$ direction. 
A phenomenologically  decoherence time of 1.5 fs is added.} \label{fig:fig5.24}
\end{figure}

So far, we have investigated how the  anomalous current behaves with respect to the intensity.  
Recently, it has been shown that the nonlinear anomalous current  in a Weyl semimetal with $n =1$
decreases with increasing the distance between the Weyl nodes~\citep{bharti2022high,avetissian2022high}. 
Let us see how the reported observation changes for m-WSMs. 
Figure~\ref{fig:fig5.21}(b) presents how the anomalous current's peak varies with  
the distance between the Weyl nodes for different $n$ values. 
The current due to the Berry curvature decreases with an increment in the distance between the nodes, 
and the decrease is different for each $n$. 
Similar observations can be made when the intensity lies within the linear-response regime as evident from Fig.~\ref{fig:fig5.23}.
The overall behavior of the anomalous current  remains  same, but the response due to the change in the distance  
is nontrivial when we compare for different $n$ values.

The normal and anomalous currents are analyzed as the optical responses in m-WSMs, which  
transit from a linear regime to a nonlinear regime as the laser's intensity increases, 
even from a perturbative regime to a nonperturbative regime.
%High-harmonic generation (HHG) is a hallmark example of a nonperturbative nonlinear optical process, which has become a method of choice to probe various 
%static and dynamical aspects of solids~\citep{schubert2014sub, mrudul2021light, mrudul2021controlling, hohenleutner2015real, zaks2012experimental, pattanayak2020influence,  pattanayak2019direct, pattanayak2022role, rana2022probing, rana2022high}.   
%In addition,  HHG from topological materials has become a center  of attention as it allows one 
%to investigate nonequilibrium topological aspects of  topological insulators, Dirac  
%and Weyl semimetals in recent years~\citep{baykusheva2021all, kovalev2020non, cheng2020efficient, bai2021high, dantas2021nonperturbative, lv2021high}. 
%Recently, it has been shown that  high-harmonic spectroscopy  can be used to probe the  
%light-induced nonlinear anomalous Hall effect in a Weyl semimetal~\citep{bharti2022high}.
%Thus, it is interesting to explore how the topological charge impacts HHG from m-WSMs. 
The high-harmonic spectra corresponding to m-WSMs are distinct for different $n$'s as can be seen  from Fig.~\ref{fig:fig5.24}. 
Odd-order harmonics are only generated as m-WSMs exhibit 
inversion symmetry in the present case. 
The harmonic cutoff not only increases drastically but 
also the yield of the  harmonics is boosted by several orders  as $n$ increases. 
The energy cutoff increases from seven to thirteen as $n$ changes from $n = 1$ to $n = 2$. 
Moreover, the yield of the seventh harmonic is two-orders in magnitude  boosted 
as $n$ transits from 1 to 2 as is evident from Figs.~\ref{fig:fig5.24}(a) and (b).  
Similar observations can be made for the harmonics shown in Fig.~\ref{fig:fig5.24}(c) for $n = 3$.
The Berry-curvature-driven anomalous current  in m-WSMs  
results in anomalous odd harmonics along the $y$ direction. 
The presence of the anomalous harmonics is a signature of the light-induced anomalous Hall effect, and 
the strength of these harmonics gives the measure of the Hall effect~\citep{bharti2022high}.

The characteristic dependence of the relative yield of the normal and anomalous harmonics on the harmonic-order 
offers a route to tailor the polarization of the emitted harmonics, which carry information about 
the topology of the Berry curvature in m-WSMs. 
The ellipticity of the first, third, and fifth harmonics for $n =1$ reads as  0.85, 0.17, and 0.55, respectively.
As the value of the topological charge changes to $n = 2(3)$, the ellipticity of the first, third, and fifth harmonics  changes drastically as 0.41(0.2), 0.25(0.11), and 0.72(0.45), respectively. 
The reduction of the ellipticity of a given harmonic with an increase in  the topological charge can be attributed to a significant change in the yield of the  anomalous harmonic. 
Thus,  the ellipticity of the emitted harmonic for a given  harmonic order is significantly different for different  $n$ values and  
can be potentially used as a characterization tool for the  topological charge in m-WSMs. 

\section{Conclusion}
In summary, we have investigated  the role of TRS breaking in strong-field driven high-harmonic generation 
 in Weyl semimetals.  
It is found that the 
non-trivial topology of the TRS-broken Weyl semimetal leads to the generation  of the anomalous odd harmonics, which are anisotropic and appear only when the driving laser has non-zero components along $y$ or $z$ direction. 
Non-trivial symmetry of the Berry curvature's components of the 
TRS-broken Weyl semimetal is responsible for 
the anisotropic nature of the anomalous  current and resultant harmonics. 
Moreover, the strength of the Berry curvature dictates 
the strength of the anomalous odd harmonics. 
Furthermore, non-trivial topology properties of the Berry curvature and its strength 
can be probed by measuring the the polarization of the emitted anomalous odd harmonics.

To go beyond ideal Weyl semimetal with degenerate,  untilted Weyl nodes, 
we studied high-harmonic generation from Weyl semimetal  that mimics the realistic conditions. 
We found that by suitably selecting the Weyl semimetal, with a moderate to large tilting, yields of higher harmonics can be improved by a few orders of magnitude with given parameters of driving laser pulse.  
Alternatively, the energy split between Weyl nodes can also significantly improve the yield of higher harmonics. 
Hence, it is expected that unconventional Weyl semimetal, which exhibits higher topological charge and finite energy split, would serve as a better source for higher-order harmonics~\citep{flicker2018chiral,chang2017unconventional}.

In the last part of this chapter,  we have explored the impact of the topological charge  
on laser-driven electron dynamics in  Weyl semimetals. 
For this purpose,  m-WSMs with topological charges $n =1, 2$, and 3 are considered. 
It has been found that the laser-driven electronic currents are distinct for different $n$'s.   
Moreover, the direction and amplitude of the current strongly depend on the laser polarization. 
The  parity and amplitude of the  Berry curvature determine the behavior of the 
anomalous current -- current perpendicular to the laser polarization. 
As the laser's intensity increases, the scaling of the anomalous current with $n$ 
deviates drastically from the linear response theory, i.e., $\mathbf{J}_\mathbf{\Omega}  \propto  n$. 
Owing to the large linear response from the Berry curvature, 
the anomalous current dominates at  lower intensity. 
However, the ratio of the anomalous current from higher to lower topological charge decreases, which leads
the saturation of the anomalous current relatively at lower intensity. 
On the other hand, normal current -- current parallel to the laser polarization -- scales with the topological charge, and therefore  remains separated in magnitude at relatively higher intensity. 

As the intensity increases, the optical responses from  m-WSMs transit from a 
linear regime to 
a nonlinear regime, and  from a perturbative regime to a nonperturbative regime.  
High-harmonic spectroscopy is used 
to probe the distinct and interesting features of the currents in  m-WSMs. 
It has been observed that the harmonic yield and the energy cutoff  of the higher-order 
harmonics increase drastically as $n$ increases. 
Moreover,  the polarization of the emitted harmonics encodes 
information about the phase and magnitude of the Berry curvature's components. 
In addition, our study will be useful in generating tailored laser pulses with tunable polarization and energy~\citep{bharti2022high,wang2024table}.
\end{sloppypar}

%Therefore, present  work opens new avenue for studying strong-field driven electron dynamics and high-harmonic generation in systems with broken TRS, such as exotic magnetic and topological materials; and tailoring the polarization of the emitted harmonics. It can be anticipated that the topological charge of the multi-Weyl semimetal can be characterized by the polarization of the emitted harmonics in an all-optical way.

\cleardoublepage
\chapter{Conclusions and Future Directions}
In this thesis, we studied selected intense-laser driven phenomena in WSMs. 
Weyl semimetals are promising quantum materials that exhibit interesting topological properties. 
Lately, it has been shown that laser-driven electron dynamics have characteristic signatures in two- 
and three-dimensional Dirac semimetals.
Hence, the topological nature of WSMs makes it  attractive candidates for nonlinear optical responses. 
Through various studies, we concluded that the interaction of an intense laser with  WSMs results 
interesting and surprising responses. 
This  thesis is broadly into two parts. In the first part, we studied frequency-independent response from the WSMs. 
When interacting with intense circularly polarized pulses, a helicity-sensitive residual electronic population is generated in the conduction band of WSMs, which can be probed by angle-resolved photoemission spectroscopy. 
In parallel,  WSMs generate photocurrent in their interaction with an intense circularly polarized pulse. 
By tuning the parameters of the driving laser, the direction and magnitude of the generated photocurrent can be tailored. Moreover, we proposed a universal method to generate photocurrent in solids irrespective of their underlying symmetries. 
In the second part of the thesis, we studied the time-varying response by employing  high-harmonic spectroscopy. Our studies show an all-optical probe of the anomalous nonlinear Hall effect and method to 
tailor the polarization of the emitted harmonics. 
The interaction of intense laser  with WSMs is simulated by solving density-matrix-based Semiconductor Bloch equations numerically. Weyl semimetals are described by different variants of tight-binding Hamiltonian in this thesis.   

Circularly polarized light fails to generate currents in inversion-symmetric WSMs with degenerate Weyl nodes at the Fermi energy. 
While each node generates a current with a 
direction depending on its chirality, the two currents in the two degenerate Weyl nodes of opposite   
chirality cancel each other. By extension, it is also generally expected that the currents generated at the same Weyl node by the laser of opposite helicities should also observe mirror
symmetry and cancel. In the \textbf{Chapter 3}, surprisingly, we found that this is not the case. 
The origin of this effect lies in the nonlinear energy dispersion of WSMs, 
which manifests strongly already in the close proximity to the Weyl nodes, where linear dispersion is expected to hold, and the Weyl fermions are thus expected to be massless. 
A scheme based on tailored circular laser pulses, composed of a counterrotating 
fundamental and its second harmonic, is proposed to control the induced population 
asymmetry at a chiral node from positive to negative, including zero. 

Generating and tailoring photocurrent in topological materials has immense importance in fundamental studies and numerous applications in technology. 
In \textbf{Chapter 4}, we introduced a universal method to generate ultrafast photocurrent in
{\it both} inversion-symmetric and inversion-broken WSMs with 
degenerate Weyl nodes at the Fermi level.
Our approach harnesses the asymmetric electronic population in the conduction band induced by an intense  
{\it single-color} circularly polarized laser pulse.  
It has been found that the induced photocurrent can be tailored 
by manipulating the helicity and ellipticity of the driving laser.  
Moreover, our approach generates photocurrent in realistic situations  
when the Weyl nodes are positioned at different energies and have finite tilt along a certain direction.
Our work adds a new dimension to practical applications of  
WSMs for optoelectronics and photonics-based quantum technologies.

In the later part of \textbf{Chapter 4}, we proposed a universal method to generate photocurrent in 
``normal'' and topological materials using a pair of multicycle linearly polarized laser pulses.
The interplay of the fundamental and its second harmonic pulses is studied
for the photocurrent generation in WSMs 
by varying the angle between the polarization direction, relative intensity, and relative phase delay of $\omega$ and $2 \omega$ pulses.
It has been observed that the  presence of a comparatively weaker $2 \omega$ pulse is sufficient to generate  
substantial photocurrent. 
Moreover, significant photocurrent is generated even when polarization directions are orthogonal for certain ratios of the lasers' intensities. 
In addition, the photocurrent is susceptible to the delay between the two pulses. 
We have unequivocally illustrated that our proposed scheme is extendable to non-topological and two-dimensional materials, such as graphene and molybdenum disulfide (MoS$_2$).

In the first part of the \textbf{Chapter 5},  we demonstrated that the laser-driven electron dynamics in a 
WSM with broken time-reversal symmetry has intriguing features in its high-harmonic spectrum. 
The parity and magnitude of the non-zero Berry curvature's components are found to control the direction and strength of the ``anomalous'' current, leading to the generation of ``anomalous'' odd harmonics. 
We demonstrate that the nontrivial topology of the Berry curvature in time-reversal symmetry broken quantum materials can be probed by measuring the polarization of the emitted  harmonics.
Furthermore, the presence of the ``anomalous''  harmonics allows to tailor the polarization of the emitted harmonics.
Our findings unequivocally illustrate that laser-driven electron dynamics lead to the generation of nonlinear anisotropic anomalous Hall effect in time-reversal symmetry broken quantum materials on ultrafast timescale.

In the second part of the \textbf{Chapter 5}, we reported a systematic and detailed investigation on strong-field driven nonperturbative HHG from WSMs in various realistic environments, i.e., going beyond the idealistic situation where the energy degenerate Weyl nodes are at Fermi energy. 
Two classes of topological semimetals are considered: time-reversal broken WSM and inversion-symmetry broken WSM. 
It has been found that type-II WSM leads to significant enhancement in the yield of the higher-order harmonics. 
In addition, energy splitting between the Weyl nodes also results in a modest boost in the harmonic yield. The underlying mechanism responsible for the enhancement can be traced to a drastic increase in the 
electronic population in the conduction band, and noticeable changes in the momentum matrix amplitude. 
A combined effect of the tilt in the Weyl cones and energy separation between the nodes allows the generation of forbidden even-order harmonics in inversion-symmetric WSM.

% The successful realization of the topological Weyl semimetals has revolutionized contemporary physics. 
% In recent years, m-Weyl semimetals, a class of  topological Weyl semimetals,  has 
% attracted broad interest in condensed matter physics. 
% Multi-Weyl semimetals are emerging topological semimetals with nonlinear anisotropic energy 
% dispersion, which is characterized by higher topological charges. 

In the third and the last part of the \textbf{Chapter 5}, we investigated how higher-order topological charge affects the nonlinear optical response 
from m-WSMs. 
It has been observed that the laser-driven electronic current is characteristic of the topological charge and
the laser polarization's direction influences the current's direction and amplitude. 
In addition, the ``anomalous'' current, perpendicular to the laser's polarization, 
carries a distinct signature of the topological charges  
and encodes the information about the parity and amplitude of the nontrivial Berry curvature of m-WSMs.
We showed that the ``anomalous'' current associated with the anomalous Hall effect remains no longer proportional to the topological charge  
at relative higher intensity of the driving laser pulse --  a significant deviation from the linear response regime. 
High-harmonic spectroscopy is employed to capture the distinct and interesting features of the currents in m-WSMs where the topological charge drastically impacts the harmonics' yield and energy cutoff. 

Our work will motivate researchers to explore intense-laser interaction with WSMs intensively in the 
coming days. 
Our work has showcased the intriguing effect of an intense circularly polarized laser on the underlying physics and exciting avenues for technological applications of WSMs in ultrafast photonics, optoelectronic devices and lightwave-driven quantum technologies at Petahertz rate. 
Moreover, we have presented a comprehensive
picture of HHG from WSMs in a conclusive and meaningful way. 
The tilting, energy shifts, the distance between Weyl nodes, and topological charges affect HHG from WSMs distinctively.  
The ingenuity of our works lie in the comprehensive selection
of parameters and in bringing easily observable conclusions. 
 Work presented in this thesis can be extended to another emerging classes of WSMs, such as chiral WSM, dipolar WSM and non-abelian topological WSM.   
We firmly believe that this thesis will have a profound impact  for both experimentalists and theoreticians.

\cleardoublepage
%\include{Chapter7}
%\cleardoublepage
\end{spacing}
%%%%%% ADD SUPPLEMENtary results

\begin{spacing}{1.3}	
\addcontentsline{toc}{chapter}{Bibliography}
\bibliographystyle{abbrvnat.bst}
\bibliography{photocurrent_comb}

\cleardoublepage \addcontentsline{toc}{chapter}{List of Publications}
\include{Publication}\cleardoublepage
\cleardoublepage
\end{spacing}

\begin{spacing}{1.25}  % Line spacing
	\appendix  % This changes the chapter numbering to A, B, C...
	% Appendix A
	\chapter*{Appendices}
	\chapter*{Appendix A: Berry Connections for Linear and Nonlinear Energy Dispersion}  % Asterisk (*) to suppress chapter numbering in document
	\addcontentsline{toc}{chapter}{Appendix A: Berry Connections for Linear and Nonlinear Energy Dispersion}  % Manually add Appendix A to ToC
	\markboth{Appendix A}{Appendix A}  % Update headers
	\renewcommand{\thesection}{A.\arabic{section}}  % Custom section numbering for Appendix A
	\setcounter{figure}{0}
	\renewcommand{\thefigure}{A.\arabic{figure}}
	\setcounter{section}{0}  % Reset section numbering
	%\section {Berry Connections for Linear and Nonlinear Energy Dispersions}{\label{chap:app0}}
\newcommand{\lam}{\lambda}
\newcommand{\E}{\mathcal{E}}
%\chapter{}
\begin{sloppypar}
\section{Berry Connections  of $\mathcal{H}_{W_{1}}$}
In this section, we will evaluate the Berry connection between conduction bands of $\mathcal{H}_{W_{1}}$ near the first Weyl node using Eq.~\eqref{eq:vecH1} as 
$$\mathcal{A}_{cc} = i \left\langle c\left|\pdv{}{k_j}\right|c \right\rangle.$$ 
Let us first compute the first step as 
\begin{eqnarray}
	\pdv{}{k_x}|c\rangle_{W_{1}} & = & \pdv{}{k_x}\left[\frac{1}{\sqrt{2\lam(\lam - k_z)}} \begin{pmatrix}
		k_z - \lam \\
		k_x - i k_y
	\end{pmatrix}  \right]\\\nonumber
	& = & \frac{1}{\sqrt{2\lam(\lam - k_z)}} \begin{pmatrix}
		\pdv{k_z}{k_x} - \pdv{\lam}{k_x} \\
		\pdv{k_x}{k_x} - i \pdv{k_y}{k_x}
	\end{pmatrix} + \begin{pmatrix}
		k_z - \lam \\
		k_x - i k_y
	\end{pmatrix} \pdv{}{k_x} \frac{1}{\sqrt{2\lam(\lam - k_z)}}.
\end{eqnarray}
Now, let us estimate  the complete term as 
\begin{eqnarray}
	\nonumber
	\left \langle c\left|\pdv{}{k_x}\right|c\right\rangle_{W_{1}} &  = & 
	\frac{1}{\sqrt{2\lam(\lam - k_z)}} \begin{pmatrix}
		k_z - \lam &	k_x + i k_y
	\end{pmatrix} \\ \nonumber 
	&& \times
	\left[ \frac{1}{\sqrt{2\lam(\lam - k_z)}} \begin{pmatrix}
		-k_x/\lam \\
		1
	\end{pmatrix} +\frac{-2k_x + \frac{k_x k_z}{\lam} }{\{2\lam(\lam -k_z) \}^{3/2}  }
	\begin{pmatrix}
		k_z - \lam \\
		k_x - i k_y
	\end{pmatrix} \right].\\ \nonumber
	& = & \frac{(-k_x k_z/\lam)+k_x +k_x +i k_y }{2\lam(\lam-k_z)} + \underbrace{\frac{k_z^2  + k_x^2 + k_y^2+ \lam^2 - 2\lam k_z }{{2\lam(\lam -k_z)}}}_{\beta=1} \left(-2k_x + \frac{k_x k_z}{\lam} \right). \\ \nonumber
\end{eqnarray}
Thus, the $x$-component  of the Berry connection can be written as 
\begin{equation}
	\mathcal{A}_{cc}^x = i \left \langle c\left|\pdv{}{k_x}\right|c\right\rangle_{W_{1}} = i \frac{(-k_x k_z/\lam) + 2k_x + i k_y }{2\lam(\lam-k_z)}  -i\left[2k_x - (k_x k_z/\lam)\right].
\end{equation}
Similarly, the $y$- and $z$-components of the Berry connection are, respectively, written as 
\begin{equation}
	\mathcal{A}_{cc}^y =  i\left \langle c \left|\pdv{}{k_y} \right|c \right\rangle_{W_{1}} = i \frac{(-k_y k_z/\lam) + 2k_y - i k_x }{2\lam(\lam-k_z)}  -i\left[2k_y - \frac{k_y k_z}{\lam}\right], 
\end{equation}
and 
\begin{equation}
	\mathcal{A}_{cc}^z ~~~ = ~~~i \left \langle c \left|\pdv{}{k_z} \right|c \right\rangle _{W_{1}}~~~ = ~~~ i \frac{k_z -\lam -\frac{k_z^2}{\lam}   }{\lam(\lam-k_z)}  -i\left[2 k_z-\frac{k_z^2}{\lam}-\lam\right].
\end{equation}

\section{Berry Connections of  $\mathcal{H}_{W_{2}}$}
In this section, we estimate the components of the Berry connection between the conduction bands of $\mathcal{H}_{W_{2}}$. Following the similar step as in the previous section, we can write the $x$-component as 
\begin{eqnarray}
	\nonumber
	\left \langle c\left|\pdv{}{k_x}\right|c\right\rangle_{W_{2}} & = & 
	\frac{1}{\sqrt{2\lam(\lam + k_z)}} \begin{pmatrix}
		-k_z - \lam &	k_x + i k_y
	\end{pmatrix}  \\ \nonumber
	&&\times 
	\left[ \frac{1}{\sqrt{2\lam(\lam + k_z)}} \begin{pmatrix}
		-k_x/\lam \\
		1
	\end{pmatrix} -\frac{2k_x + \frac{k_x k_z}{\lam} }{\{2\lam(\lam -k_z) \}^{3/2}  }
	\begin{pmatrix}
		-k_z - \lam \\
		k_x - i k_y
	\end{pmatrix} \right].\\ \nonumber
	& = & \frac{(k_x k_z/\lam)+ k_x + k_x +i k_y }{2\lam(\lam+k_z)} - \underbrace{\frac{k_z^2  + k_x^2 + k_y^2+ \lam^2 + 2\lam k_z }{{2\lam(\lam +k_z)}}}_{\beta=1} \left( 2k_x + \frac{k_x k_z}{\lam} \right). \\ \nonumber
\end{eqnarray}
Thus, we have the compact  expression as 
\begin{equation}
	\mathcal{A}_{cc}^x = i \left \langle c\left|\pdv{}{k_x}\right|c\right\rangle_{W_{2}} = i \frac{(k_x k_z/\lam) + 2k_x + i k_y }{2\lam(\lam+k_z)}  -i\left[2k_x + (k_x k_z/\lam)\right]. 
\end{equation}

Similarly, the $y$- and $z$-components are, respectively,  written as 
\begin{equation}
	\mathcal{A}_{cc}^y = i\left \langle c \left|\pdv{}{k_y} \right|c \right\rangle_{W_{2}} = i \frac{(k_y k_z/\lam) + 2k_y - i k_x }{2\lam(\lam+k_z)}  -i\left[2k_y + (k_y k_z/\lam)\right],
\end{equation}
and
\begin{equation}
	\mathcal{A}_{cc}^z = i\left \langle c \left|\pdv{}{k_z} \right|c \right\rangle_{W_{2}} =  i \frac{k_z +\lam +\frac{k_z^2}{\lam}   }{\lam(\lam+k_z)}  -i\left[2 k_z+\frac{k_z^2}{\lam}+\lam\right].
\end{equation}

The dipole matrix elements corresponding to $\mathcal{H}_{W_{1}}$ and $\mathcal{H}_{W_{2}}$ are related by mirror symmetry of $k_z$. 
If we replace $k_z\to-k_z$ in the matrix elements of $\mathcal{H}_{W_{1}}$, 
then it is the same as those for $\mathcal{H}_{W_{2}}$. 
Thus, if one part of the Weyl node, say along $+k_z$, is selectively excited by one helicity of CPL, 
then the other Weyl node is excited in $-k_z$ by the same helicity of CPL.
However, the selective excitation is valid in the linear dispersion regime only. 
In order to show that it is true in general, 
we consider the next higher-order correction to the energy band dispersion in the following section. 

The overall findings are tabulated as follows:
{\def\arraystretch{2}\tabcolsep=20pt
	\begin{table}[ht]
		\centering
		\begin{tabular}{|c|c|c|c|}
			\hline
			&$\mathcal{H}_{W_{1}}$&$\mathcal{H}_{W_{2}}$ \\
			\hline
			$	\mathcal{A}_{vc}^{x}$ & $\frac{1}{2\lam(\lam^2 - k_z^2)^{1/2}}\left( \frac{-i k_x k_z}{\lam} - k_y \right) $	& 	$ \frac{1}{2\lam(\lam^2-k_z^2)^{1/2}}\left( \frac{i k_x k_z}{\lam} - k_y \right)$\\
			\hline
			$\mathcal{A}_{vc}^{y}$  &  $\frac{1}{2\lam(\lam^2 - k_z^2)^{1/2}}\left(\frac{-ik_z k_y}{\lam} + k_x \right)$& $ \frac{1}{2\lam(\lam^2-k_z^2)^{1/2}}\left( \frac{i k_y k_z}{\lam} + k_x \right)$ \\
			\hline
			$\mathcal{A}_{vc}^{z}$& $\frac{i (\lam^2 - k_z^2)}{2\lam^2(\lam^2 - k_z^2)^{1/2}}$ & $  \frac{i\left(  k_z^2 - \lam^2 \right)}{2\lam^2(\lam^2-k_z^2)^{1/2}} $ \\
			\hline
			$\mathcal{A}_{cc}^x$& $ i \frac{(-k_x k_z/\lam) + 2k_x + i k_y }{2\lam(\lam-k_z)}  -i\left[2k_x - \frac{k_x k_z}{\lam}\right]$ & $i \frac{(k_x k_z/\lam) + 2k_x + i k_y }{2\lam(\lam+k_z)}  -i\left[2k_x + \frac{k_x k_z}{\lam}\right]$ \\
			\hline
			$\mathcal{A}_{cc}^y$ & $i \frac{(-k_y k_z/\lam) + 2k_y - i k_x }{2\lam(\lam-k_z)}  -i\left[2k_y - \frac{k_y k_z}{\lam}\right]$ & $ i \frac{(k_y k_z/\lam) + 2k_y - i k_x }{2\lam(\lam+k_z)}  -i\left[2k_y + \frac{k_y k_z}{\lam}\right]$ \\
			\hline
			$\mathcal{A}_{cc}^z$& $i \frac{k_z -\lam -\frac{k_z^2}{\lam}   }{\lam(\lam-k_z)}  -i\left[2 k_z-\frac{k_z^2}{\lam}-\lam\right]$ & $i \frac{k_z +\lam +\frac{k_z^2}{\lam}   }{\lam(\lam+k_z)}  -i\left[2 k_z+\frac{k_z^2}{\lam}+\lam\right]$ \\
			\hline
			$\mathcal{A}_{vv}^x$&$i \frac{(k_x k_z/\lam) + 2k_x + i k_y }{2\lam(\lam+k_z)}  -i\left[2k_x + \frac{k_x k_z}{\lam}\right]$ &$i \frac{(-k_x k_z/\lam) + 2k_x + i k_y }{2\lam(\lam-k_z)}  -i\left[2k_x - \frac{k_x k_z}{\lam}\right]$ \\
			\hline
			$\mathcal{A}_{vv}^y$&$ i \frac{(k_y k_z/\lam) + 2k_y - i k_x }{2\lam(\lam+k_z)}  -i\left[2k_x + \frac{k_y k_z}{\lam}\right]$ &$i \frac{(-k_y k_z/\lam) + 2k_y - i k_x }{2\lam(\lam-k_z)}  -i\left[2k_x - \frac{k_y k_z}{\lam}\right]$ \\
			\hline
			$\mathcal{A}_{vv}^z$& $i \frac{2k_z +\frac{k_z^2}{\lam}+\lam  }{2\lam(\lam+k_z)}  -i\left[ 2k_z + \frac{k_z^2}{\lam} + \lam \right]$ &$i \frac{2k_z -\frac{k_z^2}{\lam}-\lam  }{2\lam(\lam-k_z)}  +i\left[ -2k_z + \frac{k_z^2}{\lam} + \lam \right]$ \\
			\hline
		\end{tabular}
\end{table}}

\section{Beyond Linear-Energy Dispersion}
As we go away beyond linear regime,  energy band dispersion becomes nonlinear (quadratic). 
We introduce a parameter $\Delta E$, which measures the deviation of the energy dispersion from the linear one as depicted in Fig.~\ref{fig:deviation}. 
If the energy dispersion is linear, then  $\Delta E=0$.

Let us proceed with previous approach and expand the  Hamiltonian in Eq.~\eqref{eq:weylhamtb} 
to the next higher-order term in $\k$, which is quadratic in nature. 
In addition, we calculate the Berry connection by following the same procedure as in the previous section. 
The resultant Berry connections will help us 
to understand how CPL couples to Weyl nodes away from the linear dispersion region.

\begin{figure}[!h]
\centering
\includegraphics[width=0.7\linewidth]{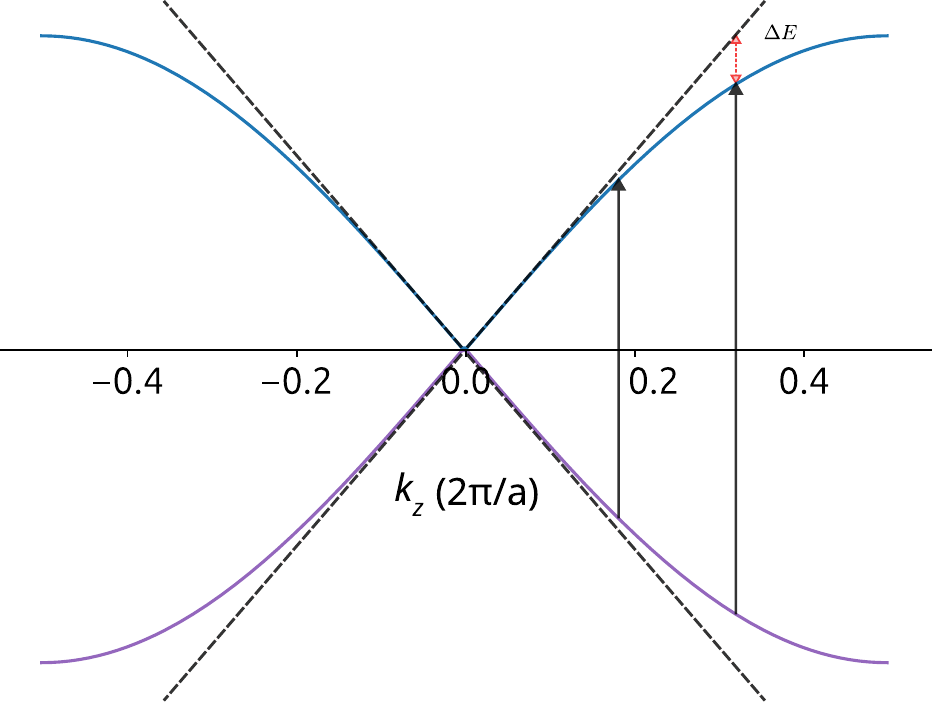}
\caption{Deviation of the energy band dispersion beyond the linear behaviour around a Weyl node.}
\label{fig:deviation}
\end{figure}

By including the next higher-order term in $\k$ n the expansion of the Eq.~\eqref{eq:weylhamtb}, 
the resultant Hamiltonian  near the first Weyl point at $k_z = \pi/2a$ is expressed  as 
\begin{equation}
\mathcal{\tilde{H}}_{W_{1}} = \nu[-k_x \sigma_x + k_y  \sigma_y - k_z^{\prime}   \sigma_z].
\end{equation}
Similarly, the Hamiltonian at the second  Weyl point at $k_z = -\pi/2a$ is   
\begin{equation}
\mathcal{\tilde{H}}_{W_{2}} = \nu[-k_x \sigma_x + k_y \sigma_y + k_z^{\prime} \sigma_z],
\end{equation}
where  $b=\alpha a$, $\nu = 2ta$ and $k_z^{\prime} = k_z - \frac{b}{2} (k_x^{2} + k_y^{2}).$	

The two eigenvalues associated with $\mathcal{H}_{W_{1}}^{\prime}$ are
\begin{equation}
\mathcal{E}^{W_{1}}_{v/c} = \pm \nu \sqrt{k_x^2 + k_y^2 + {k_z^\prime}^2 } = \pm \lambda,
\end{equation} 
where $-\lambda$ ($+\lambda$) corresponds to valence (conduction)
band.  The corresponding eigenstates can be written as 
\begin{equation}
	\ket{\tilde{v}}_{W_{1}} = \frac{1}{\sqrt{2\lambda^2 + 2\lambda k_z^\prime}}\begin{pmatrix}
		k_z^\prime + \lambda\\
		k_x - i k_y
	\end{pmatrix},~~~\textrm{and}~~
	\ket{\tilde{c}}_{W_{1}} = \frac{1}{\sqrt{2\lambda^2 - 2\lambda k_z^\prime}}\begin{pmatrix}
		k_z^\prime - \lambda\\
		k_x - i k_y
	\end{pmatrix}.
\end{equation}

In order to estimate the components of the 
Berry connection at the first Weyl node, we need to estimate the following derivative as
\begin{eqnarray}
	\pdv{}{k_x} |\tilde{c}\rangle_{W_{1}} & = &  \pdv{}{k_x}\left[\frac{1}{\sqrt{2\lam(\lam - k_z^{\prime})}} \begin{pmatrix}
		k_z^\prime - \lam \\
		k_x - i k_y
	\end{pmatrix}  \right].\\\nonumber
	& = & \frac{1}{\sqrt{2\lam(\lam - k_z^{\prime})}} \begin{pmatrix}
		\pdv{k_z^{\prime}}{k_x} - \pdv{\lam}{k_x} \\
		\pdv{k_x}{k_x} - i \pdv{k_y}{k_x}
	\end{pmatrix} + \begin{pmatrix}
		k_z^{\prime} - \lam \\
		k_x - i k_y
	\end{pmatrix} \pdv{}{k_x} \frac{1}{\sqrt{2\lam(\lam - k_z^{\prime})}}. 
\end{eqnarray}
In the case of linear dispersion, we have seen that the second term in the above equation goes to zero because of the inner product $\langle v|c \rangle$. 
Moreover, following quantities are straightforward to obtain as 
\begin{eqnarray}
	\pdv{k_z^\prime}{k_x} = -b k_x, ~~ \pdv{k_z^\prime}{k_y} = -b k_y, ~~~ \pdv{k_z^\prime}{k_z} = 1, \nonumber
\end{eqnarray}
\begin{eqnarray}	
	\pdv{\lambda}{k_x} = \frac{k_x (1- b k_z^{\prime})}{\lam},~\pdv{\lambda}{k_y} = \frac{k_y (1- b k_z^{\prime})}{\lam},~\textrm{and}~\pdv{\lambda}{k_z} = \frac{k_z}{\lam}.
\end{eqnarray}

Now, let us proceed to calculate $\mathcal{A}_{vc}$ using above relations. 
The $x-$component of the Berry connection is obtained as 
\begin{eqnarray}
	\nonumber
	\left\langle \tilde{v} \left|\pdv{}{k_x}\right| \tilde{c}\right\rangle_{W_{1}} & = & 
	\frac{1}{\sqrt{2\lam(\lam + k_z^{\prime})}} \begin{pmatrix}
		k_z^{\prime} + \lam &	k_x + i k_y
	\end{pmatrix} \cdot 
	\left[ \frac{1}{\sqrt{2\lam(\lam - k_z^{\prime})}} \begin{pmatrix}
		-bk_x - k_x(1-b k_z^{\prime})/\lam \\
		1
	\end{pmatrix}  \right] \\ \nonumber
	& = & \frac{(k_z^{\prime} + \lam)  \left[-bk_x - k_x(1-b k_z^{\prime})\right]
		+k_x +i k_y }{2\lam(\lam^2-{k_z^{\prime}}^2)^{1/2}}. \nonumber
\end{eqnarray}

Similarly, the $y$- and $z$-components are, respectively, written as 
\begin{eqnarray}
	\nonumber
	\left\langle \tilde{v}\left|\pdv{}{k_y}\right| \tilde{c}\right\rangle_{W_{1}} & = & 
	\frac{1}{\sqrt{2\lam(\lam + k_z^{\prime})}} \begin{pmatrix}
		k_z^{\prime} + \lam &	k_x + i k_y
	\end{pmatrix} \cdot 
	\left[ \frac{1}{\sqrt{2\lam(\lam - k_z^{\prime})}} \begin{pmatrix}
		-bk_y - k_y(1-b k_z^{\prime})/\lam \\
		-i
	\end{pmatrix}  \right] \\ \nonumber
	& = & \frac{\left(k_z^{\prime} + \lam \right)  \left[-bk_y - k_y(1-b k_z^{\prime})\right]-i k_x + k_y }{2\lam(\lam^2-{k_z^{\prime}}^2)^{1/2}}, \nonumber
\end{eqnarray}
and 
\begin{eqnarray}
	\nonumber
	\left\langle \tilde{v}\left|\pdv{}{k_z}\right| \tilde{c}\right\rangle_{W_{1}} & = & 
	\frac{1}{\sqrt{2\lam(\lam + k_z^{\prime})}} \begin{pmatrix}
		k_z^{\prime} + \lam &	k_x + i k_y
	\end{pmatrix} \cdot 
	\left[ \frac{1}{\sqrt{2\lam(\lam - k_z^{\prime})}} \begin{pmatrix}
		1-k_z/\lam \\
		0
	\end{pmatrix}  \right] \\ \nonumber
	& = & \frac{k_z - b \left(\frac{k_x^2}{2} + \frac{k_y^2}{2}\right) + \lam + \left(-k_z^2/\lam + b(k_z/\lam) \left(\frac{k_x^2}{2} + \frac{k_y^2}{2}\right) - k_z \right) }{2\lam(\lam^2-{k_z^{\prime}}^2)^{1/2}}. \nonumber
\end{eqnarray}
On comparing the above components of $\mathcal{A}_{vc}$ with those obtained in the previous section for the linear dispersion, we observed that the mirror symmetry of $k_z \to -k_z$ is broken.

By following the above procedures, the two eigenvalues associated with the second Weyl node are
\begin{equation}
\mathcal{E}^{W_{2}}_{v/c} = \pm \nu \sqrt{k_x^2 + k_y^2 + {k_z^\prime}^2 } = \pm \lambda,
\end{equation}
and the corresponding eigenstates are 
\begin{equation}
	\label{eq:1.71}
	\ket{ \tilde{v}}_{W_{2}} = \frac{1}{\sqrt{2\lambda^2 - 2\lambda k_z^\prime}}\begin{pmatrix}
		-k_z^\prime + \lambda\\
		k_x - i k_y
	\end{pmatrix}~~~\textrm{and}~~
	\ket{ \tilde{c}}_{W_{2}} = \frac{1}{\sqrt{2\lambda^2 + 2\lambda k_z^\prime}}\begin{pmatrix}
		-k_z^\prime - \lambda\\
		k_x - i k_y
	\end{pmatrix}.
\end{equation} 

The $x-$component of the Berry connection can be written as 
\begin{eqnarray}
	\nonumber
	\left\langle \tilde{v}\left|\pdv{}{k_x}\right| \tilde{c}\right\rangle_{W_{2}} & = & 
	\frac{1}{\sqrt{2\lam(\lam - k_z^{\prime})}} \begin{pmatrix}
		-k_z^{\prime} + \lam &	k_x + i k_y
	\end{pmatrix} \cdot 
	\left[ \frac{1}{\sqrt{2\lam(\lam + k_z^{\prime})}} \begin{pmatrix}
		bk_x - k_x(1-b k_z^{\prime})/\lam \\
		1
	\end{pmatrix}  \right] \\ \nonumber
	& = & \frac{(-k_z^{\prime} + \lam)  \left[bk_x - k_x(1-b k_z^{\prime})\right]
		+k_x +i k_y }{2\lam(\lam^2-{k_z^{\prime}}^2)^{1/2}}. \nonumber
\end{eqnarray}

Similarly, the $y$- and $z$-components are, respectively, written as 
\begin{eqnarray}
	\nonumber
	\left\langle \tilde{v}\left|\pdv{}{k_y}\right| \tilde{c}\right\rangle_{W_{2}} & = & 
	\frac{1}{\sqrt{2\lam(\lam - k_z^{\prime})}} \begin{pmatrix}
		-k_z^{\prime} + \lam &	k_x + i k_y
	\end{pmatrix} \cdot 
	\left[ \frac{1}{\sqrt{2\lam(\lam + k_z^{\prime})}} \begin{pmatrix}
		bk_y - k_y(1-b k_z^{\prime})/\lam \\
		-i
	\end{pmatrix}  \right] \\ \nonumber
	& = & \frac{\left(-k_z^{\prime} + \lam \right)  \left[-bk_y - k_y(1-b k_z^{\prime})\right]-i k_x + k_y }{2\lam(\lam^2-{k_z^{\prime}}^2)^{1/2}}, \nonumber
\end{eqnarray}
and
\begin{eqnarray}
	\nonumber
	\left\langle \tilde{v}\left|\pdv{}{k_z}\right| \tilde{c}\right\rangle_{W_{2}} & = & 
	\frac{1}{\sqrt{2\lam(\lam - k_z^{\prime})}} \begin{pmatrix}
		-k_z^{\prime} + \lam &	k_x + i k_y
	\end{pmatrix} \cdot 
	\left[ \frac{1}{\sqrt{2\lam(\lam + k_z^{\prime})}} \begin{pmatrix}
		-1-k_z/\lam \\
		0
	\end{pmatrix}  \right] \\ \nonumber
	& = & \frac{k_z - b \left(\frac{k_x^2}{2} + \frac{k_y^2}{2}\right) - \lam + \left(k_z^2/\lam - b(k_z/\lam) \left(\frac{k_x^2}{2} + \frac{k_y^2}{2}\right) - k_z \right) }{2\lam(\lam^2-{k_z^{\prime}}^2)^{1/2}}. \nonumber
\end{eqnarray}
\end{sloppypar}
  % Include content of App1 with sections
	\clearpage
	\pagestyle{empty}
	\cleardoublepage
	
	% Appendix B
	\chapter*{Appendix B: Berry Curvature Calculations I}  % Asterisk (*) to suppress chapter numbering in document
	\addcontentsline{toc}{chapter}{Appendix B: Berry Curvature Calculations I}  % Manually add Appendix B to ToC
	\markboth{Appendix B}{Appendix B}  % Update headers
	\renewcommand{\thesection}{B.\arabic{section}}  % Custom section numbering for Appendix B
	\setcounter{section}{0}  % Reset section numbering
	
\section{Berry Curvature of Time-Reversal Symmetry Broken Weyl Semimetal}
\begin{sloppypar}
Let us again revisit the calculation of Berry curvature for WSM presented in Chapter~\ref{chp:Chapter2}. For the calculations, we follow the formula presented in Eq.~\eqref{eq:berry}, rewritten below for the sake of completeness.

\begin{equation}
	\Omega(\mathbf{k})_{\mathbf{k}, \pm, i} = \pm \epsilon_{ijl} \frac{\mathbf{d}(\mathbf{k})_\mathbf{\alpha} \cdot \left( \frac{\partial{\mathbf{d}(\mathbf{k})_{\mathbf{\alpha}}}}{\partial{k_j}} \times  \frac{\partial{\mathbf{d}(\mathbf{k})_{\mathbf{\alpha}}}}{\partial{k_l}}  \right)} 
	{4|\mathbf{d}(\mathbf{k})_{\mathbf{\alpha}}|^3},
\end{equation}
where $+(-)$ corresponds to conduction (valence) energy bands,  $\epsilon_{ijl}$  
is the Levi-Civita tensor, $\mathbf{d}(\mathbf{k})_\mathbf{\alpha = 1, 2, 3}$ are  the components of $\mathbf{d}(\mathbf{k})_{\mathbf{\alpha}}$ given as $[t_x \{ \cos(k_x a) - \cos(k_0 a)\} + t_y \{\cos(k_y b) -1\} + t_z \{\cos(k_z c) -1\}, 
t_y \sin(k_y b), t_z \sin(k_z c)]$ with 
$t_{x,y,z}$ as the hopping parameters and $a, b, c$ are lattice parameters. 

After substituting the expressions of $\mathbf{d}(\mathbf{k})_{\mathbf{\alpha}}$ and performing simple mathematical steps, the components of the  Berry curvature are written as 

\begin{eqnarray}
	\Omega(\mathbf{k})_{k_{x}, +} & = & \frac{\mathcal{N}_1}{4|\mathbf{d}(\mathbf{k})_{\mathbf{\alpha}}|^3}  \left[  \cos (a k_x) \cos(b k_y) \cos(c k_z)
	+ \cos^2 (b k_y)  \cos(c k_z)  \right.  \nonumber \\  
	&& \left. + \cos(b k_y) \cos^2(c k_z) + \sin^2 (b k_y)  \cos(c k_z) + \cos(b k_y) \sin^2(c k_z)  \right],   \nonumber \\
	&& \\
	\Omega(\mathbf{k})_{k_{y}, +} & = &\frac{\mathcal{N}_2}{4|\mathbf{d}(\mathbf{k})_{\mathbf{\alpha}}|^3} \left[\sin(a k_x)  \sin(b k_y) \cos(c k_z)   \right], \nonumber  \\
	&&\\
	\Omega(\mathbf{k})_{k_{z}, +} & =& \frac{\mathcal{N}_3}{4|\mathbf{d}(\mathbf{k})_{\mathbf{\alpha}}|^3} \left[\sin(a k_x)  \cos(b k_y) \sin(c k_z)   \right].  \nonumber  \\
\end{eqnarray}
Here, $\mathcal{N}$'s depend on $t$'s, lattice parameters ($a, b, c$), and $k_{0}$. 
$|\mathbf{d}_{\mathbf{k}}|^3$ is an even function as   band dispersion is even. 
Therefore, $\Omega_{k_{x}, +}$ is an even function of all components of \textbf{k}, 
whereas  $\Omega_{k_{y}, +}$ is an odd function of $k_x$ and $k_y$, and $\Omega_{k_{z}, +}$ is an odd function of $k_x$ and $k_z$. 
\end{sloppypar}  % Include content of App2 with sections
	\clearpage
	\pagestyle{empty}
	\cleardoublepage
	
	% Appendix C
	\chapter*{Appendix C: Berry Curvature Calculations II}  % Asterisk (*) to suppress chapter numbering in document
	\addcontentsline{toc}{chapter}{Appendix C: Berry Curvature Calculations II}  % Manually add Appendix C to ToC
	\markboth{Appendix C}{Appendix C}  % Update headers
	\renewcommand{\thesection}{C.\arabic{section}}  % Custom section numbering for Appendix C
	\setcounter{section}{0}  % Reset section numbering
	\section{Berry Curvature of Multi-Weyl Semimetals}
\begin{sloppypar}
 
Let's calculate the Berry curvature corresponding to $\mathcal{H}^{(n)}(\mathbf{k})$ given by 
Eqs.~\eqref{eq:n_1}-\eqref{eq:n_3} in the Chapter~\ref{chp:chapter5} for $n = 1, 2$, and 3. 
\vspace{1cm}
\subsection{Berry curvature for the topological charge $n = 1 $}
For the sake of completeness, we again write the components of $\mathbf{d}(\mathbf{k})_\mathbf{\alpha}$, given in Eq.~\eqref{eq:n_1} as
\begin{eqnarray}\label{eq:an_1}
	\mathbf{d}^{(1)}(\mathbf{k}) & =  & \left[t\sin(k_x a), t\sin(k_y a)\right., \nonumber \\
	&&\left. t\{\cos(k_z a) - \cos(k_0 a) +2- \cos(k_x a) - \cos(k_y a)\}\right]. 
\end{eqnarray}

The components of the Berry curvature can be written as
\begin{eqnarray}
	&\Omega(\k)_{k_x} = &-a^2t^3\sin(a k_x)\cos(ak_y)\sin(ak_z),\\
	&\Omega(\k)_{k_y} = &-a^2t^3\cos(ak_x)\sin(a k_y)\sin(ak_z),\\
	&\Omega(\k)_{k_z} = & -a^2t^3 \big[\cos(ak_x)\cos(a k_y) 
	\{\cos(ak_x)+\cos(ak_y)+\cos(ak_0) -\cos(ak_z)-2\} \nonumber \\
	&&+ \sin^2(ak_x)\cos(ak_y) + \cos(ak_x)\sin^2(ak_y)\big].
\end{eqnarray}
%\begin{eqnarray}
%	\Omega(\k)_{k_z} =  -a^2t^3 \bigg[\cos(a k_y) 
%	+ \cos(ak_x)\big[1 + \cos(a k_y)  \{\cos(ak_0) -\cos(ak_z)-2\}\big] \bigg].
%\end{eqnarray}

As evident from the above equations,  $\Omega(\k)_{k_{x}/k_{y}}$ is an odd functions of $k_z, k_y$ and $k_x$. 
This implies that the  anomalous current $\mathbf{J}_\Omega \propto  \int_{\k} \dd{\k}  (\mathbf{E}\times \mathbf{\Omega})$ yields a non-zero contribution only from $\Omega(\k)_{k_z}$. 
Thus, a linearly polarized laser pulse along  $x$- or $y$- direction  generates anomalous Hall effect.
\clearpage
\subsection{Berry curvature for the topological charge $n = 2 $}
Components of $\mathbf{d}(\mathbf{k})_\mathbf{\alpha}$ for $n = 2$ 
as given in the Eq.~\eqref{eq:n_2}  are
\begin{eqnarray}\label{eq:an_2}
	\mathbf{d}^{(2)}(\mathbf{k}) & = & \left[t\{\cos(k_x a)-\cos(k_y a)\}, t\sin(k_y a)\sin(k_y a),\right. \nonumber \\
	&&  \left. t\{\cos(k_z a) - \cos(k_0 a) +2- \cos(k_x a) - \cos(k_y a)\}\right]. 
\end{eqnarray}

The corresponding components of the Berry curvature are 
\begin{eqnarray}
	&\Omega(\k)_{k_x} =  &a^2t^3\left[\sin(ak_x)\sin^2(ak_y)\sin(ak_z)  - \sin(ak_y)\sin(ak_z)
	\{\cos(ak_x) - \cos(ak_y)\}\right], \nonumber\\
	&&\\
	&\Omega(\k)_{k_y} =  &a^2t^3\left[\sin^2(ak_x)\sin(ak_y)\sin(ak_z) 
	+ \cos(ak_x)\sin(ak_y)\sin(ak_z)  \{\cos(ak_x)- \cos(ak_y)\}\right], \nonumber \\
	&&\\
	&\Omega(\k)_{k_z} = & a^2t^3 [ 2 \sin^{2}(ak_x)\sin^{2}(ak_y)\nonumber\\ &&+\{\cos(ak_y)-\cos(ak_x)\}\{\cos(ak_y)\sin^{2}(ak_x)+\cos(ak_x)\sin^{2}(ak_y)\}   \nonumber \\
	&& + \left\{ \cos(ak_x)+ \cos(ak_y) + \cos(ak_0) -\cos(ak_z) - 2\right\} \nonumber\\ 
	&&\times\left\{\cos(ak_y)\sin^{2}(ak_x) + \cos(ak_x)\sin^{2}(ak_y) \right\}].
\end{eqnarray}
%\begin{eqnarray}
%&\Omega(\k)_{k_z} = & a^2t^3 \bigg[2\cos^2(a k_y)\sin^2(a k_x) 
%+ \sin^2(a k_y) \{\cos(ak_0) -\cos(ak_z)-2\} \nonumber \\
%&&+ \cos(a k_y)\sin^2(a k_x) \{\cos(ak_0) -\cos(ak_z)-2\}\bigg].
%\end{eqnarray}

As evident from the above components, the non-zero contribution to the anomalous current  arises from 
$\Omega(\k)_{k_z}$ only, similar to previous case for $n=1$.
\clearpage
\subsection{Berry curvature for the topological charge $n = 3 $}
Components of $\mathbf{d}(\mathbf{k})_\mathbf{\alpha}$ for $n = 3$ from Eq.~\eqref{eq:n_3} 
can be written as 
\begin{eqnarray}\label{eq:an_3}
	\mathbf{d}^{(3)}(\mathbf{k}) & = & \left[t\sin(k_x a)\{3\cos(k_y a)-\cos(k_x a) -2\},\right. \nonumber \\
	&&\left. t\sin(k_y a)\{3\cos(k_x a)-\cos(k_y a) - 2\}, \right. \nonumber \\
	&&\left. t\{\cos(k_z a) - \cos(k_0 a) +2- \cos(k_x a) - \cos(k_y a)\}\right].
\end{eqnarray}

The corresponding components of the Berry curvature are
\begin{eqnarray}
	&\Omega(\k)_{k_x} = & [ a^2 t^3 \sin (a k_x) \{ 3 \cos (a k_y) -\cos (a k_x)-2 \} 
	\left\{ 2  \sin (ak_z) \cos (a k_y)  \right. \nonumber \\
	&&\left. +  \sin (ak_z) \cos ^2(ak_y) -  \sin (ak_z) \sin ^2(a k_y)  -3  \cos (a k_x) \sin (ak_z) \cos (ak_y) \right \} ] \nonumber \\
	&& + [3 a  t^3 \sin (a k_x) \sin (ak_z) \sin ^2(ak_y) \{3 \cos (a k_x)-\cos (a k_y)-2\}],\nonumber\\
	&&\\
	&\Omega(\k)_{k_y} = & [ a^2 t^3 \sin (a k_y)
	\left \{   \cos ^2(a k_x) \sin (a k_z)+2  \cos (ak_x) \sin (a k_z)  \right.\nonumber \\
	&& \left. -  \sin ^2(a k_x) \sin (a k_z) -3   \cos (ak_x) \cos (ak_y) \sin (a k_z) \right\} 
	\left\{3 \cos (a k_x)-\cos (a k_y)-2\right\} ] \nonumber \\
	&& + [3 a^2 t^3 \sin ^2(a k_x) \sin (ak_y) \sin (a k_z) \left\{3 \cos (ak_y)-\cos (a k_x)-2\right\}],\nonumber\\
	&&\\
	&\Omega(\k)_{k_z} =& [ t \{ \cos (a k_z)-\cos (ak_0)\} - t \{ \cos (a k_x) + \cos (ak_y) - 2 \} \nonumber \\
	&&  \times a^2 t^2  \{ 10  \cos ^2(a k_x) \cos ^2(ak_y) - 8 \sin ^2(a k_x) \sin ^2(a k_y) 
	-3  \cos ^3(ak_x) \cos (a k_y)  \nonumber \\
	&& + 4  \cos (ak_x) \cos (ak_y) -4  \cos ^2(a k_x) \cos (ak_y)-3  \cos (a k_x)\cos ^3(a k_y) \nonumber \\
	&&-4  \cos (a k_x) \cos ^2(ak_y)-  \cos ^2(ak_x) \sin ^2(ak_y) - \sin ^2(a k_x) \cos ^2(a k_y) \nonumber \\
	&&+3  \sin ^2(a k_x) \cos (a k_x) \cos (a k_y) +3   \cos (a k_x) \sin ^2(a k_y) \cos (a k_y) \nonumber \\
	&&\left.-2   \cos (ak_x) \sin ^2(a k_y) -2  \sin ^2(a k_x) \cos (a k_y)\right \} ] \nonumber \\
	&&+ [a^2 t^3 \sin^{2}(a k_y) \{ 2 \cos (a k_x) +  \cos ^2(a k_x)  - 4 \sin ^2(ak_x)   -3  \cos (a k_x) \cos (a k_y) \} 
	  \nonumber \\
	&& \left\{3 \cos (a k_x)-\cos (ak_y)-2\right\} ] \nonumber \\
	&&+[ a^2 t^3 \sin^{2} (ak_x) 
	\{ 2  \cos (a k_y) + \cos ^2(ak_y)  - 4 \sin ^2(a k_y) -3  \cos (a k_x) \cos (a k_y) \}  \nonumber \\
	&&  \left\{3 \cos (a k_y) -\cos (ak_x)-2\right\} ]. \nonumber\\
\end{eqnarray}

%\begin{eqnarray}
%	&\Omega(\k)_{k_z} = & a^2t^3 \Bigg[ 
%	\sin(a k_y) \{3\cos(a k_x) - \cos(a k_y) -2\}\times\nonumber\\
%	&&\times\big[2\cos(ak_x)\sin(ak_y) + \cos^2(ak_x)\sin(ak_y)\nonumber\\
%	&&- 4 \sin^2(ak_x)\sin(ak_y) -3 \cos(ak_x)\cos(ak_y)\sin(ak_y)
%	\big]\nonumber \\
%	&&+ \sin(a k_x) \{3\cos(a k_y) - \cos(a k_x) -2\}\times\nonumber \\
%	&&\times\big[2\cos(ak_y)\sin(ak_x) + \cos^2(ak_y)\sin(ak_x)\nonumber \\
%	&&- 4 \sin^2(ak_y)\sin(ak_x) -3 \cos(ak_x)\cos(ak_y)\sin(ak_x)
%	\big]\nonumber \\
%	&& + \{2-\cos(ak_x)-\cos(ak_y)-\cos(ak_0)+\cos(ak_z)\}\times\nonumber \\
%	&&\times\bigg[ 4\cos(ak_x)\cos(ak_y)-4\cos^2(ak_x)\cos(ak_y)-3 \cos^3(ak_x)\cos(ak_y)\nonumber \\ 
%	&&-4\cos(ak_x)\cos^2(ak_y) + 10 \cos^2(ak_x)\cos^2(ak_y) \nonumber\\
%	&& -3\cos(ak_x)\cos^3(ak_y) -2 \cos(ak_y)\sin^2(ak_x)-\cos^2(ak_y)\sin^2(ak_x)\nonumber \\
%	&& + 3 \cos(ak_x)\cos(ak_y)\{\sin^2(ak_x)+\sin^2(ak_y)\} - \cos(ak_x)\sin^2(ak_y) - \cos^2(ak_x)\sin^2(ak_y)\nonumber \\
%	&& - 8 \sin^2(ak_x)\sin^2(ak_y) 
%	\bigg]
%	\Bigg].
%\end{eqnarray}

Further, from the calculations of Berry curvature for the $n=3$, shown above, 
we can confirm that only $\Omega(\k)_{k_z}$ give rise to non-zero contribution to anomalous Hall current in all three cases. 
\end{sloppypar}  % Include content of App3 with sections
	\clearpage
	\pagestyle{empty}
	\cleardoublepage
	
\end{spacing}

\cleardoublepage \cleardoublepage
\addcontentsline{toc}{chapter}{List of Publications}
\markboth{List of Publications}{List of Publications}
\newpage
\thispagestyle{empty}
\begin{center}
\vspace*{-0.4cm} {\LARGE {\textbf{List of Publications}}}
\end{center}
{\setlength{\baselineskip}{8pt} \setlength{\parskip}{2pt}
\begin{spacing}{1.5}
\vspace*{.7cm} \noindent{\bf \large A. Part of this thesis} \vspace*{.7cm}
\begin{enumerate}
	\item \textbf{Amar Bharti}, M. S. Mrudul, and Gopal Dixit: High Harmonic spectroscopy of nonlinear anomalous Hall effect in Weyl semimetals, Physical Review B \textbf{105}, 155140 (2022).
	\item  \textbf{Amar Bharti}, and Gopal Dixit: Role of topological charges in the nonlinear optical response from Weyl semimetals: Physical Review B \textbf{107}, 224308 (2023).
	\item  \textbf{Amar Bharti}, Misha Ivanov, and Gopal Dixit: How massless are Weyl fermions in Weyl semimetals: Physical Review B \textbf{108}, L020305 (2023).
	\item  \textbf{Amar Bharti}, and Gopal Dixit: Tailoring photocurrent in Weyl semimetals via intense laser irradiation: Physical Review B \textbf{108}, L161113 (2023).
	\item  \textbf{Amar Bharti}, and Gopal Dixit: Photocurrent generation in solids via linearly polarized laser: Physical Review B \textbf{109}, 104309 (2024).
	\item  \textbf{Amar Bharti}, and Gopal Dixit: Non-perturbative nonlinear optical responses in Weyl semimetals: Applied Physics Letter \textbf{125}, 051104 (2024).		

\end{enumerate}

\vspace*{.7cm} \noindent{\bf \large B. Not part of this thesis} \vspace*{.7cm}
\begin{enumerate}
	\item \textbf{Amar Bharti}, Margarita Khokhlova, Misha Ivanov, and Gopal Dixit: Chiral sensitive valleytronics in Weyl semimetals, \textbf{under preparation.}
	\item  Navdeep Rana, \textbf{Amar Bharti}, Sucharita Giri and Gopal Dixit: Nonlinear spectroscopy of nodal line semimetals, \textbf{under preparation.}
\end{enumerate}

\newpage
\thispagestyle{empty}
\begin{center}
	\vspace*{-0.4cm} {\LARGE {\textbf{Conferences Attended}}}
\end{center}

\vspace*{.7cm} \noindent{\bf \large A. International}
\vspace*{.7cm}

\begin{enumerate}
	\item  \textbf{Control of Ultrafast (Attosecond and Strong Field) Processes Using Structured Light}, MPIPKS Dresden, Germany (26 June - 2 July 2023), Contributed Talk: ``All-optical probe of anomalous Hall effect''.
	\item  \textbf{ATTO VIII: 8\textsuperscript{th} International Conference On Attosecond Science And Technology}, University of Central Florida (11-15 July 2022), Contributed Talk: ``High-Harmonic spectroscopy of nonlinear anisotropic anomalous Hall effect in topological materials''.

\end{enumerate}

\vspace*{.7cm} \noindent{\bf \large B. National}
\vspace*{.7cm}

\begin{enumerate}
	\item  \textbf{SYMPHY 2024}, IIT Bombay (8-10 March 2024), Contributed Talk: ``Intense circularly polarized laser on Weyl semimetals''.
	\item  \textbf{9\textsuperscript{th} Topical conference on Ultrafast Photonics and Quantum Science}, Physical Research Laboratory, Ahmedabad (15-17 February, 2024), Contributed Talk: ``Intense circularly polarized laser on Weyl semimetals''.	
	\item \textbf{Quantum Materials 2023},  NISER, Bhubaneswar (27 - 30 November 2023), Presented Poster: ``Attosecond Electron Dynamics Driven Photocurrent in Inversion Symmetric Weyl Semimetals''.	
	\item \textbf{Ultrafast Sciences 2022}, IISER Thiruvananthapuram (03-05 November 2022), Presented Poster: ``High-Harmonic spectroscopy of nonlinear  anomalous Hall effect in Weyl semimetal''. 

\end{enumerate}

\end{spacing}

\cleardoublepage
\end{document}